\documentclass[a4paper,11pt]{book}
\usepackage[utf8]{inputenc}
\usepackage[T1]{fontenc} 
\usepackage{lmodern} 
\usepackage[margin=28mm,includeheadfoot,bindingoffset=5mm]{geometry}[2010/03/13]

\usepackage{graphicx}
\usepackage{amsmath}
\usepackage{bbold}
\usepackage{amssymb} % pour le signe \lesssim
\usepackage{textcomp} % \textdegree
\usepackage[most]{tcolorbox} 
\usepackage{enumitem} 
\usepackage{xcolor}
\usepackage{float}
\usepackage{lscape}
\usepackage{physics} 
\usepackage{stmaryrd}
\usepackage{wasysym} 
\usepackage{tikz}
\usepackage{cite}
\usepackage{hyperref}
 \usepackage{textgreek}

\usepackage{stmaryrd}
\usepackage{trimclip}

\makeatletter
\DeclareRobustCommand{\shortto}{%
  \mathrel{\mathpalette\short@to\relax}%
}

\newcommand{\short@to}[2]{%
  \mkern2mu
  \clipbox{{.5\width} 0 0 0}{$\m@th#1\vphantom{+}{\shortrightarrow}$}%
  }
\makeatother

\usepackage{thcover}

\tcbset{enhanced,colback=red!5!white, colframe=red!75!black,fonttitle=\bfseries}
\graphicspath{{figures/}} %Setting the graphicspath
\usepackage[explicit, clearempty]{titlesec}
\titleformat{\chapter}[display]{\bfseries\filright}{\huge\chaptername~\thechapter}{20pt}{\Huge#1}
\titleformat{name=\chapter, numberless}[display]{\bfseries\filright}{}{0pt}{\Huge#1}[\addcontentsline{toc}{chapter}{#1}]

\begin{document}

\frontcover
%\documentclass[a4paper,11pt]{book}
%\usepackage[utf8]{inputenc}
%\usepackage[T1]{fontenc} 
%\usepackage{lmodern} 
%\usepackage[margin=28mm,includeheadfoot,bindingoffset=5mm]{geometry}[2010/03/13]

%\usepackage{graphicx}
%\usepackage{amsmath}
%\usepackage{bbold}
%\usepackage{amssymb} % pour le signe \lesssim
%\usepackage{textcomp} % \textdegree
%\usepackage[most]{tcolorbox} 
%\usepackage{enumitem} 
%\usepackage{xcolor}
%\usepackage{float} 
%\usepackage{physics}
%\usepackage{stmaryrd}
%\usepackage{wasysym} 
%\usepackage{tikz}
%\usepackage{hyperref}
%\usepackage{cite}
%\usepackage{thcover} % Pour la couverture UPMC

%\newcommand*\circled[1]{\tikz[baseline=(char.base)]{
%          \node[shape=circle,draw,inner sep=2pt] (char) {#1};}}
%\renewcommand{\thesubsubsection}{\roman{subsubsection}}
%\tcbset{enhanced,colback=red!5!white, colframe=red!75!black,fonttitle=\bfseries}
%\graphicspath{{figures/}} %Setting the graphicspath
%\setcounter{tocdepth}{3}
%\setcounter{secnumdepth}{3}

%\begin{document}

% \tableofcontents

\chapter*{Abstract}

Exciton-polaritons are quasi-particles arising from the strong coupling regime between excitons and photons. In planar microcavitites, phenomena such as superfluidity or Bose-Einstein condensation can be observed. Those systems have demonstrated to be very efficient in the hydrodynamic generation of topological excitations, such as vortex-antivortex pairs or dark solitons. However, the lifetime and motion of those excitations were limited by the driven dissipative nature of the system.

\paragraph{}
In this thesis, we present a rich variety of results about the generation and control of
such topological excitations. Taking advantage of the optical bistability present in our system, we were able to greatly enhanced the propagation length of vortices and solitons generated in the wake of a structural defect, revealing in the mean time an unexpected binding mechanism of the solitons which propagate parallel.
This behaviour was recovered in  a specifically designed  experiment, where we artificially imprint dark soliton pairs on demand on a polariton superfluid. The adaptability of our technique allowed for a detailed study of this phenomenon, that we directly connected to the driven-dissipative nature of our system.
Finally, confined dark solitons were generated within guided intensity channels on a static polariton fluid. The absence of flow lead to the development of transverse snake instabilities of which we studied the interesting properties.

%\end{document}
%\documentclass[a4paper,11pt]{book}
%\usepackage[utf8]{inputenc}
%\usepackage[T1]{fontenc} 
%\usepackage{lmodern} 
%\usepackage[margin=28mm,includeheadfoot,bindingoffset=5mm]{geometry}[2010/03/13]

%\usepackage{graphicx}
%\usepackage{amsmath}
%\usepackage{bbold}
%\usepackage{amssymb} % pour le signe \lesssim
%\usepackage{textcomp} % \textdegree
%\usepackage[most]{tcolorbox} 
%\usepackage{enumitem} 
%\usepackage{xcolor}
%\usepackage{float} 
%\usepackage{physics}
%\usepackage{stmaryrd}
%\usepackage{wasysym} 
%\usepackage{tikz}
%\usepackage{hyperref}
%\usepackage{cite}
%\usepackage{thcover} % Pour la couverture UPMC

%\newcommand*\circled[1]{\tikz[baseline=(char.base)]{
%          \node[shape=circle,draw,inner sep=2pt] (char) {#1};}}
%\renewcommand{\thesubsubsection}{\roman{subsubsection}}
%\tcbset{enhanced,colback=red!5!white, colframe=red!75!black,fonttitle=\bfseries}
%\graphicspath{{figures/}} %Setting the graphicspath
%\setcounter{tocdepth}{3}
%\setcounter{secnumdepth}{3}

%\begin{document}

% \tableofcontents

\chapter*{Résumé}

Les polaritons excitoniques sont des quasi-particules créées par le régime de couplage fort entre des excitons et des photons. Dans des microcavités planaires, des phénomènes tels que la superfluidité ou la condensation de Bose-Einstein ont pu être observés.
Ces systèmes ont démontrés être particulièrement appropriés pour la génération hydrodynamique d'excitations topologiques comme des paires de vortex-antivortex ou de solitons sombres.
Cependant, le temps de vie et la propagation de ces excitations étaient limités par la dissipation du système.

\paragraph{}
Dans cette thèse, nous présentons une succession de résultats sur la génération et le contrôle de telles excitations topologiques. 
En utilisant la bistabilité optique de notre système, nous avons fortement augmenté la distance de propagation de vortex et de solitons formés dans le sillage de défauts structurels, ce qui a également révélé un comportement inattendu des solitons liés qui restent parallèles.
Ce comportement a été confirmé par l’expérience suivante, où nous avons artificiellement imprimé des paires de solitons sombres dans des superfluides de polaritons. 
La flexibilité de notre technique nous a permis d'étudier ce phénomène en détails et de le relier directement à la dissipation du système.
Enfin, des solitons sombres ont été produits dans des canaux d'intensités dans un fluide statique. L'absence de flux a permis le développement d'instabilités transverses ou \textit{snake instabilities} dont nous avons étudié les propriétés.

%\end{document}
%\input{Publications.tex}
%\documentclass[a4paper,11pt]{book}
%\usepackage[utf8]{inputenc}
%\usepackage[T1]{fontenc} 
%\usepackage{lmodern} 
%\usepackage[margin=28mm,includeheadfoot,bindingoffset=5mm]{geometry}[2010/03/13]

%\usepackage{graphicx}
%\usepackage{amsmath}
%\usepackage{bbold}
%\usepackage{amssymb} % pour le signe \lesssim
%\usepackage{textcomp} % \textdegree
%\usepackage[most]{tcolorbox} 
%\usepackage{enumitem} 
%\usepackage{xcolor}
%\usepackage{float} 
%\usepackage{physics}
%\usepackage{stmaryrd}
%\usepackage{wasysym} 
%\usepackage{tikz}
%\usepackage{hyperref}
%\usepackage{cite}
%\usepackage{thcover} % Pour la couverture UPMC

%\newcommand*\circled[1]{\tikz[baseline=(char.base)]{
%          \node[shape=circle,draw,inner sep=2pt] (char) {#1};}}
%\renewcommand{\thesubsubsection}{\roman{subsubsection}}
%\tcbset{enhanced,colback=red!5!white, colframe=red!75!black,fonttitle=\bfseries}
%\graphicspath{{figures/}} %Setting the graphicspath
%\setcounter{tocdepth}{3}
%\setcounter{secnumdepth}{3}

%\begin{document}

% \tableofcontents

\chapter*{Remerciements}

\paragraph{}
Avant de rentrer dans le vif du sujet de ma th\`{e}se, je voudrais prendre le temps de remercier tous ceux qui ont également participé à ce travail et sans qui mes résultats n’auraient certainement pas été ce qu’ils sont aujourd’hui.

\paragraph{}
Tout d’abord, j’ai eu la chance d’être encadré par Alberto Bramati, qui a très rapidement su m’intégrer à l’équipe et me mettre en confiance sur mon sujet de thèse. 
Sa porte a toujours été ouverte et il a su nous guider sur la bonne voie à chaque étape de nos projets, prenant toujours en compte nos remarques et nos intuitions. 
J’aimerais en particulier le remercier pour sa patience, sa gentillesse et sa bonne humeur qui ont grandement participé à garder ma motivation à son maximum durant ces trois années.

\paragraph{}
J’ai également été énormément conseillé par Élisabeth Giacobino et Quentin Glorieux, qui tout en suivant nos projets d’un peu plus loin, ont toujours été disponibles et ont su prendre le temps de nous aider à résoudre nos problèmes. 
J’aimerais également remercier particulièrement Simon Pigeon, qui a travaillé avec nous pendant presque toute la durée de ma thèse sur la partie théorique des projets de polaritons, et dont je suis très heureuse qu’il ait accepté de faire partie de mon jury. 
Il a lui aussi toujours su garder sa porte ouverte et a énormément participé à ma compréhension de la théorie en répondant patiemment à mes questions plus ou moins pertinentes.

\paragraph{}
I would also probably not have been able to achieve this work if it was not for Giovanni Lerario. 
During his time as a post-doc with us, he taught me how to build and run an optical experiment, how to get information on your system from tiny details you wouldn’t have notice, but also useful Italian words, quite interesting psychedelic music and many mysterious chess moves. 
I am very glad to have had the opportunity to work with him and to rely on his knowledge to make me understand the bigger picture, even after he left the group and until the very last night before my defense.

\paragraph{}
J’ai eu aussi l’occasion d’interagir régulièrement avec nos collaborateurs de l’Institut Pascal à Clermont-Ferrand, et en particulier avec Sergei Koniakhin, qui a réalisé sa thèse théorique simultanément à la mienne, a réalisé beaucoup de simulations de nos travaux et est même venu passer quelques jours en salle de manip pour découvrir avec nous la partie expérimentale de nos recherches. 
Je laisse maintenant les expériences de polaritons entre de bonnes mains, celles de Ferdinand, qui a également beaucoup participé à ce travail, aujourd’hui accompagné de Maxime et dont je suis sûre qu’ils obtiendront encore de beaux résultats. 
J’ai également une pensée pour tous les membres du groupe qui se sont succédés durant ma thèse, et qui ont tous contribué à la bonne ambiance qui y régnait : Maxime, Tom, Popi, Stefano, Lorenzo, Chengjie, Rajiv, Murad, Huiqin, Thomas, Marianna, Wei et Tanguy.

\paragraph{}
Une thèse expérimentale dépend aussi beaucoup de l’environnement dans lequel elle est réalisée, et nous sommes très bien entourés au laboratoire Kastler-Brossel. 
Je voudrais donc remercier tous ceux qui contribuent à nos recherches d’un peu plus loin, au service administratif, à l’atelier mécanique ou à l’atelier d’électronique. 
Enfin, je pense que ma motivation a aussi été renforcée par les amitiés que j’ai pu créer au-delà de notre équipe, et c’est pourquoi j’ai également une pensée pour Tom, Tiphaine, Yohann, Ferhat, Michaël, Arthur, Paul, Thomas, Adrien, Jérémy, Félix, et tous ceux que j’ai croisé un peu moins régulièrement, pour ces longues discussions animées autour d’un café ou de quelques bières.

\paragraph{}
Enfin, je sais que ma thèse a parfois occupé mes pensées en dehors des murs du laboratoire, et j’ai été très heureuse de la patience et de la très grande curiosité de mes proches et amis quant à mes recherches (pas encore de sabre laser, ce sera pour la prochaine fois). 
Je tiens donc à les remercier tous pour leur soutien, que ce soit les anciens mini ou les habitués du BDF qui ont suivi ça très régulièrement, mais aussi les petits merdeux, les crazy Wednesdays ou encore Solène, Manu et Jeanne, avec qui les emplois du temps sont plus difficiles à accorder mais les retrouvailles toujours aussi intenses.

\paragraph{}
Bien évidemment, je terminerai par remercier ma famille, mes grandes sœurs, mon père ainsi que mes adorables neveux qui ont chacun à leur manière activement participé à ma rédaction confinée.

%\end{document}

\setcounter{tocdepth}{-1}
%\mainmatter
\tableofcontents
\addtocontents{toc}{\protect\setcounter{tocdepth}{2}}%

%\documentclass[a4paper,11pt]{book}
%\usepackage[utf8]{inputenc}
%\usepackage[T1]{fontenc} 
%\usepackage{lmodern} 
%\usepackage[margin=28mm,includeheadfoot,bindingoffset=5mm]{geometry}[2010/03/13]

%\usepackage{graphicx}
%\usepackage{amsmath}
%\usepackage{bbold}
%\usepackage{amssymb} % pour le signe \lesssim
%\usepackage{textcomp} % \textdegree
%\usepackage[most]{tcolorbox} 
%\usepackage{enumitem} 
%\usepackage{xcolor}
%\usepackage{float} 
%\usepackage{physics}
%\usepackage{stmaryrd}
%\usepackage{wasysym} 
%\usepackage{tikz}
%\usepackage{hyperref}
%\usepackage{cite}
%\usepackage{thcover} % Pour la couverture UPMC

%\newcommand*\circled[1]{\tikz[baseline=(char.base)]{
%          \node[shape=circle,draw,inner sep=2pt] (char) {#1};}}
%\renewcommand{\thesubsubsection}{\roman{subsubsection}}
%\tcbset{enhanced,colback=red!5!white, colframe=red!75!black,fonttitle=\bfseries}
%\graphicspath{{figures/}} %Setting the graphicspath
%\setcounter{tocdepth}{3}
%\setcounter{secnumdepth}{3}

%\begin{document}

% \tableofcontents

\chapter*{Introduction}

\paragraph{}
The concepts of light and matter are fascinating topics whose description strongly evolved over history.
Light in particular remained an enigma for physicists for centuries, before the debates over its wave or particle nature were definitely settled by Louis de Broglie in 1923, who theorized the light matter-wave duality \cite{DEBROGLIE1923}.
It became the starting point of quantum mechanics, which the basis were developed in the following years \cite{VonNeumann1955}.

\paragraph{}
This new mechanics allows both light and matter to have similar behaviour. In particular, the light quanta or \textit{photons} follow a bosonic statistic, identical to the one describing atoms with integer spins. 
They can have therefore identical properties, like the ability to condense in a single state, leading to a macroscopic quantum state where all particles are indistinguishable, called a Bose-Einstein condensate.

\paragraph{}
An important difference still remains between light and matter in the fact that photons do not interact in vacuum. However, photons can interact with themselves when propagating in a non linear medium. 
Such effects have been observed with the discovery of the laser: a sufficiently high intensity beam passing through a material medium can perturb it; and this perturbation also induces modifications on the light field.
In particular, nonlinear effects such as self focusing and self defocusing can be observed , which are equivalent respectively to an effective photon-photon attractive or repulsive interaction.  

\paragraph{}
This interactions between light and matter can be strongly enhanced by correctly adapting the environment. One possibility is for instance to confine an electromagnetic mode in order to couple it to the energy transition of the material emitter. This can be achieved by placing the emitter inside an optical cavity, whose resonance corresponds to the transition frequency \cite{Dovzhenko2018}. If the system losses are low enough, a weak coupling is established when the confined light changes the emission properties of the transition, known as the Purcell effect \cite{Purcell1995}.
The coupling can even be enhanced to the point where the rate of coherent exchange of energy between light and matter is higher than their decay rates \cite{Tormo2015}. The system enters then the strong coupling regime where the initial states splits into two: this characteristic energy \textit{anticrossing} is proportional to the coupling strength and is called the Rabi splitting.

\paragraph{}
Such optical cavities can be developed in many geometries, although we have focused our work on a particular type of material known as semiconductors. 
Those insulators become conductive when electrically or optically excited. Electrical excitation have been deeply studied and developed in the electronic domain; but what interests us in this work is the optical excitation and the fast response and strong control it offers. In particular, the development of optically controlled semiconductor heterostructures opened the way to many possibilities of light-matter interaction.

\paragraph{}
In the end of the twentieth century, new techniques of heterostructures growth reached the nanometer scale of layers thickness. This allowed to design optical microcavities, in which were embedded planar quantum wells.
It is in such a structure that Claude Weisbuch observed for the first time in 1992 the strong coupling between light and matter in semiconductors \cite{Weisbuch1992}. This configuration, previously theorized by John J. Hopfield in 1958 \cite{Hopfield1958}, results in the appearance of bosonic quasiparticles known as \textit{exciton polaritons}, made of a superposition of cavity photons on one hand and electron-hole pairs on the other hand.

\paragraph{}
Polaritons are of particular interest as they inherit properties from both their constituents. Their photonic part gives them a low effective mass, while the excitonic one ensures strong interactions and non-linearity.
They have been intensively studied in the past few decades, so that their essential properties are well known, such as their dispersion curve \cite{Bloch1997}, lifetime \cite{Houdre1994} or relaxation \cite{Stanley1996}.
All those properties allow these systems to exhibit interesting phenomena, like Bose Einstein condensation at high temperature \cite{Kasprzak2006b, Tsintzos2008}, four wave mixing \cite{Romanelli2005, Baumberg2000} or squeezing \cite{Karr2004, Boulier2014}.

\paragraph{}
Excitons polaritons also appear as ideal candidates to study the hydrodynamic properties of quantum fluids. Indeed, many parameters are easily accessible and controlled in those systems, such as momentum, density and phase which have already lead to the observation of superfluidity \cite{Carusotto2004a, Amo2009, Amo2009a, Carusotto2013a} and of topological excitations \cite{Lagoudakis2008, Amo2011, Grosso2011, Nardin2011}.

\paragraph{}
Those last ones in particular are the guiding line of this thesis work. The different experiments indeed followed each others in the study of those phenomena, and more importantly on their sustainability and control. This thesis is therefore organized as followed.

\paragraph{}

The \textbf{first chapter} focuses on the description of our system. We first study independently its two components: the light part with the photons confined within a planar microcavity, and the matter part represented by the quantum well excitons.
We then combined these two together by embedding a quantum well inside a microcavity, which allows to reach the strong coupling between the two oscillators, and therefore to obtain the polariton quasiparticles.

We focus then on the two main ways of exciting such a system, namely resonantly or non-resonantly, and describe how they influence the polaritons properties and the fluid behaviour.

\paragraph{}

The \textbf{second chapter} details the experimental tools available in the lab that we used to implement our work. The first part lists the devices used in the excitation part in order to create the desired polariton fluid, while the second part focuses on the detection equipment and describes the process of data analysis.

\paragraph{}
The experimental results are divided in three different experiments, reported in the last chapters.

\paragraph{}
\textbf{Chapter 3} begins by reporting a theoretical proposal \cite{Pigeon2017} which is the starting point of the experimental work of this thesis. It suggests to use the property of optical bistability, present in our system due to a strong non linearity, to get rid of the phase constraint usually imposed by the pump in a quasi-resonant excitation scheme.
To exploit this phenomenon, a novel implementation is suggested, using two excitation beams and leading to a large fluid area within the bistability cycle.

This idea was put in application experimentally to generate and sustain topological excitations, such as vortex-antivortex pairs or dark solitons \cite{Lerario2020, Lerario2020a}, depending on the hydrodynamic conditions of the system.
They form spontaneously in the wake of a structural defect hit by a nonlinear flow of polaritons.
A comparable configuration had previously been implemented \cite{Amo2011}, but the solitons were generated within the decaying tale of the excited polaritons, resulting in a short propagation distance through an intensity decaying fluid. 
Our configuration not only greatly enhances the propagation distance, but also highlights an unexpected behaviour of the dark soliton pairs, which under the influence of the driving field align to each other and propagate parallel.
This phenomenon is in total opposition to all former results in atomic quantum fluids as well as in polariton systems. Indeed, the only observation of an attractive behaviour between dark solitons has been realized in a thermo-optic medium with non-local interactions, whereas polariton interaction are based on the exciton exchange interaction and are fully local.

\paragraph{}
Those surprising results still relies on the spontaneous generations of such turbulence, under specific conditions around a structural defect on which we do not have any control. 
The next step of our work, presented in \textbf{chapter 4}, was therefore to get rid these constraints and to artificially imprint dark solitons, to have full control over them and to lead deeper studies.
A smart design of the excitation beam allowed us to artificially create dark solitons within a polariton flow, and to let them propagate freely through a bistable fluid. 

Once again, the soliton propagate parallel and experience a binding mechanism, leading to the formation of a dark soliton molecule \cite{Maitre2020}.
The scalability of our new technique allowed us to further investigate this unexpected behaviour.
In particular, we observed that the solitons always reach the same equilibrium separation distance, no matter the initial conditions. We managed to understand this phenomenon as a consequence of the driven-dissipative nature of our system.
The implementation of this new method opens the way to a fine control and manipulation of collective excitations, which could lead to the generation of multiple soliton pattern or a quantitative study of quantum turbulence phenomena.

\paragraph{}
We followed this direction in the \textbf{fifth chapter}, where we once again artificially create solitons in our system, but within a static fluid. They are this time generated inside a low density channel, stabilized by the pressure from the high density walls \cite{Claude2020}.
This configuration allows the observation of transverse instabilities called \textit{Snake instabilities}, inhibited by the fast flow in the previous implementations. Moreover, the behaviour of those solitons strongly depends on the ratio of intensities between the high and low density regions, and on the connection of the channel to low density fluid surrounding the excited domain. A careful implementation can therefore leads to interesting phenomena such as the resolution of a maze pattern.

%\bibliographystyle{unsrt}
%\bibliography{bibs/LKB-bibs-bib_thesis-Intro}  

%\end{document}
%\documentclass[a4paper,11pt]{book}
%\usepackage[utf8]{inputenc}
%\usepackage[T1]{fontenc} 
%\usepackage{lmodern} 
%\usepackage[margin=28mm,includeheadfoot,bindingoffset=5mm]{geometry}[2010/03/13]

%\usepackage{graphicx}
%\usepackage{amsmath}
%\usepackage{bbold}
%\usepackage{amssymb} % pour le signe \lesssim
%\usepackage{textcomp} % \textdegree
%\usepackage[most]{tcolorbox} 
%\usepackage{enumitem} 
%\usepackage{xcolor}
%\usepackage{physics} 
%\usepackage{wasysym} 
%\usepackage{tikz}
%\usepackage{hyperref}
%\usepackage{textgreek}
%\usepackage{cite}
%\usepackage{wrapfig}

%\newcommand*\circled[1]{\tikz[baseline=(char.base)]{
%          \node[shape=circle,draw,inner sep=2pt] (char) {#1};}}
%\renewcommand{\thesubsubsection}{\roman{subsubsection}}
%\tcbset{enhanced,colback=red!5!white, colframe=red!75!black,fonttitle=\bfseries}
%\graphicspath{{figures/}} %Setting the graphicspath
%\setcounter{tocdepth}{3}
%\setcounter{secnumdepth}{3}

%\begin{document}

%\tableofcontents

\chapter{Exciton-polaritons quantum fluid of light}
\label{chap:Polaritons}

\paragraph{}
Exciton-polaritons are quasiparticles arising from the strong coupling between light and matter. 
They are indeed a superposition of cavity photons and quantum well excitons. 
In order to discover all the tools of our system, we start this chapter by independently describing the two elements of the polaritons.
We will focus on particularities of photon within a planar microcavity, then study the consequences of bidimensional confinement on semiconductor excitons.
We will then be able to strongly couple those elements together and describe this new system of excitons-polaritons.

\paragraph{}
On the second part of this chapter, we will extend our field of view to the ensemble of the polariton fluid: we will see how the excitation configuration influences the system properties, and describe the main regimes of such a fluid.

\newpage

\section{Light matter coupling in semiconductor microcavities}

\paragraph{}
Polaritons are made of light and matter strongly coupled.
To figure out how those quasi-particles arise, it is essential to study their components independently and how their environment influences their behaviour.
Then, by combining them together, we will be able to understand their interactions and how to describe our new system.

\subsection{Photons in a planar microcavity}

\paragraph{}
This section focuses on the photonic part of the polaritons. 
As we ultimately want to achieve a strong coupling with some matter particle, the electromagnetic field needs to be enhanced, which is done by confining it within a microcavity.

\subsubsection{Optical microcavity}

\paragraph{Microcavity properties}

A cavity is an optical resonator where a standing wave is amplified.
The light is confined within the cavity through a particular mirrors arrangement and follows a well defined path: interferences take place and lead to the amplification of some specific modes.

A microcavity is a particular case of optical cavity, which dimension is of the order of the light wavelength. 

Two parameters are mainly used to describe the optical cavity properties: the quality factor and the finesse.
The \textbf{quality factor} or \textit{Q-factor} characterizes the frequency width of the field enhancement. 
It is defined as the ratio between the resonant frequency $\omega_{cav}$ and the full width at half maximum (FWHM) of the cavity mode $\delta \omega_{cav}$:

\begin{equation}
    Q = \dfrac{\omega_{cav}}{\delta \omega_{cav}}
\end{equation}

The Q-factor describes the rate of the energy decay, which can be due to losses of the mirrors, absorption, or scattering on the cavity imperfections: $Q^{-1}$ is the fraction of energy lost in one round-trip of the cavity \cite{Kavokin2007a}. Consequently, the photon population decays exponentially and its lifetime can be defined as

\begin{equation}
    \tau = \frac{Q}{\omega_{cav}}
\end{equation}

\paragraph{}
The \textbf{finesse} connects the FWHM of the cavity mode to the free spectral range (FSR), \textit{i.e.} the frequency separation between two consecutive longitudinal resonant modes. It characterizes the spectral resolution of the cavity.
It can also be defined using the total reflectivity $R$ of the cavity:

\begin{equation}
    F = \dfrac{\Delta \omega_{cav}}{\delta \omega_{cav}} = \dfrac{\pi \sqrt{R}}{1-R}
\end{equation}

The frequency separation is connected to the total length $L$ of the cavity: $\Delta \omega_{cav} = \frac{2 \pi c}{L}$. Therefore, in the particular case of a microcavity, it is of the same order of magnitude than the cavity mode frequency, so the finesse and the Q-factor are close.

\paragraph{}
As the finesse is usually dominated by the mirror losses, the field enhancement within the cavity can be written as:

\begin{equation}
    \dfrac{I_{intracavity}}{I_{incident}} \approx \dfrac{1}{1-R} = \dfrac{F}{\pi \sqrt{R}}
\end{equation}

In a standing wave microcavity, the electromagnetic field is distributed in the form of an interference pattern, creating therefore localized maxima and minima. Consequently, in order to couple an emitter within the microcavity, it has to be placed at a maximum of the field to enhance the coupling, as we will describe more in detail later.

\paragraph{Planar microcavity}

The type of cavity that we considered in this thesis is a \textbf{Fabry-Perot cavity}. It is one of the simplest optical cavity and was invented by Charles Fabry and Alfred Perot in 1899.
It consists in two planar mirrors, parallel and facing each other: the light with  specific wavelengths is enhanced by going back and forth between them, due to constructive interferences.

Fabry-Perot microcavities are usually built using Distributed Bragg Reflectors (DBR). Those mirrors consist in a pile of alternate layers of two materials with different refractive indices $n_{1}<n_{2}$.
They are designed for a specific wavelength $\lambda_{0}$: each layer has a thickness $d_{i} = \frac{\lambda_{0}}{4n_{i}}$ ($i=(1, 2)$).
This structure then acts like a high reflectivity mirror for a frequency range called stop-band $\Delta \lambda$, centered at $\lambda_{0}$ \cite{Macleod2001}:

\begin{equation}
    \Delta \lambda = \dfrac{4 \lambda_{0}}{\pi} \arcsin \Bigg( \dfrac{n_{2}-n_{1} }{n_{2} + n_{1} } \Bigg)
\end{equation}

\begin{figure}[h]
    \centering
    \includegraphics[width=0.6\linewidth]{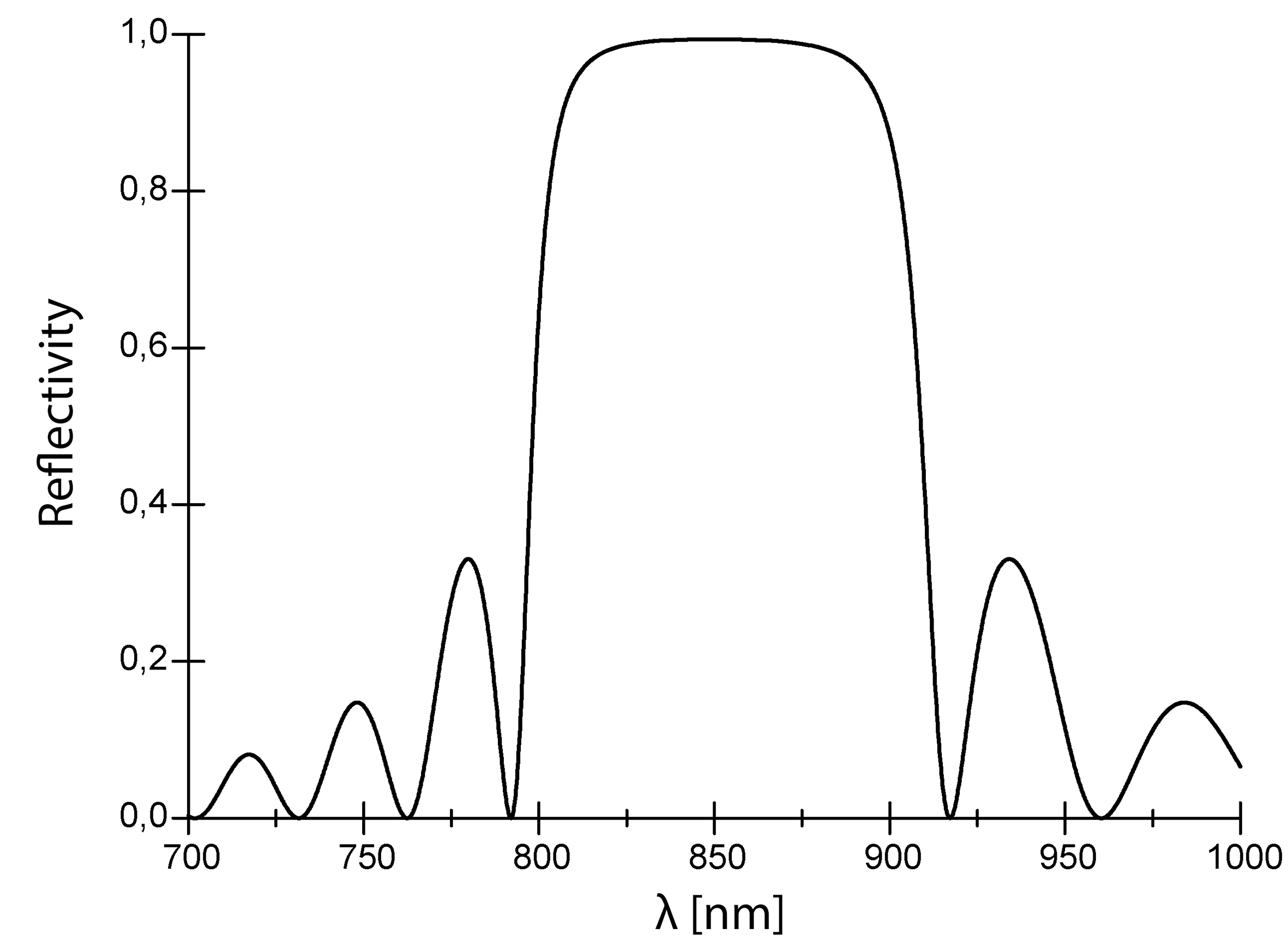}
    \caption{\textbf{DBR reflectivity}. Reflectivity of a Bragg mirror at normal incidence made of 20 pairs of Ga\textsubscript{0.9}Al\textsubscript{0.1}As/AlAs and centered in $\lambda_{0} = 850$ nm. From \cite{Hivet2013}}
    \label{fig:ReflDBR}
\end{figure}

Figure \ref{fig:ReflDBR} presents the reflectivity of a DBR made of 20 pairs of Ga\textsubscript{0.9}Al\textsubscript{0.1}As/AlAs with a stop band centered at $\lambda_{0} = 850$ nm. The optical indices of the materials are $n_{Ga_{0.9}Al_{0.1}As} = 3.48$ and $n_{AlAs} = 2.95$.
The reflectivity is very close to one over a hundred nanometers range.

\begin{figure}[h]
    \centering
    \includegraphics[width=0.6\linewidth]{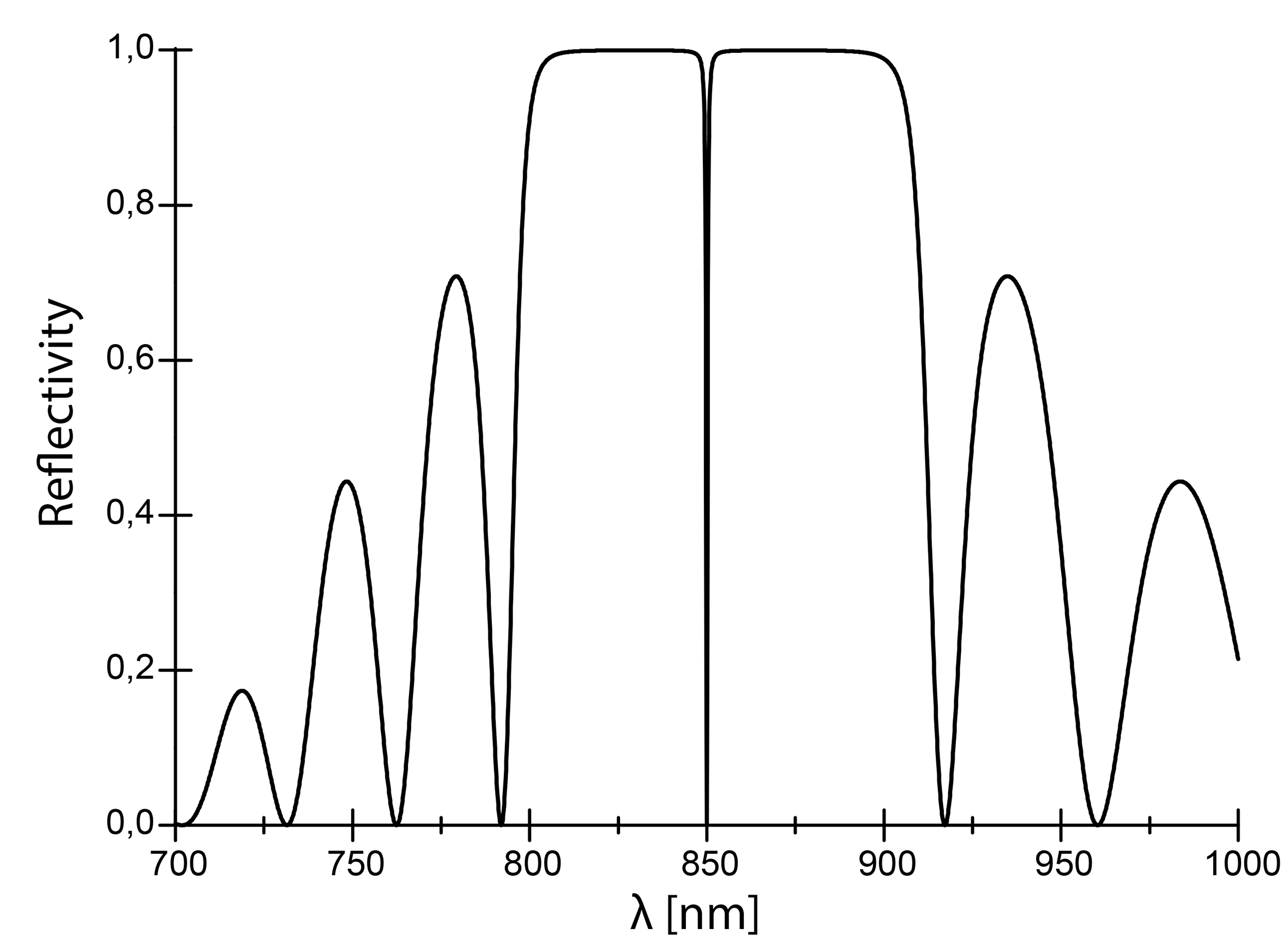}
    \caption{\textbf{Microcavity reflectivity}. Reflectivity at normal incidence of a Fabry Perot made with two DBR identical to the one presented in figure \ref{fig:ReflDBR} and separated by a distance $L_{cav} = 2 \frac{\lambda_{0}}{2 n_{cav}}$. From \cite{Hivet2013}}
    \label{fig:ReflMicrocav}
\end{figure}

The Fabry Perot cavity is realized by placing two DBRs in front of each other, at a distance $L_{cav} = m \frac{\lambda_{0}}{2n_{cav}}$, where $n_{cav}$ is the refractive index of the intracavity medium and $m$ an integer.
For instance, figure \ref{fig:ReflMicrocav} shows the reflectivity of a microcavity as a function of the wavelength, at normal incidence. The cavity consists of two DBR identical to the one of figure \ref{fig:ReflDBR}, parallel to each other and at a distance $L_{cav} = \frac{2\lambda_{0}}{m_{cav}}$.
A very sharp dip is visible at $\lambda = \lambda_{0} = 850$ nm, which corresponds to the cavity resonance.

\subsubsection{Cavity photons}

\paragraph{}

The term \textbf{cavity photons} is due to the particular properties imposed by the cavity to the magnetic field, that we are going to develop.

\paragraph{Dispersion relation}

\begin{figure}[h]
    \centering
    \includegraphics[width=0.75\linewidth]{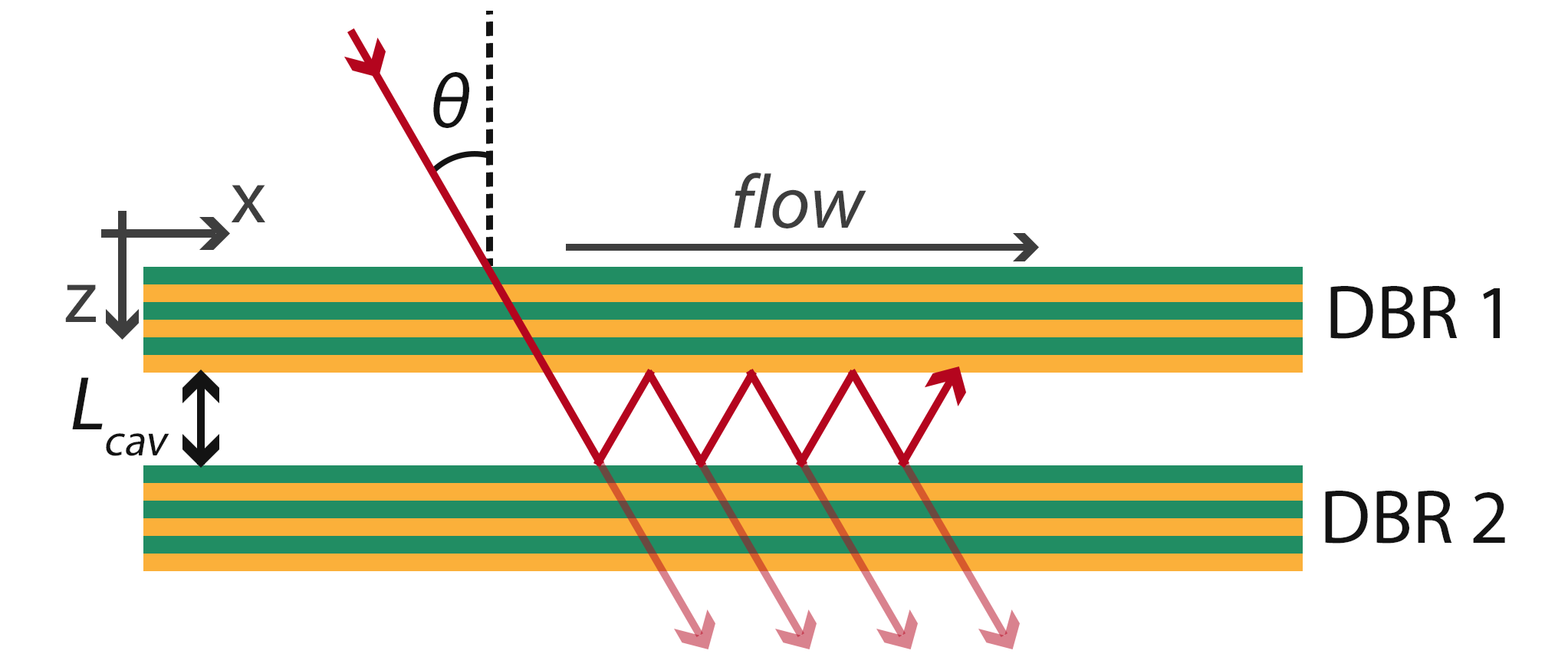}
    \caption{\textbf{Photon flow inside a planar cavity}. The photons travel a distance $L_{\theta} = \frac{L_{cav}}{\cos{\theta}}$ between the mirrors, which corresponds to another resonance energy.}
    \label{fig:CavityFlow}
\end{figure}

The dispersion relation arises from the fact that the effective length of the cavity felt by the photons depends on their angle of incidence, as shown in figure \ref{fig:CavityFlow}. 
If the photons enter the cavity with an incident angle $\theta$, they feel a cavity of length $L_{\theta} = \frac{L_{cav}}{\cos{\theta}}$, which shifts the resonance energy.

\paragraph{}
More quantitatively, the cavity imposes a quantization condition on the \textit{z} component of the photon wavevector $\mathbf{k}^{\gamma} = (\mathbf{k}^{\gamma}_{||}, k^{\gamma}_{z})$:

\begin{equation}
    k^{\gamma}_{z} = \dfrac{2 \pi n_{cav}}{\lambda_{0}}
\end{equation}

There is no constraint on the parallel component of the wavevector, so the energy of the photons can be written as:

\begin{equation}
    E_{\gamma}(\mathbf{k}^{\gamma}_{||}) = \dfrac{\hbar c}{n_{cav}}
    \sqrt{\bigg( \dfrac{2 \pi n_{cav}}{\lambda_{0}} \bigg)^{2} + (k^{\gamma}_{||})^2 }
\end{equation}

Considering that the orthogonal component of the wavevector is significantly larger than the parallel one ($k^{\gamma}_{||} \ll k^{\gamma}_{z}$), this expression can be approximated with a parabolic dispersion:

\begin{equation}
    E_{\gamma} \simeq \dfrac{hc}{\lambda_{0}} 
    \Bigg( 1 + \dfrac{1}{2} 
    \bigg( \dfrac{\lambda_{0}k^{\gamma}_{||}}{2 \pi n_{cav}} \bigg)^{2}
    \Bigg)
\end{equation}

\paragraph{}
This equation leads to the definition of an \textbf{effective mass of the cavity photons} $m^{*}_{\gamma}$, connected to the curvature of the parabola:

\begin{equation}
    \dfrac{1}{m^{*}_{\gamma}} = \dfrac{1}{\hbar^{2}} 
    \dfrac{\partial^{2} E}{\partial k^{2}}
\end{equation}

which gives us the expression:

\begin{equation}
    m^{*}_{\gamma} = \dfrac{n_{cav}^{2} \hbar}{\lambda_{0}c}
\end{equation}

Typically, in semiconductor microcavities, it is of the order of 10\textsuperscript{-5} times the free electron mass.

\paragraph{Photon lifetime}
The lifetime of the cavity photons depends on the reflectivity of the mirrors.
We saw that the finesse is indeed associated with the number of round-trips a photon makes before getting out of the cavity.
However, the photon lifetime of such Fabry-Perot microcavity can also be properly defined by the relation \cite{Savona1999}:

\begin{equation}
    \tau_{cav} = L_{eff} \dfrac{n_{cav}}{\pi} \dfrac{F}{c}
\end{equation}

where $L_{eff}$ is the effective length of the cavity. Indeed, the fact that the electromagnetic field effectively enters the DBR in the form of an evanescent field has to be taken into account, hence the definition of $L_{eff} = L_{cav}\Big( 1 + 2 \dfrac{n_{1}n_{2}}{n_{2}-n_{1}} \Big) $.

\paragraph{Leaky modes}

\begin{figure}[h]
    \centering
    \includegraphics[width=0.6\linewidth]{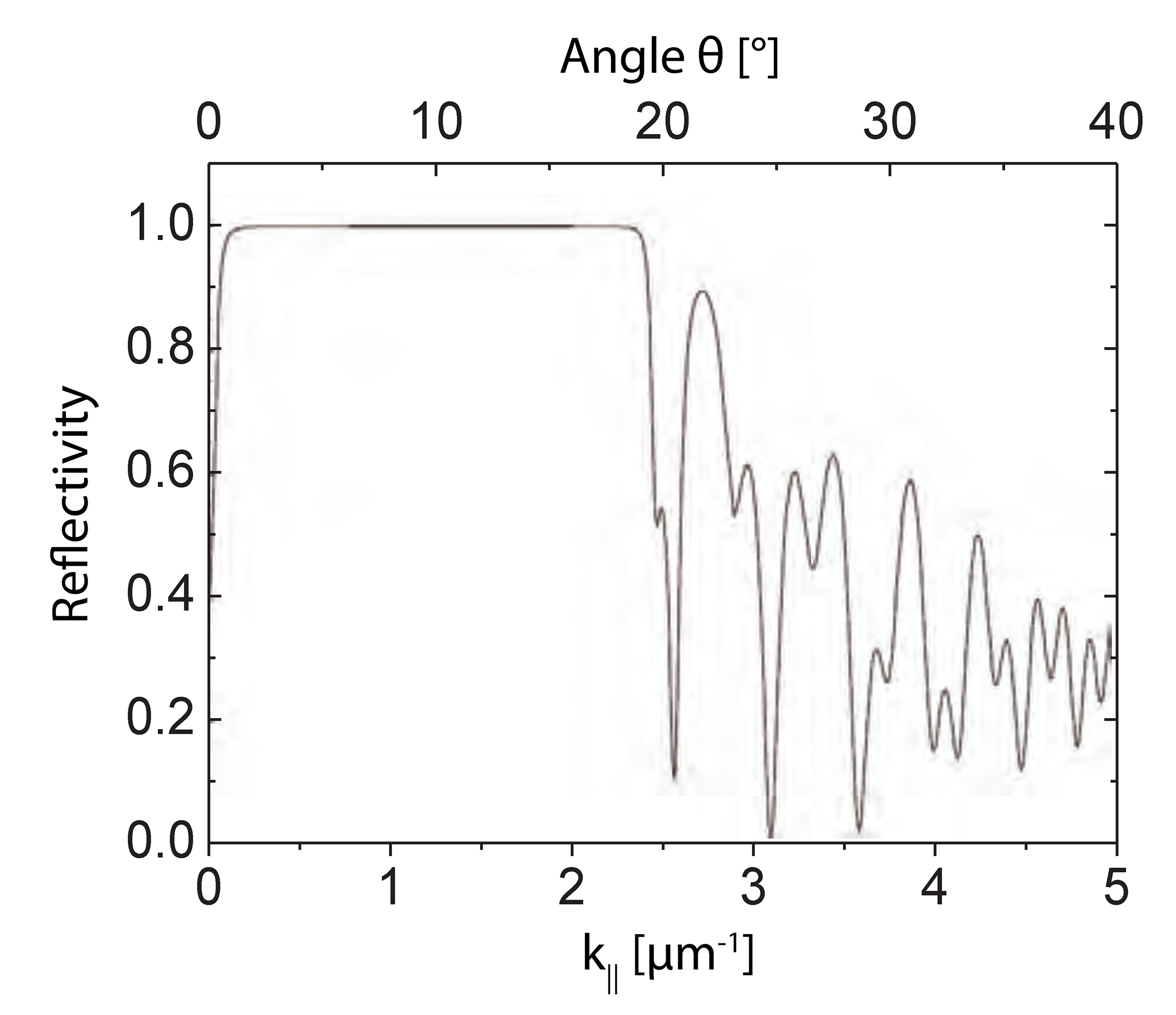}
    \caption{\textbf{Leaky modes}. For small angles of incidence ($\theta < 18^{\circ}$), the reflectivity is equal to one except for the sharp dip at $k^{\gamma}_{||} = 0$ \textmu m\textsuperscript{-1} which coincides with the resonance. But for higher angles, the reflectivity decreases and oscillates: is shows some dips corresponding to the leaky modes of the cavity, where a strong coupling can not be sustained anymore. From \cite{Leyder2007}}
    \label{fig:LeakyModes}
\end{figure}

A further investigation about the cavity field dependence on the incidence angle is given in figure \ref{fig:LeakyModes}. Here is presented the reflectivity of the system as a function of the in-plane wavevector $k^{\gamma}_{||}$ and its correspondence with the incident angle $\theta$, for an incident light of wavelength $\lambda = \lambda_{0}$.

\paragraph{}
The reflectivity shows different behaviours along the wavevector scan. At $k^{\gamma}_{||} = 0$ \textmu m\textsuperscript{-1}, the reflectivity sharply decreases down to zero, and illustrates the resonance of the cavity for a light with $\lambda_{0}$. It corresponds to the main mode.

By increasing the incident angle, the reflectivity stays constant at 1. The mirrors work properly and the light is trapped inside the cavity.
But when the angle becomes too large, the reflectivity decreases and some oscillations are visible, with large dips.
It shows that the field of those modes escapes easily outside of the cavity, hence the name \textit{leaky modes}. The linewidth of the dips are larger than the one of the main mode, which shows that they have a much shorter lifetime.

In the case of our cavity, the leaky modes appear for $k = 2.5$ \textmu m\textsuperscript{-1}, which corresponds to an angle of 18$^{\circ}$ \cite{Messin2000}. But our cavity is made in GaAs, for which the total reflection angle is 16.6$^{\circ}$: the leaks do not happen at the interface with air.

\subsection{Quantum well excitons}
\label{sec:excitons}

\subsubsection{Excitons in a bulk semiconductor}

\paragraph{Band theory}

\begin{figure}[h]
    \centering
    \includegraphics[width=0.9\linewidth]{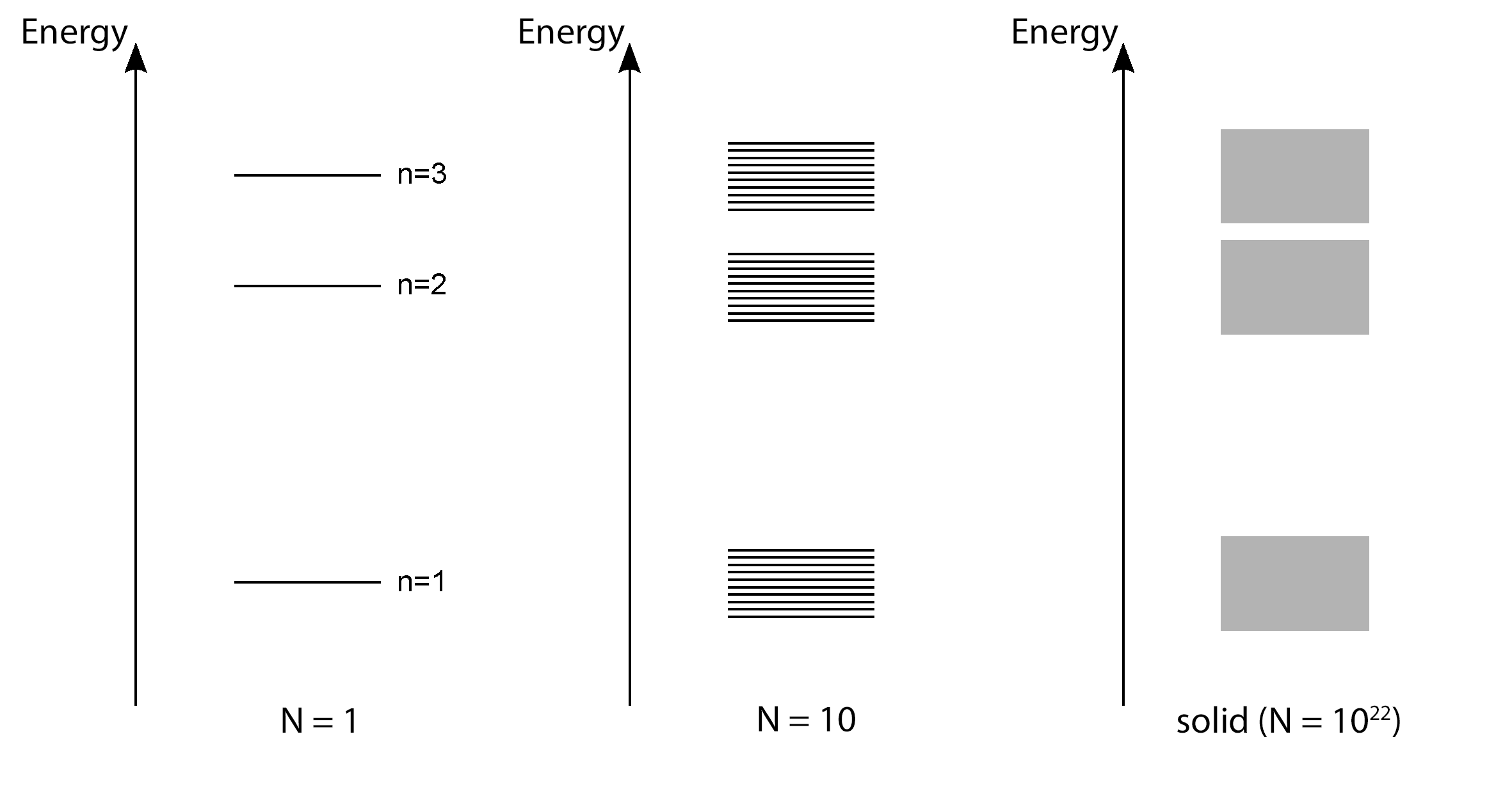}
    \caption{\textbf{Development of conduction bands}. Energy levels of a chain of N atom. The isolated atom (N=1) possess a few number of states, which increases with the number of atoms (N=10) until forming continuum in a solid. From \cite{Hivet2013}}
    \label{fig:EnergLev}
\end{figure}

Quantum physics is based on the quantification of the energy in an isolated atom: its electron can only access discrete energy levels, contrary to the energy continuum of a free electron \cite{Klimov2004}.
This property is modified when several atoms are put together, as illustrated in figure \ref{fig:EnergLev}. If a chain of N atoms are placed close together, and separated by a distance of the order of their Bohr radius, a coupling takes place between them and the degeneracy of their energy levels is removed. The electrons are not only connected to one atom anymore but are delocalized over the whole chain, leading to new energy levels closed to the ones of the isolated atom.

Finally, the case of a solid material can be considered as a chain of a very large number of atoms: the density of energy levels becomes so high that it can be seen as a band of accessible energy, hence the name \textit{band theory}.

\paragraph{}
The distribution of the electrons within these bands at T = 0 K imposes the electronic properties of a material.
At this temperature, the maximum level of occupation of the electrons is given by the Fermi energy. It actually corresponds to the chemical potential at 0 Kelvin of the material.
The position of the Fermi energy compared to the energy bands determines the conduction properties of a solid. 

The first energy band entirely below the Fermi energy is called the \textbf{valence band}; at T = 0 K, it is the completely filled band with the highest energy.
The band just above the valence band is called the \textbf{conduction band}. 
If the Fermi energy is located within the conduction band, the electrons are free to evolve in the whole band, as it is not entirely full, and can thus propagate through the material: it is the case of a \textbf{conductor}.
On the other hand, if the Fermi energy is in the gap between the bands, then all the states of the valence band are filled while all the ones of the conduction band are empty: the material is an insulator.

However, this situation can be different at room temperature. If the gap between the valence and the conduction band is small enough, typically $E_{g} \approx 1$ eV, then the electronic density of the conduction band at room temperature is not negligible, and the material becomes conductor, even though is is an insulator at low temperatures. Such type of material is called \textbf{semiconductor}.

\paragraph{Direct gap semiconductor}

The energy of the valence and conduction bands also depends on the momentum. In particular, two types of semiconductors can be defined: the direct gap semiconductors, where the maximum of the valence band is reached at the same wavevector than the minimum of the conduction band, and the indirect gap semiconductors, where the maximum of the valence band and the minimum of the conduction band do not correspond to the same \textbf{k}.

\begin{figure}[h]
    \centering
    \includegraphics[width=0.9\linewidth]{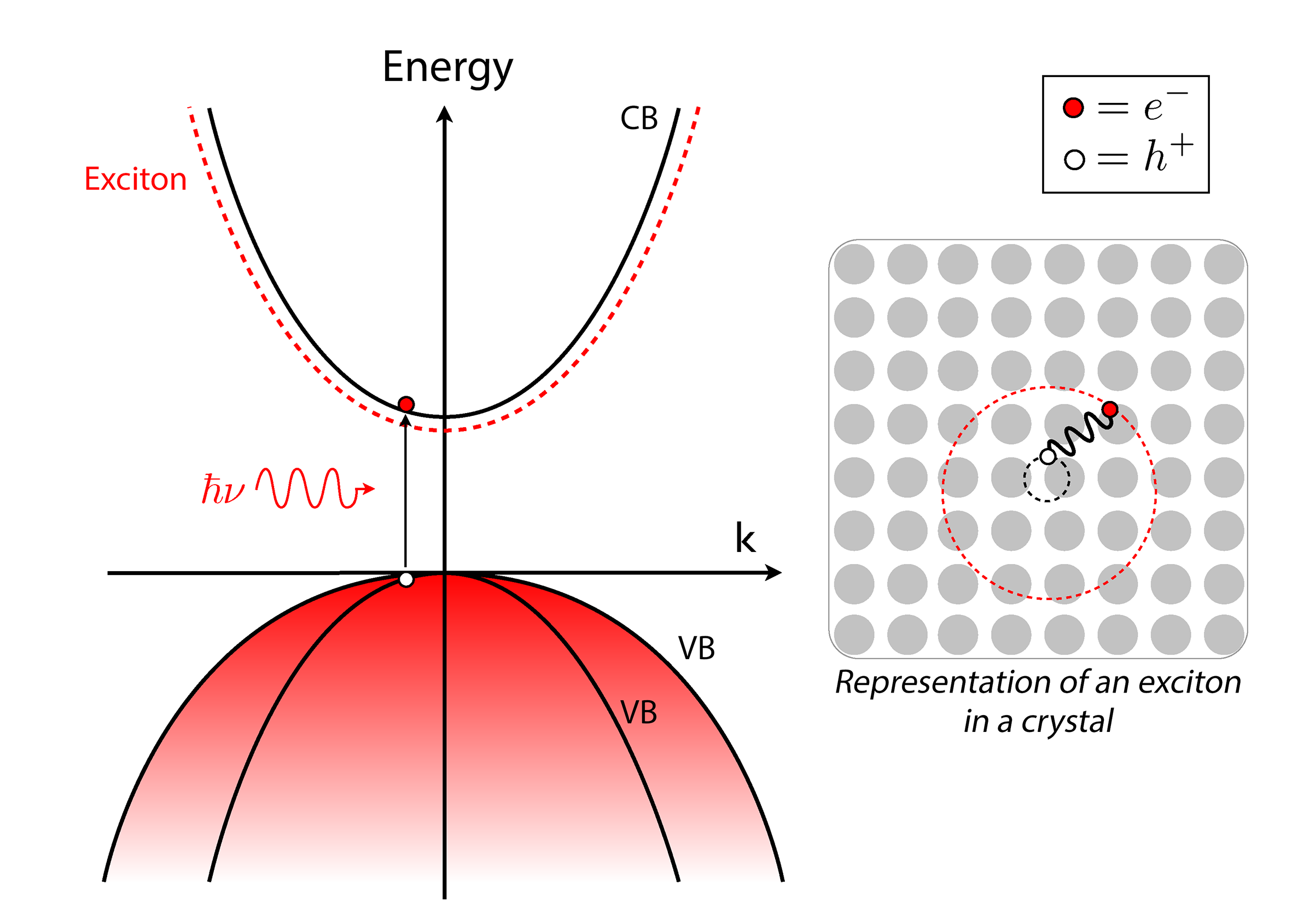}
    \caption{\textbf{Direct gap semiconductor dispersion}. The bottom parabola in red corresponds to the valence band filled with electrons, while the upper parabola is the empty conduction band. Their extrema all take place at the same wavevector. The red dashed line illustrates the energy of the exciton, slightly lower than the one of the conduction band: it is created by the absorption of a photon which energy coincides with the one of the gap. On the right, a symbolic representation of an exciton in a crystal, extending over several crystal cells. From \cite{Pigeon2011phd}}
    \label{fig:DirectGapDisp}
\end{figure}

\paragraph{}
In the case of direct gap semiconductor, as illustrated in figure \ref{fig:DirectGapDisp}, a photon can be absorbed if its energy is larger than the gap energy: $\hbar \omega > E_{g}$. An electron of the valence band is then excited to the conduction band, leaving a hole in the valence band. The excited electron is negatively charged while the presence of the hole imposes a positive charge, and the whole system is called an \textbf{electron-hole pair}.
However, if the incoming photon has an energy slightly lower than the gap energy, this one can still be absorbed. In that case, the excited electron jumps just below the conduction band, and thus stays coupled to the hole: the electron remains only within a couple crystal cells (see inset of figure \ref{fig:DirectGapDisp}), and the bound pair is called an \textbf{exciton}. 

\paragraph{}
An exciton is described as a single neutral quasi-particle, even though it is constituted of two charged particles. Its effective mass is the sum of the ones of both the electron and the hole ($m_{X} = m_{e} + m_{h}$), themselves extracted from their respective dispersion curvature.
Its energy follows the relation \cite{yamamoto2003}:

\begin{equation}
    E_{X}(\mathbf{k}, n) = E_{g} - E_{b}(n) + \dfrac{\hbar^{2} \mathbf{k}^{2}}{2 m_{X}}
\end{equation}

where $E_{b}$ is the binding energy coming from the coulombian interaction and $n$ the electronic density. The binding energy can be written as \cite{Kavokin2007a}:

\begin{equation}
    E_{b}(n) = 
    \dfrac{\hbar^{2}}{2m_{X} a_{X}^{*2} n^{2}} = \dfrac{Ry^{*}}{n^{2}}
\end{equation}

with $Ry^{*}$ an effective Rydberg constant, an $a_{X}^{*}$ the Bohr radius of the exciton defined by analogy with the hydrogen atom as:

\begin{equation}
    a^{*}_{X} = \dfrac{\hbar^{2} \epsilon}{e^{2}m_{X}}
\end{equation}

with $\epsilon$ the dielectric constant of the medium and $e$ the elementary charge.
Semiconductors like GaAs have typically a high dielectric constant, while the effective mass of the exciton is usually one order of magnitude lower than the mass of a free electron in vacuum. 
We obtain $a^{*}_{X} \approx 50$ \AA, which is much bigger than the crystalline cell: the exciton extends over several cells, as pictured in the inset of figure \ref{fig:DirectGapDisp}.

Therefore, the binding energy in such dielectric medium reaches a few meV ($E_{b} \approx 4.5$ meV for GaAs). However, at room temperature, the thermal energy is $E_{th} = k_{B} T = 25$ meV, which means that excitons can be created only at cryogenic temperature. 

This type of exciton are called the \textit{Wannier-Mott excitons}, in opposition to the \textit{Frenkel excitons} that are created in material with lower dielectric constant, and which have consequently a higher binding energy and a smaller size, and are stable at room temperature.

\paragraph{}
As we saw that the excitons can be considered as particles, a hamiltonian operator $H_{X}$ can be defined to describe them, with an associated creation operator $\hat{a}^{\dagger}_{\mathbf{k}}$ and annihilation one $\hat{a}_{\mathbf{k}}$. We thus have:

\begin{equation}
    H_{X} = \sum_{\mathbf{k}} \hbar \omega_{X}(\mathbf{k}) \hat{a}^{\dagger}_{\mathbf{k}} \hat{a}_{\mathbf{k}}
\end{equation}

\paragraph{}
We now have all the elements to describe the excitons and their life cycle: creation through a photon absorption and destruction by a photon emission. However, we have described here \textit{bulk excitons}, \textit{i.e.} excitons within a three dimensional material, where no direction is privileged. The emitted photons are therefore randomly distributed in all directions. 
In order to obtain polaritons, we want to strongly couple the excitons with the light field: an efficient way to improve the coupling is to confine the excitons in two dimensions, so that the photons are emitted in a single direction, as we will now describe.

\subsubsection{Bidimensional confinement in a quantum well}
\label{sec:2DXconfined}

\paragraph{Bidimensional confinement}

Three dimensional bulk semiconductors possess a full translation invariance, which imposes the momentum conservation. A recombination of an exciton of wavevector $\mathbf{k}^{X}$ thus leads to the emission of a photon with the same wavevector: $\mathbf{k}^{ph} = \mathbf{k}^{X}$. 
Therefore, the exciton can only be coupled to one mode of the electromagnetic field.
Moreover, the energy conservation must also be verified: only the excitons which the total energy is the same than the one of the electromagnetic field can realize a radiative relaxation.

\begin{figure}{h}
    \centering
    \includegraphics[width=0.5\textwidth]{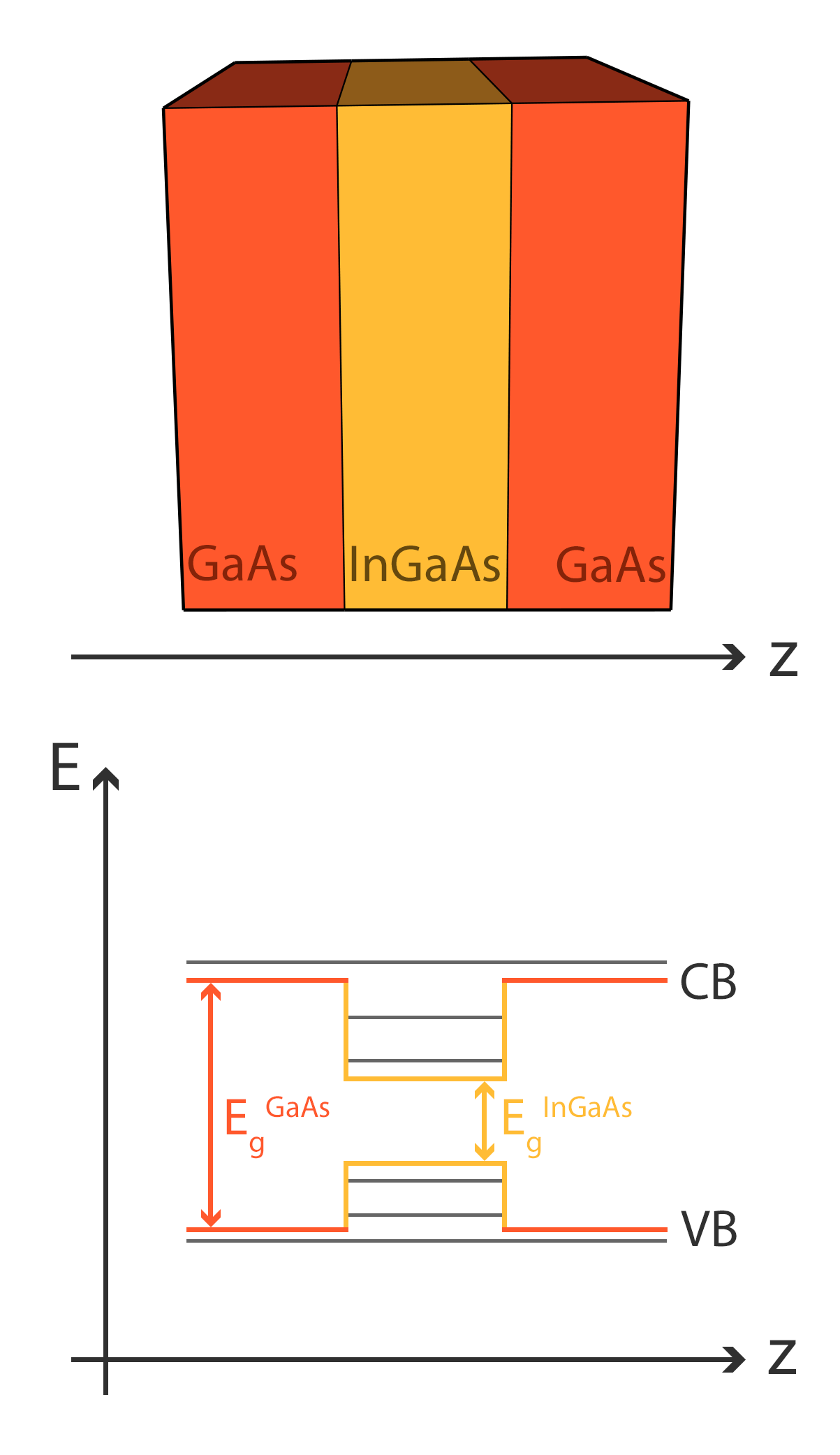}
    \caption{\textbf{Spatial and energetic structures of GaAs-InGaAs quantum well}. A layer of InGaAs is placed between two layers of GaAs. As $E^{InGaAs}_{g} < E^{GaAs}_{g}$, the energy profile of the conduction band presents a dip (respectively a bump for the valence band) where some new energy levels are accessible (represented by the gray lines). The excitons created between these levels are confined along the \textit{z} axis.}
    \label{fig:QutWellGaAs}
\end{figure}

This translational invariance is broken by a bidimensional confinement. It is easily realized by placing a layer of semiconductor in between two layers of another semiconductor with a different gap energy.
For instance, in the case of our sample, a layer of InGaAs is sandwiched between two layers of GaAs, as illustrated in figure \ref{fig:QutWellGaAs}.

\paragraph{}
The interesting case takes place when the central semiconductor has a smaller gap energy than the outer one. 
It indeed induces a dip in the conduction band energy profile and a bump in the one of the valence band, as well as the existence of energy levels only accessible within the central space layer but forbidden everywhere else, plotted in grey in figure \ref{fig:QutWellGaAs}.
An exciton created between two of these levels is therefore confined along the \textit{z} axis, but free to travel on the plane of the layer.

\paragraph{Properties of 2D excitons}
Such a configuration breaks the translational invariance in the $z$ direction: the momentum conservation is only valid on the in-plane directions (\textit{x}, \textit{y}).
Therefore, a photon emitted after the recombination of an exciton of wavevector $\mathbf{k}^{X} = (\mathbf{k}^{X}_{||}, k^{X}_{z})$ has to satisfy only the condition $\mathbf{k}^{ph}_{||} = \mathbf{k}^{X}_{||}$ while its \textit{z} component remains free.
The confinement of the exciton also influences the Bohr radius of the exciton which gets reduced (by half in the ideal case \cite{Kavokin2007a}): this leads to an increase of the binding energy and an enhancement of the coupling.

\paragraph{}
Again, the exciton can be treated as a quasi-particle and described with a Hamiltonian operator. It can be written as:

\begin{equation}
    \hat{H}^{2D}_{X} = \hat{H}_{\mathbf{k}^{X}_{||}} + \hat{H}^{2D}_{b}
\end{equation}

with $\hat{H}_{\mathbf{k}^{X}_{||}}$ the kinetic Hamiltonian and $\hat{H}^{2D}_{b}$ the bidimensional equivalent to the bulk excitons Hamiltonian.

The kinetic Hamiltonian has a parabolic continuum of eigenenergies:

\begin{equation}
    E(\mathbf{k}^{X}_{||}) = E_{g} + 
    \dfrac{\Big(\hbar \mathbf{k}^{X}_{||}\Big)^{2} }{2 m^{*}_{X}}
\end{equation}

while $\hat{H}^{2D}_{b}$ has discrete eigenvalues. Our cavity has been designed so that the ground state is energetically well separated from the other states; by analogy with the hydrogen atom, it is called the bidimensional 1s exciton level. $E_{1s}^{2D}$ can thus be considered the fundamental state of the exciton, and the eigenenergies of $\hat{H}^{2D}_{b}$ can be written:

\begin{equation}
    E^{2D}_{X} = E_{g} + E^{2D}_{1s} + 
    \dfrac{\Big(\hbar \mathbf{k}^{X}_{||}\Big)^{2}}{2 m^{*}_{X}}
\end{equation}

\paragraph{}
The bidimensionnal confinement imposes a last condition on the exciton: the momentum conservation during a photon-excitation interaction. This can be written as:

\begin{equation}
    E_{X} + \dfrac{\Big(\hbar \mathbf{k}_{||}^{X}\Big)^{2}}{2 m_{X}^{*}}
    = \dfrac{\hbar c}{n_{cav}^{2}} 
    \sqrt{\Big(\mathbf{k}_{||}^{\gamma} \Big)^{2} + \Big(k_{z}^{\gamma} \Big)^{2}}
\end{equation}

here the left hand side is the exciton energy, where we have associated $E_{g} + E_{1s}^{2D} = E_{X}$, and the right hand side the photon one, with its wavevector decomposed as $\mathbf{k^{\gamma}} = (\mathbf{k}^{\gamma}_{||}, k^{\gamma}_{z})$.
The interaction takes place when we have $\mathbf{k}^{X}_{||} = \mathbf{k}^{\gamma}_{||}$, thus leading to:

\begin{equation}
    E_{X} + \dfrac{\Big( \hbar \mathbf{k}_{||}^{X} \Big)}{2 m^{*}_{X}} \geqslant \dfrac{\hbar c}{n_{cav}} |\mathbf{k}_{||}^{X}|
\end{equation}

Which for small $|\mathbf{k}_{||}^{X}|$ simplifies to:

\begin{equation}
    |\mathbf{k}_{||}^{X}| \lesssim \dfrac{n_{cav} E_{X}}{\hbar c}
\end{equation}

Therefore, the excitonic modes that does not verify this relation are non radiative. In the type of semiconductor we used, this usually corresponds to $|\mathbf{k}_{||}^{X}| \lesssim $ 30 \textmu m\textsuperscript{-1}.

\paragraph{}
Quantum well excitons in free space are coupled to a continuum of optical modes. Therefore, the radiative emission of a photon by exciton recombination is an irreversible process and the probability of finding an exciton in the excited state decreases exponentially with time. 
Typically, the associated timescale is of the order of 10 ps, which corresponds to the radiative broadening of the exciton linewidth.

\paragraph{}
The linewidth of the exciton also broadens due to non-radiative relaxation process in the quantum wells, which consequently limit the exciton lifetime.
They can happen because of disorder within the semiconductor structure, because of interactions with crystal lattice phonons or even because of exciton-exciton interaction. 
Taking this phenomena into account, the typical non-radiative exciton lifetimes are of the order of 100 ps.

\subsection{Excitons polaritons}

\paragraph{}

We described in the previous sections two different components.
First, the optical cavity breaks the symmetry of the electromagnetic field along the \textit{z} axis, which leads to a discretization of the \textit{z} component of the incoming photons.
On the other hand, the quantum well confinement plays a similar role on the excitons: the symmetry which induces the momentum conservation along \textit{z} is also broken.

The starting point of this section is to combine those elements to enhance the light-matter interactions, in order to reach the strong coupling regime.

\subsubsection{Quantum well embedded in a microcavity}

\begin{figure}[h]
    \centering
    \includegraphics[width=0.9\linewidth]{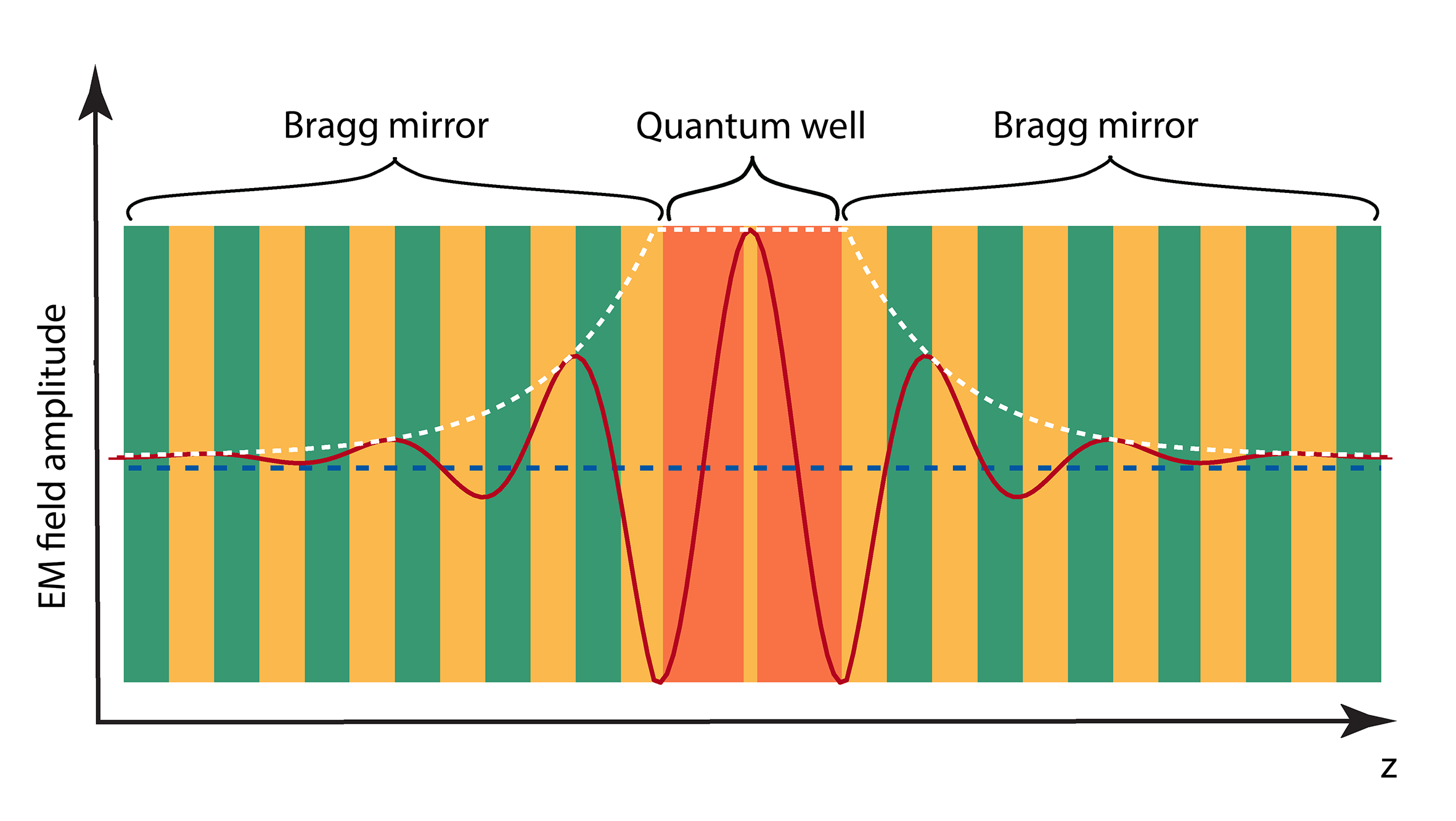}
    \caption{\textbf{Quantum well embedded in a Fabry-Perot microcavity}. The semiconductor microcavity is constituted by two piles of green and yellow layers, while the quantum well, in orange and yellow, is placed inside the cavity. The red line shows the electromagnetic field distribution, and the white dashed line its envelope with an exponential decay within the DBR. Note that the quantum well is placed at an antinode of the electromagnetic field. Adapted from \cite{Boulier2014}}
    \label{fig:CavityQW}
\end{figure}

\paragraph{Geometry of the samples}
In order to strongly couple excitons and photons, the 2D quantum well is placed inside a microcavity (see figure \ref{fig:CavityQW}).
The cavity consists of two DBRs (piles of yellow and green layers) designed to be quasi-resonant with the excitonic transition. 

To maximize the interactions, the quantum well (thin yellow layer inserted in the orange material) is placed at an antinode of the electromagnetic field, whose amplitude is plotted in red. The white dashed line shows its envelope and the exponential decay within the DBRs.

\begin{figure}[h]
    \centering
    \includegraphics[width=0.95\linewidth]{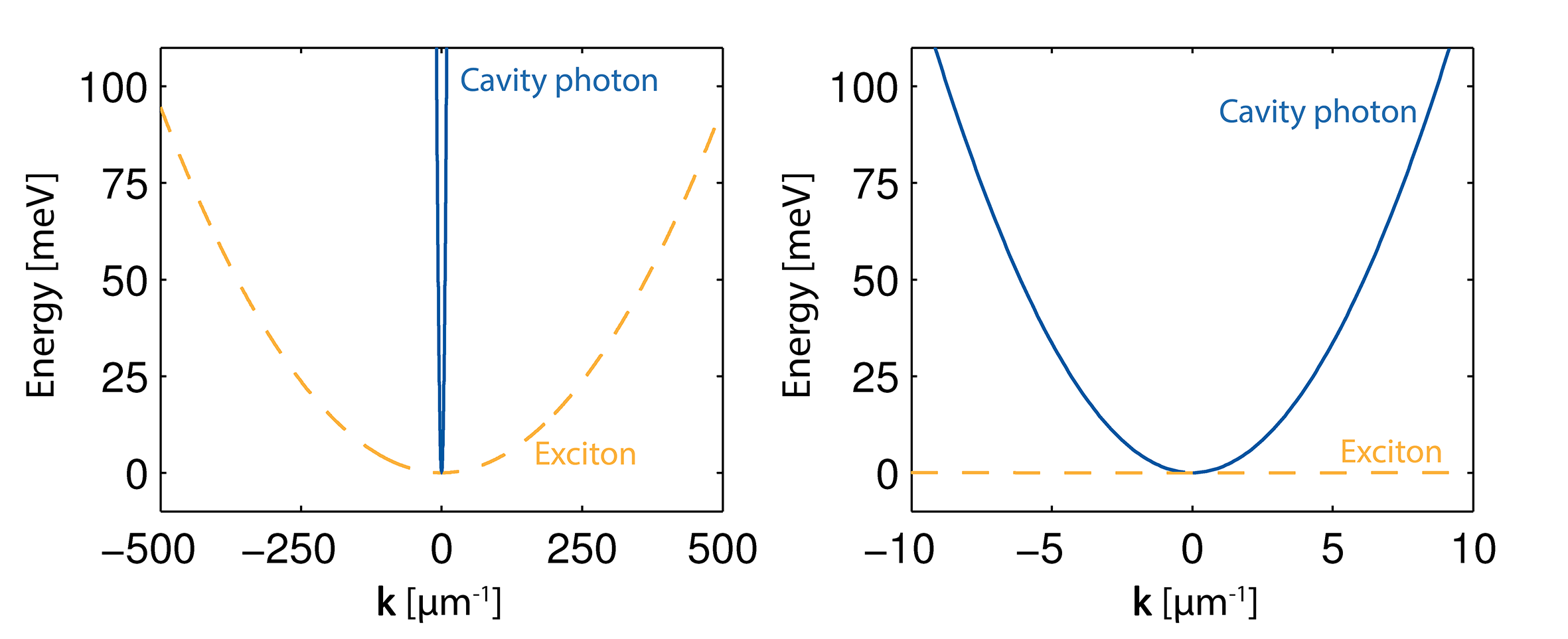}
    \caption{\textbf{Comparison of photonic and excitonic resonances}. Photonic and excitonic numerical dispersions for different momentum ranges. As we work with wavevectors of the order of 1 \textmu m\textsuperscript{-1}, the picture on the right is the one that we consider, and the excitonic resonance is considered constant from now on.}
    \label{fig:disptheoXcav}
\end{figure}

\paragraph{}
At this point, an important comparison has to be made between the effective masses of both components.
The cavity photons have an effective mass 4 orders of magnitude smaller than the quantum well excitons.
It results in the fact that, even though they both have a parabolic dispersion, they take place in a different scale of wavevectors, as shown numerically in figure \ref{fig:disptheoXcav}.

The left image shows both dispersions scaled so that the curvature of the excitonic one is visible. The photonic dispersion is in this case very narrow around zero.
The right image is the same, but for a much shorter range of wavevectors: the photonic curvature is clear but the excitonic resonance seems flat.
As it corresponds to the wavevectors considered in this work, the excitonic curvature is neglected in the following discussion and its resonance energy is considered constant.

\paragraph{}
The interactions can be modified by putting more quantum wells inside the cavity, all of them at some maxima of the electromagnetic field. In that case, the density is shared between all the quantum wells while the coupling is maintained, which allows to reach high densities that can lead to a condensation phenomenon as we will see later. A typical configuration consists of 12 quantum wells, distributed as 3 packs of 4 wells.

The quasi-resonance between the exciton and the cavity photon can also be tuned by design. The excitonic resonance energy is fixed by the quantum well materials, but the photonic one depends on the length of the cavity. 
By implementing a slight angle (wedge) between the two DBRs, the detuning exciton-cavity photon $\Delta E_{Xcav} = E_{X} - E_{\gamma}$ depends on the position on the sample and can be chosen on will.

\paragraph{}
Let us consider that the medium outside of the cavity is air, and neglect the linewidth broadening.
This configuration allows the selective excitation of one exciton of in-plane momentum $\mathbf{k}_{||}^{X} = (k_{x}^{X}, k_{y}^{X})$. It is done by choosing the incident photon $\mathbf{k}_{||}^{\gamma} = (k_{x}^{\gamma}, k_{y}^{\gamma})$ with an angle $\theta^{\gamma} = (\theta_{x}^{\gamma}, \theta_{y}^{\gamma})$ that verifies:

\begin{align}
    \label{eq:Xcavexcitationx}
    k_{x}^{X} &= k_{x}^{\gamma , cav} = \dfrac{E n_{cav}}{\hbar c} \sin \theta_{x}^{\gamma, cav} = \dfrac{E}{\hbar c} \sin \theta_{x}^{\gamma} \\
    \label{eq:Xcavexcitationy}
    k_{y}^{X} &= k_{y}^{\gamma , cav} = \dfrac{E n_{cav}}{\hbar c} \sin \theta_{y}^{\gamma, cav} = \dfrac{E}{\hbar c} \sin \theta_{y}^{\gamma}
\end{align}

where $\mathbf{k}_{||}^{\gamma, cav} = (k_{x}^{\gamma, cav}, k_{y}^{\gamma, cav})$ is the photon momentum inside the cavity, $E$ is the excitation energy, $n_{cav}$ the index of the cavity and $\theta^{\gamma, cav} = (\theta_{x}^{\gamma, cav}, \theta_{y}^{\gamma, cav})$ the photon angle inside the cavity obtained by the Snell-Descartes law.

Reciprocally, the detection of a photon with an angle $\theta^{\gamma} = (\theta_{x}^{\gamma}, \theta_{y}^{\gamma})$ out of the cavity comes from the recombination of an exciton $\mathbf{k}_{||}^{X} = (k_{x}^{X}, k_{y}^{X})$ that verifies equations \ref{eq:Xcavexcitationx} and \ref{eq:Xcavexcitationy}.

\paragraph{Spontaneous emission}

A parameter plays a crucial role in the light matter interactions within our system: the spontaneous emission of photons by an exciton. This process of energy dissipation is described by the \textit{Fermi golden rule}, which gives the spontaneous emission rate of a photon from an initial state \textit{i} to a final state \textit{f}. It can be written as \cite{cohen2012}:

\begin{equation}
    \Gamma_{i \rightarrow f} = \dfrac{2 \pi}{\hbar} 
    \big| \langle f | V | i \rangle \big|^{2}
    \delta(E_{i} - E_{f}) \rho (E_{f})
\end{equation}

where $\Gamma_{i \rightarrow f}$ is the transition rate from state \textit{i} to \textit{f} per time unit, $\big| \langle f | V | i \rangle \big|^{2}$ is the probability of transition from states \textit{i} to \textit{f}, $\delta(E_{i} - E_{f})$ is the energy conservation condition and $\rho (E_{f})$ the number of final states \textit{f} of energy $E_{f}$. This process thus directly depends 
on the density of final states.
The coupling matrix element is also depicted as the quantity $\Omega_{R}$ or \textit{Vacuum Rabi frequency} \cite{Adrados2011}.

\paragraph{Coupling regimes}

Both cavity photons and quantum well excitons have limited lifetimes $\tau$ which result into spectral broadening $\gamma = 1/ \tau$.
Various intrinsic factors are responsible for the homogeneous broadening of both elements.
For instance, the cavity mode can couple to other empty electromagnetic modes of the cavity, or to the continuum of the extra-cavity modes, which are all contained in the cavity photon broadening $\gamma_{cav}$.
On the other hand, the excitonic broadening $\gamma_{X}$ includes the coupling of the excitons with each other \cite{Ciuti1998}, with the phonons \cite{Karr2001} or with the disorder of the crystal \cite{Savona1997}.

\paragraph{}
Different coupling regimes can be defined by comparing the two lifetimes with the coupling strength through the Rabi frequency $\Omega_{R}$.
When $\Omega_{R} \ll \gamma_{cav}, \gamma_{X}$, the oscillator strength is not strong enough to reabsorb a photon spontaneously emitted which will be lost in the cavity: the emission is irreversible, it is the regime of \textbf{weak coupling}. 
However, if $\Omega_{R} \gg \gamma_{cav}, \gamma_{X}$, then the lifetimes are long enough so that an emitted photon stays in the cavity and can be absorbed again, which corresponds to the \textbf{strong coupling} regime.
The phenomenon is reversible and many coherent exchange of energy can take place before the photon leaves the cavity, which are called the \textbf{Rabi oscillations}.

\subsubsection{Strong coupling regime}
\label{sec:StrongCplgReg}

\paragraph{}
The strong coupling regime is the one that we are interested in as it results in the apparition of new eigenmodes called \textbf{cavity polaritons}.
The first observation of such strong coupling in a quantum well embedded in a semiconductor microcavity has been realized in 1992 in the group of C. Weisbuch \cite{Weisbuch1992}.

\paragraph{}
In order to study the optical properties of the strong coupling, a simple model consists in considering the coupled elements as harmonic oscillators, to which is associated a creation operator $\hat{a}^{\dagger}$ and an annihilation one $\hat{a}$.
An important condition of this description is however that both elements are bosonic, meaning that they follow the commutation law $[\hat{a}_{m}, \hat{a}^{\dagger}_{n}] = \delta_{mn}$, with $m$ and $n$ two quantum states of the system.

In our case, the coupling takes place between the cavity photons and the quantum well excitons. 
By nature, photons are bosons, but the excitons are made of fermionic elements. 
Let us first then verify if they can be treated as bosons.

\paragraph{Excitons as bosons}

It is possible to associate to excitons a creation $\hat{b}^{\dagger}_{1s}$ and an annihilation $\hat{b}_{1s}$ operators \cite{Usui1960, Rochat2000}. In that case, the commutation relation can be written as \cite{Karr2001}:

\begin{equation}
    \big[ \hat{b}_{1s}, \hat{b}_{1s}^{\dagger} \big] = 1 - \mathcal{O} \Big( n \big(a_{X}^{*2D} \big)^{2} \Big)
\end{equation}

where $a_{X}^{*2D}$ is the Bohr radius of the exciton in the quantum well as defined in section \ref{sec:excitons}, and $n$ the excitonic density per surface unit. 
It is therefore possible to consider the exciton as a boson if this density is small enough ($n \big(a_{X}^{*2D} \big)^{2} \ll 1$), \textit{i.e.} in a regime of low light excitation.
Typically, the Bohr radius of a 2D exciton is around 5 nm, which allows a maximal excitonic density of the order of $4 \cdot 10^{4}$ \textmu m\textsuperscript{-2}.
In our case, we always work with lower densities, which allows us to use the bosons description for the excitons.

\paragraph{Linear hamiltonian of the strong coupling regime}

Let us simplify the notation as we only consider the in-plane wavevector: from now on, the $||$ notation is forgotten and $\mathbf{k} = \mathbf{k}_{||}$.

As a first step, we place the system in a configuration of low excitation. The exciton-exciton interactions are therefore negligible: the resulting hamiltonian is purely linear and can be written as:

\begin{equation}
\label{eq:linHamilt}
    \hat{H}_{lin} = \sum_{\mathbf{k}} \hat{H}_{\mathbf{k}} = \sum_{\mathbf{k}} 
    \bigg( E_{X} \hat{b}_{\mathbf{k}}^{\dagger} \hat{b}_{\mathbf{k}}
    + E_{\gamma} (\mathbf{k}) \hat{a}_{\mathbf{k}}^{\dagger} \hat{a}_{\mathbf{k}}
    + \dfrac{\hbar \Omega_{R}}{2} \Big( \hat{a}_{\mathbf{k}}^{\dagger} \hat{b}_{\mathbf{k}} + \hat{b}_{\mathbf{k}}^{\dagger} \hat{a}_{\mathbf{k}} \Big) \bigg)
\end{equation}

with $\hat{a}_{\mathbf{k}}^{\dagger}, \hat{a}_{\mathbf{k}}$ and $\hat{b}_{\mathbf{k}}^{\dagger}, \hat{b}_{\mathbf{k}}$ are the creation and annihilation operators for a cavity photon and a quantum well exciton with an in-plane wavevector \textbf{k}, respectively.
$E_{X}$ is the excitonic energy, $E_{\gamma}(\mathbf{k})$ the photonic one, and $\Omega_{R}$ the Rabi frequency.

The last term of equation \ref{eq:linHamilt} is particularly interesting. It contains the superposition of two events: on one side, the creation of a photon and the annihilation of an exciton, \textit{i.e.} the recombination of the exciton, and on the other side the creation of an exciton and the annihilation of a photon, so the absorption of a photon. 
They are connected by the coefficient $\frac{\hbar \Omega_{R}}{2}$ which represents the \textbf{coupling energy}.

\paragraph{}
Equation \ref{eq:linHamilt} is very close to the coupling of a two level system by an external electromagnetic field. 
In this case, the system oscillates between its ground state and the excited state, by absorbing a photon to acquire energy or by emitting a similar one to relax.
The polariton case is analog if we consider the exciton as the two level system, with one particularity: the ground state in his case is the absence of exciton. It is indeed the lowest energy state to which the system tends after a long enough time.
The first excited state is in our case the presence of the \textit{1s} exciton.

In time, polaritons can therefore be seen as the succession of cavity photon, absorbed by creating an exciton, which later recombines by emitting an identical photon, that can be absorbed once again, and so on.
This image describes the polaritons as a coherent linear combination of an electromagnetic field (the photonic part) and a polarization field (the excitonic one).

\subsubsection{Polariton basis}

\paragraph{Eigenmodes and Hopfield coefficients}

The hamiltonian given by the equation \ref{eq:linHamilt} can be diagonalized. It results in new eigenstates of the system that contain both the photonic and excitonic part: the upper and lower \textbf{exciton-polaritons}, denoted as UP and LP respectively. The hamiltonian can be rewritten in this basis:

\begin{equation}
    \hat{H}_{lin} = \sum_{\mathbf{k}} \hbar \omega_{LP} (\mathbf{k})
    \hat{p}_{\mathbf{k}}^{\dagger} \hat{p}_{\mathbf{k}}
    + \sum_{\mathbf{k}} \hbar \omega_{UP} (\mathbf{k})
    \hat{u}_{\mathbf{k}}^{\dagger} \hat{u}_{\mathbf{k}}
\end{equation}

with $\hat{p}_{\mathbf{k}}^{\dagger}, \hat{p}_{\mathbf{k}}$ and $\hat{u}_{\mathbf{k}}^{\dagger}, \hat{u}_{\mathbf{k}}$ the creation and annihilation operators of the upper and lower polaritons, respectively.
They are bosonic operators and therefore follow the commutation rule: $\big[ \hat{p}_{\mathbf{k}}^{\dagger}, \hat{p}_{\mathbf{k}'} \big] = \delta_{\mathbf{k}, \mathbf{k}'}$ and $\big[ \hat{u}_{\mathbf{k}}^{\dagger}, \hat{u}_{\mathbf{k}'} \big] = \delta_{\mathbf{k}, \mathbf{k}'}$.

$\omega_{LP} (\mathbf{k})$ and $\omega_{UP} (\mathbf{k})$ express the dispersions associated to both of these polaritonic modes:

\begin{equation}
    \omega_{UP/LP} = 
    \dfrac{\omega_{X} (\mathbf{k}) \pm \omega_{\gamma}(\mathbf{k}) }{2}
    \pm \dfrac{1}{2}
    \sqrt{\big( \omega_{X}(\mathbf{k}) - \omega_{\gamma} (\mathbf{k}) \big)^{2} + 4 \Omega_{R}^{2}}
\end{equation}

\begin{figure}[H]
    \centering
    \includegraphics[width=0.98\linewidth]{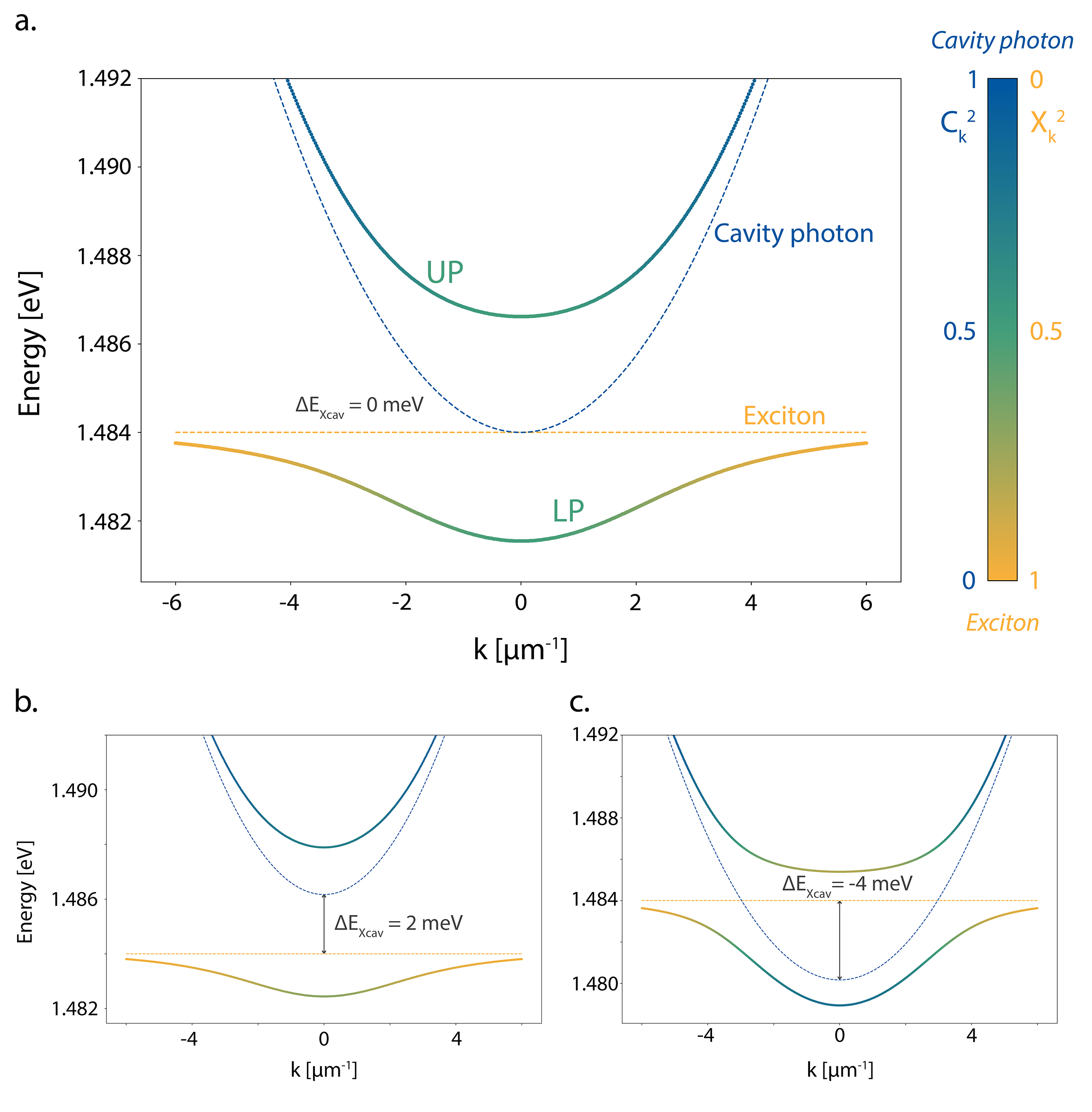}
    \caption{\textbf{Polariton dispersions and Hopfield coefficients}. a. Upper (UP) and lower (LP) polariton dispersions. The color gradient illustrates the excitonic ($X_{\mathbf{k}}^{2}$ in yellow) and photonic ($C_{\mathbf{k}}^{2}$ in blue) fractions. The dotted lines show the photon and exciton dispersion in the weak coupling case. The detuning exciton-cavity photon is chosen to be zero.
    b. and c. Polariton dispersions with the same color scale, for an energy detuning $\Delta E_{Xcav} = 2$ meV and $\Delta E_{Xcav} = -4$ meV, respectively.}
    \label{fig:DispHopfCoeff}
\end{figure}

The two dispersions are plotted in figure \ref{fig:DispHopfCoeff} by the solid lines: the strong coupling greatly modifies the system (the dashed lines illustrate the system dispersions in the weak coupling regime).

\paragraph{}

The two polaritonic operators can be expressed through the photonic and excitonic ones by the relation:

\begin{equation}
    \begin{pmatrix}
    \hat{p}_{\mathbf{k}} \\ \hat{u}_{\mathbf{k}}
    \end{pmatrix}
    =
    \begin{pmatrix}
    X_{\mathbf{k}} & C_{\mathbf{k}} \\
    -C_{\mathbf{k}} & -X_{\mathbf{k}}
    \end{pmatrix}
    \begin{pmatrix}
    \hat{b}_{\mathbf{k}} \\ \hat{a}_{\mathbf{k}}
    \end{pmatrix}
\end{equation}

$X_{\mathbf{k}}^{2}$ and $C_{\mathbf{k}}^{2}$ represent the excitonic and the photonic fraction of the polaritons, respectively, and are called the \textbf{Hopfield coefficients} \cite{Hopfield1958}. 
The transformation is therefore unitary: $X_{\mathbf{k}}^{2} + C_{\mathbf{k}}^{2} =1$.
They are plotted in figure \ref{fig:DispHopfCoeff} through the color scale of the dispersions, highlighting the light and matter part of the polaritons.
They can also be quantified more precisely by the expressions \cite{Boulier2014}:

\begin{align}
    X_{\mathbf{k}}^{2} &= \dfrac{\sqrt{\Delta E_{Xcav}^{2} 
    + \hbar^{2} \Omega_{R}^{2}} + \Delta E_{Xcav} }
    {2 \sqrt{\Delta E_{Xcav}^{2} + \hbar^{2} \Omega_{R}^{2}}}\\
    C_{\mathbf{k}}^{2} &= \dfrac{\sqrt{\Delta E_{Xcav}^{2} 
    + \hbar^{2} \Omega_{R}^{2}} - \Delta E_{Xcav} }
    {2 \sqrt{\Delta E_{Xcav}^{2} + \hbar^{2} \Omega_{R}^{2}}} 
\end{align}

These coefficients strongly depend on the detuning between the bare exciton energy and the cavity photon resonance $\Delta E_{Xcav} = \hbar \omega_{\gamma}(\mathbf{k}) - \hbar \omega_{X}(\mathbf{k})$, as illustrated in the figures \ref{fig:DispHopfCoeff}.b. and c.
The polaritons are fully hybrid typically if $\Delta E_{Xcav} = 0$ and at zero wavevector: $C_{\mathbf{k} = 0}^{2} = X_{\mathbf{k}=0}^{2} = \frac{1}{2}$.

\paragraph{}
An interesting interpretation of the Hopfield coefficients and the Rabi frequency connects them with the state of the system. 
As we saw, polaritons are continuously oscillating between their photonic and excitonic states, which are coupled through the Rabi frequency: it thus represents the average conversion rate between those two states. 
As for the Hopfield coefficients, they can be seen as the time ratio the system passes in the photonic or excitonic state.

\paragraph{}
The eigenenergies $E_{UP}(\mathbf{k})$ and $E_{LP}(\mathbf{k})$ of the polaritonic states are given by:

\begin{align}
    E_{UP}(\mathbf{k}) &=  E_{X}(\mathbf{k}) + E_{\gamma}(\mathbf{k})
    + \sqrt{ \Delta E_{Xcav}^{2} + \big(\hbar \Omega_{R}\big)^{2} } \\
    E_{LP}(\mathbf{k}) &= E_{X}(\mathbf{k}) + E_{\gamma}(\mathbf{k})
    - \sqrt{ \Delta E_{Xcav}^{2} + \big(\hbar \Omega_{R}\big)^{2}}
\end{align}

\paragraph{}
If we now take into account the finite lifetimes of both the cavity photon and the exciton, the eigenstates are modified. 
A common approximation consists in adding the decay rates $\gamma_{X}$ and $\gamma_{cav}$ to the particle energies as an imaginary part:

\begin{align}
    E_{X}^{*} (\mathbf{k}) &= E_{X}(\mathbf{k}) - i \hbar \gamma_{X} \\
    E_{\gamma}^{*} (\mathbf{k}) &= E_{\gamma}(\mathbf{k}) - i \hbar \gamma_{cav}
\end{align}

It is an approximation as it neglects the influence of the exciton-photon coupling on relaxation, and does not take into account the shape of the dispersion and in particular the dependence on \textbf{k} of the decay rates.
However, as we focus our study on the small wavevectors, those hypothesis are acceptable.

\paragraph{}

The introduction of the decay rates also influences the expression of the upper and lower polariton energies, which are now complex as well:

\begin{align}
    E_{UP}^{*} &= \dfrac{E_{X}(\mathbf{k}) + E_{\gamma}(\mathbf{k})}{2}
    - i \hbar \dfrac{\gamma_{X} + \gamma_{cav}}{2}
    + \sqrt{ \dfrac{\big( \Delta E_{Xcav} - i \hbar (\gamma_{cav} - \gamma_{X}) \big)^{2} + \big( \hbar \Omega_{R} \big)^{2} }{2} } \\
    E_{LP}^{*} &= \dfrac{E_{X}(\mathbf{k}) + E_{\gamma}(\mathbf{k})}{2}
    - i \hbar \dfrac{\gamma_{X} + \gamma_{cav}}{2}
    - \sqrt{ \dfrac{\big( \Delta E_{Xcav} - i \hbar (\gamma_{cav} - \gamma_{X}) \big)^{2} + \big( \hbar \Omega_{R} \big)^{2} }{2} } 
\end{align}

The last term of the previous equations defines the coupling between photons and excitons. In particular, for $\Delta E_{Xcav} = 0$, it simplifies into $\dfrac{\hbar}{2} \sqrt{\Omega_{R}^{2} - (\gamma_{cav} - \gamma_{x})^{2}}$.
In the case of the strong coupling regime that we are concerned about, it results in an energy difference between the two branches, known as the \textbf{anticrossing}, and characteristic of the strong coupling.

\begin{figure}[h]
    \centering
    \includegraphics[width=0.95\linewidth]{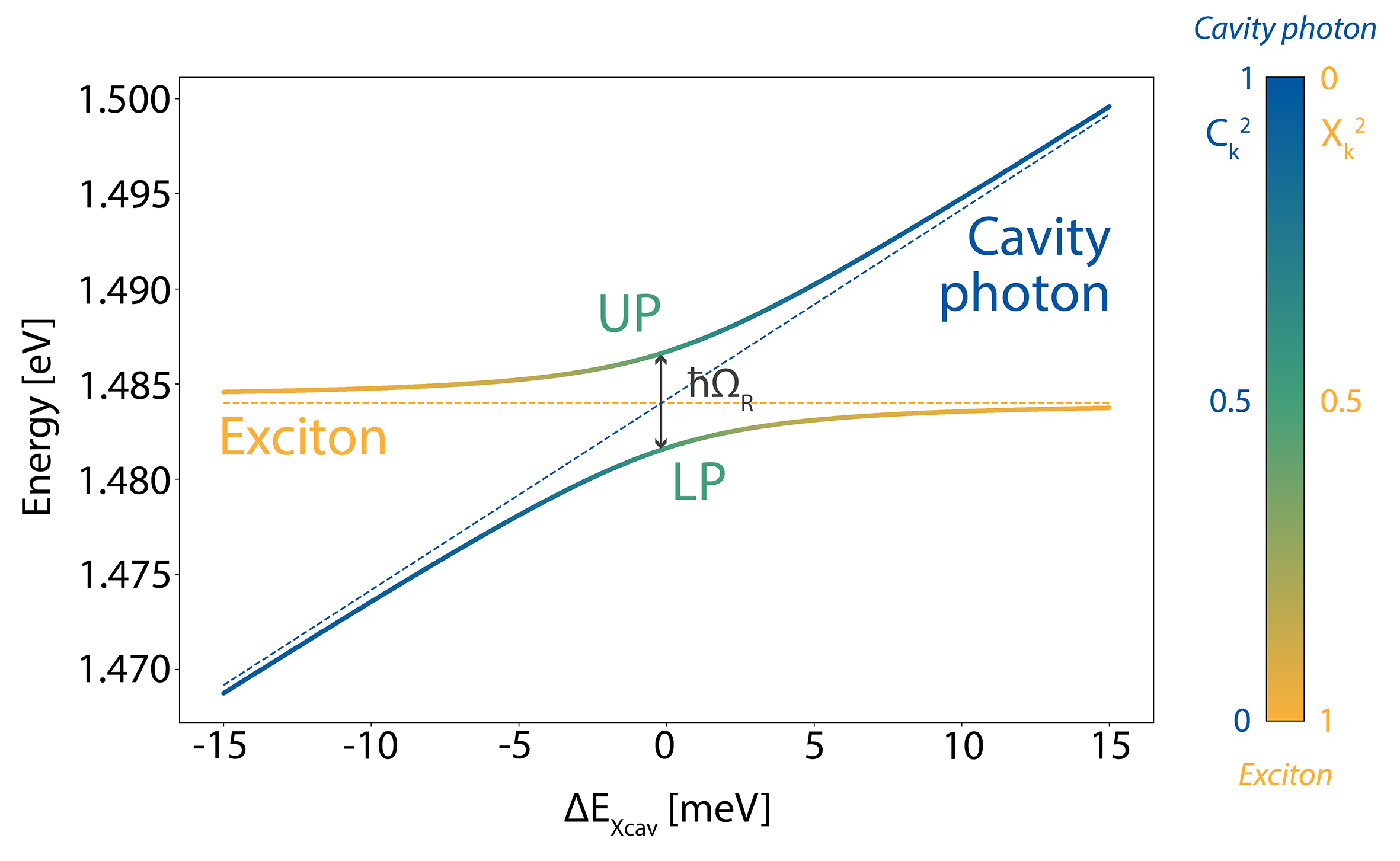}
    \caption{\textbf{Anticrossing of the energy levels}. Energies of the upper (UP) and lower (LP) polariton branches, at $\mathbf{k} = 0$ \textmu m\textsuperscript{-1}, depending on the exciton - cavity photon detuning $\Delta E_{Xcav}$, with the Hopfield coefficients indicated through the colorscale. The dashed lines are the energies of the bare photon and exciton. At $\Delta E_{Xcav} = 0$ meV, there is an energy gap equals to the Rabi energy $\hbar \Omega_{R}$.}
    \label{fig:Anticross}
\end{figure}

This behaviour is highlighted in figure \ref{fig:Anticross}, where are plotted the upper and lower polariton energies as a function of the detuning $\Delta E_{Xcav}$. As previously, the colorscale indicates the excitonic and photonic coefficients. 
The anticrossing is visible for $\Delta E_{Xcav} = 0$ meV, where the energy gap between the branches is equal to the Rabi energy $\hbar \Omega_{R}$.

\paragraph{}
Finally, the linewidth of the two polariton branches can be expressed as a superposition of the excitonic and photonic ones:

\begin{equation}
    \begin{pmatrix}
    \gamma_{UP}(\mathbf{k}) \\ \gamma_{LP}(\mathbf{k})
    \end{pmatrix}
     = 
     \begin{pmatrix}
     C_{\mathbf{k}}^{2} & X_{\mathbf{k}}^{2} \\
     X_{\mathbf{k}}^{2} & C_{\mathbf{k}}^{2}
     \end{pmatrix}
     \begin{pmatrix}
     \gamma_{X} \\ \gamma_{cav}
     \end{pmatrix}
\end{equation}

Thus, the polaritons usually have short lifetimes, typically of the order of some tens of picoseconds.
Indeed, the polariton population is characterized by a continuous decay due to the photon escape out of the cavity. 
The system is therefore out-of-equilibrium and needs a constant pumping to ensure a stable population density: we talk about its \textbf{driven-dissipative} nature.

This particularity is actually a great advantage for polaritonic systems: not only are they perfectly suited to study out-of-equilibrium quantum gases, but the photon escape allows for the simplest optical detection. Indeed, the photons leave the cavity with all the properties of the polaritons they are issued from: the analysis of the cavity emission gives access to all the interesting features of the polaritons, as we will discuss in the next section.

\subsubsection{All-optical control}

\paragraph{}
Let us summarize the correspondence between the optical excitation and detection of the system and the properties of the polaritons. 
We saw that the selection of a specific angle and energy of the excitation field leads to the excitation of a single resonant polariton mode within the dispersion curve. 
Reciprocally, the emission of a photon with a specific wavevector and energy comes from a polariton with the same properties. 

\paragraph{}
We therefore have an equivalence between the following quantities:
\begin{itemize}
    \item the excitation intensity and the polariton density
    \item the incident angle of the excitation field and the in-plane wavevector of the created polaritons
    \item the in-plane wavevector of the polaritons and their velocity
    \item the density of polaritons and the emitted intensity
    \item the polariton phase and the phase of the emitted field
    \item the polariton wavevector and the emission angle
    \item the polariton energy and the emission energy
    \item the polariton lifetime and the emission linewidth
\end{itemize}

\subsubsection{Polaritons-polaritons interactions}

\paragraph{Nonlinearity}

Since the beginning of section \ref{sec:StrongCplgReg}, the description of the exciton-polaritons system has been done in the case of low density, thus neglecting any interactions between the particles. 
However, when the incident intensity is increased, this approximation can not be sustained anymore and the different interactions must be taken into account.
The interactions take place in particular between the electrons and the holes of the excitons, leaving the system in a particular regime where the excitons are still considered as bosons, but where their fermionic composition can not be neglected.

\paragraph{}
The linear hamiltonian previously used needs to be completed by some second order terms \cite{Karr2001, Ciuti2003}:

\begin{itemize}
    \item the hamiltonian of the coulombian interaction between electrons and holes of two colliding excitons of wavevectors $\mathbf{k}$ and $\mathbf{k}'$, resulting in the creation of two new excitons of wavevectors $\mathbf{k} + \mathbf{q}$ and $\mathbf{k}' - \mathbf{q}$ to respect the momentum conservation. It is also know as the exchange interaction \cite{Ciuti1998a} and can be expressed as:
    \begin{equation}
    \label{eq:4wavemixX}
        \hat{H}_{XX} = \dfrac{1}{2} 
        \sum_{\mathbf{k}, \mathbf{k}', \mathbf{q}} V_{0}^{XX}(\mathbf{q})
        \hat{b}_{\mathbf{k}+\mathbf{q}}^{\dagger} \hat{b}_{\mathbf{k}'-\mathbf{q}}^{\dagger}
        \hat{b}_{\mathbf{k}} \hat{b}_{\mathbf{k}'}
    \end{equation}
    
    \item the hamiltonian of the coupling saturation between exciton and photon, leading to the saturation of the optical transition with the excitation intensity:
    \begin{equation}
        \hat{H}_{Xcav}^{sat} = -\dfrac{1}{2}
        \sum_{\mathbf{k}, \mathbf{k}', \mathbf{q}}
        V_{sat} \Big(
        \hat{a}_{\mathbf{k} + \mathbf{q}}^{\dagger} 
        \hat{b}_{\mathbf{k}' - \mathbf{q}}^{\dagger}
        \hat{b}_{\mathbf{k}} \hat{b}_{\mathbf{k}'} + 
        \hat{a}_{\mathbf{k} + \mathbf{q}}
        \hat{b}_{\mathbf{k} - \mathbf{q}}
        \hat{b}_{\mathbf{k}}^{\dagger} \hat{b}_{\mathbf{k}'}^{\dagger}
        \Big)
   \end{equation}
\end{itemize}

A few comments can be done on those two terms.
First, the coulombian interaction of a bidimensional system is defined as:

\begin{equation}
    V_{0}^{XX}(\mathbf{q}) = \dfrac{2 \pi e^{2}}{\epsilon A q}
\end{equation}

with $A$ the quantification area, \textit{i.e.} the quantum well surface in our case. Under the Bohr approximation, stating that $\frac{2 \pi}{q} \gg a_{X}^{*2D}$, which means, for our excitonic Bohr radius of 5 nm: $q \ll 1000$ \textmu m\textsuperscript{-1}, the previous potential can be considered independent of $q$ \cite{Karr2001}, as it is the case in our work.

\paragraph{}
As for the saturation potential, it appears for a certain regime of excitation, for which the repulsion between the interacting excitons becomes so high that they can not stay in the same state, which results in a fermion-like behaviour.
The excitons are then described as \textit{hard-core} bosons, as the N-excitons wavefunction stays symmetric for the exchange of two excitons even with such a Pauli-like repulsion.

It can be expressed by applying the Usui transformation \cite{Usui1960} to quantum wells \cite{Rochat2000}:
\begin{equation}
    V_{sat} = \dfrac{\hbar \Omega_{R}}{2 n_{sat} A}
\end{equation}
where $n_{sat}$ is the exciton density from which the exciton-photon coupling saturation takes place \cite{Houdre1995, Ciuti2003}, typically of the order of $10^{3}$ \textmu m\textsuperscript{-2} in our cavities \cite{Adrados2011}.
In the case of a gaussian excitation spot, the saturation starts from the higher intensity, resulting in a flat density in the center of the spot.

\paragraph{}

The two previous potentials can be combined in the polariton basis under the single potential $V_{\mathbf{k}, \mathbf{k}', \mathbf{q}}^{pol-pol}$, which depends on the Hopfield coefficients, on $V_{0}^{XX}$ and on $V_{sat}$.
Considering now only the lower branch of the polaritons, the associated hamiltonian is defined as:

\begin{equation}
    \hat{H}_{int} = \dfrac{1}{2} \sum_{\mathbf{k}, \mathbf{k}', \mathbf{q}}
    V_{\mathbf{k}, \mathbf{k}', \mathbf{q}}^{pol-pol}
    \hat{p}_{\mathbf{k} + \mathbf{q}}^{\dagger}
    \hat{p}_{\mathbf{k}' - \mathbf{q}}^{\dagger}
    \hat{p}_{\mathbf{k}} \hat{p}_{\mathbf{k}'}
\end{equation}

\paragraph{}
This nonlinear hamiltonian, as well as the one expressed in \ref{eq:4wavemixX}, describes an interaction between four polaritonic modes: the annihilation of two polaritons of wavevectors $\mathbf{k}$ and $\mathbf{k}'$ combined with the creation of two polaritons of wavevectors $\mathbf{k} + \mathbf{q}$ and $\mathbf{k}' - \mathbf{q}$.
This type of scattering is commonly referred to as \textbf{polariton four-wave mixing} and leads to the spontaneous generation of new populations of polaritons when the conservation of momentum and energy is satisfied. 
It is typically responsible for the polariton optical parametric oscillator (or OPO), analog to the nonlinear optics one \cite{Baumberg2000, Stevenson2000, Romanelli2005}.

\paragraph{}
Finally, the total hamiltonian of the lower polariton branch corresponds to the sum of the linear one and the interaction one:

\begin{equation}
    \hat{H} = \hat{H}_{lin} + \hat{H}_{int}
     = \sum_{\mathbf{k}} \bigg( \hbar \omega_{LP} (\mathbf{k})
     \hat{p}_{\mathbf{k}}^{\dagger} \hat{p}_{\mathbf{k}} 
     + \dfrac{1}{2}
     \sum_{\mathbf{k}', \mathbf{q}}
     V_{\mathbf{k}, \mathbf{k}', \mathbf{q}}^{pol-pol}
     \hat{p}_{\mathbf{k}+\mathbf{q}}^{\dagger}
     \hat{p}_{\mathbf{k}'-\mathbf{q}}^{\dagger}
     \hat{p}_{\mathbf{k}} \hat{p}_{\mathbf{k}'} \bigg)
\end{equation}

\paragraph{Energy renormalization}

In order to study the effect of nonlinearity on the polariton dispersion, let us consider two modes of the lower polariton branch $\mathbf{k}_{1}$ and $\mathbf{k}_{2}$. 
The interaction hamiltonian $\hat{H}_{int}$ of these modes is thus the sum of the two terms $\hat{p}_{\mathbf{k}_{2}}^{\dagger} \hat{p}_{\mathbf{k}_{1}}^{\dagger} \hat{p}_{\mathbf{k}_{2}} \hat{p}_{\mathbf{k}_{1}}$ and $\hat{p}_{\mathbf{k}_{1}}^{\dagger} \hat{p}_{\mathbf{k}_{2}}^{\dagger} \hat{p}_{\mathbf{k}_{1}} \hat{p}_{\mathbf{k}_{2}}$. 
But as $\hat{p}_{\mathbf{k}_{1}}$ and $\hat{p}_{\mathbf{k}_{2}}$ are eigenmodes of the system, they commute, therefore the two terms are the same.

The evolution of the $\mathbf{k}_{1}$ mode is given by the Heisenberg equation:

\begin{equation}
\label{eq:Evolmode1}
    i \hbar \dfrac{d}{dt} \hat{p}_{\mathbf{k}_{1}} = 
    \big[ \hat{p}_{\mathbf{k}_{1}}, \hat{H} \big] = 
    \big[ \hat{p}_{\mathbf{k}_{1}}, \hat{H}_{lin} \big] + 
    \big[ \hat{p}_{\mathbf{k}_{1}}, \hat{H}_{int} \big]
\end{equation}

and a complete calculation tells us that:

\begin{equation}
    \big[ \hat{p}_{\mathbf{k}_{1}}, \hat{H}_{int} \big] =
    \hbar g_{LP} \hat{p}_{\mathbf{k}_{2}}^{\dagger}
    \hat{p}_{\mathbf{k}_{2}} \hat{p}_{\mathbf{k}_{1}} = 
    \hbar g_{LP} \hat{N}_{2} \hat{p}_{\mathbf{k}_{1}}
\end{equation}

where $g_{LP} = f(V_{\mathbf{k}_{1}, \mathbf{k}_{2}, \mathbf{k}_{2} - \mathbf{k}_{1}})$ is the polariton-polariton interaction constant of the modes $\mathbf{k}_{1}$ and $\mathbf{k}_{2}$, and $\hat{N}_{2} = \hat{p}_{\mathbf{k}_{2}}^{\dagger} \hat{p}_{\mathbf{k}_{2}}$ the polariton population in the mode $\mathbf{k}_{2}$.

Now if we consider the $\mathbf{k}_{2}$ mode to be macroscopically populated, its density can be replaced by its mean field value $\langle \hat{N}_{2} \rangle$.
And if the previous equations are normalized by the quantization area $A$, we can even introduce the $\mathbf{k}_{2}$ mode density $n_{2} = \dfrac{\langle \hat{N}_{2} \rangle}{A}$, which works as an external potential for the polaritons in the $\mathbf{k}_{1}$ mode, as the equation \ref{eq:Evolmode1} becomes:

\begin{equation}
    \dfrac{d}{dt} \hat{p}_{\mathbf{k}_{1}} = -i \bigg(
    \dfrac{E_{LP}(\mathbf{k}_{1})}{\hbar} + g_{LP} n_{2}
    \bigg) \hat{p}_{\mathbf{k}_{1}}
\end{equation}

where $E_{LP}$ comes from the linear hamiltonian. We see here that the interaction of the $\mathbf{k}_{1}$ mode with a pump of wavevector $\mathbf{k}_{2}$ introduces a shift of its energy toward the blue, therefore commonly known as \textit{blue shift}:

\begin{equation}
    \Delta E_{\mathbf{k}_{1}} = \hbar g_{LP} n_{2}
\end{equation}

This shift is an effect of energy renormalization of the $\mathbf{k}_{1}$ mode by the pump mode ($\mathbf{k}_{2}$) and thus a multimode effect. It works the same way for a continuum of modes, even though in that case the contributions from all the modes have to be taken into account.
It happens as well on the pump mode itself, which then can be related to the self Kerr effect known in nonlinear optics.

\paragraph{}
This energy renormalization has to be taken into account while considering the pumped mode dynamics. 
Indeed, a typical way of creating a polariton fluid, and which will be the case in the present work, is to excite it quasi-resonantly, which means at an energy slightly blue-detuned from its resonance (the lower polariton branch in our case).
In order to evaluate the effective detuning between the pump and the polaritons, the previously described blue shift must not be forgotten, and we have:

\begin{equation}
    \Delta E_{lasLP}^{eff} = E_{las} - (E_{LP}{\mathbf{k}_{las}} - \hbar g_{LP} n_{las})
\end{equation}

where the index $las$ corresponds to the pumped mode of the laser.

\paragraph{}

This renormalization is also responsible for the \textbf{bistability} phenomenon, where for a high enough detuning between the pump and the lower polariton branch, a certain range of input intensities gives access to two stable output states.
However, we will not focus on it here as it will be discussed in more details in section \ref{sec:OpticalBist}.

\paragraph{Spin}

\paragraph{}
Until here, the description of our polariton system was entirely scalar. However, the polariton do inherit a spin degree of freedom from their excitonic part. The spin of the exciton can indeed take two possible values of their total internal momentum, $J = 1$ and $J = 2$. 
Yet only the $J=1$ excitons can couple to single photons, while $J=2$ excitons are called dark-excitons. Which means that the only possible projections of the polaritons angular momentum along the \textit{z} axis orthogonal to the cavity are $J_{z} = \pm 1$.
Polaritons and out-of-cavity photons have a one-to-one relation which allows a well defined polarization of the emitted light.

\paragraph{}
Polaritons have therefore a circular polarization. Therefore, an excitation with a linearly polarized light, as it will mainly be used in this work, creates two polariton populations with opposite circular polarization.
As a first approximation, the coupling between the different polariton polarization is negligible: the two polariton populations are blind to each other \cite{Ciuti1998a}. 
Therefore, as this work focuses on phenomena not influenced by the spin, we will neglect it in the following.

\subsubsection{Other geometries}

\paragraph{}
The samples used in the present work are all planar microcavities as described previously. However, they are not the only configuration in which the strong coupling between a microcavity photon and a quantum well exciton can take place.
This section will give a brief overview of some other geometries.

\begin{figure}[h]
    \centering
    \includegraphics[width=0.8\linewidth]{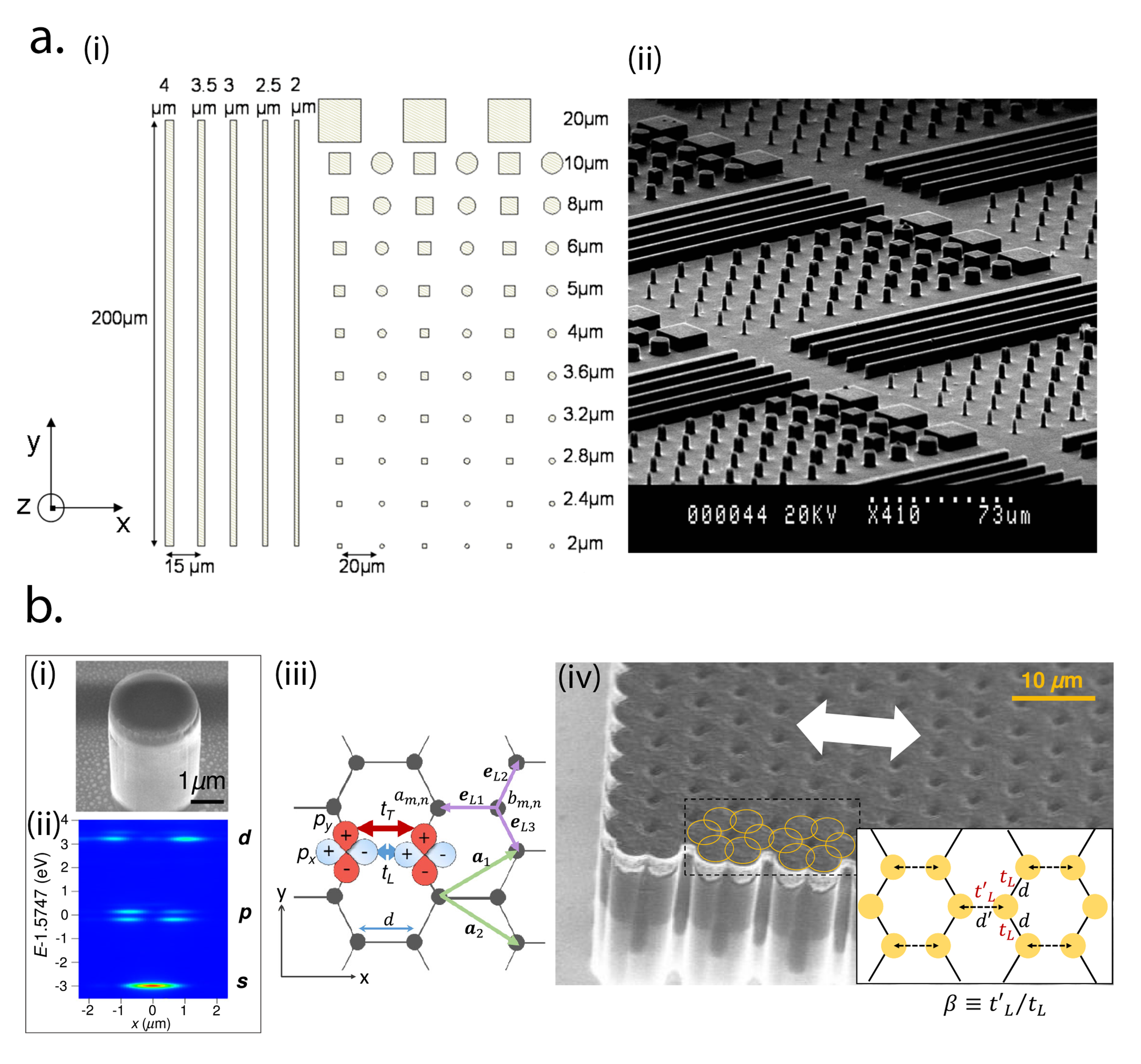}
    \caption{\textbf{Examples of other geometry microcavitites}.
    a. Pattern (i) and Scanning Electron Microscopy image (ii) of a sample combining microwires and micropillars of different shape and size. From \cite{Wertz2010}.
    b. Structure of a honeycomb polariton lattice. (i) Electron microscopy image of a single micropillar, whose discrete dispersion is plotted on (ii). The pillars can also be combined as a honeycomb lattice, as sketched on (iii) and pictured on (iv). From \cite{Milicevic2019}.}
    \label{fig:OtherGeom}
\end{figure}

\paragraph{}
Once the planar cavity is grown, it is indeed possible to chemically etch it in an order pattern, to create new confinement geometries. 
Typically, one could create microwires, as presented in figure \ref{fig:OtherGeom}.a, to get an unidimensional confinement of the excitons, and therefore study different behaviour of polariton fluids \cite{Wertz2010, Nguyen2013, Goblot2016b, Su2018}.

\paragraph{}
It is also possible to lower again the dimensionality and to etch micropillars, as presented in figure \ref{fig:OtherGeom}, which can for instance be used to study the squeezing of polaritons \cite{Boulier2014a}.
Those pillars can also be combined together, for example two-by two as polariton molecules \cite{MichaelisdeVasconcellos2011}, or as lattices, as picture in figure \ref{fig:OtherGeom}.b., typically used to observe topological phenomena \cite{Klembt2018a, Milicevic2019, Jamadi2020}.

\section{Superfluidity and condensation in polariton systems}

\paragraph{}
In the previous section, we have discovered the necessary conditions for polaritons to exist and the tools to describe them.
We can now explore the different techniques to generate them and how they influence their dynamics.

\subsection{Resonant excitation: polariton superfluid}

\paragraph{}
The most direct way to get a strong coupling in our cavity is to inject photons at the cavity resonance. 
It results in a population of polaritons that we will describe in this section.

\subsubsection{Mean-field theory}

\paragraph{}
To consider the dynamics of the polariton fluid, we now need to take into account the ensemble of particles and not just one of them.
To do so, we use the well known \textbf{mean-field approximation} \cite{Carusotto2013a}, where we do not consider the particle operators anymore.
Indeed, as the total number of particles $N$ is high, we have $N \sim N+1$: we can replace the particle quantum operators by their expectation values.
Therefore, the creation and annihilation operators $\hat{a}^{\dagger}(\mathbf{r}, t)$ and $\hat{a}(\mathbf{r}, t)$ of a single particle at the position $\mathbf{r}$ and instant $t$ can be replaced in the equations of motion by their mean field terms $\Psi (\mathbf{r}, t) = \langle \hat{a}(\mathbf{r}, t) \rangle$ and $\Psi^{\dagger} (\mathbf{r}, t) = \langle \hat{a}^{\dagger}(\mathbf{r}, t) \rangle$.
$\Psi$ represents then the wavefunction of the fluid and its modulus square the density of particles: $|\Psi(\mathbf{r}, t)|^{2} = n(\mathbf{r}, t)$.

\paragraph{}
The first description of the dynamics of a quantum gas of material particles was given in 1961 by Gross \cite{Gross1961} and Pitaevskii \cite{Pitaevskii1961} while describing quantum vortices in liquid Helium.
Adapted in our case by taking into account the Bogoliubov theory of the dilute Bose gas \cite{Carusotto2013a}, the equation describing the system can be written in the form:

\begin{equation}
\label{eq:GPE}
    i \hbar \dfrac{d}{dt} \Psi (\mathbf{r}, t) = \Big(
    - \dfrac{\hbar^{2}}{2m} \nabla^{2}_{r} + V_{ext}(\mathbf{r}) 
    + \hbar g n(\mathbf{r}, t) \Big) \Psi (\mathbf{r}, t)
\end{equation}

with $V_{ext}$ an external potential in which the particles are confined. This equation is referred to as the \textbf{Gross Pitaevskii equation} (GPE) or nonlinear Schr\"{o}dinger equation.
It states in particular that the macroscopic ensemble of particles that forms the fluid behaves collectively and that the quantum matter field $\hat{\Psi}(\mathbf{r}, t)$ acts as a classical field $\Psi (\mathbf{r}, t)$ \cite{Leggett2001, Pitaevskii2003a, Carusotto2013a}.

\paragraph{}
The previous description has been realized for a Bose gas of cold atoms in weak interaction, but it can be adapted to polariton fluids.
If the excitonic density is low, polaritons act as bosons, therefore the picture is coherent. Furthermore, their low effective mass (of the order of 10\textsuperscript{-5} times the one of the free electron) ensures a de Broglie wavelength higher than the polariton-polariton distance, at cryogenic temperatures.
Indeed, for a typical excitonic density of 40 \textmu m\textsuperscript{-1}, the mean distance between particles is about 0.1 \textmu m, while the de Broglie wavelength has the expression:

\begin{equation}
    \lambda_{dB} = \sqrt{\dfrac{\hbar^{2}}{2 m^{*}_{LP}k_{B}T}}
\end{equation}

where $k_{B}$ is the Boltzmann constant and $T$ the temperature, and reaches about 1 \textmu m for $T = 5$ K.

Finally, polaritons have coulombian interactions through their excitonic part, that we saw in section \ref{sec:2DXconfined} are weak compared to the one of an hydrogenoid atom.

\paragraph{}
However, a few differences need to be noticed. First of all, the definition of the phase of the wavefunction $\Psi(\mathbf{r}, t)$ of the system.
For a standard Bose gas, the phase is defined by the chemical potential $\mu$ of the gas.
In the case of resonantly pumped polaritons however, the thermodynamic equilibrium is never reached, therefore the phase is not intrinsically defined by the fluid but depends on the pump properties. However, it is enough for the Gross Pitaevskii theory to be applied \cite{Domb1972, Adrados2011}.

A second particularity of polaritons is the fact that they are composed of two types of bosons: the cavity photon and the quantum-well excitons. The two particles feel different external potentials, $V_{c}(\mathbf{r})$ and $V_{X}(\mathbf{r})$ respectively. 
Therefore the GPE should be written for both particles, taking into account the respective potentials and the fact that the interactions only happen between the excitons.
The full Gross-Pitaevskii theory applied to polariton fluids have been detailed by Iacopo Carusotto and Cristiano Ciuti \cite{Ciuti2005, Carusotto2013a}, and is further presented in the next section.

\subsubsection{Driven-dissipative Gross-Pitaevskii equation}

\paragraph{}
In order to apply the GPE to microcavity polariton fluids, lets us first describe the system in the exciton-photon basis by considering the associated fields $\Psi_{X}(\mathbf{r}, t)$ and $\Psi_{\gamma}(\mathbf{r}, t)$.
The system losses are represented by the two linewidths $ \gamma_{X}$ and $ \gamma_{cav}$.
The continuous pumping $F_{p}(\mathbf{r}, t)$ only applies to the photonic field while the interactions defined by $g$ happen between excitons. 
The external potentials $V_{\gamma}(\mathbf{r})$ and $V_{X}(\mathbf{r})$ come from the intrinsic dissipation phenomena of the microcavity, usually related to its defects.

Those statements lead us to the coupled equations:

\begin{multline}
\label{eq:GPEXcav}
    i \hbar \dfrac{d}{dt} 
    \begin{pmatrix}
    \Psi_{\gamma}(\mathbf{r}, t) \\ \Psi_{X}(\mathbf{r}, t)
    \end{pmatrix} = 
    \begin{pmatrix}
    \hbar F_{p}(\mathbf{r}, t) \\ 0
    \end{pmatrix} + \\
    \Bigg[ H_{lin} (\mathbf{r}) + 
    \begin{pmatrix}
    V_{\gamma}(\mathbf{r}) - i \hbar \gamma_{cav} & 0 \\
    0 & V_{X}(\mathbf{r}) - i \hbar \gamma_{X}
    + \hbar g n(\mathbf{r}, t)
    \end{pmatrix}
    \Bigg]
    \begin{pmatrix}
    \Psi_{\gamma}(\mathbf{r}, t) \\ \Psi_{X}(\mathbf{r}, t)
    \end{pmatrix}
\end{multline}

where $H_{lin}(\mathbf{r})$ is the linear hamiltonian written in the real space, \textit{i.e.} obtained by changing $\mathbf{k}$ to $-i \nabla_{\mathbf{r}}$ in $H_{lin}(\mathbf{k})$ : 

\begin{equation}
    H_{lin}(\mathbf{r}) = 
    \begin{pmatrix}
    E_{X}^{0} & \dfrac{\hbar}{2} \Omega_{R} \\
    \dfrac{\hbar}{2} \Omega_{R} & 
    E_{\gamma}(-i \nabla_{\mathbf{r}})
    \end{pmatrix}
\end{equation}

\paragraph{}
The equation \ref{eq:GPEXcav} can also be written in the polariton basis, and in particular if we focus on the lower branch, we obtain the \textbf{driven-dissipative Gross-Pitaevskii equation} for the polariton field:

\begin{equation}
\label{eq:ddGPE}
    i\hbar \dfrac{\partial}{\partial t} \Psi_{LP}(\mathbf{r}, t) = 
    \bigg( - \dfrac{\hbar^{2}}{2m_{LP}^{*}} \nabla_{\mathbf{r}}^{2}
    + V_{LP}(\mathbf{r})
    - i \hbar \gamma_{LP} +
     \hbar g_{LP} n(\mathbf{r}, t) \bigg)
    \Psi_{LP}(\mathbf{r}, t) + \hbar F_{p}(\mathbf{r}, t)
\end{equation}

with $m_{LP}^{*}$ the effective mass of the lower polariton, $V_{LP} = |X_{\mathbf{k}}|^{2} V_{X} + |C_{\mathbf{k}}|^{2} V_{\gamma}$ the external potential felt by the lower polaritons and $g_{LP} = |X_{\mathbf{k}}|^{4} g$ the interaction constant between the lower polaritons in the same modes.

\paragraph{}
Until the end of this section, the study is done only on the lower branch of the polariton. Therefore, in order to lighten the notation, the indices will be removed: $\Psi_{LP}(\mathbf{r}, t) = \Psi(\mathbf{r}, t)$, $\gamma_{LP} = \gamma$, $g_{LP} = g$ and $m_{LP}^{*} = m^{*}$.

\paragraph{}
As expected, this equation is indeed very similar to \ref{eq:GPE}, except for the loss term $-i \hbar \gamma$ and its necessary compensation with the pump term $F_{p}(\mathbf{r}, t)$.
It is indeed an important difference between our system and cold atom gases for which the theory was first written: the lifetime of our particles $\tau = \dfrac{\hbar}{2 \pi \gamma}$ is of the order of some tens of picosecond, compared to the second for atoms. 
Hence the fact that the losses need to be continuously compensated: our system is \textbf{out of equilibrium}.

\paragraph{}
Let us now focus on the mean-field stationary solutions of equation \ref{eq:ddGPE} in the homogeneous case, \textit{i.e.} for an external potential equals to zero.
We need to remember that the system is driven by the pump field $F_{p}(\mathbf{r}, t) = F_{p}(\mathbf{r}) e^{i(\mathbf{k}_{p}\mathbf{r} - \omega_{p}t)}$.
Therefore, the solutions can be written $\Psi(\mathbf{r}, t) = \Psi_{0}(\mathbf{r}) e^{i(\mathbf{k}_{p} \mathbf{r} - \omega_{p}t)}$.
It results in the mean field stationary equation:

\begin{equation}
\label{eq:statsolddGPE}
    \bigg( - \dfrac{\hbar^{2} \mathbf{k}_{p}^{2}}{2m^{*}}
    - i \hbar \gamma 
    + \hbar g n_{0}(\mathbf{r}) \bigg)
    \Psi_{0}(\mathbf{r}) + \hbar F_{p}(\mathbf{r}) = 0
\end{equation}

\paragraph{}
This equation is responsible for the \textbf{bistability} phenomenon observed in our system. As it is the starting point of most of the experimental work presented in this thesis, it will be further detailed in chapter \ref{chap:SpontChap}, and only quickly introduced here for the sake of the discussion.

\paragraph{}
The bistability indeed comes from the quasi-resonant pumping of the system. We need therefore to take into account the detuning between the pump and the lower polariton branch $\Delta E_{lasLP} = \hbar \omega_{p} - \hbar \omega_{LP}$.
Multiplying the stationary equation by its complex conjugate, we obtain the equation in intensity:

\begin{equation}
\label{eq:bistabilityIntensity}
    I(\mathbf{r}) = \bigg( (\hbar \gamma)^{2} 
    + \Big( \Delta E_{lasLP} - \hbar g n_{0}(\mathbf{r}) \Big)^{2}
    \bigg) n_{0}(\mathbf{r})
\end{equation}

with $I(\mathbf{r})$ the intracavity intensity.
Now this equation can be derived into:

\begin{equation}
    \dfrac{\partial I}{\partial n_{0}} = 3 g^{2} n_{0}^{2}
    - 4 \Delta E_{lasLP} g n_{0} + (\hbar \gamma)^{2}
    + \Delta E_{lasLP}^{2}
\end{equation}

which discriminant is $(2g)^{2} ( \Delta E_{lasLP}^{2} - 3 ( \hbar \gamma)^{2} ) $. Therefore, the previous equation can have two distinct roots if the detuning validates the condition $\Delta E_{lasLP} > \sqrt{3} \hbar \gamma$.

\begin{figure}[h]
    \centering
    \includegraphics[width=0.9\linewidth]{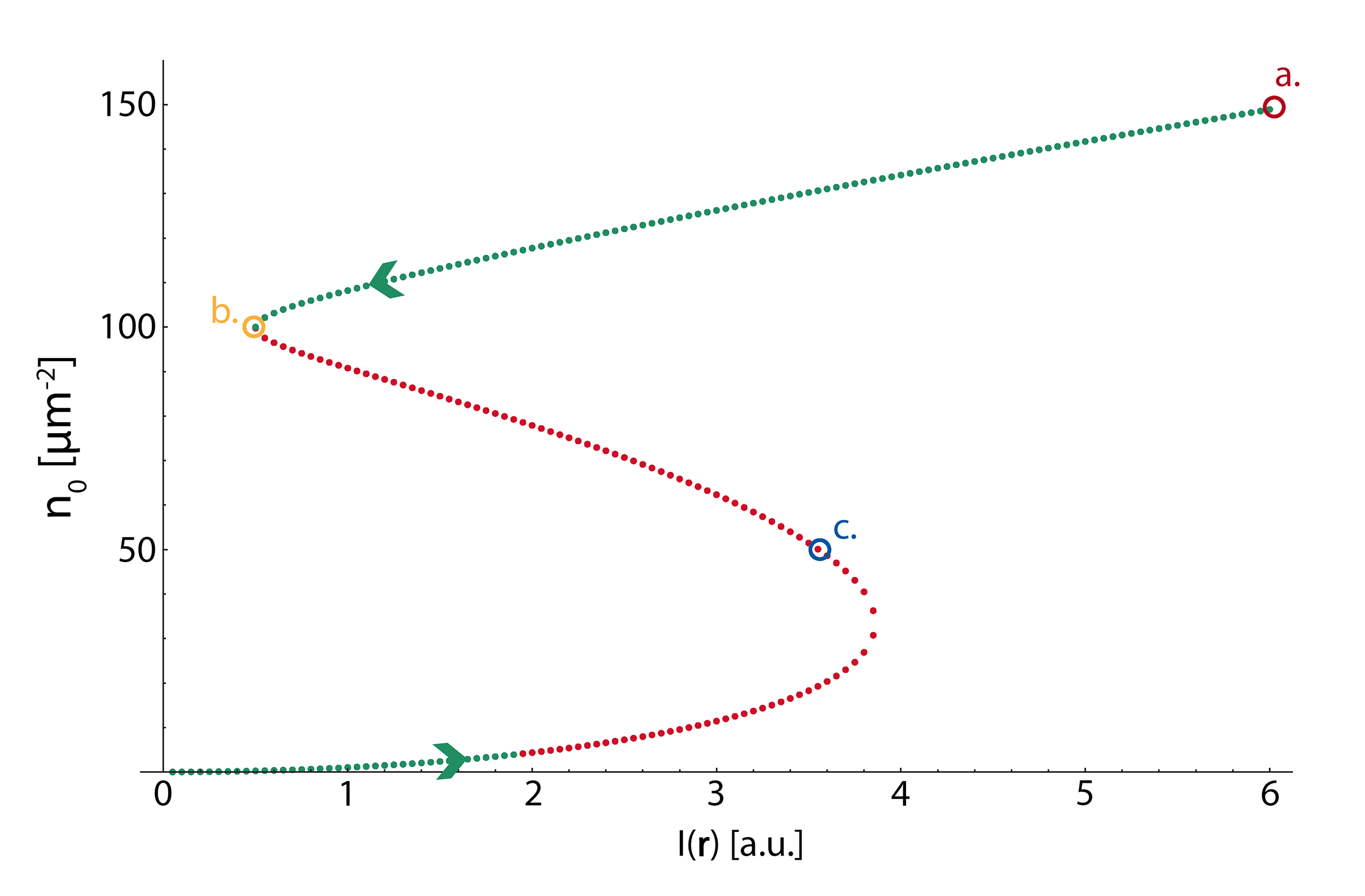}
    \caption{\textbf{Bistability}. Numerical resolution of equation \ref{eq:bistabilityIntensity} for $\hbar \gamma = 0.1$ meV and $\Delta E_{lasLP} = 1$ meV. The green region shows the stable solutions while the red one indicates the unstable ones \cite{Ciuti2005}. The density jumps from one branch to the other at the frontier between the green and red regions: we have a hysteresis cycle illustrated by the arrows. The a, b and c points corresponds to the conditions of the curves later presented in figure \ref{fig:ExcSpectrDisp}.
    Adapted from \cite{Boulier2014}}
    \label{fig:BistThom}
\end{figure}

\paragraph{}
In that case, the system presents a bistable behaviour: the curve $n_{0}(I)$ displays a range of intensities for which the system has two possible outputs, as plotted in figure \ref{fig:BistThom}.
The system is stable on the green parts of the curve, while the red region shows non-physical solutions \cite{Ciuti2005}. 
The system therefore jumps from one branch to another at the limits between the red and green regions. 
We observe a bistability cycle: within the bistable range, the upper branch is only accessible by lowering the system intensity from a higher one, while the lower branch can only be reached by increasing the intensity from a lower one, as indicated by the arrows.

\subsubsection{Landau criterion}
\label{sec:LandauCrit}

\paragraph{}
The superfluidity is a particular state of matter that was first observed in 1937 by John F. Allen, A. Don Misener and Piotr Kapitza \cite{Allen1938, Kapitza1938}. 
They independently studied liquid Helium at low temperature and observed a transition point at 2.17 K, called the \textit{lambda point}, below which the thermal conductivity of the Helium becomes very high and the viscosity abruptly drops.

A year later, Fritz London put in relation for the first time this phenomenon and the Bose Einstein condensation, earlier theorized for atomic gases, starting from the fact that the two transition temperatures are of the same order of magnitude - 2.17 K for the Helium lambda point and 3.2 K for the condensation of a Bose gas of the same density \cite{London1938}.
However, the main issue of this theory comes from the interaction between Helium atoms that are too strong to consider it as a perfect gas.
Indeed, later studies have shown that only 10\% of superfluid \textsuperscript{4}He is actually condensed \cite{Sosnick1990}.

The best explanation of the phenomenon was given by Lev Landau in 1941 \cite{Landau1941}, following an idea expressed by Tisza in 1938 \cite{Tisza1938}: he developed a model based on two fluids in interaction, a classical one and a condensed one. It can explain many of the observed behaviours in such fluids, that were not yet modeled due to the complex interaction between He atoms.
The understanding of superfluidity was completed by N. N. Bogoliubov in 1946 \cite{Bogoliubov1946}, who established a non phenomenological model of the dispersion of elementary excitations.

\paragraph{}
Landau proposed several experiments to determine if a system is superfluid and to find out its superfluid fraction, as the rotating fluid experiment, later realized by Hess and Fairbank \cite{Hess1967}, or the study of persistent currents, experimentally observed in Helium \cite{Reppy1964} and much later in atomic condensates \cite{Ryu2007}. 
However, a most used one and related to the commonly called the \textbf{Landau criterion} is based on the collective thermal excitations in a superfluid.

\paragraph{}
The Landau criterion is based on the elementary excitation spectrum that we develop here.
Let us thus consider a fluid of total mass $m$ in a Galilean reference frame \textit{R}, with an energy $E$ and a momentum $\mathbf{P}$.
In the reference frame \textit{R\textsubscript{0}} moving at a speed $\mathbf{v}$ with respect to \textit{R}, its energy $E'$ and momentum $\mathbf{P}'$ are given by:

\begin{align}
    E' &= E - \mathbf{P} \cdot \mathbf{v} + \dfrac{1}{2} m \mathbf{v}^{2} \\
    \mathbf{P}' &= \mathbf{P} - m \mathbf{v}
\end{align}

\paragraph{}
Now considering the fluid is actually flowing in a tube at a constant speed $\mathbf{v}$ and energy $E_{0}$, we add an elementary excitation of momentum $\mathbf{p}$ and energy $\epsilon (\mathbf{p})$.
The total energy of the fluid becomes $E_{0} + \epsilon (\mathbf{p})$, and in the tube reference frame, we have:

\begin{align}
    E' &= E_{0} + \epsilon(\mathbf{p}) + \mathbf{p} \cdot \mathbf{v} + \dfrac{1}{2} m \mathbf{v}^{2} \\
    \mathbf{p}' &= \mathbf{p} + m \mathbf{v}
\end{align}

Comparing these relations with the previous ones, we see that the excitation energetic contribution is $\epsilon (\mathbf{p}) + \mathbf{p} \cdot \mathbf{v}$. Now if this contribution is negative, \textit{i.e.} if $|\mathbf{v}| > \dfrac{\epsilon (\mathbf{p})}{|\mathbf{p}|}$, it means that the system looses energy by getting excited: the fluid is unstable and excitations are spontaneously generated, which leads to an increase of the viscosity.

On the contrary, below this critical velocity, the system is very stable in its ground state, as the presence of an excitation would induce an energetic increase. Therefore the excitations are strongly inhibited.
There is no viscosity in this case and the system is superfluid.
The Landau criterion is thus expressed as:

\begin{equation}
    |\mathbf{v}| < v_{c} = \min_{\mathbf{p}} \dfrac{\epsilon (\mathbf{p})}{|\mathbf{p}|}
\end{equation}

\subsubsection{Polariton superfluid}

\paragraph{}
In order to apply the previous theory to the polariton fluid case, we consider a small perturbation propagating in the fluid with a momentum $\mathbf{p} = \hbar \mathbf{k}$. 
It can be added to the polariton stationary solutions given by equation \ref{eq:statsolddGPE}, which leads to, using the same notations:

\begin{equation}
    \Psi(\mathbf{r}, t) = \bigg( \Psi_{0}(\mathbf{r})
    + A e^{i (\mathbf{kr} - \omega t)} + 
    B^{*}e^{-i(\mathbf{kr} - \omega t)} \bigg)
    e^{i(\mathbf{k}_{p}\mathbf{r} - \omega_{p}t)}
\end{equation}

with $A$ and $B^{*}$ the amplitudes of the perturbation.
The previous wavefunction can be inserted into the driven-dissipative Gross Pitaevskii equation (equation \ref{eq:ddGPE}). Then, keeping only the linear terms, and as the second order terms cancel out, we obtain the coupled equations:

\begin{align}
     A \Bigg( \dfrac{\hbar^{2} (\mathbf{k}_{p} 
     + \mathbf{k})^{2}}{2m^{*}}
    - i \hbar \gamma + 2 \hbar g n_{0}
    - \hbar (\omega_{p} + \omega) \Bigg) 
    + \hbar g n_{0} B &= 0 \\
    B \Bigg( \dfrac{\hbar^{2} (\mathbf{k}_{p} 
    - \mathbf{k})^{2}}{2m^{*}}
    + i \hbar \gamma + 2 \hbar g n_{0}
    - \hbar (\omega_{p} - \omega) \Bigg)
    + \hbar g n_{0} A &= 0
\end{align}

resulting in the new dispersion relation:

\begin{equation}
\label{eq:DispElemPert}
    \hbar \omega_{\pm} = 
    \dfrac{\hbar^{2} \mathbf{k}_{p}  \mathbf{k} }{m^{*}}
    - i \hbar \gamma
    \pm \sqrt{ \Bigg( 
    \dfrac{\hbar^{2} \big(\mathbf{k}_{p}^{2}+\mathbf{k}^{2} \big)}
    {2m^{*}} - \hbar \omega_{p} + 2 \hbar g n_{0} \Bigg)^{2} 
    + \Big( \hbar g n_{0} \Big)^{2} }
\end{equation}

\begin{figure}[h]
    \centering
    \includegraphics[width=0.9\linewidth]{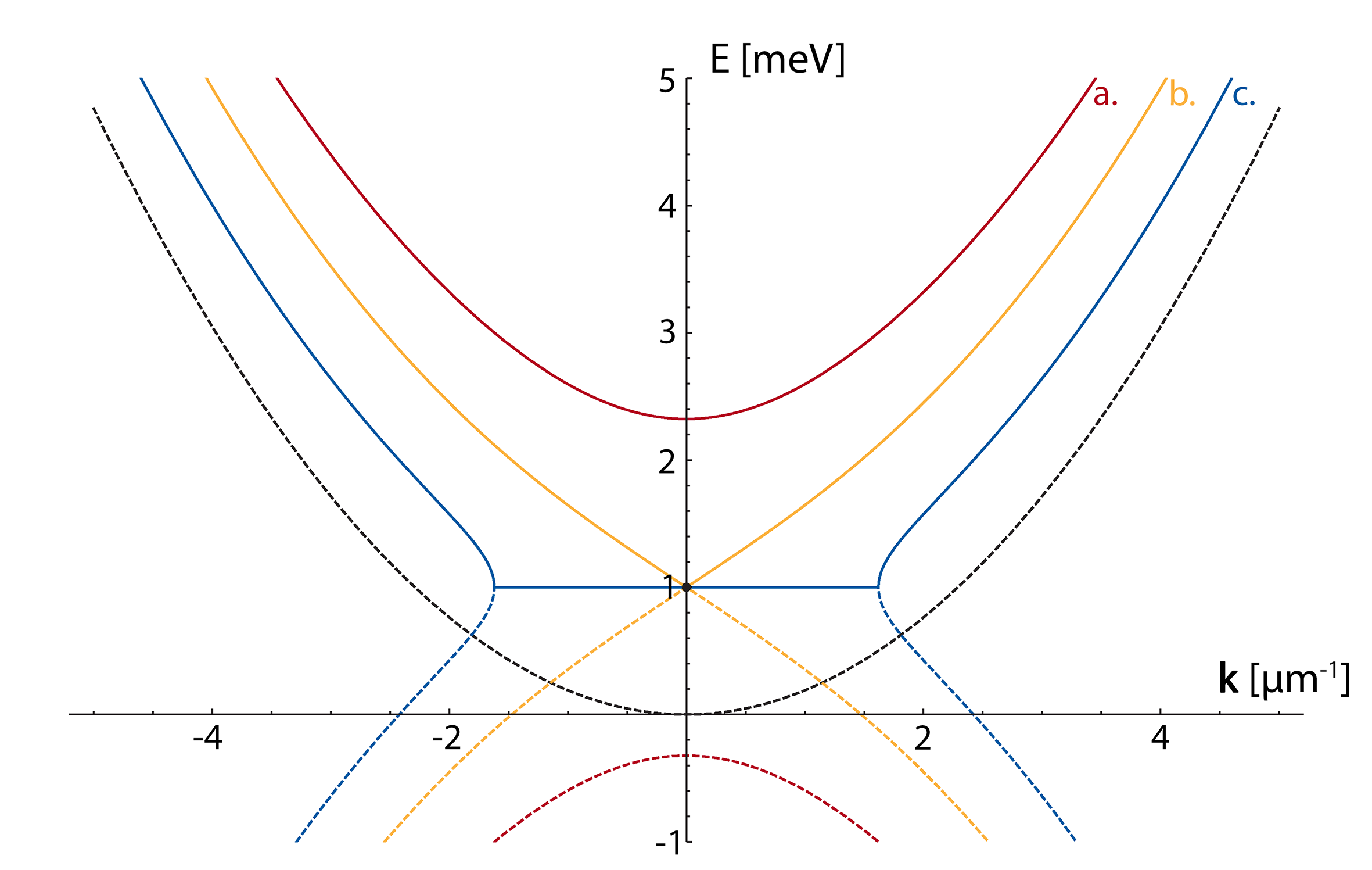}
    \caption{\textbf{Elementary excitation dispersion}. Real part of the solutions of equation \ref{eq:DispElemPert} of a static fluid ($\mathbf{k}_{p} = 0$ meV) pumped quasi-resonantly ($\Delta E_{lasLP} = 1$ meV) for different densities. The solid lines correspond to the $\omega_{+}$ solutions while the $\omega_{-}$ ones are plotted with the dashed lines. The black one indicates the simplified parabolic dispersion of non interacting polaritons, setting the energy origin.
    An energy shift appears at high density (a. in red, $\hbar gn_{0} > 2 \Delta E_{lasLP}$), while for exactly $\hbar gn_{0} = \Delta E_{lasLP}$ (case b. in yellow) we get the Bogoliubov dispersion with a linearization for the small wavevectors. Case c. in blue is unstable \cite{Ciuti2005} and corresponds to $\hbar gn_{0} < \frac{2}{3} \Delta E_{lasLP}$.
    Adapted from \cite{Hivet2013}.}
    \label{fig:ExcSpectrDisp}
\end{figure}

The real part of this equation is plotted in figure \ref{fig:ExcSpectrDisp}, for different polariton densities $n_{0}$ and for zero wavevector of the pump $\mathbf{k}_{p} = 0$ \textmu m\textsuperscript{-1}.
The solid lines illustrate the $\omega_{+}$ solutions while the $\omega_{-}$ is plotted in dashed lines.

\paragraph{}
The black dashed line shows the polariton dispersion without interaction, simplified as a parabola. We saw that different regimes appear depending on the polariton density, directly connected also to the energy detuning between the pump and the lower polariton branch $\Delta E_{lasLP} = \hbar \omega_{p} - E_{LP}(\mathbf{k}_{p})$.

\paragraph{}
Depending on the position of the system within the bistability curve presented earlier, different behaviour can be observed. The cases correspond to the ones indicated in figure \ref{fig:BistThom}.
The red curve corresponds to the highest density, with $\hbar g n_{0} > 2 \Delta E_{lasLP}$, above the bistability cycle: we observe the energy renormalization and an energy gap compared to the initial dispersion. 
The yellow curve illustrates the particular case where the renormalization and the detuning coincide: $\hbar g n_{0} = \Delta E_{lasLP}$, which happens at the turning point of the bistability cycle. We are in the particular case of the Bogoliubov spectrum, with a linear dispersion for small wavevector. Finally, when $\hbar g n_{0} < \Delta E_{lasLP}$ as for the blue curve, we observe the apparition of a flat region around $\mathbf{k} = 0$ \textmu m\textsuperscript{-1}, which corresponds to the behaviour of the system in the unstable branch of bistability.

\paragraph{}
In the interesting case where we have exactly $\hbar g n_{0} = \Delta E_{lasLP}$, equation \ref{eq:DispElemPert} can be much simplified into:

\begin{equation}
    \omega = |\mathbf{k}| \sqrt{\dfrac{\hbar g n_{0}}{m^{*}}}
\end{equation}

An analogy with the atomic condensates can then be made \cite{Pitaevskii2003a}, with the definition of a sound velocity 
\begin{equation}
    c_{s} = \sqrt{\dfrac{\hbar g n}{m^{*}}}
\end{equation}

which actually corresponds to the critical velocity of the Landau criterion defined in section \ref{sec:LandauCrit}: the system keeps superfluid properties until reaching a propagation speed $v_{f} = c_{s}$.
If the fluid speed gets above $c_{s}$, it leaves the superfluid regime to enter the Cerenkov one, where elastic scattering takes place.

\begin{figure}[H]
    \centering
    \includegraphics[width=0.9\linewidth]{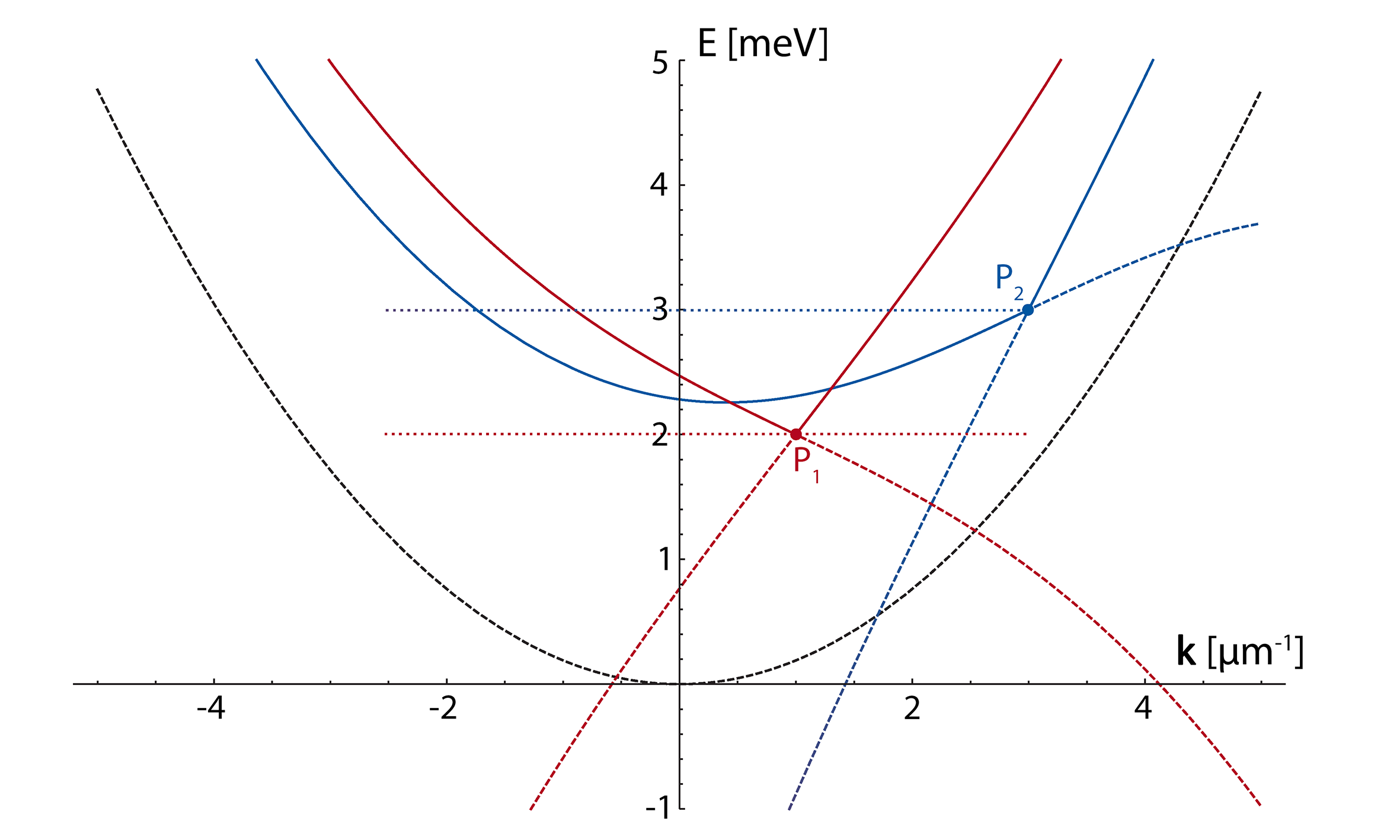}
    \caption{\textbf{Superfluid and Cerenkov regimes}. Elementary excitation spectra for a flowing fluid pumped quasi-resonantly. Again, the solid lines corresponds to the $\omega_{+}$ solutions of equation \ref{eq:DispElemPert} while the $\omega_{-}$ are plotted in dashed lines. The black one is the parabolic dispersion of the bare polaritons. The densities are set so that $\hbar g n_{0} = \Delta E_{lasLP}$. The red curve P\textsubscript{1} shows the case where $\mathbf{k}_{p} = 1$ \textmu m\textsuperscript{-1} and $\Delta E_{lasLP} = 2$ meV so that the Landau criterion is satisfied. Only one state is available: the system can not scatter and is therefore superfluid. On the contrary, the P\textsubscript{2} case plotted in blue illustrates the configuration where $\mathbf{k}_{p} = 3$ \textmu m\textsuperscript{-1} and $\Delta E_{lasLP} = 3$ meV: this time, an other state is available at the same energy to which the system can scatter, it is the Cerenkov regime.
    Adapted from \cite{Hivet2013}.}
    \label{fig:SuperfldCerenkov}
\end{figure}

\paragraph{}
Both cases dispersions are illustrated in figure \ref{fig:SuperfldCerenkov}. It corresponds again to the solutions of equation \ref{eq:DispElemPert} but for a moving polariton fluid ($\mathbf{k}_{p} \neq 0$). The $\omega_{+}$ solutions are plotted in solid lines and the $\omega_{-}$ in dashed lines, while the dashed black one is the dispersion of the non interacting lower polariton branch, to which is set the reference of energy.
The densities have been chosen so that the relation $\hbar g n_{0} = \Delta E_{lasLP}$ is satisfied.

\paragraph{}
The P\textsubscript{1} case plotted in red shows the configuration where $\mathbf{k}_{p} = 1$ \textmu m\textsuperscript{-1} and $\Delta E_{lasLP} = 2$ meV, which means that the Landau criterion is verified. 
We see that at the energy of the system, only one state is available: no scattering is therefore possible, the system is superfluid.

On the other hand, we have in blue the situation where $\mathbf{k}_{p} = 3$ \textmu m\textsuperscript{-1} and $\Delta E_{lasLP} = 3$ meV, so that the system is supersonic. In that case, a second state is available at the energy of the excitation: the system can thus scatter to it, we are in the Cerenkov regime.

\paragraph{}
The model that we considered here simplifies the polaritonic dispersion to be parabolic, which is valid for small wavevectors. However, a more complete model considering the full polariton dispersion can be found in \cite{Ciuti2005}.

\begin{figure}[H]
    \centering
    \includegraphics[width=0.9\linewidth]{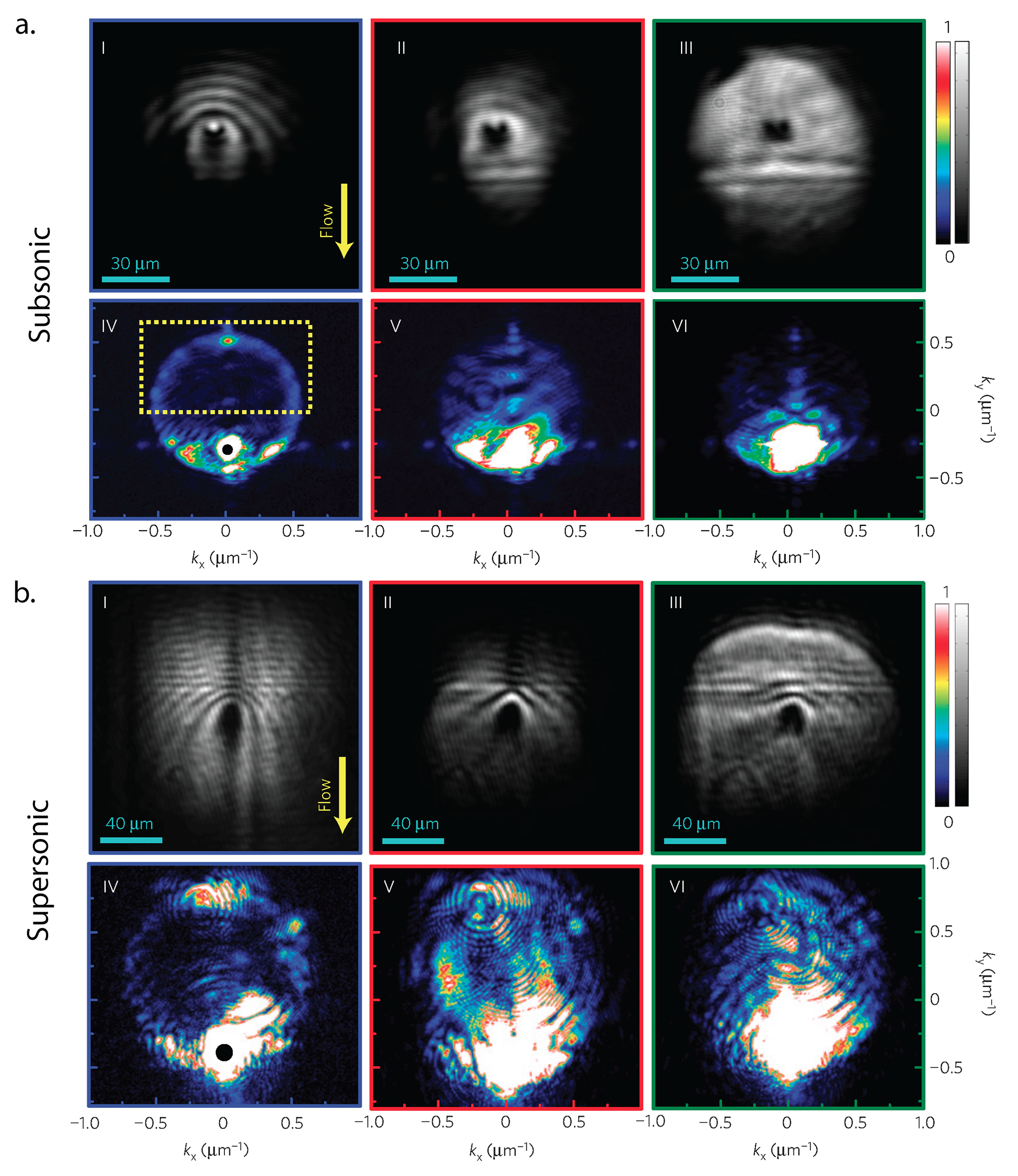}
    \caption{\textbf{Superfluidity in polariton fluid}. a. Experimental results of a polariton fluid transition to superfluidity. Top images are real space while bottom ones are the momentum space. The flow goes from top to bottom hitting a defect, the in-plane wavevector is $\mathbf{k}_{p} = -0.337$ \textmu m\textsuperscript{-1} and the detuning $\Delta E_{lasLP} = 0.10$ meV. The polariton density increases from left to right: at low density, the fluid scatters on the defect, as shown by the scattering ring in the far field and the parabolic wavefront in the density map. Those gradually vanish while the density increases, until they completely disappear in the superfluid regime on the right images.
    b. Same configuration in the supersonic regime ($\mathbf{k}_{p} = -0.521$ \textmu m\textsuperscript{-1} and $\Delta E_{lasLP} = 0.11$ meV). This time, the increase of density leads to a linearisation of the scattering fringes around the defect, also known as the Cerenkov cone.
    From \cite{Amo2009}.}
    \label{fig:SuperfldCerenkExp}
\end{figure}

\paragraph{}
These two regimes have been experimentally observed for the first time in 2009 in the group \cite{Amo2009}, and reported in figure \ref{fig:SuperfldCerenkExp}.

The experimental transition to the superfluid regime is pictured in figure \ref{fig:SuperfldCerenkExp}.a. The upper line presents the real space images and the bottom line the momentum space.

The polariton fluid is sent toward a defect, with an in-plane wavevector $\mathbf{k}_{p} = -0.337$ \textmu m\textsuperscript{-1} and a flow from top to bottom. The detuning between the laser and the lower polariton branch is $\Delta E_{lasLP} = 0.10$ meV.
The blue-circled images correspond to low excitation power, \textit{i.e.} low polariton density. The fluid elastically scatters on the defect: parabolic wavefronts are visible in real space while a ring appears in momentum space.

By increasing the excitation power (red-circled images), the fluid behaviour changes: the wavefronts start to fade away as well as the scattering ring, eventually reaching the superfluid regime in the last green-circled images: the viscosity has completely disappeared, the fluid flows around the defect without any scattering, as the single excitation spot in the momentum space confirms.

\paragraph{}
The Cerenkov effect is illustrated by the images of figure \ref{fig:SuperfldCerenkExp}.b.
Again, the upper line shows the real space experimental results and the lower one the momentum space, with a polariton flow from top to bottom hitting a defect.
The conditions are this time different: the pump wavevector is $\mathbf{k}_{p} = -0.521$ \textmu m\textsuperscript{-1} for a detuning $\Delta E _{lasLP} = 0.11$ meV, in order to be able to reach the Cerenkov regime.

\paragraph{}
As in the previous case, the excitation power increases in the images from left to right. 
At low density (left blue-circled image), one can again observe the elastic scattering of the fluid on the defect, with its characteristic parabolic fringes in real space and scattering ring in the far field.

However this time, the increase of density leads to a different behaviour of the fluid: instead of vanishing, the scattering fringes change shape: the wavefronts become linear around the defect and we have the appearance of the so-called Cerenkov-Mach cone.
At much higher density (density on the right images is 5 times higher than the one on the left), the linear shape of the fringes is maintained but widens, along with the suppression of the elastic scattering ring.

\subsection{Non-resonant excitation: polariton condensate}

\paragraph{}
The previous section described the dynamics of a polariton fluid when excited resonantly or quasi-resonantly. 
The incoming photons were therefore sent with an energy close to the one of the lower polariton branch, and the properties of the polariton fluid were inherited from the ones of the pump.

This excitation scheme is the one that was mostly used in the results of this thesis, however it is not the only way to create a polariton fluid. This section gives an overview of the non-resonant injection which can lead to a condensation of the polaritons.

\subsubsection{Non-resonant pumping}

\paragraph{}

One of the particularity of the coherent, quasi-resonant pumping of a microcavity is evidently the coherence of the generated fluid with the driving fluid, leading to the fixation of many properties of the fluid to the pump ones.
To seek for a spontaneous formation of coherence in a polariton fluid, several configurations can be implemented, as working with a large angle optical drive \cite{Deng2003} or by electrically injecting polaritons \cite{Bajoni2008, Tsintzos2008}.
However, the most common scheme is to  pump the microcavity out of resonance, \textit{i.e.} using a far blue detuned light, where the coherence of the pumping laser is lost during the relaxation process towards the lower polariton branch.
The detuning of the excitation light must match a dip of the cavity reflectivity, so that enough signal can enter the cavity and relax into the ground state polariton population.

\begin{figure}[h]
    \centering
    \includegraphics[width=0.8\linewidth]{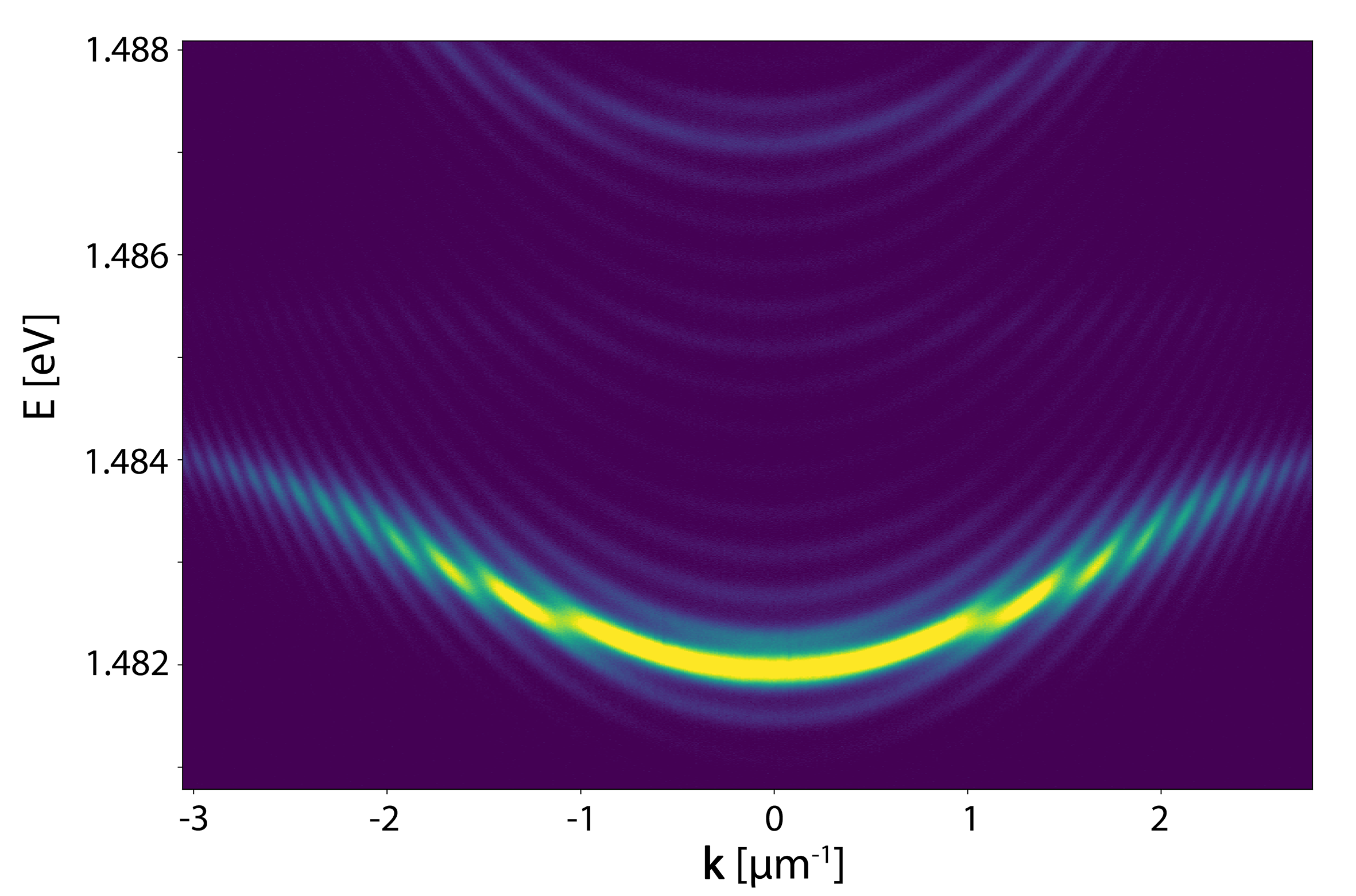}
    \caption{\textbf{Photoluminescence image of the polariton dispersions}. Both branches are visible, even though there is more signal coming from the lower branch since it is of lower energy. The replica fringes come from multireflections within the sample substrate.}
    \label{fig:DispExpPhotolum}
\end{figure}

\paragraph{}
First, the non-resonant pumping leads to a cloud of electrons and holes that thermalizes to its own temperature through exciton-exciton interactions \cite{Kavokin2007a}.
The system then reduces its kinetic energy by interacting with phonons, resulting in a relaxation along the polariton branches.
It allows to obtain a photoluminescence image of the dispersion curves, where both polariton branches are visible, as shown in figure \ref{fig:DispExpPhotolum}.
However, the signal is not homogeneously distributed and is much stronger at the bottom of the lower polariton branch of lower energy.
The replica fringes come from multiple reflections within the sample substrate.

%\paragraph{}
%Those incoherent pumping configurations typically lead to the accumulation of incoherent polariton population in the bottleneck region of the momentum space, close to the inflection point of the lower polariton branch. 
%Indeed, the additional relaxation down to the very bottom of the lower branch dispersion is slower due to the fact that the density of states of the final states is low.
%The detailed theory of these mechanisms can be found in \cite{Porras2002}.

\paragraph{}
However, for high enough polariton densities, a new relaxation phenomenon can take place, based on polariton-polariton collisions \cite{Carusotto2013a}.
Two polaritons can collide and respectively scatter to the bottom of the LP branch and to the large wavevector region, mostly excitonic and with a large density of states.
As polaritons have a bosonic statistics, as soon as the polariton population already at the bottom of the dispersion starts to have a phase-space density of order one, this relaxation scheme is enhanced by the bosonic stimulation process.
Therefore, if the stimulation overcomes the losses, an accumulation of a macroscopic coherent polariton population takes place in the final state which can be defined as a condensate, that we will further study in the next section.
The irreversibility of this process is ensured by the fact that the phase-space density of the excitonic region always stays much lower than one.

\subsubsection{Bose-Einstein condensation}

\paragraph{}
As we already saw in the previous sections, polaritons can be considered as bosons in the low density regime. 
We can therefore wonder if they possess all the characteristics of this type of particles, and in particular a very typical one known as the \textbf{Bose-Einstein condensation} and initially though for cold atoms.
We will first describe the theoretical approach of this phenomena, before looking at the polariton case.

\paragraph{}
In 1924, Satyendranath Bose proposed a theoretical description of the photons statistics which took into account their indistinguishability and the possibility to have several particles in the same state \cite{Bose1924a}.
The year later, Albert Einstein generalized its statistics to bosons in general, \textit{i.e.} to all particles with an integer spin, as what is now known as the Bose-Einstein distribution and can be written as:

\begin{equation}
    n(\mathbf{k}, T, \mu) = \dfrac{1}
    {exp \bigg(\dfrac{E(\mathbf{k}) - \mu}{k_{B} T} \bigg) - 1}
\end{equation}

where $\mathbf{k}$ is the particle wavevector, $E(\mathbf{k})$ their dispersion function, $k_{B}$ the Boltzmann constant, $T$ the temperature and $\mu$ the chemical potential, which corresponds to the amount of energy needed to add a particle to the system.
In particular, Einstein predicted that at low temperature and at thermodynamic equilibrium, non-interacting identical bosons should all condense in the same state and become indistinguishable: it is the Bose-Einstein Condensation.

\paragraph{}
At low temperature, the particles can not be considered classic anymore, and they are described by their wavepacket, characterized by the thermal de Broglie wavelength $\lambda_{dB}$:

\begin{equation}
    \lambda_{dB} = \sqrt{\dfrac{2 \pi \hbar^{2}}{m k_{B}T}}
\end{equation}

with $m$ the particle mass.
If the temperature is sufficiently low so that the de Broglie wavelength becomes of the same order of magnitude than the interparticle distance $d$, the wavepackets of the particle start to overlap. A part of the gas can therefore be described through a macroscopic wavefunction, which is equivalent to consider that a fraction of it is condensed and the other part not. 
However, at zero temperature, all of the wavepackets merged and the gas is described by a single macroscopic wavefunction, it is a pure Bose Einstein condensate.
The first experimental observation of such a condensate has been realized by M. H. Anderson \textit{et al.} \cite{Anderson1995a} in 1995 using ultracold Rubidium atoms. They managed to reach temperatures of a few hundreds nanokelvins, and the atomic gas was diluted enough so that the interaction could be neglected.

\paragraph{}
However, the full description of the condensate dynamics must consider the particles interaction. 
To do so, the mean field approximation can also be used and the global wavefunction of a condensate of $N$ atoms be written $\hat{\Phi}(\mathbf{x}_{1}, \mathbf{x}_{2}, ..., \mathbf{x}_{N}) = \prod_{N} \hat{\Psi}_{i} (\mathbf{x}_{i})$, where $\hat{\Psi}_{i}$ is the individual particle wavefunction.
It is based on this approximation that Lev P. Pitaevskii and Eugene Gross established their famous \textit{Gross-Pitaevskii equation}, initially developed for such atomic gases:

\begin{equation}
    i \hbar \dfrac{\partial \hat{\Psi}(\mathbf{x})}{\partial t}
    = \bigg( - \dfrac{\hbar^{2}}{2m} \nabla^{2} + V(\mathbf{x})
    + \hbar g |\hat{\Psi}(\mathbf{x})|^{2} \bigg) \hat{\Psi}(\mathbf{x})
\end{equation}

with $m$ the particle mass, $V$ an external potential and $g$ the coupling constant. $g$ is proportional to the diffusion length between particles $a_{s}$ such that $g = 4 \pi \hbar a_{s}/m$.
The minimization of the energy leads to the definition of a chemical potential $\mu = \partial E / \partial N = \hbar gN$ in the homogeneous case.

\subsubsection{Polariton condensates}

\paragraph{}
As we saw in the previous sections, exciton-polaritons can be considered as bosons in the low density regime. One can thus expect to observe a phase transition similar to the Bose-Einstein condensation in polariton fluid as well. Imamoglu and Ram were the first to suggest in 1996 \cite{Imamoglu1996} that the bosonic properties of the polaritons could lead to a condensate state emitting a coherent laser light. They described a coherent population of the ground state coming from an incoherent excitonic reservoir; that could be interpreted as a phase transition, in that case similar to the Bose-Einstein condensation, or as a lasing from bosonic stimulated scattering.

\paragraph{}
The first observation of polariton population with spontaneous coherence was realized in 2000 \cite{Savvidis2000, Baumberg2000}, using a quasi-resonant scheme, which prevents it to be considered a polariton laser. Above a certain threshold intensity, a phenomenon similar to an Optical Parametric Oscillator takes place \cite{Ciuti2000, Ciuti2001, Whittaker2001}, leading to the appearance of signal and idler modes with long-range coherence in both space and time \cite{Baas2006}. 
This kind of polariton population can be seen as a non-equilibrium BEC \cite{Carusotto2005}, as its coherence is not directly inherited from the pump but comes from a spontaneous symmetry breaking.

\begin{figure}[h]
    \centering
    \includegraphics[width=0.8\linewidth]{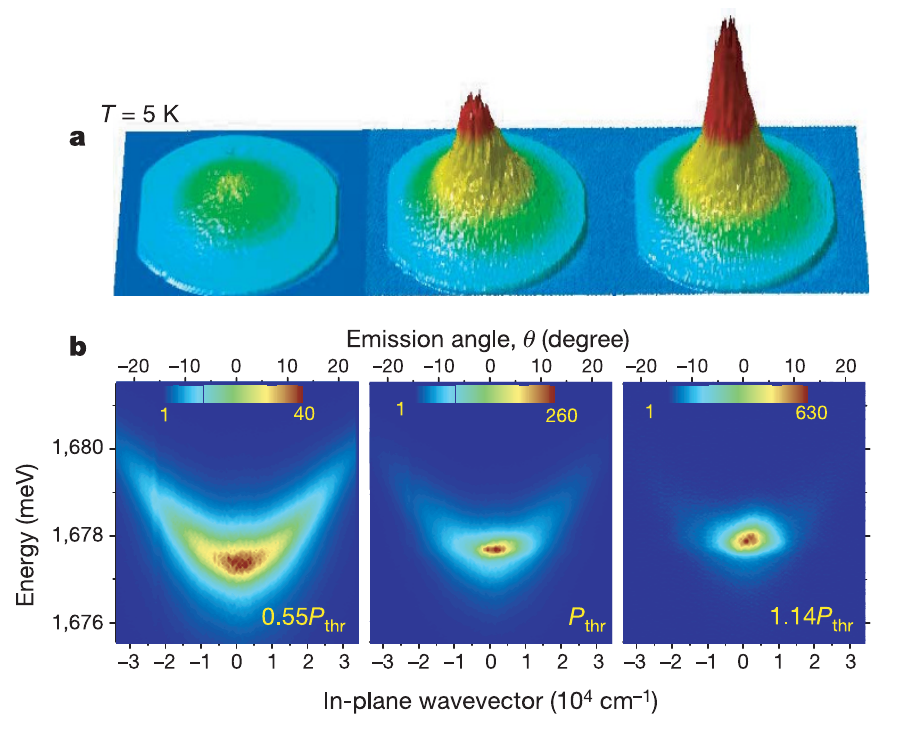}
    \caption{\textbf{Polariton Bose-Einstein Condensation}. Real space density (a.) and dispersion (b.) of a polariton BEC, obtained by increasing the intensity of the incoherent pump from left to right. A long-range coherence is reached in the real space while the momentum distribution shows an accumulation in the lower energy state. From \cite{Kasprzak2006b}. }
    \label{fig:PolaritonsBEC}
\end{figure}

\paragraph{}
The actual equivalent for Bose-Einstein condensate in polariton has however been realized under non-resonant pumping in 2006 by Jacek Kasprzak \textit{et al.} \cite{Kasprzak2006b}. An important advantage of exciton-polaritons is their low effective mass, several order of magnitude smaller than the exciton mass, which allows condensation at much higher temperatures than for the atoms (a few Kelvins while the atoms only condense at the nanokelvin scale). 
By exciting the system with a sufficiently high power, the incoherently injected polaritons reach a higher density than the critical one for Bose Einstein condensation, which results in the creation of an actual coherent polariton condensate, as presented in figure \ref{fig:PolaritonsBEC}.
The long range coherence is reached in the real space while the momentum space shows an accumulation of the population in the lower energy state.

%\bibliographystyle{unsrt}
%\bibliography{bibs/LKB-bibs-bib_thesis-PolaritonsChap}

%\end{document}
%\documentclass[a4paper,11pt]{book}
%\usepackage[utf8]{inputenc}
%\usepackage[T1]{fontenc} 
%\usepackage{lmodern} 
%\usepackage[margin=28mm,includeheadfoot,bindingoffset=5mm]{geometry}[2010/03/13]

%\usepackage{graphicx}
%\usepackage{amsmath}
%\usepackage{bbold}
%\usepackage{amssymb} % pour le signe \lesssim
%\usepackage{textcomp} % \textdegree
%\usepackage[most]{tcolorbox} 
%\usepackage{enumitem} 
%\usepackage{xcolor}
%\usepackage{physics} 
%\usepackage{float} 
%\usepackage{wasysym} 
%\usepackage{tikz}
%\usepackage{hyperref}
%\usepackage{textgreek}
%\newcommand*\circled[1]{\tikz[baseline=(char.base)]{
%          \node[shape=circle,draw,inner sep=2pt] (char) {#1};}}
%\renewcommand{\thesubsubsection}{\roman{subsubsection}}

%\tcbset{enhanced,colback=red!5!white, colframe=red!75!black,fonttitle=\bfseries}
%\graphicspath{{figures/}} %Setting the graphicspath
%\setcounter{tocdepth}{3}
%\setcounter{secnumdepth}{3}

%\begin{document}

%\tableofcontents
%\setcounter{chapter}{1}

\chapter{Experimental tools}
\label{chap:Setup}

\paragraph{}
The previous chapter detailed the theory of our system, where we established our tools to describe the polariton evolution.
Before discussing the experimental results in the following chapters, we will report here the tools we used in the lab to implement our experiment.

\paragraph{}
The first section focuses on the excitation part of the experiment: after a detailed characterization of the sample used in this work, we will describe the two cryostats we have in the lab and the laser source we used, as well as the Spatial Light Modulator or SLM, a device shaping the wavefront of the light beam which was very useful for the following results.

\paragraph{}
On the second part of the chapter, we will discover which detection tools were used to retrieve the system information. The main devices will be described, the CCD camera and the spectrometer, and an explanation of the data analysis will be given.

\newpage

\section{Excitation}

\subsection{Microcavity sample}

\paragraph{Fabrication}
The sample used for this work is a planar microcavity realized at the Ecole Polytechnique F\'{e}d\'{e}rale de Lausanne (EPFL) by Romuald Houdr\'{e}. This semiconductor sample is very robust and has been used in the group since many years. 

It has been fabricated by \textbf{Molecular Beam Epitaxy (MBE)}: this technology is used to grow ultrathin films with a very good control of their thickness and composition. Realized under ultra high vacuum, target materials are heated to their sublimation points and produce molecular beams. They can then chemically react with other species before condensing as a layer on the single crystal substrate. This technique has a nanometer scale precision as the atomic layers can be grown one by one \cite{Lakhtakia2013}. 

\begin{figure}
    \centering
    \includegraphics[width=\linewidth]{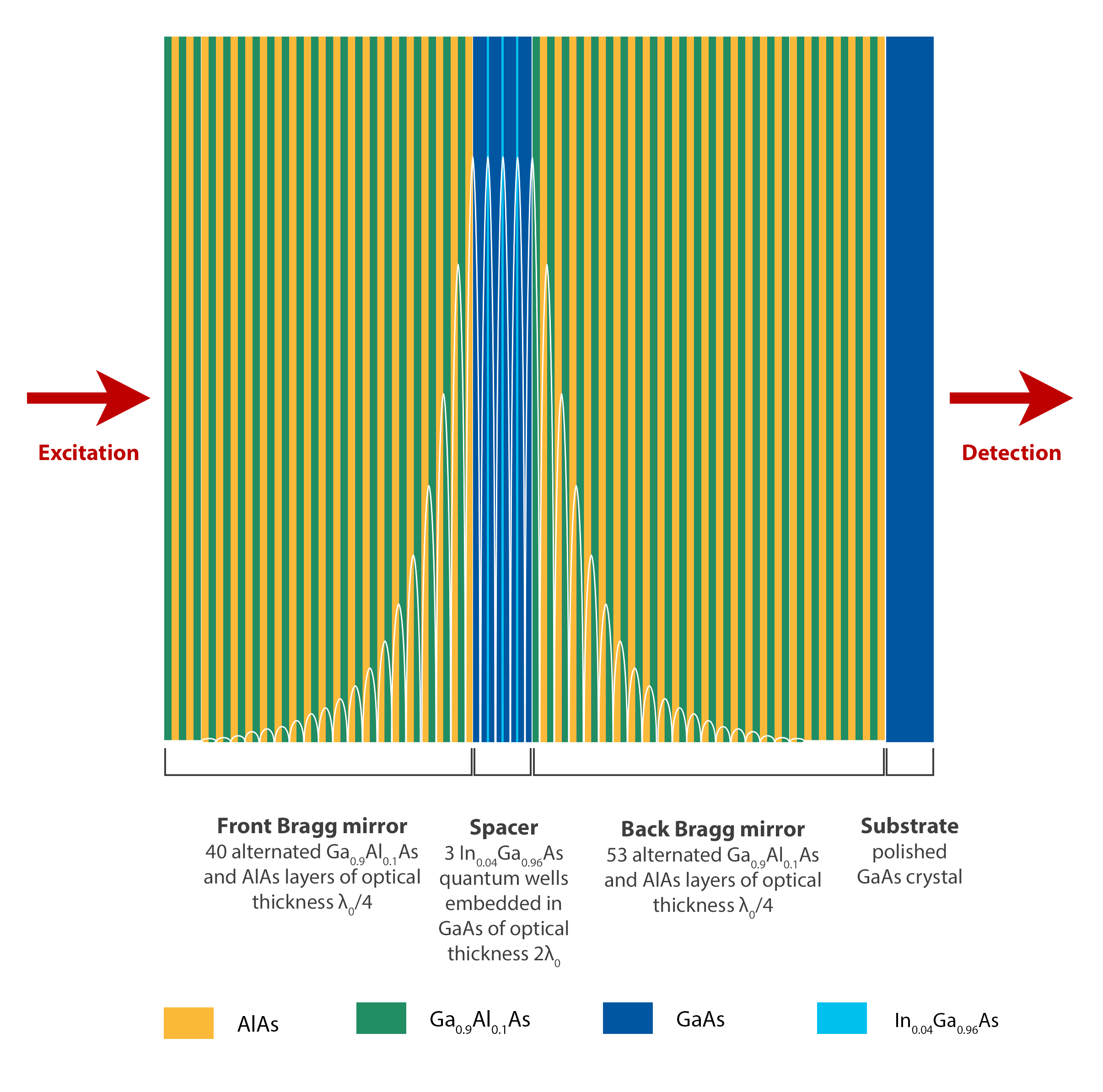}
    \caption{\textbf{Semiconductor microcavity} Two Bragg mirrors of alternated layers of Ga\textsubscript{0.9}Al\textsubscript{0.1}As and AlAs of optical thickness $\lambda_{0}/4$ (40 and 53 layers respectively), surrounding a spacer of GaAs of optical thickness $2\lambda_{0}$, in which three quantum wells of In\textsubscript{0.04}Ga\textsubscript{0.96}As are embedded at the antinodes of the electromagnetic field, which amplitude is presented with the white line. The cavity is grown on a substrate of GaAs, previously polished to allow working in transmission. The excitation takes place on the thin mirror side, while the detection is done on the substrate side.}
    \label{fig:cavity_layers}
\end{figure}

\paragraph{}
In our case, the sample is a stack of layers of four different compounds:
Ga\textsubscript{0.9}Al\textsubscript{0.1}As, AlAs, GaAs and In\textsubscript{0.04}Ga\textsubscript{0.96}As (see figure \ref{fig:cavity_layers}).
The substrate is a thick layer of GaAs, polished in order to allow working in transmission. Then the
first Distributed Bragg Mirror (DBR) is grown: it is a pile of 53 alternate Ga\textsubscript{0.9}Al\textsubscript{0.1}As and AlAs layers, each of them being $\lambda_{0}/4n_{i}$ thick, with $\lambda_{0}$ the wavelength of the excitonic resonance ($\lambda_0 = 836$ nm) and $n_{i}$ the respective refractive index of the considered compound ($n_{Ga_{0.9}Al_{0.1}As}=3.48$ and $n_{AlAs}=2.95$). This ensures that all layers have the same optical thickness, leading to constructive interferences in reflection and destructive ones in transmission.

The intracavity spacer is composed of GaAs, thick of $2\lambda_{0}/n_{GaAs}$, in which are embedded three quantum wells of In\textsubscript{0.04}Ga\textsubscript{0.96}As at the antinodes of the electromagnetic field. The thickness of the cavity defines the resonance energy of the cavity photons, and therefore the detuning between the excitonic resonance and the photonic one. In order to be able to tune it experimentally, a small wedge is introduced between the two mirrors, in the order of $10^{-6}$ radians. This way, the photonic resonance scales linearly with the position on the sample, with an energy gradient of 710 $\mu$eV.mm\textsuperscript{-1} leading to detunings from +8 meV to -4 meV.

Finally, the front mirror is grown, similar to the back one except that it only contains 40 alternated dielectric layers.

{}

\paragraph{Reflectivity and finesse}

The reflectivity of a DBR is defined with the formula \cite{Sheppard1995}:
\begin{equation}
    R = \frac{n_{0}(n_{H})^{2N} - n_{1}(n_{L})^{2N}}{n_{0}(n_{H})^{2N} + n_{1}(n_{L})^{2N}}
\end{equation}
where 
\begin{itemize}
    \item $n_{0}$ is the refractive index of the medium before the mirror; for us, the inside of the cavity in GaAs: $n_0 = n_{GaAs} = 3.54$
    \item $n_{1}$ is the refractive index of the medium after the mirror; i.e. the air for the front mirror ($n_1 = 1$) and the GaAs substrate for the back mirror ($n_1 = n_{GaAs} = 3.54$)
    \item $n_H$ and $n_L$ the refractive indices of the two media constituting the DBR, with $n_L<n_H$. For our sample, $n_L = n_{AlAs} = 2.95$ and $n_H = n_{Ga_{0.9}Al_{0.1}As} = 3.48$
    \item $N$ is the number of pairs of alternated layers in the DBR: 21 for the front mirror, 24 for the back one
\end{itemize}

This leads to a reflectivity of 99,85\% for the front mirror and 99,93\% for the back one. From that we can extract the finesse of the cavity \cite{Kavokin2007a}:
\begin{equation}
    F = \frac{\pi \sqrt{R}}{1-R}
\end{equation}
with $R = \sqrt{R_{front}R_{back}}$ the total power reflectivity. For our sample, the finesse is close to 3000.

\paragraph{Anticrossing}

\begin{figure}[h]
    \centering
    \includegraphics[width=0.8\linewidth]{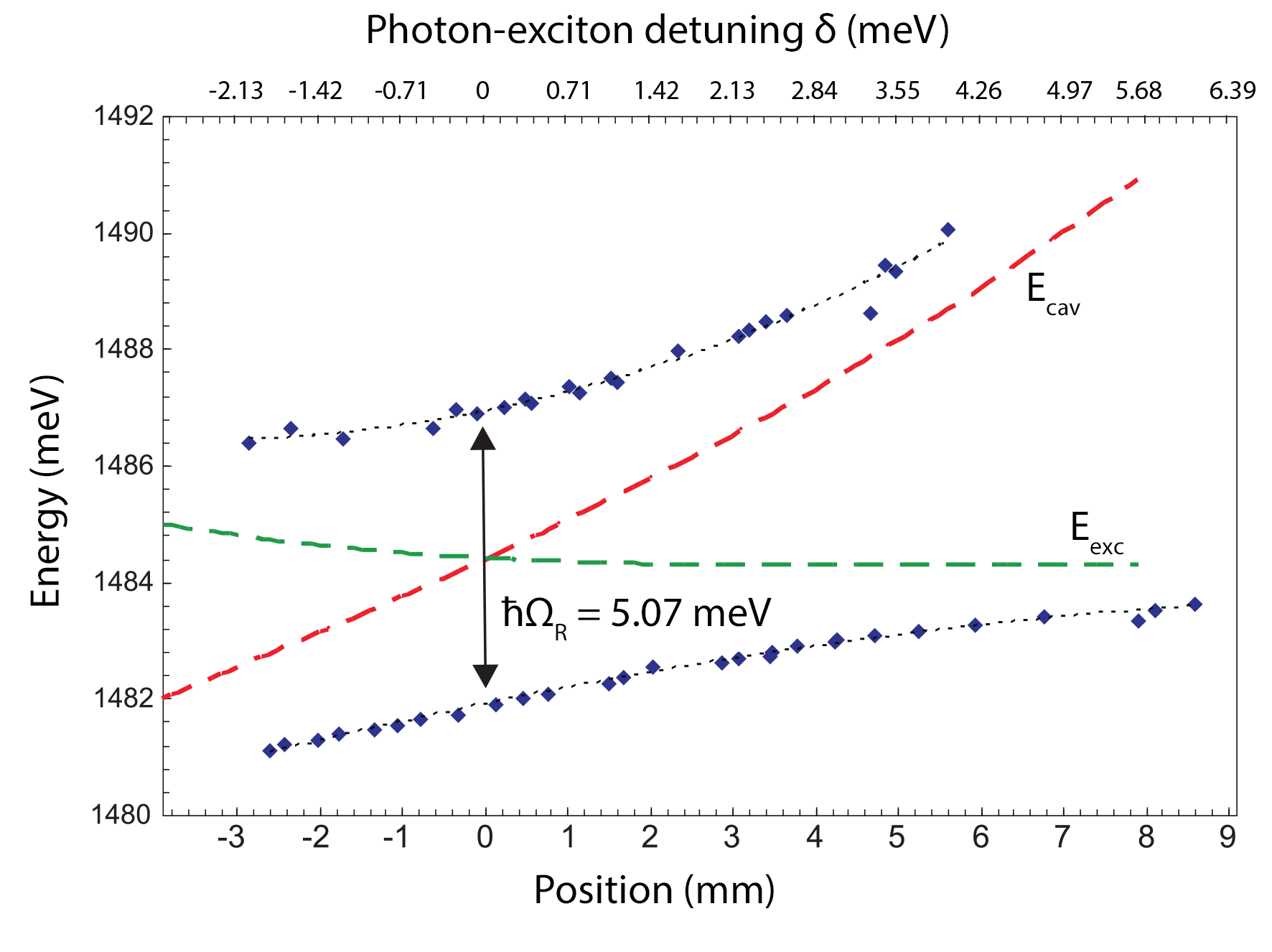}
    \caption{\textbf{Experimental anticrossing}. The blue dots represent the upper and lower polariton branches, obtained by non resonant pumping at \textbf{k} = 0 $\mu$m\textsuperscript{-1}. Changing the position changes the photon-exciton detuning $\Delta E_{Xcav}$ (horizontal axis); the exciton and photon energies are plotted in green and red dashed line, respectively. The experimental value for the Rabi splitting is extracted at zero detuning: $\hbar \Omega_R = 5.07$ meV. From \cite{Hivet2013}}
    \label{fig:exp_anticrossing}
\end{figure}

The measurement of  the anticrossing of our cavity is presented in figure \ref{fig:exp_anticrossing}. By pumping the sample out of resonance and at \textbf{k} = 0 $\mu$m\textsuperscript{-1}, the energies of the upper and lower polariton branches can be extracted for several positions on the sample. This leads to the reconstruction of the photonic and excitonic resonances, in red and green dashed lines in figure \ref{fig:exp_anticrossing}. The Rabi energy corresponds to the difference between the upper and lower branches for an exciton-photon detuning equals to zero: $E_R = \hbar \omega_R = 5.07$ meV for this sample.

\paragraph{Sample inhomogeneities}

\begin{figure}[h]
    \centering
    \includegraphics[width=0.9\linewidth]{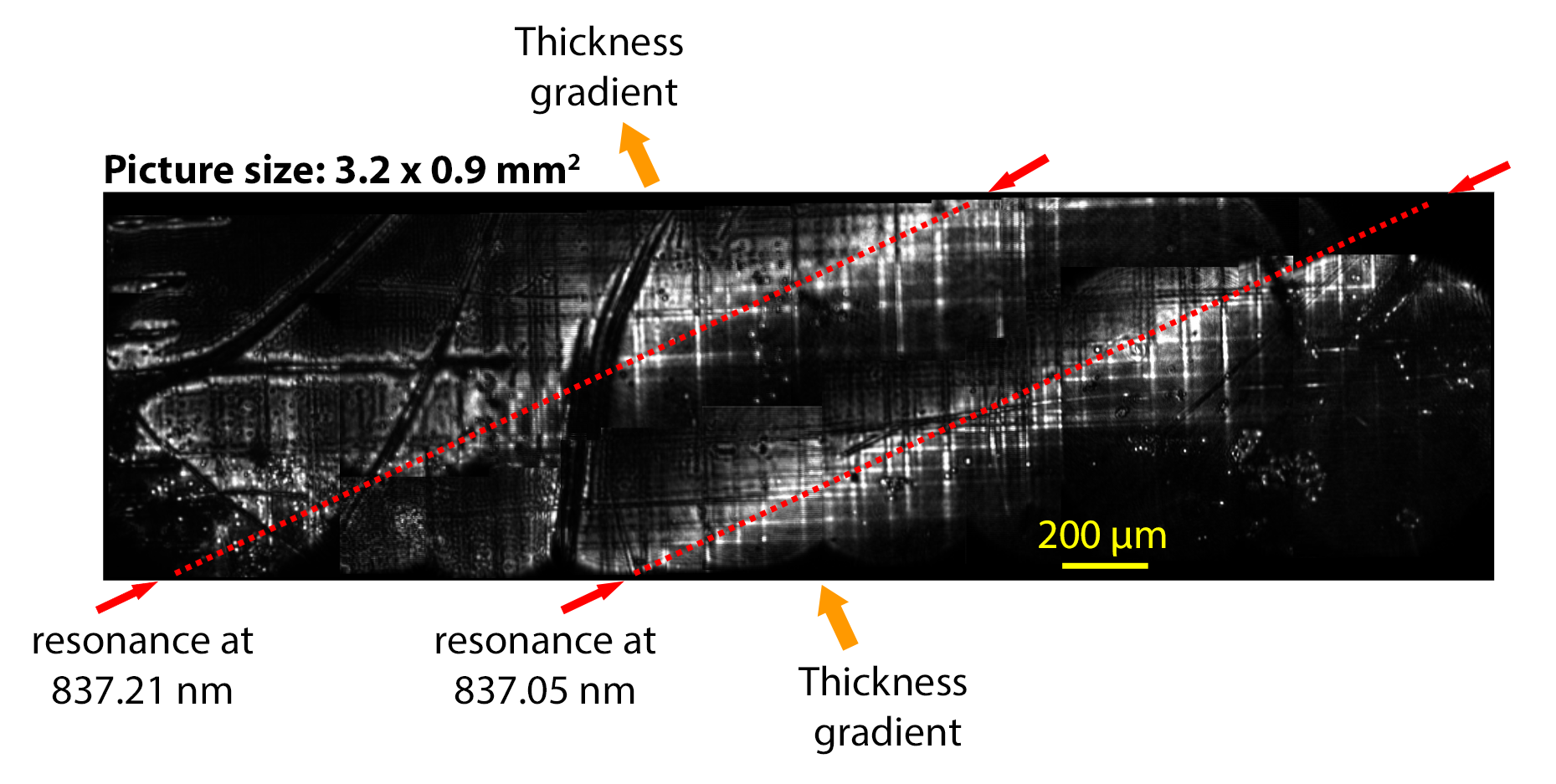}
    \caption{\textbf{Cavity transmission}. Picture of a sample area of 3.2x0.9 mm\textsuperscript{2}, excited at two different wavelengths (837.05 and 837.21 nm). The resonances are clearly visible orthogonal to the wedge, as well as the different types of defects of the cavity: the regular pattern of mosaicity, some elongated dislocations and point-like defects. From \cite{Adrados2011}}
    \label{fig:cav_transmission}
\end{figure}

Figure \ref{fig:cav_transmission} presents a picture of a large area of the sample, taken from \cite{Adrados2011}. Reconstructed from several images, this picture was realized at normal excitation with two different wavelengths of the laser, 837.21 and 837.05 nm. Both resonances are clearly visible on the picture as bright lines: they indicate regions of the cavity with the same effective length, orthogonal to the direction of the thickness gradient.

Different types of irregularities are showing up in figure \ref{fig:cav_transmission}. 
First, a thin regular pattern of an orthogonal lattice: this indicates the mosaicity phenomenon, due to mechanical constraints of the crystal \cite{Holy1993, Abbarchi2012}. 
Some irregular defects are also visible, elongated and thick, mainly on the left side of the picture: those dislocations can be attributed to mechanical shocks on the sample \cite{Marchetti2006, Kavokin2007a, Liew2009}. 
Finally, point-like defects are randomly distributed over the sample. They can be bright or dark as they correspond to potential barriers or wells: a barrier shifts the resonance locally, preventing the nearby polaritons to excite the states of the defect, while a well traps the polaritons, leading to a stronger emission \cite{Hegarty1982, Haacke1997}.

\paragraph{}
The sample is glued on the holder using silver paste. The sample holder is a plate of copper, to insure good thermal conductivity, with a hole in its center to allow working in transmission. 

\subsection{Cryostats}

We have in the lab two different cryostats that will be described separately in this section.

\subsubsection{Oxford Instrument Microstat}
\begin{figure}[h]
    \centering
    \includegraphics[width=\linewidth]{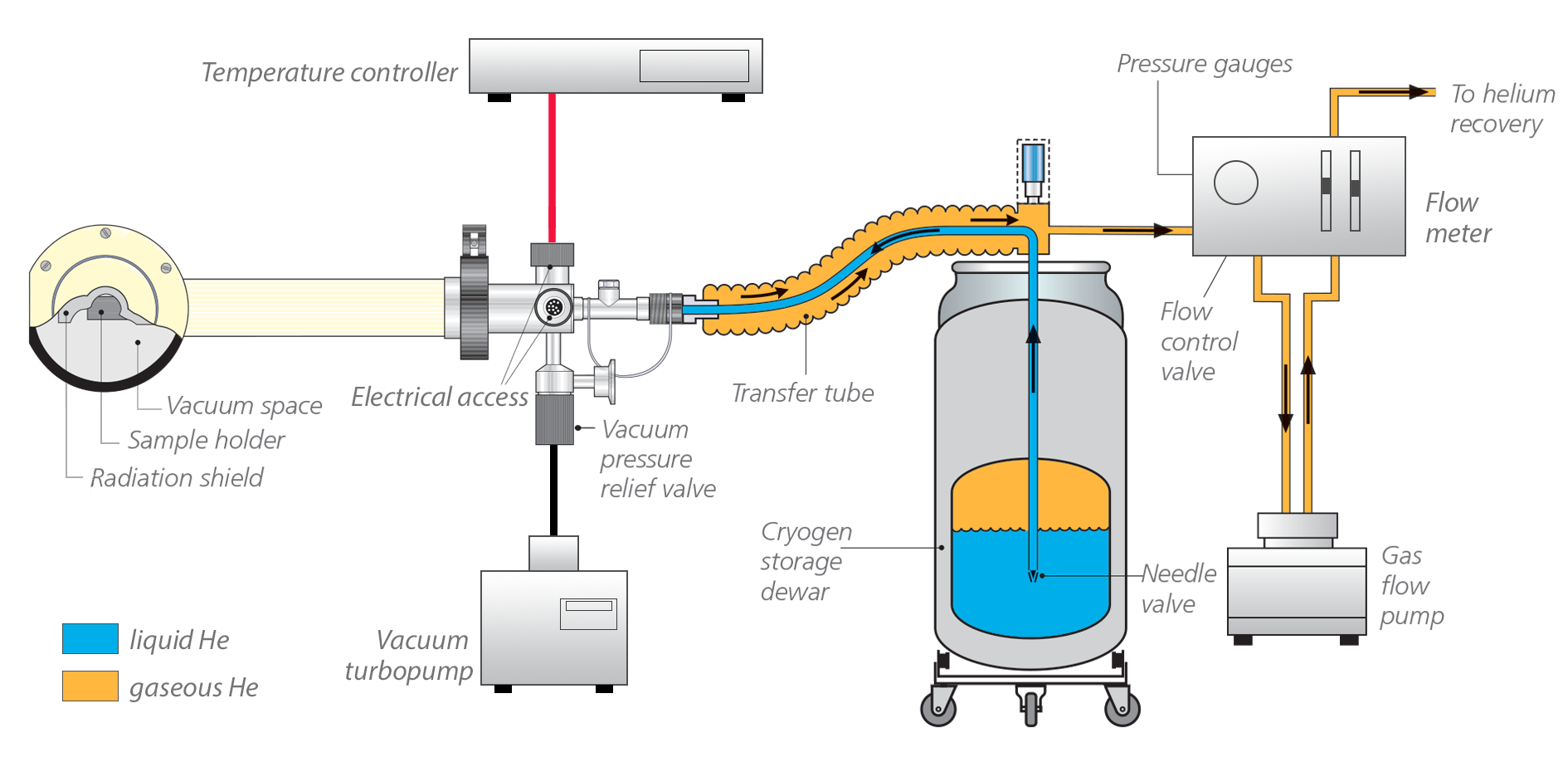}
    \caption{\textbf{Cryogenic setup}. The liquid Helium stored in the Dewar tank is dragged by the pump along the transfer tube. The flow goes along the cryostat, in direct contact with the cold finger and consequently with the sample. The heated Helium flow is then pumped to the recovery network of the university. The flow can be controlled with a flow meter, adjusting the sample temperature. This one is displayed on the temperature controller, connected to a sensor on the sample holder. Adapted from \cite{OxfordInstr2017}}
    \label{fig:oldcryo_setup}
\end{figure}

The Oxford Instrument cryostat was used for most of the results presented in this thesis. The whole cryogenic setup is sketched in figure \ref{fig:oldcryo_setup}. 

The cryostat has two circular windows on both sides of the cold finger end, on which the sample holder is screwed. The vacuum inside the cryo is done using a primary pump and a turbopump and reaches 10\textsuperscript{-4} mbar. As the cryostat is quite old and has a few leaks around the cold finger, we usually keep the pump running while doing experiment, even though it induces small vibrations.

The cooling process is realized with a continuous Helium flow. The liquid Helium is stored in a Dewar tank, connected to the cryostat with a transfer tube. This one is inserted inside the cold finger of the cryostat, and plugged to the university Helium network through a gas flow pump (see figure \ref{fig:oldcryo_setup}). The Helium cool down the cold finger and thus the attached sample. Its temperature is displayed using a sensor on the sample holder, and can be adjusted by tuning the Helium flow. We can cool the sample down to 3.8 K, but we usually work a bit higher, around 4.5 K, to not use to much Helium.

This system works well, but has the drawback of a high Helium consumption. Fortunately, the university has a Helium recovery network and its own liquefier, which ensure us an easy supply. However, we still do not let it run constantly, which limits the measurements to one-day experiments.

\subsubsection{MyCryoFirm Optidry}

A second cryostat has been recently installed in the group and has been implemented in parallel of the reported work.
This closed circuit cryostat is an Optidry from MyCryoFirm: this company offers cryostats with great flexibility to match the customer needs. 

The vacuum of the whole system is again ensured by a turbopump, preceded by a primary pump, and can also reach 10\textsuperscript{-4} mbar. The vacuum is well conserved in this cryostat so we do not keep the pump running when the vacuum is reached.

The MyCryoFirm products are working in closed circuit, i.e. they do not need Helium refueling but are connected to a compressor which ensures the Helium re-compression. The cooling of the compressor is managed by a heat exchanger, itself connected to the cold water network of the university. 

This system leads to a great stability of the cryostat: once the target temperature is reached, it can stay on for months with very small fluctuations of the conditions. It is much more convenient for a continuity of the experiment from one day to the next.

\begin{figure}[h]
    \centering
    \includegraphics[width=0.9\linewidth]{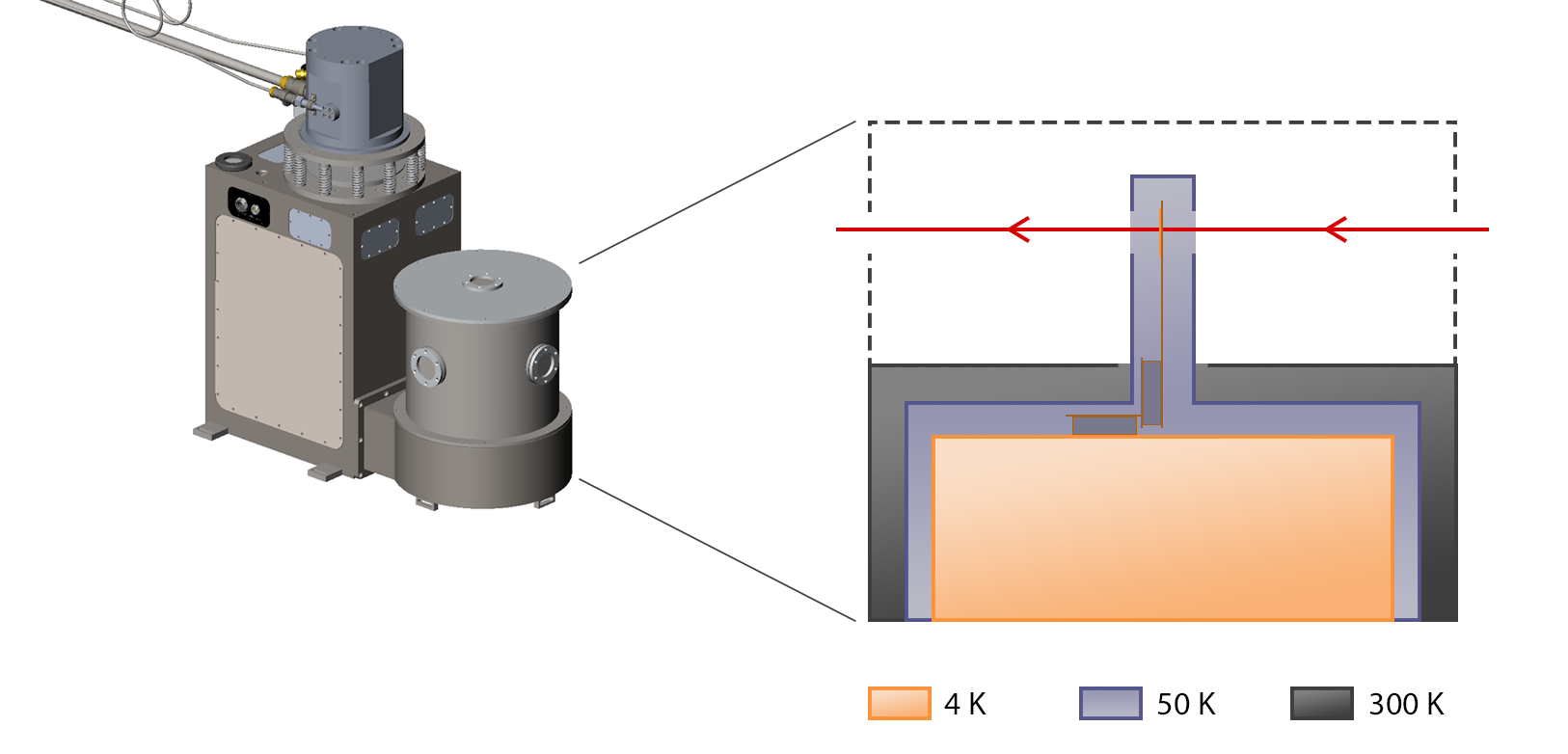}
    \caption{\textbf{New cryostat system}. The left image, taken from \cite{MyCryoFirmwebsite}, is a 3D model of the outside of the cryostat. The working place corresponds to its right part, composed of three plates at different temperatures detailed on the right picture.
    The lower plate, in orange, is the coldest one and can be cooled down to less than 4 K. It is the one where is attached the sample, mounted on piezo-based nanopositioners to be able to move it, and placed at the height of the windows to be optically accessible. A second plate surrounds the first one, in blue, at 50 K to ensure a thermal shield. Finally, the upper one in grey is a breadboard at ambient temperature, as it is directly connected to the outside wall of the cryostat. This is where the closest optical components can be mounted. The red line correspond to the optical axis}
    \label{fig:newcryo_plates}
\end{figure}

Our cryostat consists of three plates at different temperatures, as illustrated in figure \ref{fig:newcryo_plates}. The coldest one is the deepest, plotted in orange: it reaches the target temperature, that can be set to less than 4K. The sample is in direct contact with it and placed at the height of the optical axis (in red). It is mounted on two piezo-based nanopositioners from Attocube to be scanned along the x and y axis. The temperature is controlled through a sensor screwed on the sample holder.

A second plate, in blue in the figure, is at 50 K and works as a thermal shield. The third plate, in grey, corresponds also to the walls of the cryostat, and is therefore at room temperature. In the cryo, the horizontal plate is a breadboard where optical components can be mounted to be as close as possible to the sample.

\subsection{Laser source}
    % \label{subsec:laser}
    
To excite the polariton fluid resonantly, the light field needs to validate several requirements. First of all, the light needs to enter the cavity: the laser should be at the resonance energy, around 835 nm in  our case. However, we also want to be able to transfer an in-plane wavevector to the fluid, and therefore explore the dispersion curve. To do that, the laser should be continuously tunable over the corresponding energy range, in the order of 10 nm. Finally, the spectral bandwidth of the laser should be thin enough to excite only one excitonic mode.

The more suitable laser type for those requirements are the Titanium-Sapphire lasers. The active medium of these solid state lasers is a sapphire crystal doped with titanium ions ($Ti^{3+}:Al_{2}O_{3}$), optically pumped. In the continuous wave configuration, they can be tuned from about 700 to 1200 nm.
In our laboratory, we use a commercial TiSa laser from Spectra Physics, the Matisse 2 TR. Tunable from 730 to 930 nm, it is pumped by a commercial doubled Nd:Yag green laser Verdi from Coherent, delivering 10 W at 532 nm. This way, the TiSa can emit up to 1.2 W in the $TEM_{00}$ mode for a linewidth thinner than 50 kHz.

\subsection{Spatial Light Modulator}
\label{sec:SLM}

\begin{figure}[h]
    \centering
    \includegraphics[width=\linewidth]{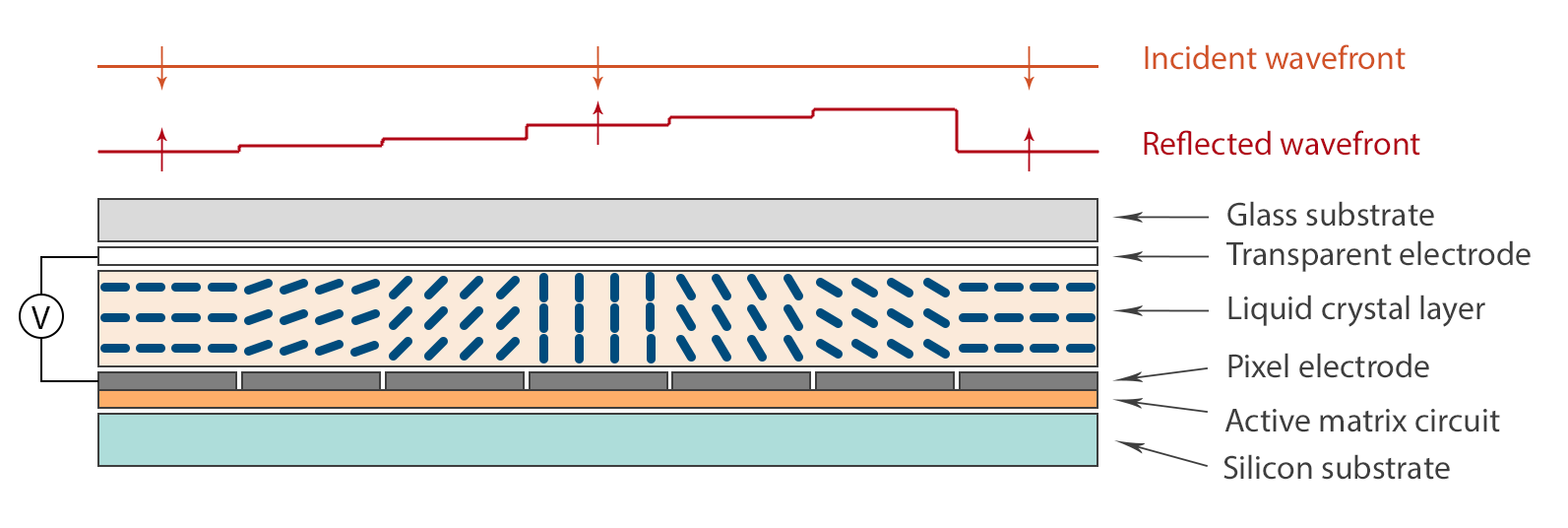}
    \caption{\textbf{Transverse view of the Hamamatsu LCOS-SLM screen}. The silicon substrate supports an active matrix circuit on which pixels electrodes are connected. A specific voltage is applied between each pixel and a transparent grounded electrode. This voltage induces a certain orientation of the liquid crystals in between the electrodes, and thus a specific optical refraction index due to their birefringence. A glass substrate finally covers the top. An incident plane wave consequently follows a specific optical path on each pixel, resulting in a modulated reflected wavefront.}
    \label{fig:SLMprinc}
\end{figure}

A Spatial Light Modulator or SLM is a device that can shape the wavefront of a light beam. 
In our lab, we are using an LCOS model from Hamamatsu, which stands for Liquid Crystal On Silicon, and whose structure is presented on figure \ref{fig:SLMprinc}.

Liquid crystals are a particular phase of matter, which flows and takes the shape of its container, but whose particles still have a positional order, and therefore a preferred orientation \cite{Collings1990}. This preferred orientation can be modulated by applying an electric field, which induces in the same time a change in the refractive index of the medium as liquid crystals are birefringent.

To use this property, LCOS SLM screen has a liquid crystal layer embedded between electrodes (see figure \ref{fig:SLMprinc}): 

\begin{itemize}
    \item a grounded transparent electrode on a glass substrate, on the side of the incoming light
    \item a matrix of pixels electrodes, connected to an active circuit on a silicon substrate. There are 792x600 pixels, of 9.9x7.5 mm\textsuperscript{2} each, with a pixel pitch of 12.5 \textmu m.
\end{itemize}

A specific voltage can be applied between each pixel electrode and the grounded one, defining also the orientation of the liquid crystals in between, and with it their optical index. 
A planar incident wave experiences therefore different optical paths depending on the pixels, and the reflected wavefront is shaped accordingly.

The SLM head, which contains the screen, is connected to a controller, itself connected to a DVI input sendind a 600x800 pixels image in grey shades. 
The controller translates the greyness level (from 0 to 255, bitmap image) into voltage and reproduces the image on the SLM screen, thus drawing it on the light wavefront.

\paragraph{}
The efficiency of the SLM is not perfect: the liquid crystals do not cover the whole surface of the screen, therefore a part of it only reflects on the back of the SLM which acts like a standard mirror.
According to its manufacturer \cite{Hammamatsu2019}, the light utilization efficiency of our model is 97\%. In order to clean the output image, a filtering is realized in the Fourier plane of the SLM: a grating is added to the input image, so that only the light effectively interacting with the SLM is diffracted into the first order. By blocking all the other orders, the beam is cleaned and corresponds exactly to the desired pattern.

This technique allows also to use the SLM not only to shape the phase pattern, but also the intensity profile. Indeed, the quantity of light diffracted into the first order depends on the contrast of the grating. If the whole spectrum of grey scale is used, all the available light is diffracted; if the grating only uses half of the scale, half of the light stays on the zeroth order and is therefore filtered. 
Some patterns can thus be drawn on the grating, playing with the contrast, and be translated as intensity levels in the reflected light beam.

\section{Detection}

\begin{figure}[h]
    \centering
    \includegraphics[width=0.9\linewidth]{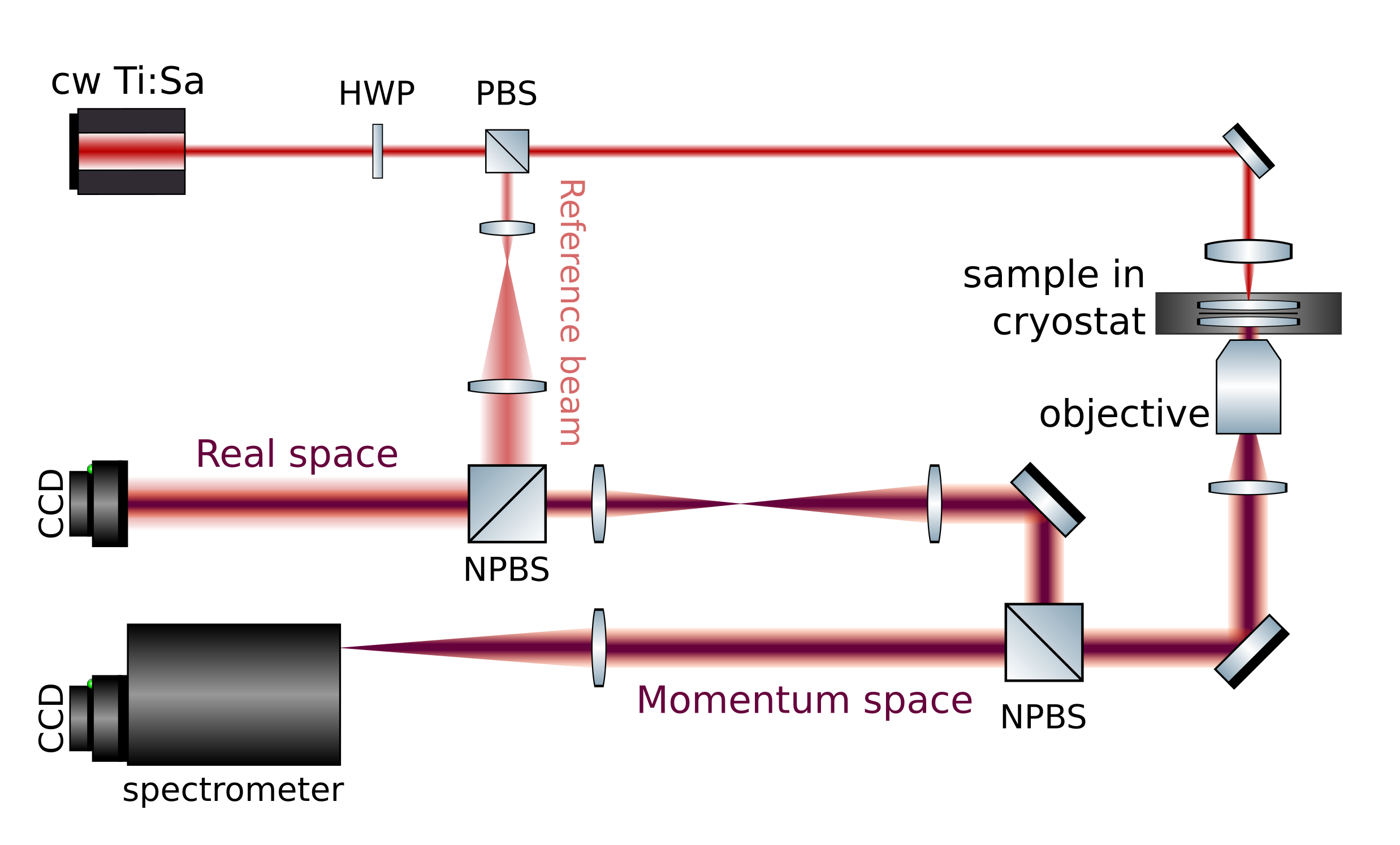}
    \caption{\textbf{Typical experimental setup}. The signal out of the sample is split into two arms which detect the real and momentum space. The real space can interfere with a reference beam, while the momentum space goes through a spectrometer to get information on the signal energy.}
    \label{fig:ExpSetup}
\end{figure}

\paragraph{}
The detection of our system is done in real space and in momentum space.
A typical experimental setup is presented in figure \ref{fig:ExpSetup}: the signal from the sample goes through an objective with a large numerical aperture (0.42), before being separated into two arms, which correspond to the real and momentum spaces.
The momentum space passes through a spectrometer while the real space interacts with a reference beam, extracted from the initial laser beam.

\subsection{CCD Camera}

\paragraph{}
The cameras used in this experiment are PIXIS1024BR eXcellon models from Princeton Instruments. This charged couple device (CCD) has a  back illuminated chip of 1024x1024 pixels, each 13 $\mu$m\textsuperscript{2}, and a quantum efficiency of 95\% at 830 nm. The reading rate of the CCD is 100 kHz, while the integration time can be on the order of the millisecond. The camera also has a mechanical shutter.

It is connected to a computer and driven through the Princeton Instruments software Winspec, which also manages the spectrometer.

\subsection{Spectrometer}

\paragraph{}
The spectrometer of the lab is also from Princeton Instruments and is an Acton Series SP2750 model. This device has a focal length of 750 mm and an interchangeable triple grating turret, on which are mounted 2 gratings and one mirror. 
The gratings respectively have 1200 and 1800 lines per millimeter, and the mirror allows to couple the entrance plane with the CCD one. The entrance plane has an adjustable slit that can be manually tuned between 10 \textmu m and 3 mm, i.e. to a minimal aperture smaller than the pixel size. 
With the 1800 lines per millimeter grating, the spectral resolution is of the order of 1 $\AA$, which is close to 0.2 meV at 835 nm.

\paragraph{}
The spectrometer is connected to the same computer as the camera, and controlled simultaneously with the Winspec software. Both were calibrated together, so that the camera images appear as a function of the wavelength. 
The exposition time or the position of the grating are defined from the computer, and some quick analysis are also available in the software, like transverse cuts or integration along a specific axis. However, the main data analysis work is done afterwards, as explained in the next section.

\subsection{Data analysis}
\label{sec:DataAnalysis}

\subsubsection{Image analysis}

\paragraph{SPE file format}
All the measurements are pictures of the CCD, taken through the Winspec software. As it manages both the camera and the spectrometer, pictures are saved in SPE format, which also stores information about the experimental conditions.
SPE files contain a matrix corresponding to the camera pixels, with for each of them indicates the associated photon counts received during the integration time. This matrix is preceded by a header filled by many parameters, as the position and calibration of the spectrometer, the date of the measurement or the integration time. Those data are easily accessible and useful for the later analysis.
The integration time is typically on the order of a few millisecond, close to the shortest time accessible by the camera. 
We have enough signal to record a clean picture during this time - we usually use filters in front of the spectrometer - and a short integration time reduces our sensitivity to vibrations of the system induced by the pump and the cryostat.

\begin{figure}[H]
    \centering
    \includegraphics[width=\linewidth]{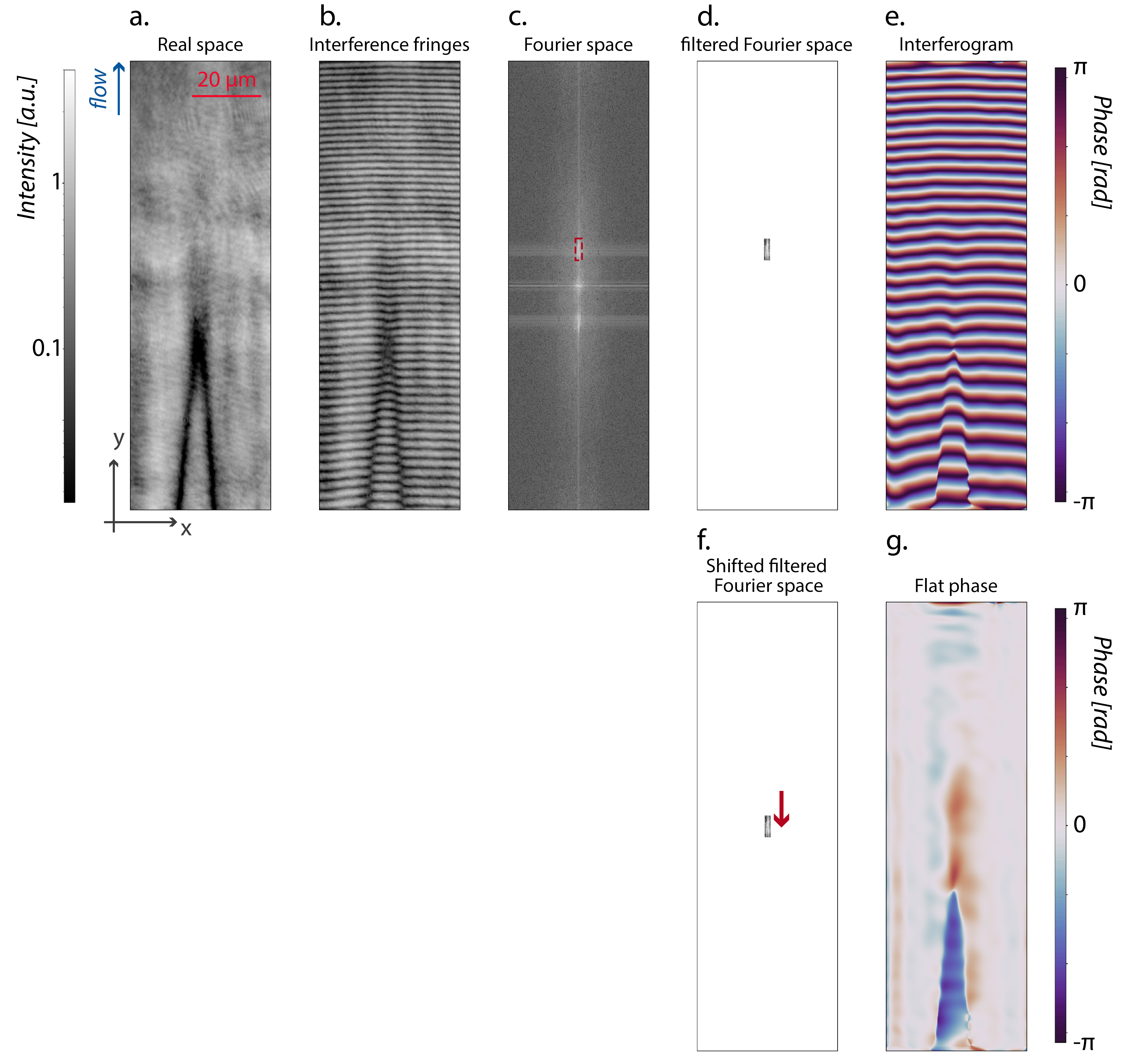}
    \caption{\textbf{Data Analysis}. a. Real space image, signal from the sample renormalized. b. Experimental fringes obtained by interfering the signal with the reference beam. c. Fourier transform of image b: the smallest spatial frequencies are in the center, while the signal outlined in red corresponds to the fringes that we want to keep. d. Filtering of image c., kept at its original position. e. Phase of the inverse Fourier transform of image d.: the fringes are cleaned and highlighted. f. Filtering of image c., shifted towards the center to erase the fringes frequency. g. Phase of the inverse Fourier transform of image f, fitted to place at 0 the phase of the area without modulation: only the phase profile is kept.}
    \label{fig:DataAn}
\end{figure}

\paragraph{}
The different steps of the data analysis are presented in figure \ref{fig:DataAn}, and explained more precisely in the following paragraphs.

\paragraph{Intensity}
The intensity image is a direct picture of the output plan of the cavity. Photons escape from the cavity accordingly to their decay rate, and are detected by the camera: their spatial distribution is directly proportional to the polariton density.

Thus the pictures do not need much analysis: after defining the Region-Of-Interest for the considered measurement, the background counts are removed, as the mean value of a region not illuminated by the polariton signal. The colorscale can also be readjusted, and chosen to be linear or logarithmic depending on what is shown.
An example is given in figure \ref{fig:DataAn}.a, where a phase modulation marked with dark lines is sent on the bottom of the picture, marked by the dark lines, with a flow from bottom to top. The intensity scale is chosen logarithmic.

\paragraph{Phase profile}
The phase profiles are extracted from interference images of the cavity signal with a reference beam of the same laser source, as shown in figure \ref{fig:DataAn}.b. The interference beam is aligned in order to have linear thin fringes to optimize the resolution of the extracted pattern. However, as the system sightly vibrates, too thin fringes also lead to a bad contrast: we usually work with fringes around 5 pixels width.

The background is subtracted and the images are renormalized. In order to enhance the interference fringes, they are filtered in the Fourier space.
The Fourier transform of the data is computed, as presented in figure \ref{fig:DataAn}.c: the signal outlined in red corresponds to the fringes spatial frequency.
The bright spot in the center of the image comes from the signal with spatial frequency equals to zero, \textit{i.e.} the background noise that we want to remove. Finally, the elongated spot symmetric to the selected one also comes from the fringes signal, considering negative frequencies.
Only this part is kept and all others frequencies are put to zero, as illustrated in figure \ref{fig:DataAn}.d.
The data are then inversely transformed, and by plotting their unwrapped phase, only remains the highlighted fringes pattern, plotted in figure \ref{fig:DataAn}.e.

Depending on the pattern we want to show, we can also choose not to keep the fringes, but to flatten the phase pattern, for example when we do not have any flow and only want to reproduce the wavefront. To do so, we still use the interference picture, but when filtering it in the Fourier space, we also translate the remaining signal towards the center of the image, i.e. to the zero frequency, as shown in figure \ref{fig:DataAn}.f. This way, by inversely transforming it, the fringes are removed and the region corresponding to a plane wave gets a constant value, shown in figure \ref{fig:DataAn}.g.

\paragraph{Momentum space}
Several pictures need to be taken in momentum space in order to access all the parameters of the experiment. The first one is the dispersion, which depends on the position of the sample. This one is carried by photoluminescence: by exciting the sample out of resonance, \textit{i.e.} with a much higher energy than the polariton branches, the relaxation of the excess energy fills all the accessible polariton states and allows for a direct detection of the polariton dispersion. 

\begin{figure}[h]
    \centering
    \includegraphics[width=0.6\linewidth]{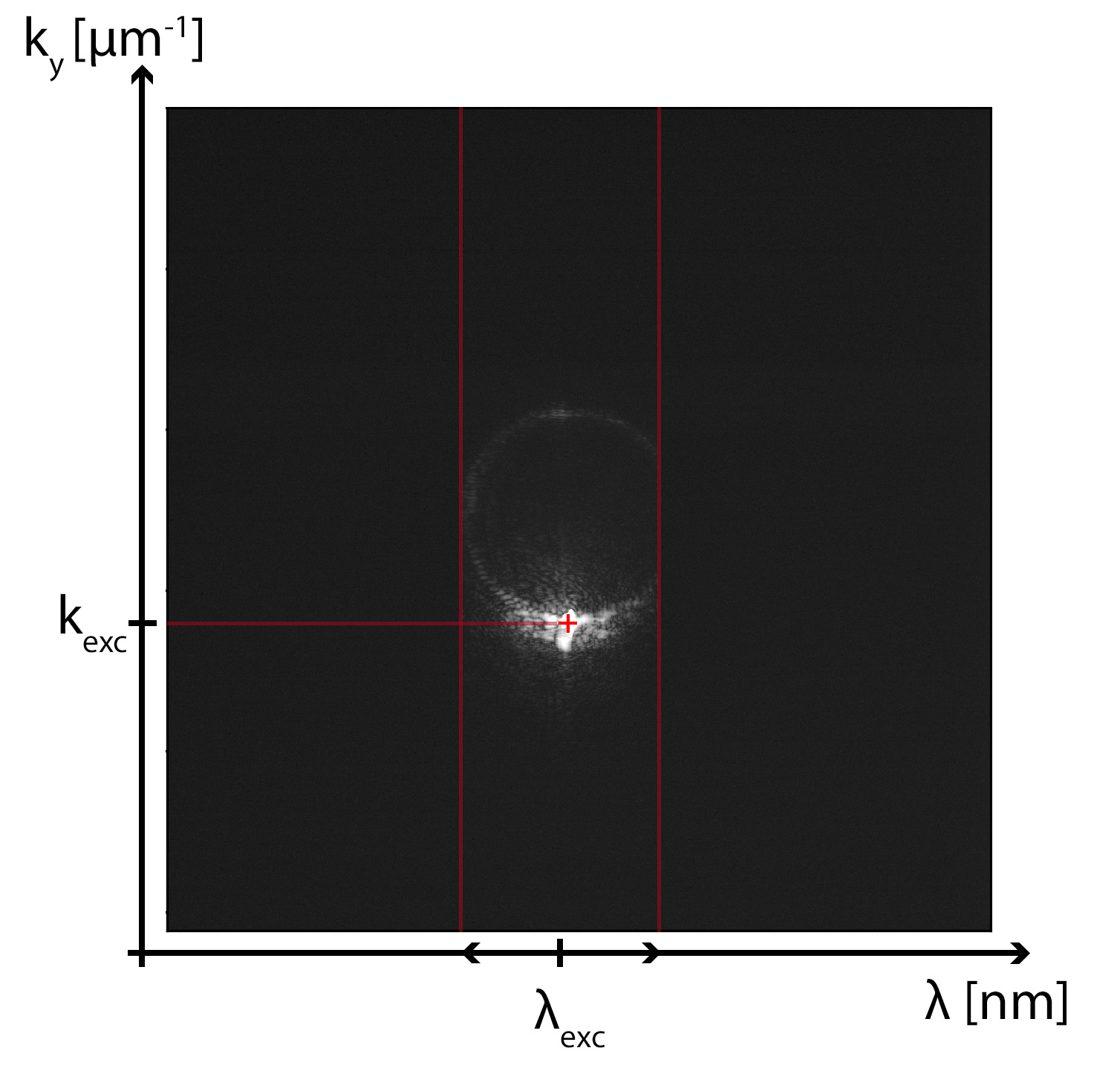}
    \caption{\textbf{Momentum space detection image}. The brightest spot, located by the red cross, indicates the excitation momentum, while the elastic scattering of the polaritons on the cavity disorder is responsible for the shallow Rayleigh ring. In between the open slit of the spectrometer, indicated by the vertical red lines, we detect a far field image which can be graduated in \textmu m\textsuperscript{-1}, while the slit position along the wavelength axis gives us the excitation energy.}
    \label{fig:kspace}
\end{figure}

The wavelength of the incoherent excitation is chosen to correspond to the first minimum of the cavity transmission, which is 810 nm in our case. The dispersion is very well resolved with a thin entry slit of the spectrometer, even though the integration time of the camera should be increased to a few hundred milliseconds as the signal is quite weak. 

\paragraph{}
Another necessary measurement to carry out in the momentum space is the far field, which is the \textbf{k}\textsubscript{x}-\textbf{k}\textsubscript{y} image, presented in figure \ref{fig:kspace}.
This picture gives us the shape and location of the excitation in the momentum space, and its position according to the dispersion.
As the excitation spot is usually collimated on the sample, it gives a main single spot in the momentum space, whose location is extracted through a bidimensional gaussian fit of the image. As the sample is not perfect but presents also some structural disorder, the polaritons can elastically scatter on the defects which results in the appearance of a ring in the momentum space, called the \textit{Rayleigh ring} which intensity is usually smaller than the signal from the pump, and visible in figure \ref{fig:kspace}.

Those images are usually taken through the spectrometer, keeping its entry sit open, as illustrated by the two vertical red lines. The position of the slit is therefore defined by the wavelength of the signal $\lambda_{exc}$, giving us the excitation energy, while inside the slit the image can be scaled in \textmu m\textsuperscript{-1}, hence the \textbf{k}\textsubscript{x}-\textbf{k}\textsubscript{y} image.

\subsubsection{Parameters extraction}

\paragraph{Excitation parameters}

The first parameters are extracted from the dispersion of the lower polariton branch:

\begin{equation}
    E_{LP} = \dfrac{E_{X} - E_{\gamma}(\mathbf{k})}{2}
    - \dfrac{1}{2} \sqrt{\big( E_{X} - E_{\gamma}(\mathbf{k}) \big)^{2}
    + 4 \Omega_{R}^2}
\end{equation}

with $E_{X}$ the excitonic resonance energy, considered constant over our range of wavevectors; $E_{\gamma}(\mathbf{k})$ the photonic resonance energy, approximated parabolic with a curvature proportional to the photonic effective mass $m_{\gamma}^{*}$, and $\Omega_{R}$ the Rabi frequency.

Four parameters need therefore to be extracted: the excitonic resonance, the cavity photon energy at normal incidence, the effective mass of the photon and the Rabi frequency, which is constant for a given sample.

\begin{figure}[h]
    \centering
    \includegraphics[width=\linewidth]{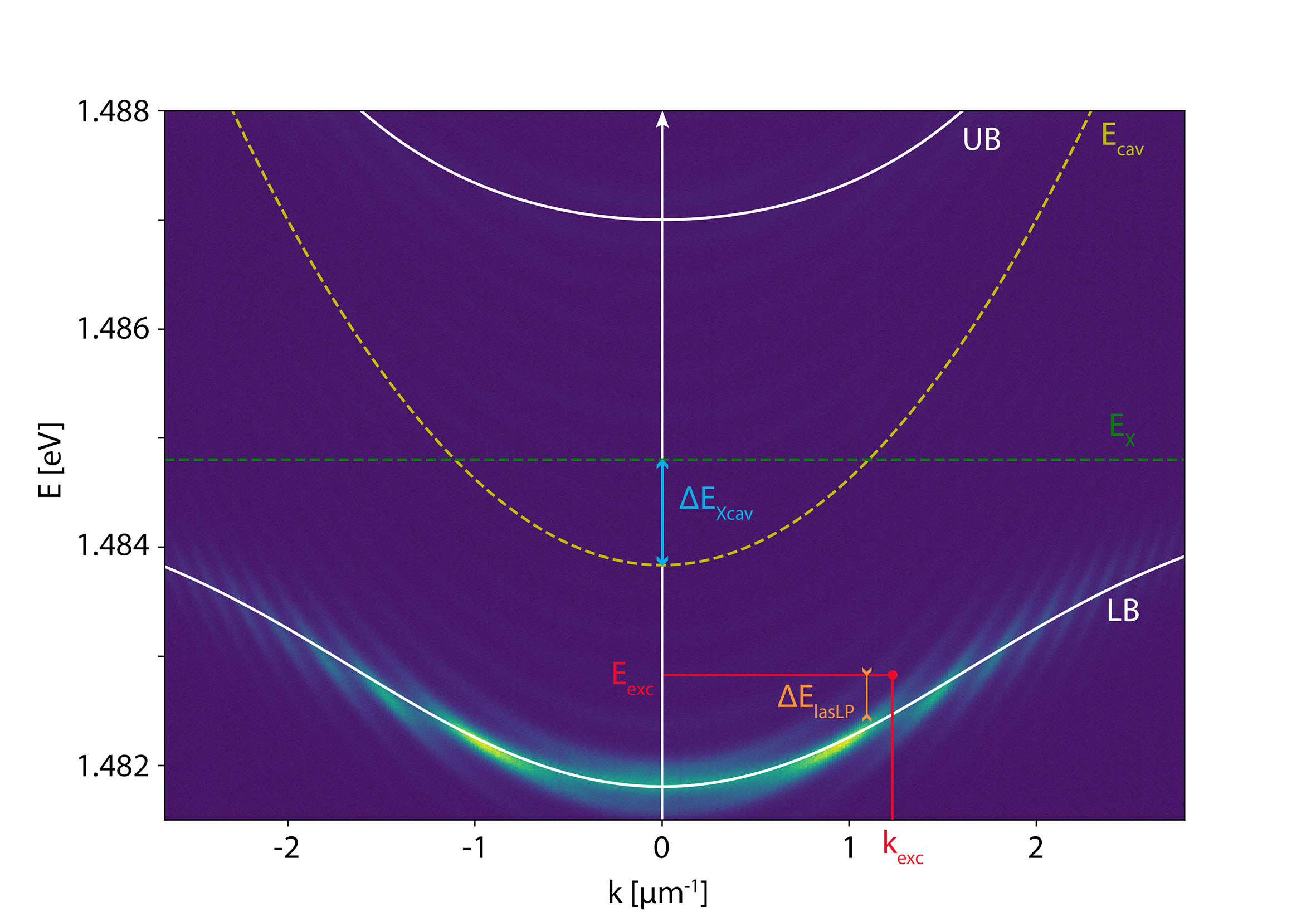}
    \caption{\textbf{Experimental dispersion and quasi resonant excitation}. The background image corresponds to photoluminescence of the sample, carried out by non resonant pumping. The relaxation of the excess energy leads to a polariton distribution along the lower branch, much brighter than the upper one. Therefore only the lower one is used for the numerical fit, illustrated by the white line. The corresponding excitonic and photonic dispersions are plotted in green and yellow dashed lines, respectively. The exciton-photon detuning $\Delta E_{Xcav}$ is presented as the blue arrow.
    The excitation spot, in red, is extracted from the \textbf{k}\textsubscript{x}-\textbf{k}\textsubscript{y} image. It is slightly blue-shifted from the polariton dispersion, from an energy detuning $\Delta E_{lasLP}$ illustrated with the orange arrow.}
    \label{fig:expdisp}
\end{figure}

Those are obtained by fitting the lower polariton branch after a background subtraction, as shown in figure \ref{fig:expdisp}: the white lines are the upper and lower polariton branches, while the yellow and green dashed lines respectively illustrate the corresponding cavity photon and bare exciton dispersions.

The polariton dispersion is necessary to calibrate the camera screen in a wavevector scale. Indeed, the optical path is realigned each day, and with it the position of the zero wavevector. With the dispersion fit, each pixel line is assigned to its corresponding wavevector, and in particular the one of the maximum of the excitation spot from the \textbf{k}\textsubscript{x}-\textbf{k}\textsubscript{y} image, giving us the excitation wavevector \textbf{k}\textsubscript{exc}, in red in figure \ref{fig:expdisp}. The excitation energy E\textsubscript{exc} is obtained as the position of the center of the spectrometer slit.

In the case of a quasi resonant excitation, mainly used in the present work and illustrated in figure \ref{fig:expdisp}, the excitation spot is blue-shifted with respect to the polariton dispersion. Its energy is thus slightly higher than the lower branch one, by a detuning $\Delta E_{lasLP}$ illustrated with the orange arrow in the figure.

\paragraph{Derived parameters}

The previously described parameters, directly extracted from the experimental data, are used to evaluate some derived parameters. 
First of all, the \textbf{fluid velocity} is defined as the group velocity of the polaritons, i.e. the first derivative of the dispersion:
\begin{equation}
    v_{fluid} = \frac{1}{\hbar}\frac{\partial E}{\partial k}
\end{equation}
In our system, the fluid velocity can be up to 1.5 $\mu$m/ps.

From the dispersion can also be extracted the \textbf{effective mass} of the polaritons $m_{LP}^{*}$ created at \textbf{k\textsubscript{exc}}, using the relation 
\begin{equation}
    \frac{1}{m_{LP}^{*}} = \frac{1}{\hbar^{2}} \frac{\partial^{2}E}{\partial \mathbf{k}^{2}}
\end{equation}

As we saw in the theoretical chapter \ref{chap:Polaritons}, the \textbf{lifetime} of the polaritons is experimentally determined with the dispersion analysis, as it is directly connected to the linewidth of the cavity:
\begin{equation}
    \delta E = \hbar \gamma = \frac{\hbar}{\tau}
\end{equation}
with $\gamma$ the cavity decay rate and $\tau$ the lifetime of the polaritons. In our case, the linewidth $\delta E$ is found to be 0.07 meV, which corresponds to a polariton lifetime of 9 ps.

Finally, the \textbf{speed of sound} $c_{sound}$ also derived from the excitation parameters:
\begin{equation}
    c_{sound} = \sqrt{\frac{\hbar g |\psi|^{2} }{m_{LP}^{*^2}}}  = \sqrt{\frac{\Delta E_{lasLP}}{m_{LP}^{*^2}}}
\end{equation}
where $g$ is the interaction constant and $\psi$ the polariton wavefunction. 
The second equality is valid under resonant excitation within the polariton linewidth and in the nonlinear regime.
The speed of sound in our system is typically of the order of 0.5 $\mu$m/ps.
%The later equality is theoretically valid only for the S point 

\paragraph{}
From the speed of the fluid and the speed of sound can be defined the \textbf{Mach number} of the system, $M = v_{fluid}/c_{sound}$. By comparison with unity, this quantity defines sub- or supersonic flow conditions and will be particularly used in the following work.

%An other interesting quantity in fluid of light systems is the \textbf{healing length} $\xi$

%\bibliographystyle{unsrt}
%\bibliography{bibs/LKB-bibs-bib_thesis-ExpChap}  
        
%\end{document}
        
%\documentclass[a4paper,11pt]{book}
%\usepackage[utf8]{inputenc}
%\usepackage[T1]{fontenc} 
%\usepackage{lmodern} 
%\usepackage[margin=28mm,includeheadfoot,bindingoffset=5mm]{geometry}[2010/03/13]

%\usepackage{graphicx}%
%\usepackage{amsmath}
%\usepackage{bbold}
%\usepackage{amssymb} % pour le signe \lesssim
%\usepackage{textcomp} % \textdegree
%\usepackage[most]{tcolorbox} 
%\usepackage{enumitem} 
%\usepackage{float}  
%\usepackage{xcolor}
%\usepackage{stmaryrd}
%\usepackage{physics} 
%\usepackage{wasysym} 
%\usepackage{tikz}
%\usepackage{cite}
%\usepackage{hyperref}
%\usepackage{textgreek}
%\newcommand*\circled[1]{\tikz[baseline=(char.base)]{
%           \node[shape=circle,draw,inner sep=2pt] (char) {#1};}}
%\renewcommand{\thesubsubsection}{\roman{subsubsection}}
%\tcbset{enhanced,colback=red!5!white, colframe=red!75!black,fonttitle=\bfseries}
%\graphicspath{{figures/}} %Setting the graphicspath
%\setcounter{tocdepth}{3}
%\setcounter{secnumdepth}{3}

%\begin{document}

%\tableofcontents
%\setcounter{chapter}{2}

\chapter{Spontaneous generation of topological defects}
\label{chap:SpontChap}

\paragraph{}

The first two chapters introduced all the necessary tools to understand our system: the theoretical description first, then the experimental devices available in the lab.

\paragraph{}
We present in this chapter the first part of our experimental results, the spontaneous generation of topological excitations. To do so, we use the property of bistability of our system: this one is detailed in the first part of the chapter.

\paragraph{}
The experiment was implemented following a theoretical proposal published in 2017 \cite{Pigeon2017}, suggesting to split the excitation into two beams of different intensities, the seed and the support. 
This configuration allows to obtain a large area of the fluid on the upper branch of bistability. 
Then, by giving it a flow and sending it toward a defect, we can observe the generation of quantum turbulence, and in particular of vortex-antivortex pairs in the subsonic case.
Not only does it show that the bistable regime releases the phase constraint of the quasi-resonant excitation, but also that the generated excitations are sustained for hundreds of micrometers.

\paragraph{}
Increasing the Mach number of the fluid also increases the generation rate of the vortices. 
At some point close to the subsonic-supersonic transition, vortices on one hand and antivortices on the other merge together and form a pair of dark solitons.
Those ones where already observed in a different configuration \cite{Amo2011}, without the presence of the driving field, which lead to oblique solitons vanishing over a few tens of microns.
Our implementation greatly enhances the propagation distance, and exhibits a new behaviour of the solitons under the influence of the driving field: the dark solitons align to each other and propagate parallel, as long as they can be sustained.

\newpage

\section{Optical bistability}
\label{sec:OpticalBist}

\paragraph{}
The optical bistability is the ability of a system, over a certain range of input intensities, to possess two possible output values.
Such a property requires nonlinearity: the output intensity can not be linked to the input one only by a multiplicative constant. 
However, nonlinearity itself is insufficient to explain bistability: knowing an input intensity inside the bistability range is not enough to determine the output one \cite{Gibbs1985}.

\paragraph{}
The first observation of optical bistability in a passive medium have been realized in 1969 by A. Sz\"{o}ke and coworkers \cite{Szoke1969}. They used a ring cavity filled with a saturable two-level medium: a "saturable resonator". 
They explain how the saturation of the absorption leads to bleaching, and by combination with the resonator feedback induces a hysteresis cycle. This phenomenon is mainly due to the intensity-dependent absorption and therefore called \textbf{absorptive optical bistability} \cite{Meystre1999}.

\paragraph{}
Another configuration was observed a few years later by Gibbs \textit{et al.} \cite{Gibbs1976}, using this time a purely dispersive medium, with a nonlinear index but no absorption or gain, typically a Kerr medium. 
This \textbf{dispersive bistability} can be studied through the first nonlinear index term $n_2$ or the susceptibility term $\chi^{(3)}$ \cite{Meystre1999}. It is the case we will focus on from now on as it corresponds to our system (see section \ref{sec:quasresPump}).

\paragraph{}
A theoretical model of such phenomenon has first been given by Luigi Lugiato and Rodolfo Bonifacio \cite{Bonifacio1978} in 1978, providing an analytical treatment of optical bistability in the absorptive and dispersive cases, that they have later develop to different configurations \cite{Lugiato1984}.

\subsection{Dispersive bistability}

\begin{figure}[h]
    \centering
    \includegraphics[width=\linewidth]{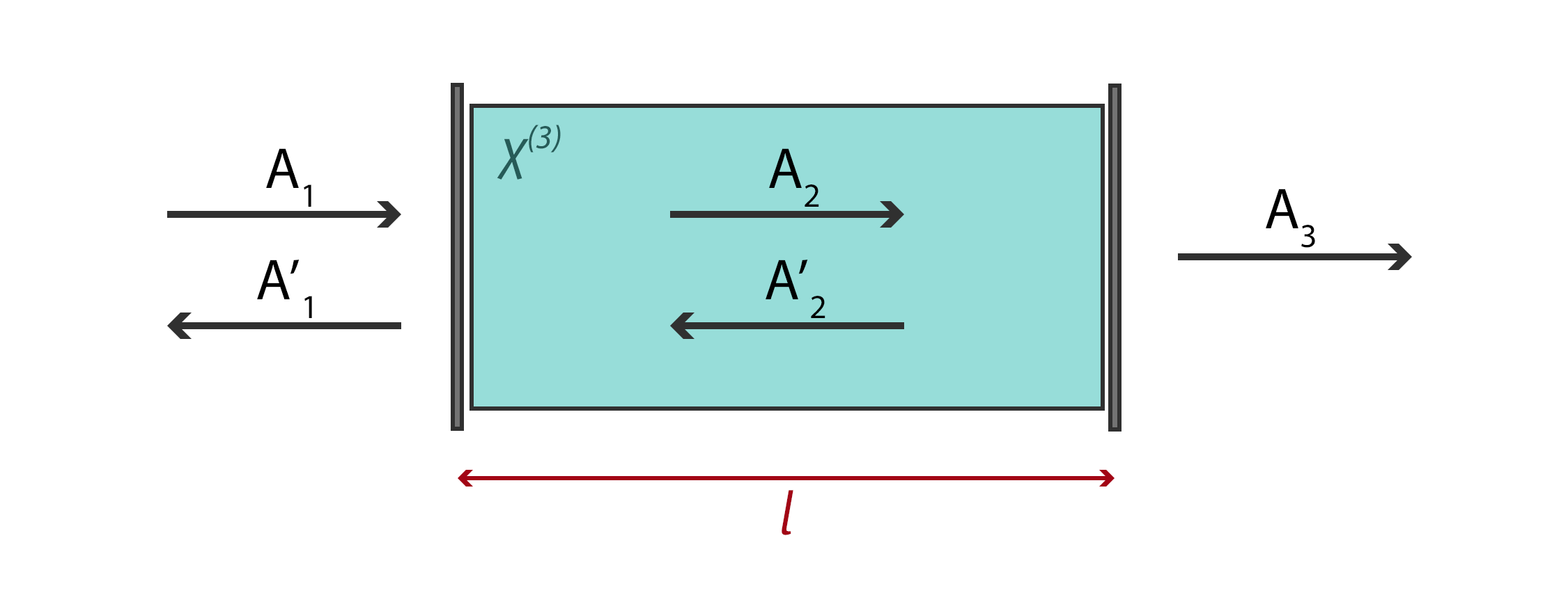}
    \caption{\textbf{Bistable optical system} Fabry-Perot interferometer of length $l$ filled by an ideal Kerr medium without absorption. }
    \label{fig:FabryPerotNonlinMed}
\end{figure}

\paragraph{}
To understand the generation of our system's bistability, let us consider a nonlinear medium inside a Fabry-Perot cavity, as displayed in figure \ref{fig:FabryPerotNonlinMed}. We will consider an ideal configuration, in which the two mirrors are identical and lossless, and follow the relations:
\begin{equation*}
    R = |\rho|^2 
    \qquad
    T = |\tau|^2 
    \qquad
    R + T = 1 
\end{equation*}
where $\rho$ and $\tau$ are reflectance and transmittance in amplitude, and $R$ and $T$ the intensity ones.

\paragraph{}
We want to focus on the case of a dispersive bistability, so we neglect any absorption. The propagation constant $k = n \omega /c$ is taken to be a real quantity with both a linear and nonlinear contributions, and spatially invariant. The different fields are thus linked by the relations \cite{Boyd2008}:

\begin{align}
    A^{'}_{2} &= \rho A_{2} e^{2ikl} \\
    A_{2} &= \tau A_{1} + \rho A^{'}_{2}
\end{align}

Solving the previous system by eliminating $A^{'}_{2}$ leads to the Airy's equation

\begin{equation}
\label{eq:Airys}
    A_{2} = \frac{\tau A_{1}}{1- \rho^{2} e^{2ikl}} = \frac{\tau A_{1}}{1+R e^{i\delta}}
\end{equation}

where $\rho ^{2} = R e^{i \phi}$ contains its amplitude and phase while $\delta = \delta_{lin} + \delta_{nonlin}$ is the total phase shift acquired in a round trip along the cavity. It consist of a linear part:
\begin{equation}
    \delta_{lin} = \phi + 2n_{lin}l \frac{\omega}{c}
\end{equation}
and a nonlinear one:
\begin{equation}
    \delta_{nonlin} = 2n_{nonlin}l \frac{\omega}{c} I
\end{equation}
with $I = I_{2}+I^{'}_{2} \approx 2I_{2}$. 

\paragraph{}
The equation \ref{eq:Airys} can be rewritten for the intensities:
\begin{equation}
    I_{2} = \frac{T I_{1} }{(1-Re^{i\delta})(1+Re^{i\delta}) }
\end{equation}

which leads after a few transformations to the relation:
\begin{equation}
\label{eq:Iratio}
    \frac{I_{2}}{I_{1}} = \frac{1/T }{1+(4R/T^{2})\sin^{2}{\delta/2} }
\end{equation}

\begin{figure}[h]
    \centering
    \includegraphics[width=\linewidth]{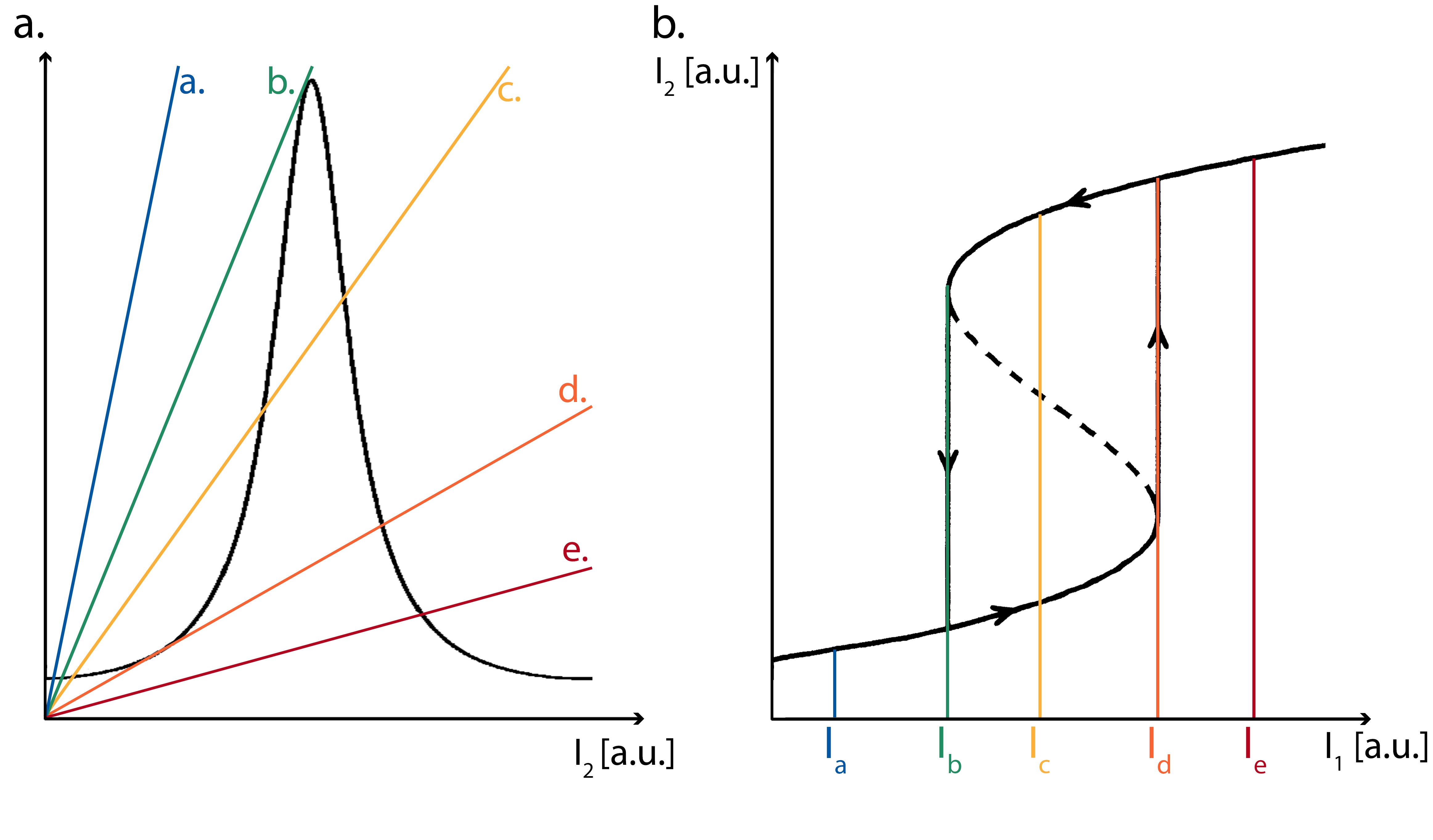}
    \caption{\textbf{Intensities relation and bistability cycle}. a. Graphical resolution of equation \ref{eq:Iratio}: the black curve represents the right hand side of the equation as a function of $I_{2}$, while the colored straight lines correspond to the ratio of intensities, for increasing values of $I_{1}$ from line a to e. Adapted from \cite{Boyd2008}. b. Evolution of the intracavity intensity as a function of the input one. Intensities $I_{a}$ to $I_{e}$ correspond to the associated lines in panel a. The dashed line is an unstable solution. The system presents a hysteresis cycle: the bistability occurs between $I_{b}$ and $I_{d}.$
    }
    \label{fig:GraphSolution}
\end{figure}

\paragraph{}
This equation is resolved graphically in figure \ref{fig:GraphSolution}.a., where both sides of the relation are plotted. The periodic black curve represents the right hand side as a function of $I_{2}$, and the colored lines the ratio $I_{2}/I_{1}$ for different values of $I_{1}$. On the right, figure \ref{fig:GraphSolution}.b., is illustrated the shape of the evolution of $I_{2}$ as a function of $I_{1}$, where the intensities corresponding to the colored lines on a. are reported.

The blue line labeled a. illustrates the lowest value of $I_{1}$. It crosses the black curve only once: only one internal intensity value $I_2$ corresponds to this input intensity, as plotted on figure \ref{fig:GraphSolution}.b.
The yellow c. line shows the case of a higher value of $I_1$, and crosses the black line three times. That it the case of bistability: on figure b., two states are accessible, while the dashed line is unstable. Increasing again the incident intensity, like in the e. case, only one value of $I_{2}$ is possible: the red curve on figure a. crosses the black one only once.

Cases b. and d. are particular: they have two intersection points with the black line on \ref{fig:GraphSolution}.a. They correspond to the turning points between which the bistability occurs. It results in a hysteresis cycle: once the output on one branch, the system remains on it. It therefore does not follow the same path by increasing or reducing the input intensity, as indicated by the arrows on figure \ref{fig:GraphSolution}.b.

\subsection{Quasi resonant pumping}
\label{sec:quasresPump}

\paragraph{}
The nonlinearity of a polariton fluid comes from the polariton-polariton interactions. They result in a third order term, equivalent to a Kerr-like medium in such a way that the polariton system behaviour is analogous to an electromagnetic wave resonant with a cavity filled by a Kerr medium, as presented in the previous section.

We induce bistability in the polariton system by pumping it quasi-resonantly with a laser slightly blue-detuned with respect to the lower polariton branch. The energy detuning between the pumping laser and the lower polariton branch $\Delta E_{lasLP} = \Delta E = E_{las} - E_{LP}$ plays an important role in the state of the system, as it induces an energy renormalization and shifts the dispersion.

In the case of a red detuning, this renormalization drives the mode away from the resonance, as an increase of the pump intensity induces a decrease of absorption. The interesting case is however the one of a blue detuned pump: this time, a higher pump intensity leads to a higher polariton density, and an energy renormalization closer to resonance. 
Thus, an intensity threshold appears, above which an increase in the polariton population self amplifies and leads to a blue shift of the dispersion up to the laser energy.

\begin{figure}[h]
    \centering
    \includegraphics[width=0.9\linewidth]{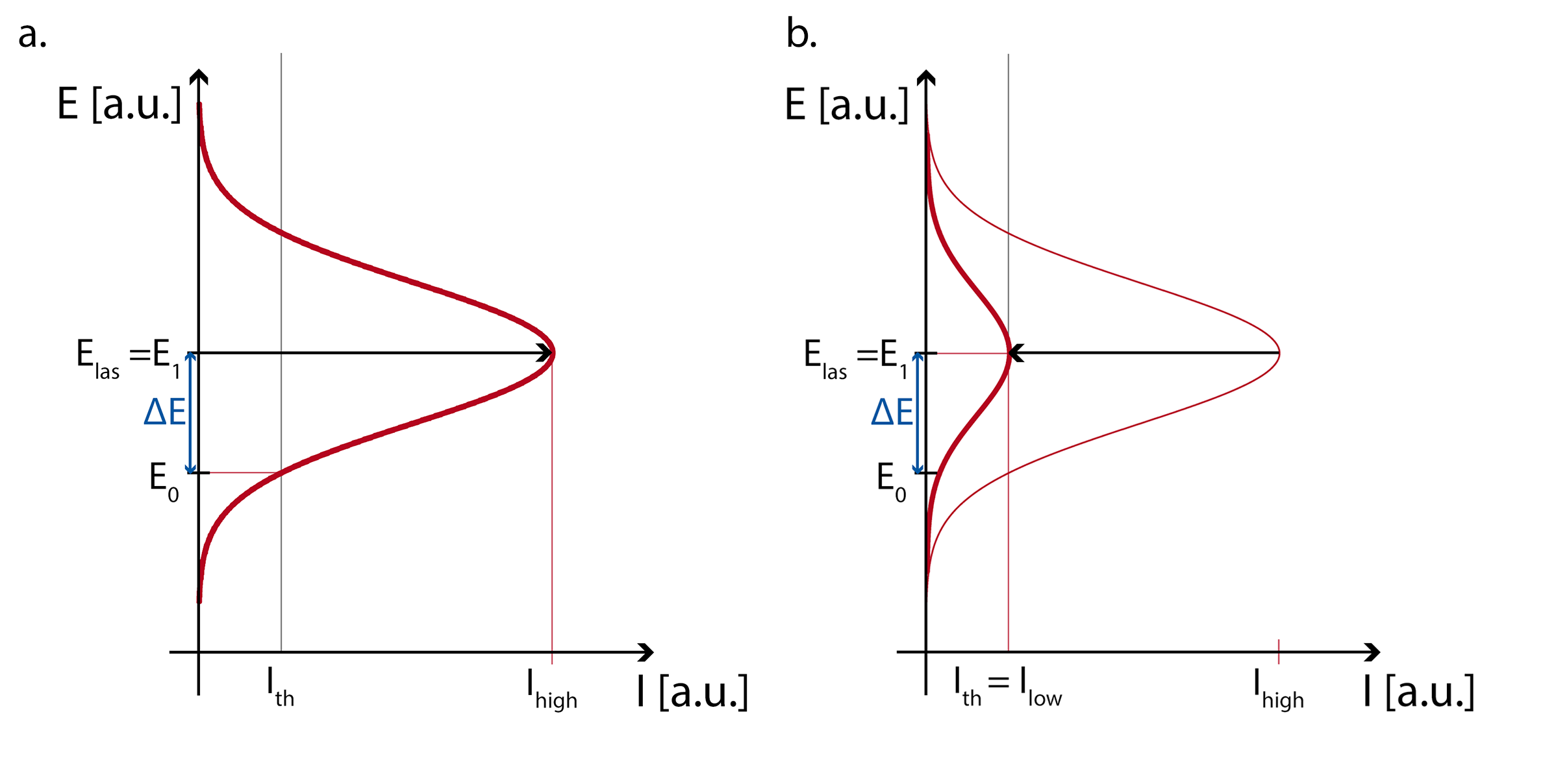}
    \caption{\textbf{Bistability in quasi-resonant pumping}. $E_{0}$ corresponds to the energy of the linear dispersion of the lower polariton branch, while the laser pumps the system at a detuned energy $E_{1} = E_{0} + \Delta E_{lasLP}$, inducing a population represented by the red gaussian curves. 
    On a. is illustrated the situation of an \textbf{increase} of the pumping intensity: initially at $E_{0}$, the system jumps to $E_{1}$ for a pump intensity of $I_{high}$, which corresponds to the threshold value at $E_{0}$. 
    Figure b. shows the inverse situation of a \textbf{decreasing} intensity: once the system as jump to $E_{1}$, it stays there until the pump intensity decreases down to $I_{th} = I_{low}$; only then will the system go back to the unperturbed dispersion energy $E_{0}$. Thus, between $I_{low}$ and $I_{high}$, depending on its initial state, the system can be either at $E_{0}$ or $E_{1}$ and is therefore bistable.}
    \label{fig:BistQuasiresScheme}
\end{figure}

\paragraph{}
This mechanism induces a bistability, sketched in figure \ref{fig:BistQuasiresScheme}.
Let us consider our system at the energy of the unperturbed lower polariton branch $E_{0}$, and a laser pump slightly blue detuned from it, at $E_{1} = E_{0} + \delta E$; its induced population it represented by the red gaussian curves.

By gradually increasing the pump intensity, the system remains at $E_{0}$ as long as the intensity felt by the system stays lower than the threshold intensity $I_{th}$.
The jumps happens thus at the configuration illustrated by the figure \ref{fig:BistQuasiresScheme}.a: the intensity at $E_{0}$ is the threshold one $I_{th}$, while the total one is actually much higher and corresponds to the upper limit of the bistability cycle $I_{high}$.
By jumping to the $E_{1}$ state, the system also strongly increases its density: it has now reached the nonlinear regime of the bistability upper branch.

Once in that state, the system stays there as long as it density stays above the threshold $I_{th}$. The figure \ref{fig:BistQuasiresScheme} shows the configuration for which the system jumps back down to the ground state $E_{0}$: it takes place when the total pump intensity decreases down to $I_{th}$, which therefore corresponds to the lower limit of the bistability cycle $I_{low}$.

However, a decrease of the pump does not induce the same path for the system. Indeed, being already in the upper state $E_{1}$, it stays there while the pump intensity is higher than the threshold.
The unlocking back to the $E_{0}$ state occurs for a pump intensity $I_{th} = I_{low}$, shown in figure \ref{fig:BistQuasiresScheme}.b.

Between $I_{high}$ and $I_{low}$, the state of the system therefore depends on its initial position, and can be either in $E_{0}$ or $E_{1}$. That is the range of intensity where the bistability occurs, resulting in a hysteresis cycle similar to the one presented in figure \ref{fig:GraphSolution}.b.

\paragraph{}
A more quantitative way of describing this phenomena in our system is to start with the steady-state equation. 
Let us place ourselves in the case of a normal incidence excitation with $\mathbf{k} =0$ \textmu m\textsuperscript{-1}, \textit{i.e.} in the case of the \textit{degenerate four-wave mixing}.
The two-by-two interaction ("four-wave mixing") indeed happens between identical and at rest polaritons, hence the term degenerate.
The steady-state equation can then be written as \cite{Karr2001}:

\begin{equation}
\label{eq:deg4wavemix}
    F_{las}(\mathbf{r}) = (i \hbar \gamma + \Delta E_{lasLP} - g n_{0}(\mathbf{r}) ) \psi_{0} (\mathbf{r})
\end{equation}

where $F_{las}$ represents the laser pump, $\gamma = \gamma^{LP}_{\mathbf{k}}$ the relaxation term of the lower polaritons, $g$ the polaritons interaction constant, and $\psi_{0}$ and $n_{0}$ the static wavefunction and density: $n_{0}(\mathbf{r}) = \psi_{0}(\mathbf{r}) \psi^{*}_{0}(\mathbf{r})$.
By multiplying equation \ref{eq:deg4wavemix} by its complex conjugate, we obtain:

\begin{equation}
\label{eq:bistability}
    I_{las}(\mathbf{r}) = \Big( (\hbar \gamma)^{2} + (\Delta E_{lasLP} - g n_{0} (\mathbf{r}))^{2} \Big) n_{0}(\mathbf{r})
\end{equation}
with $I_{las}$ the intracavity laser pump intensity.

If the detuning between the pump and the lower polariton branch $\Delta E_{lasLP}$ is high enough, this equation leads to a bistable behaviour. Indeed, this equation can be derived:

\begin{equation}
    \dfrac{\partial I_{las}}{\partial n_{0}} = 3 g^{2}n_{0}^{2}
    - 4 \Delta E_{lasLP} g n_{0} + (\hbar \gamma)^{2} + \Delta E_{lasLP}^{2}
\end{equation}

the discriminant has the expression $(2g)^{2} \big( \Delta E_{lasLP} - 3 (\hbar \gamma)^{2}\big)$. 
Therefore, the derivative of the intensity possesses two distinct roots if the detuning follows the conditions $\Delta E_{lasLP} > \sqrt{3}\hbar \gamma$.

\begin{figure}[h]
    \centering
    \includegraphics[width=0.75\linewidth]{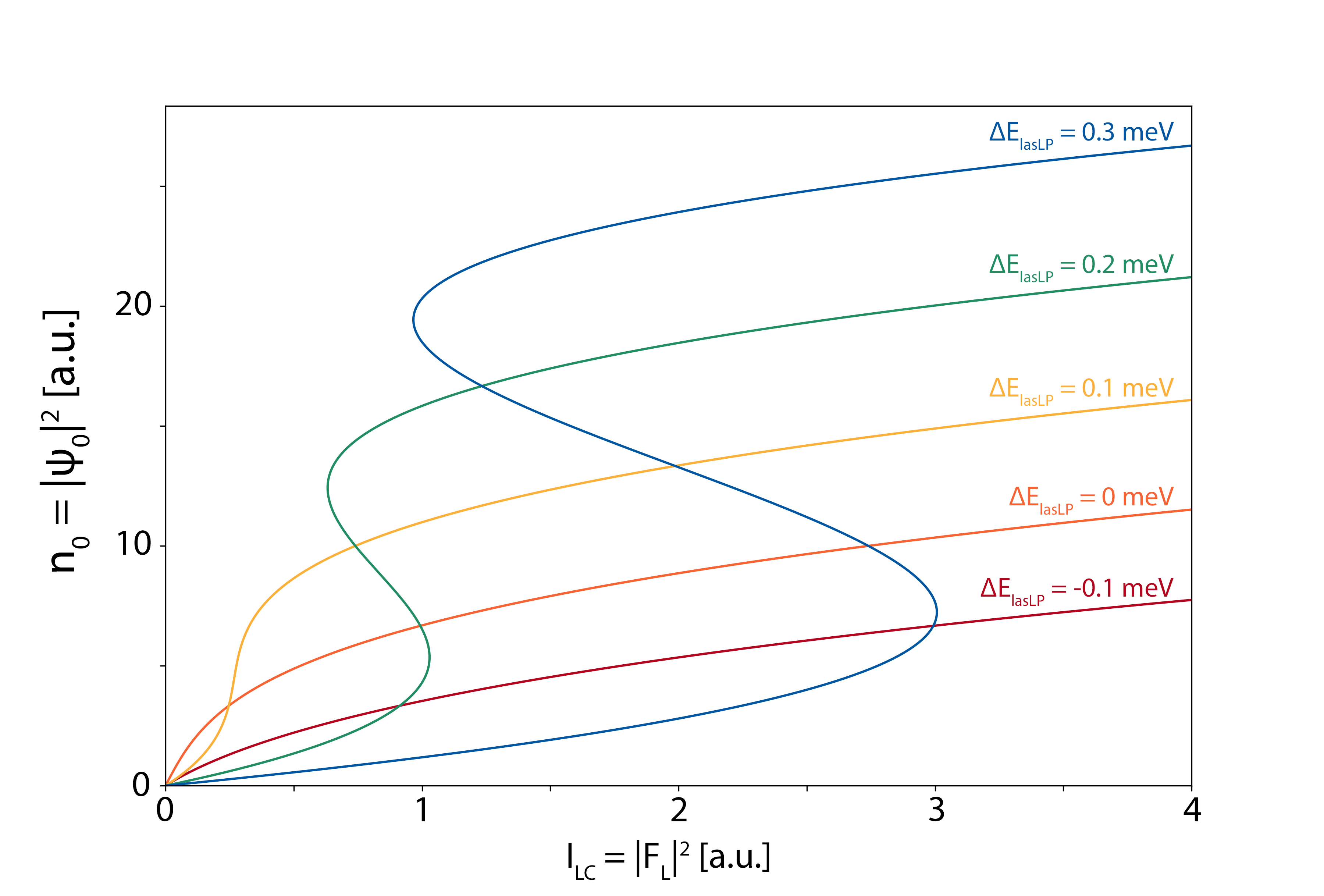}
    \caption{\textbf{Density evolution for different detunings}. Numerical resolution of equation \ref{eq:bistability} for different detunings, from -0.1 meV to 0.3 meV. The bistability only appears for detunings larger than $\sqrt{3} \hbar \gamma $.}
    \label{fig:BistDetSimu}
\end{figure}

\paragraph{}
This equation has also been solved numerically and the results are presented on figure \ref{fig:BistDetSimu} for parameters chosen to be similar to the typical experimental ones: $\hbar \gamma = 0.07$ meV,  $g = 0.015$ meV, and energy detunings $\Delta E_{lasLP}$ from 0 to 0.3 meV.

As expected, different behaviours can be observed. For a red-detuned pump ($\Delta E \leq 0$), the energy shift has a profile as a square root: increasing the excitation leads to larger shift, and with it to a larger detuning with the pump, which interacts less with the excited states. 
On the other hand, when the pump is blue detuned ($\Delta E >0 $), the profile is different. However, the bistability occurs only for large detunings compared to the linewidth of the cavity. The yellow curve, which corresponds to $\Delta E_{lasLP} = 0.1$ meV only presents an inflection point, while the S shape of bistability appears for $\Delta E_{lasLP} = 0.2$ meV and $\Delta E_{lasLP} = 0.3$ meV.
One can see that the bistability intensity range expands with the detuning; but it also shifts to higher intensities. Experimentally, the size of the bistability cycle is often limited by the available laser power.

\paragraph{}
An interesting theoretical idea about the bistability is its role in the pump influence on the system. A resonant pumping is known to impose its properties to the fluid, and in particular to give it its phase. 
From there on, the sustainability of topological excitations such as vortices or solitons is inhibited in a pumped region, as they typically induce phase modulations. However, the presence of a second energy state for the fluid to jump on could release the system of this constraint, and thus tolerate the generation of such phase discontinuities.

%\subsection{Pigeon effect: phase constraint release}

%\paragraph{}
%??

\section{Seed-support configuration: enhancement of propagation distance}
\sectionmark{Seed-support configuration}

\paragraph{}
In 2017, Simon Pigeon and Alberto Bramati \cite{Pigeon2017} suggested the use of the optical bistability of an exciton-polariton system to enhance the propagation of the superfluid. 
In particular, this theoretical proposal explained how the use of two different beams could create a high density polariton fluid, bistable and over a macroscopic scale.
This section will focus on the results of this paper, as its experimental realizations are presented afterward.

\subsection{Superfluid propagation enhancement}
\label{sec:propenhancemt}

\paragraph{}

Let us consider an exciton-polariton system, offering an optical bistability between the intracavity intensities $I_{low}$ and $I_{high}$, as presented in section \ref{sec:OpticalBist}.

The suggested configuration of Pigeon \textit{et al.} \cite{Pigeon2017} is to consider two driving fields, with the same frequency $\omega_{p}$ and the same in-plane wavevector $\mathbf{k}_{p}$.
The first one, called the seed or the reservoir, is localized in space and has a high intensity: $I_{r} > I_{high}$. It thus produces a nonlinear superfluid above the bistability cycle.
The second field is the support, much more extended in space; for now, let us consider it theoretically as an infinitely extended constant field, stationary in time and homogeneous in space. Its intensity is weaker than the intensity of the seed one, and is placed inside the bistability cycle: $I_{low} < I_{s} < I_{high}$.

\paragraph{}
The evolution of such a system is described with the Gross-Pitaevskii equation:

\begin{multline}
    i \partial_{t} 
    \begin{pmatrix} \Psi_{C}(\mathbf{x}, t) \\ \Psi_{X}(\mathbf{x}, t) \end{pmatrix}
    = \hbar \begin{pmatrix} F_{s} + F_{r}(\mathbf{x}) \\ 0 \end{pmatrix}
    e^{-i(\mathbf{k}_{p}\cdot \mathbf{x} - \omega_{p}t)} \\
    + \hbar
    \begin{pmatrix}
    \omega_{C}(\mathbf{k}) + V(\mathbf{x})-i \gamma_{C} & \Omega_{R} \\
    \Omega_{R} & \omega_{X}^{0} + g |\Psi_{X}(\mathbf{x}, t)|^{2} - i \gamma_{X}
    \end{pmatrix}
    \times \begin{pmatrix} \Psi_{C}(\mathbf{x}, t) \\ \Psi_{X}(\mathbf{x}, t) \end{pmatrix}
\end{multline}

with $\Psi_{C}(\mathbf{x}, t)$ the cavity field and $\hbar \omega_{C}(\mathbf{k})$ the cavity mode dispersion; $\Psi_{X}(\mathbf{x}, t)$ the excitonic field and $\hbar \omega_{X}^{0}$ the exciton energy considering an infinite mass. $F_{s}$ is the amplitude of the support driving field ($I_{s} = |F_{s}|^{2}$) and $F_{r}(\mathbf{x})$ the one of the seed. 
$V(\mathbf{x})$ is the photonic potential, $\Omega_{R}$ the Rabi frequency, $g$ the exciton-exciton interaction term and $\gamma_{C}$ and $\gamma_{X}$ the decay rates of the cavity and the exciton, respectively.

\paragraph{}

\begin{figure}[h]
    \centering
    \includegraphics[width=0.8\linewidth]{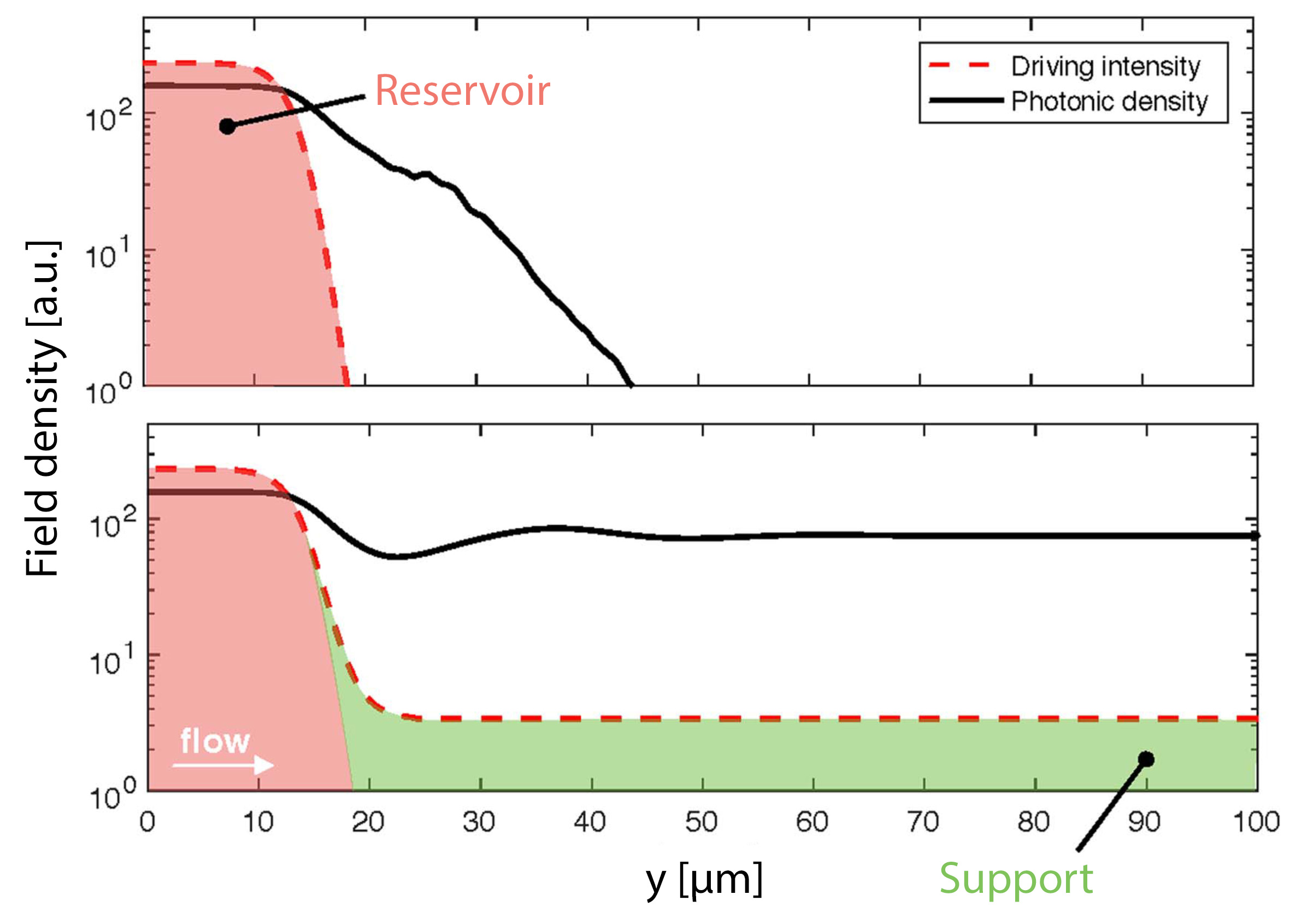}
    \caption{\textbf{Field density in seed only or seed support configuration}. The black solid line illustrates the intracavity intensity (logarithmic scale), while the red dashed one is the driving intensity sent to the system. The colored regions delimits the different driving fields: the seed (or reservoir) in red and the support in green. The polariton flow goes from left to right. On the upper panel, only the seed is sent: it creates a high density polariton fluid, decaying exponentially out of the pumping region. The propagation length is limited by the polariton lifetime. In the lower panel, the support field is added: even if its density is two order of magnitude lower than the seed one, the total polariton density is maintained at its higher level all along the presence of the support.
    From \cite{Pigeon2017}}
    \label{fig:SeedSuppDens}
\end{figure}

The goal of this configuration is to enhance the propagation and density of the polariton fluid by combining the properties of both beams. Numerical simulations were done to understand their combination, presented in figure \ref{fig:SeedSuppDens}. 
They have been realized for a cavity without any defects ($V=0$) such that: $\hbar \omega_{C}(\mathbf{k} = \mathbf{0}) = 1602$ meV, $\hbar \omega_{X}^{0} = 1600$ meV, $\hbar \gamma_{X} = \hbar \gamma_{C} = 0.05$ meV, $\hbar \Omega_{R} = 2.5$ meV and $\hbar g = 0.01$ meV/\textmu m\textsuperscript{2}. 
The driving field parameters are $\delta E = \hbar \omega_{p} - \hbar \omega_{LP}(\mathbf{k} = \mathbf{k}_{p}) = 1$ meV and $ |\mathbf{k}_{p}| = (0.1)^{T} $ \textmu m\textsuperscript{-1}.

The field intensity is shown as a function of space, plotted in logarithmic scale. The flow of particles goes from left to right. The driving intensity sent in the system is pictured with the red dashed line, while the black solid line shows the steady-state photonic intracavity density.

\paragraph{}
On the upper panel, only the seed is sent on the left, illustrated with the red highlighted region ($F_{r}(\mathbf{x}) \neq 0$; $ F_{s} =0 $). 
Given the presence of an in-plane wavevector, the photonic density expands a bit, but with an exponential decay due to the finite polariton lifetime.
Even with the present best quality samples, the propagation distance in this configuration is limited to around 50 ps. Moreover, as the density is decreasing all along propagation, all related parameters are also constantly changing.

\paragraph{}
The lower panel shows what happens when the support, in green, is added to the previous configuration.
The support intensity is two order of magnitude lower than the seed one, but infinitely extended in space for this simulations.
The presence of the support field has a strong impact on the total polariton density: its high level created from the reservoir is maintained without any decay all over the region where is the support, despite its much weaker intensity.
Some modulations can be observed at the frontier between the support and the reservoir, due to the sharp designed profile of $F_{r}(\mathbf{x})$. %; they are dispersive shock waves. 
They were not observable in the seed-only configuration due to the quick decay of the photonic density, which does not occur with a support field.

\paragraph{}
Those simulations do not exactly match an experimental situation. In particular, the extent of the support beam can obviously not be infinite: the fluid propagation length is limited by the pump available power. 
However, thanks to the localization of the seed and the low intensity of the support, it can easily reach macroscopic scale of hundreds of microns, and this independently of the polariton lifetime.

\subsection{Bistability role}

\begin{figure}[h]
    \centering
    \includegraphics[width=0.7\linewidth]{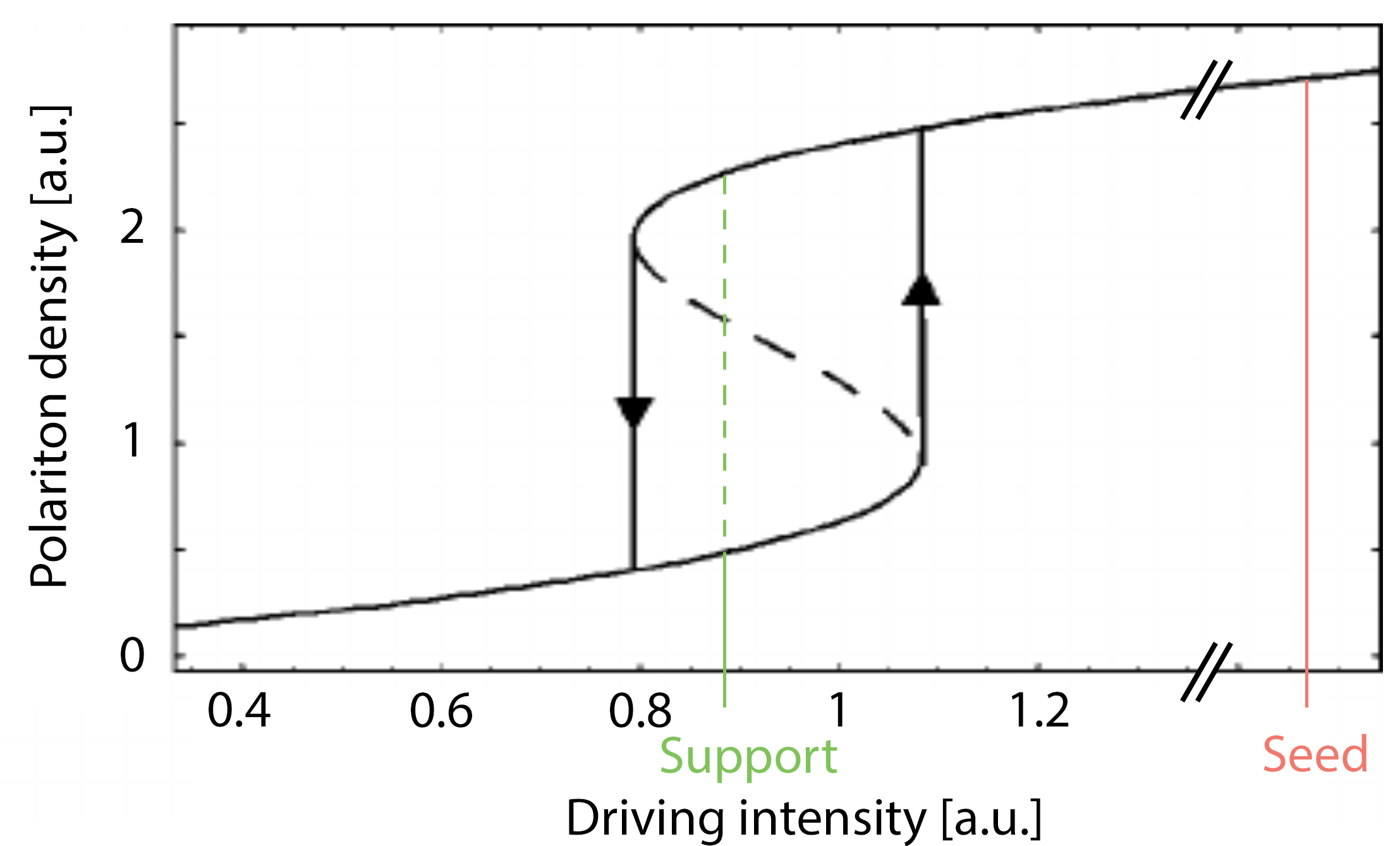}
    \caption{\textbf{Positions of seed and support intensities compared to the bistability cycle}. The seed is chosen to be above the hysteresis cycle. Only one stable state is available at this place, on the upper nonlinear branch of bistability. The support has much lower intensity, inside the bistable region: the support alone would not be enough to place the system on the nonlinear branch , and would remain linear at low density. However, if both fields are sent, the seed presence ensures the system to stay on the upper branch, and this everywhere sustained by the support. 
    }
    \label{fig:SeedSuppBist}
\end{figure}

\paragraph{}
Bistability is essential to explain this behaviour. The seed is placed outside the bistability cycle, where only one stable state is accessible (see figure \ref{fig:SeedSuppBist}). It therefore ensures the system on the upper branch of the cycle, the nonlinear branch.
On the other hand, the support is chosen to be inside the bistability cycle, where two states are reachable. The support itself can not reach the upper branch and, if alone, would only drive the system in the linear configuration. 
Yet, as the seed places the system in the nonlinear branch and touches the support illuminated region, by extension, the neighbour region jumps also on the upper branch which expands to all the support area.

\section{Subsonic flow: vortex stream generation}

\paragraph{}
The idea of the seed-support configuration is not only to obtain an extended fluid of polaritons, but also to study its properties, and in particular its ability to generate topological excitations, such as dark solitons and quantized vortex-antivortex pairs.
The different hydrodynamic regimes of a polariton fluid and their effects on the generation of topological excitations have been previously studied in the case of a single intense and localized pump, placed just upstream of a structural defect \cite{Pigeon2011, Amo2011}.
The presence of the defect creates some turbulence along the flow, which evolves differently depending on the ratio between the speed of the fluid and the sound velocity, \textit{i.e.} the Mach number of the system: $M = v_{f}/c_{s}$.
However, the speed of the fluid that needs to be taken into account is the one around the defect, therefore always a bit higher than the one extracted far from the defect. Indeed, the particles close to the defect are accelerated \cite{Frisch1992}, which induces a phase shift as the fluid phase and speed are connected.
In the case of a global subsonic flow, but locally supersonic around the defect, vortex-antivortex pairs emerge in the wake of the defect. Increasing the Mach number induces an increase in the emission rate of the vortices, which will finally merge together in a pair of dark solitons for a Mach number close to 1 or higher.

\paragraph{}
This section focuses on the generation of a vortex stream. As in the numerical case of \cite{Pigeon2011}, the vortices are spontaneously generated in the wake of a structural defect. However, the seed-support configuration previously described is used to enhance of one order of magnitude their propagation length. The first part of this section presents the numerical predictions published in \cite{Pigeon2017}, while the second part focuses on the experimental observations.

\subsection{Numerical predictions}

\begin{figure}[h]
    \centering
    \includegraphics[width=0.6\linewidth]{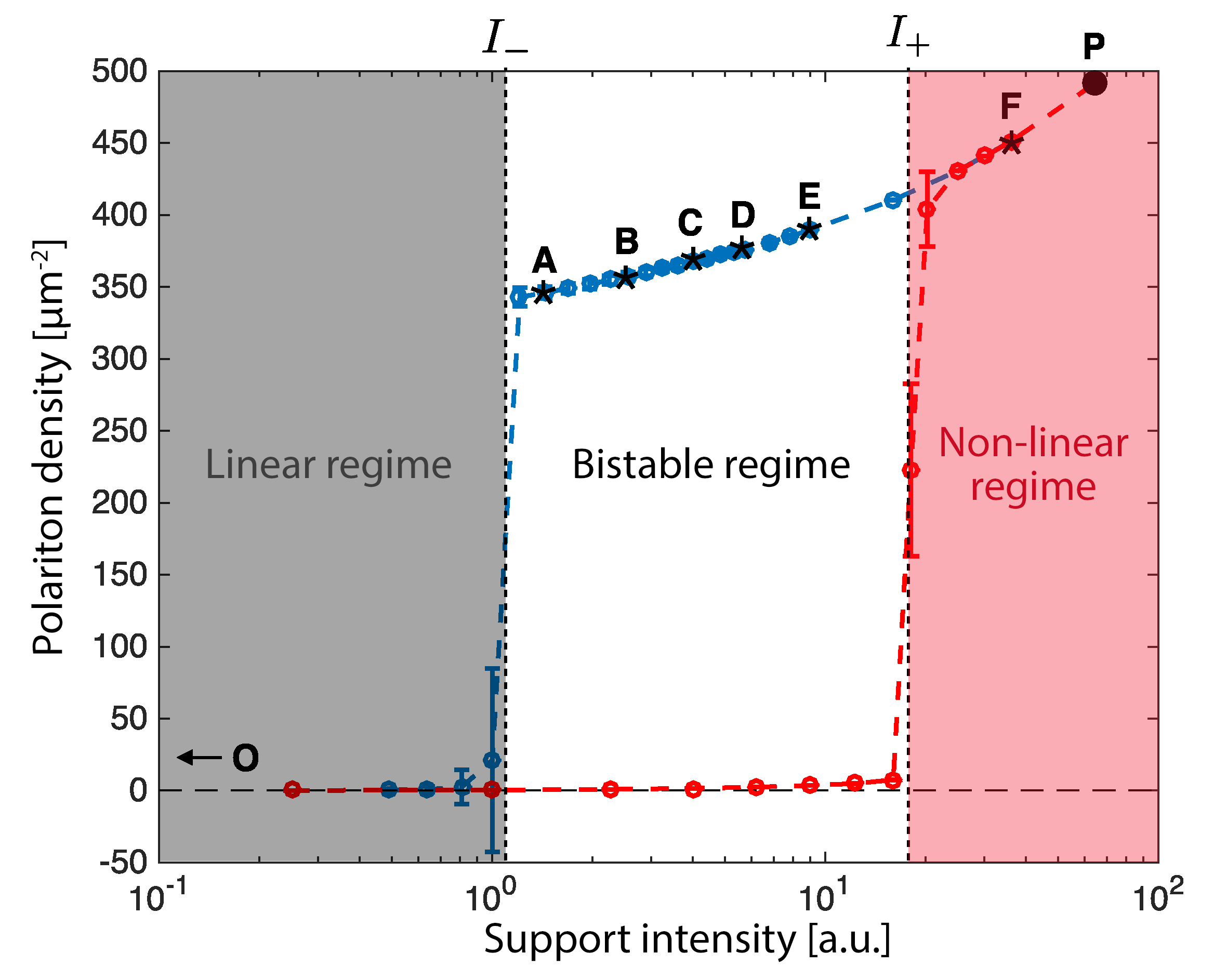}
    \caption{\textbf{Distribution of the support intensities presented in figure \ref{fig:SimuVortexSpont}}. The red dots are obtained with the support field only, while the blue ones were computed with both the seed and the support. The seed intensity is placed above the bistability cycle, represented by the black dot labeled P. The label O corresponds to a case without support, the cases A to E are in the bistable regime and the F one is just above it.
    Adapted from \cite{Pigeon2017}}
    \label{fig:SeedSuppWorkingPoints}
\end{figure}

\paragraph{}
To numerically reproduce the effect of a cavity structural defect, a large potential photonic barrier is introduced ($V \neq 0$). 
The seed is placed upstream to it, localized and with a fixed intensity above the bistability cycle, illustrated in point P in figure \ref{fig:SeedSuppWorkingPoints}.

\begin{figure}[H]
    \centering
    \includegraphics[width=\linewidth]{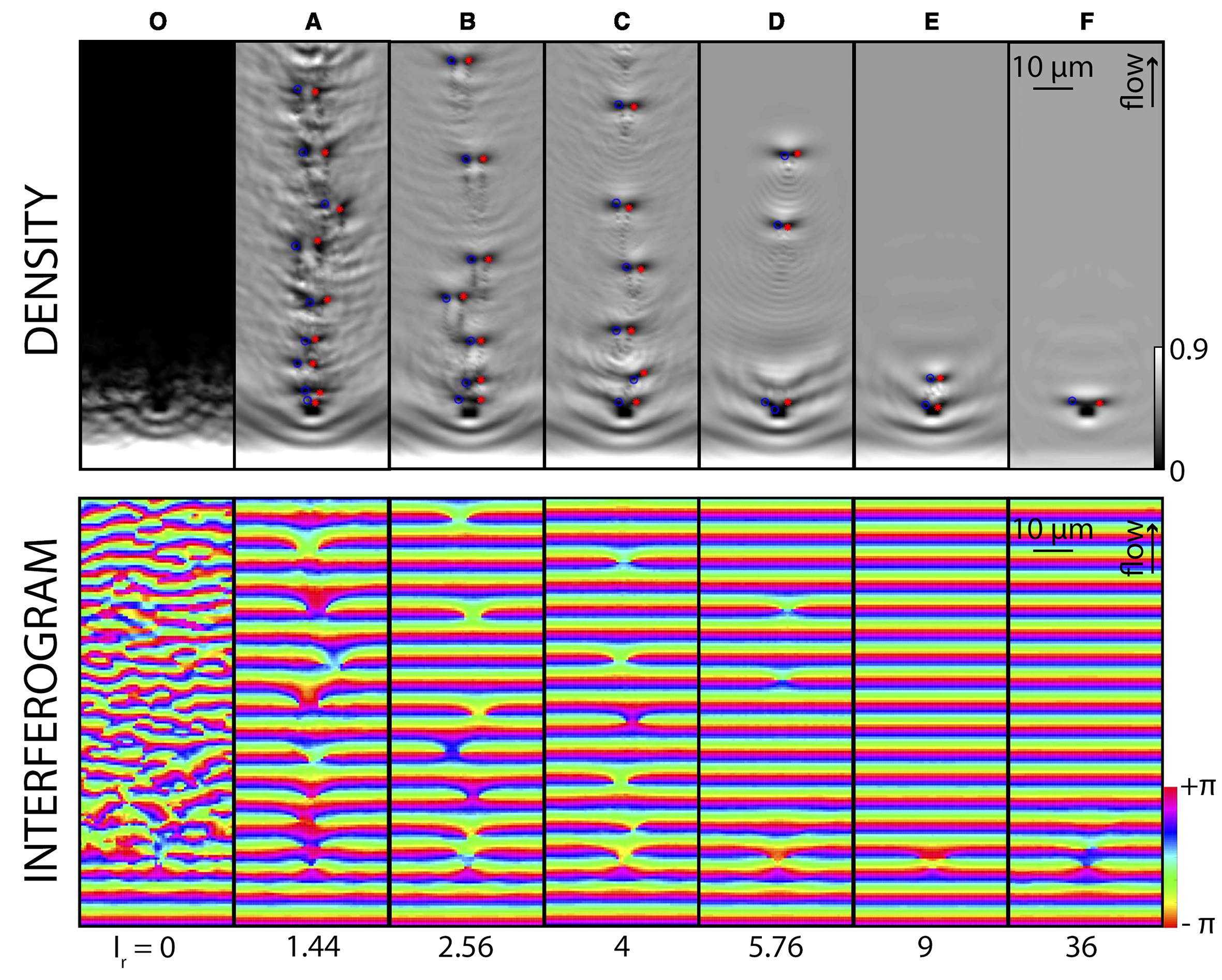}
    \caption{\textbf{Density and phase snapshots of a polariton superfluid around a photonic defect for different support intensities}. Case O corresponds to a seed without any support field: the polariton population decays exponentially. Cases A to F present different configurations for an increasing support intensity (detailed in figure \ref{fig:SeedSuppWorkingPoints}): the non-linear fluid is extended over hundreds of microns, and vortex-antivortex pairs are spontaneously generated on the wake of the defect for bistable support intensities.
    Adapted from \cite{Pigeon2017}}
    \label{fig:SimuVortexSpont}
\end{figure}

\paragraph{}
The goal of exploring the bistable regime and its influence on the vortex generation is achieved by tuning the intensity of the support field. Figure \ref{fig:SeedSuppWorkingPoints} presents the bistability system of the considered system, and the positions of the support intensities shown in figure \ref{fig:SimuVortexSpont}. The other cavity and driving parameters are identical to the one presented in the section \ref{sec:propenhancemt}.

The situation labeled O corresponds to a localized seed pump in the absence of the support one. 
As previously explained, the polariton density downstream of the pump exponentially decays due to the short lifetime of the polaritons. 
When reaching the defect, it is already too low to observe any non-linear interactions, and no topological excitations can be observed.

Intensities A to E are placed within the bistability cycle. 
The first conclusion to get from those pictures is that the non linear fluid is effectively sustained for macroscopic length, even in the region where only the support is sent, which would not be enough if it were only the driving.

As expected, the bistable regime ensures a certain release of the phase constraint imposed by a resonant driving. 
Thus, even though the support field possesses a flat phase, vortex-antivortex pairs are generated in the wake of the defect as long as the support intensity places the fluid in the bistable regime. 
Vortex and antivortex have opposite circulation, and are spotted by red and blue dots, respectively.
Vortex-antivortex can stay bounded and propagate along the flow side by side, or annihilate each other and vanish.

\begin{figure}[h]
    \centering
    \includegraphics[width=0.6\linewidth]{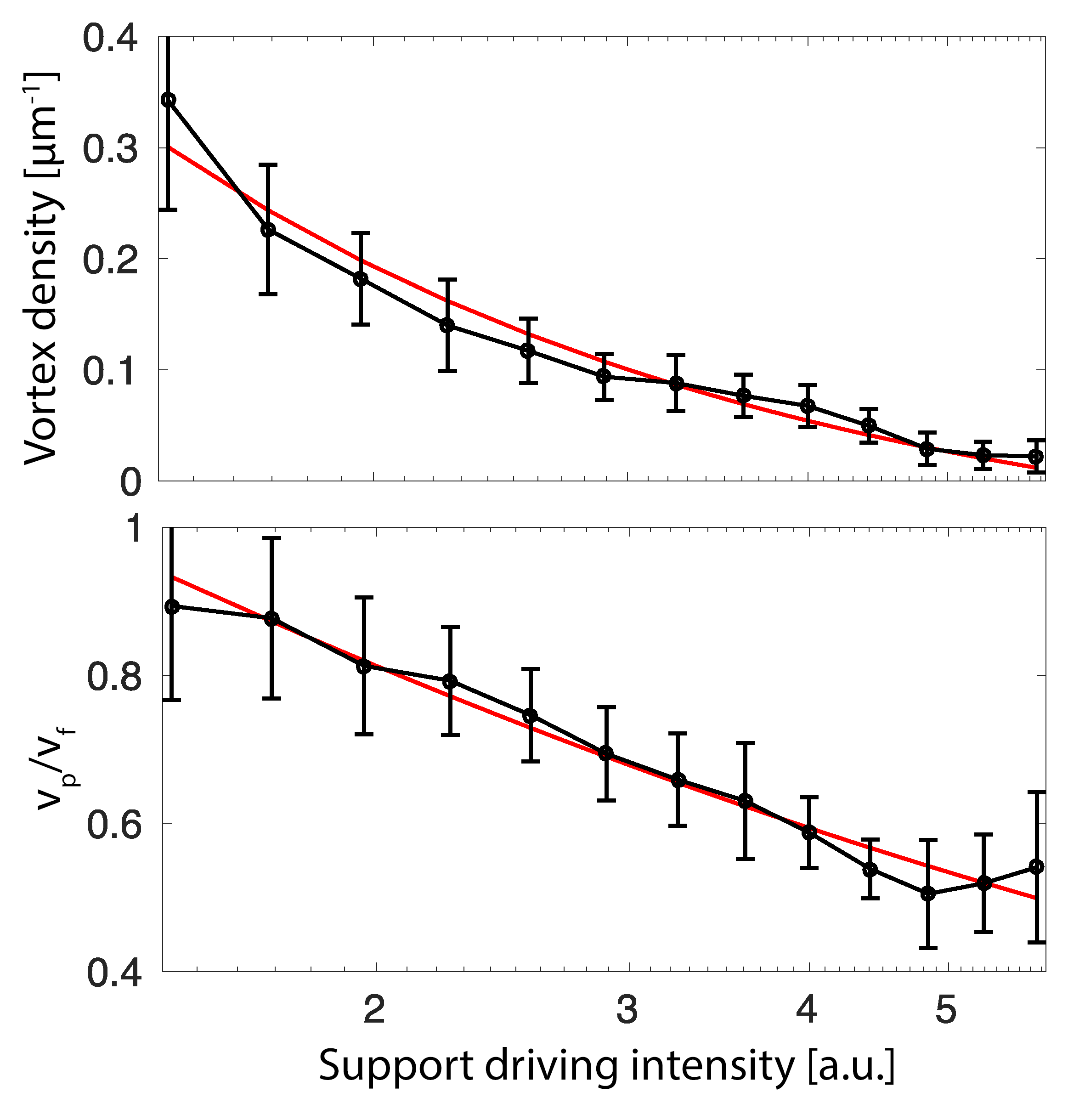}
    \caption{\textbf{Support intensity influence on the vortex stream}. The upper panel shows the vortex density in the stream decreasing inversely to the support intensity. On the lower panel is plotted the evolution of the vortex pair velocity $v_{p}$, normalized to the fluid velocity $v_{f}$, with the support field. The red curves are a linear fit of the data, depending on the support field amplitude for the upper panel and on the support intensity in the lower one.
    Adapted from \cite{Pigeon2017}}
    \label{fig:VortexDensitySpeed}
\end{figure}

An interesting parameter to study is the vortex density in the stream. In case A, close to the low threshold intensity $I_{low}$, the vortex density is quite high. But with the increase of the support intensity, from case B to E, the number of vortex pairs observable in the stream is gradually decreasing, until they almost do not propagate but annihilate quickly as in case E, even if still inside the bistability cycle.
The vortex density is plotted on the upper panel of figure \ref{fig:VortexDensitySpeed}, and actually scales inversely with the support intensity. On case E in particular, one can see that the vortex pairs are not sustained long in the stream but recombine close to the potential barrier.

The lower panel of figure \ref{fig:VortexDensitySpeed} illustrates how the velocity of the vortex pairs $v_{p}$ evolves with the driving intensity. The velocity has been renormalized to the fluid one $v_{f}$. Even at smallest support intensity, the vortex speed is a bit lower than the fluid one, but remains close. However, it strongly decreases with the support increase, down to half of its initial value. 
It actually scales as the inverse of the amplitude of the support field: this shows the coherence of the phenomenon. The vortices are slowed down by the support field which acts as a friction force on them. This could be interesting as the tuning of the support intensity could have a direct control on the properties of the vortices.

\begin{figure}[ht]
    \centering
    \includegraphics[width=0.7\linewidth]{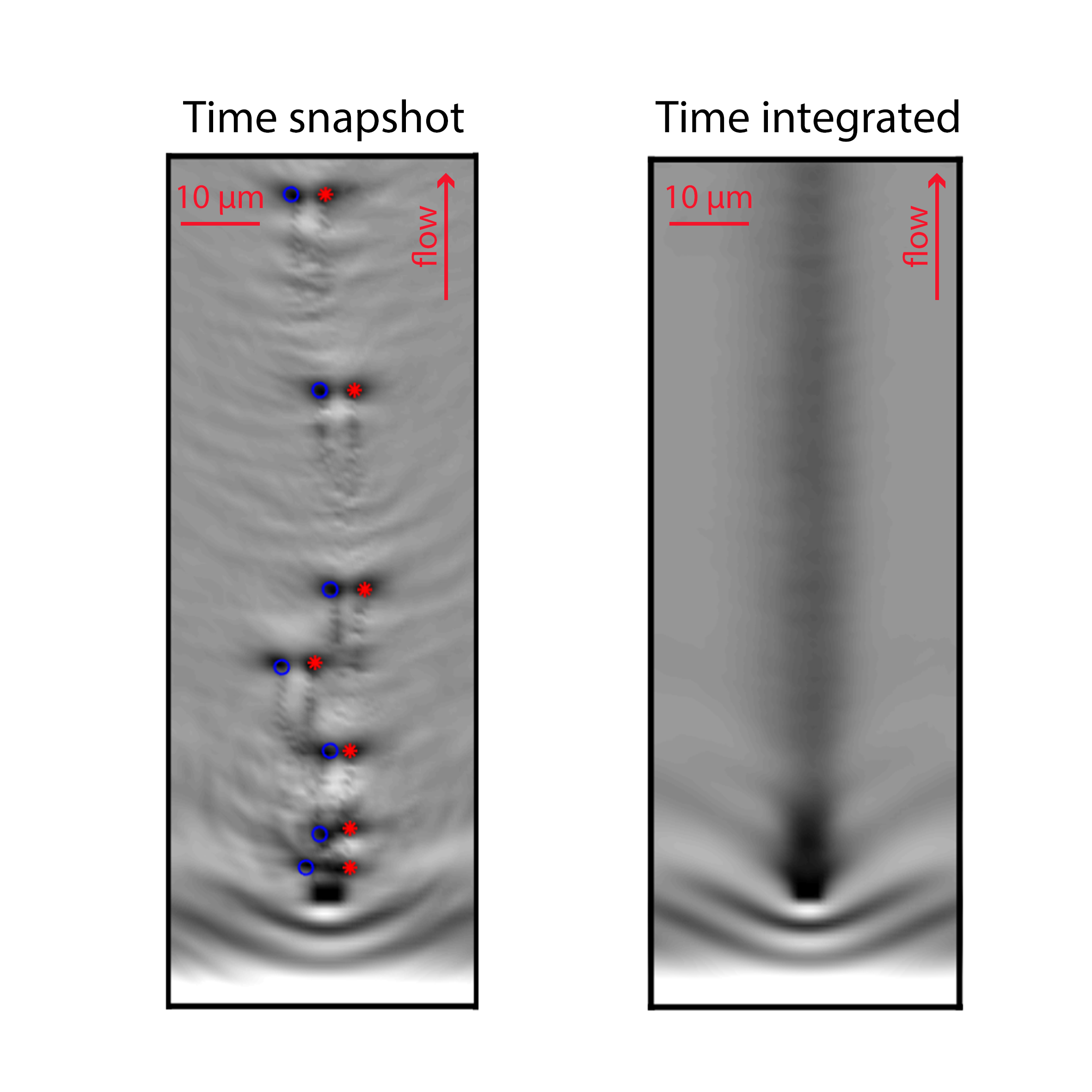}
    \caption{\textbf{Comparison between a time snapshot and a time integrated image of a vortex stream}. On the left, the time snapshot shows the position of vortex and antivortex flowing in the wake of the defect. On the right, the time integrated image has blurred the positions of the vortices and results in a thick shadow along the flow.}
    \label{fig:VortSnapInt}
\end{figure}

\paragraph{}
The images presented previously in figure \ref{fig:SimuVortexSpont} are time snapshots of a simulated polaritons fluid. They enable to precisely locate the vortices, and to observe the fork shape of the phase pattern typical of vortices.
However, those pictures can not be reproduced experimentally, as it would require a single-shot picosecond time resolution that we do not have in the lab. We only have access to millisecond resolution: numerical simulations have been made in order to predict how those phenomena would look like in a time integrated picture. Figure \ref{fig:VortSnapInt} shows the comparison between one of the previous snapshot image on the left, with the same parameters, and on the right what it would look like if the picture was taken with 1 ms integration time.

A time average flow of vortex pairs appears as a thick line of lower density, which dip height is proportional to the vortex density. The phase pattern would be blurry and a decrease of visibility would appear along the vortex stream.

\subsection{Experimental realizations}
\label{sec:ExpVortSpont}

\begin{figure}[h]
    \centering
    \includegraphics[width=0.85\linewidth]{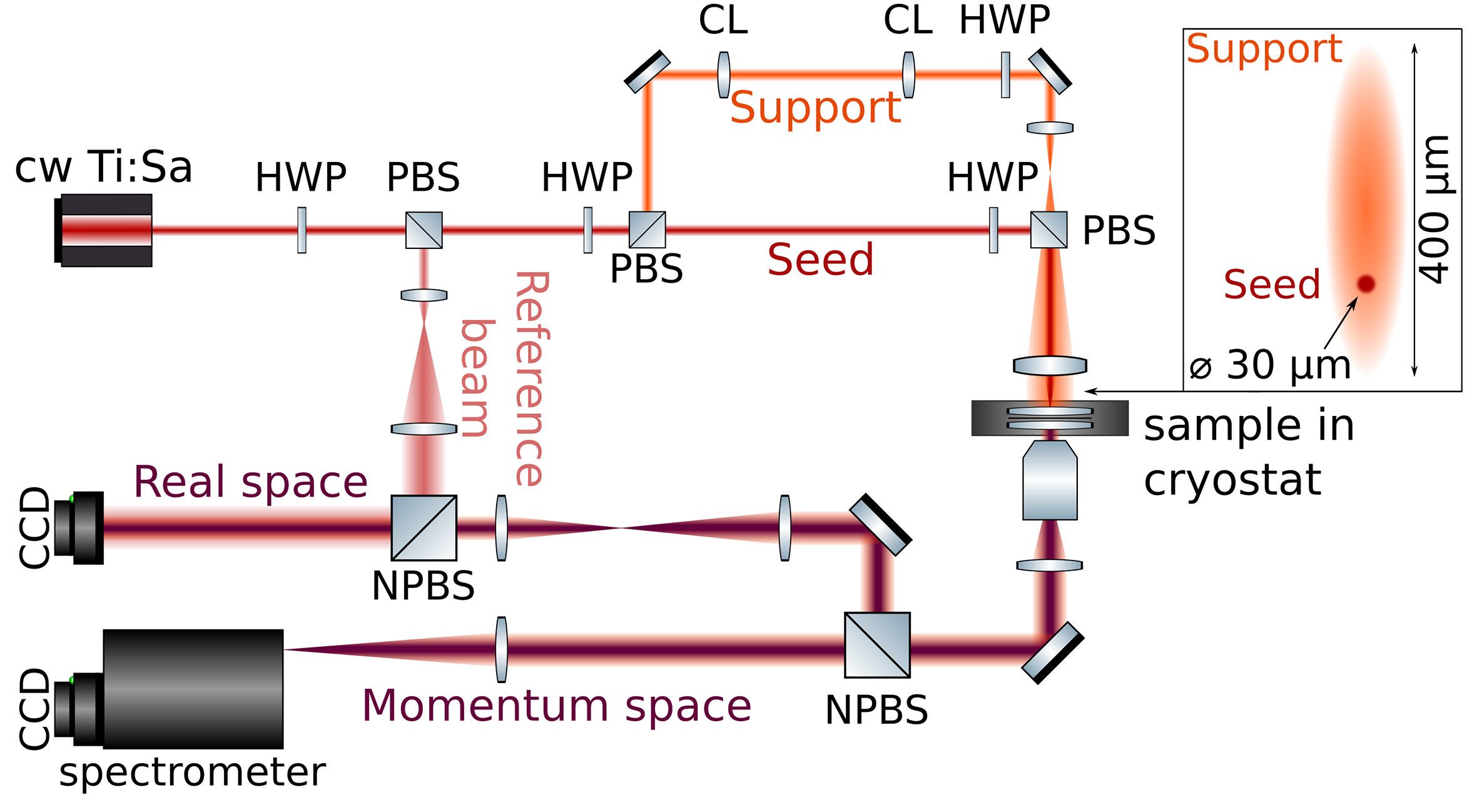}
    \caption{\textbf{Experimental setup}. The continuous Titanium-Sapphire laser beam (cw Ti:Sa) is split into three using half waveplates (HWP) and polarizing beam-splitter (PBS). A reference beam, in light red, is first set aside for interferogram on the detection part; the seed beam in red is focused on the sample to a spot of 30 microns diameter; the support in orange is extended and elongated in the vertical direction through cylindrical lenses (CL) before being collimated and sent into the microcavity. The seed and the support are not centered to one another (see inset) but share the same wavevector. The detection is done in real space, from which we can get information on the density and phase maps, and also in momentum space through the spectrometer.}
    \label{fig:SetupSeedSupp}
\end{figure}

\paragraph{Implementation}
We have implemented this proposal to experimentally verify it and our work got published in \cite{Lerario2020}: most of the figures of this section are adapted from the ones of the paper, as the setup displayed in figure \ref{fig:SetupSeedSupp}. The initial laser source is our continuous Titanium Sapphire Matisse laser. It is split up a first time into the main beam and the reference beam, later used to realize interferograms and get information on the phase. The main beam is later split a second time to generate the seed and the support beams. The seed is then focused on the sample into a spot of 30 microns diameter, for an intensity $I_{r} = 10.6$ W/mm\textsuperscript{2}. On the other hand, the support is elongated on the vertical axis through two cylindrical lenses, then collimated and sent to the sample as an elliptical spot of 400 microns length, with an intensity of 5.8 W/mm\textsuperscript{2}. The inset of figure \ref{fig:SetupSeedSupp} gives a representation of the relative position of the beams; the seed is not centered to the support so that the topological excitations can be studied on the flat part of the gaussian support beam.

The sample used for this experiment works in transmission: an objective was placed just after the cryostat with a large numerical aperture of 0.42. The detection is done in two spaces: the real space gives a copy of the polariton density in the plane of the cavity. The reference beam previously extracted from the initial laser beam can also interact with the signal of the cavity to get interferograms, used to reconstruct the phase map. All the other parameters of the excitation are coming from the momentum space data, acquired through a spectrometer. Details about the data extraction and analysis have been given in section \ref{sec:DataAnalysis}.

\paragraph{Vortex stream generation}

\begin{figure}[H]
    \centering
    \includegraphics[width=0.65\linewidth]{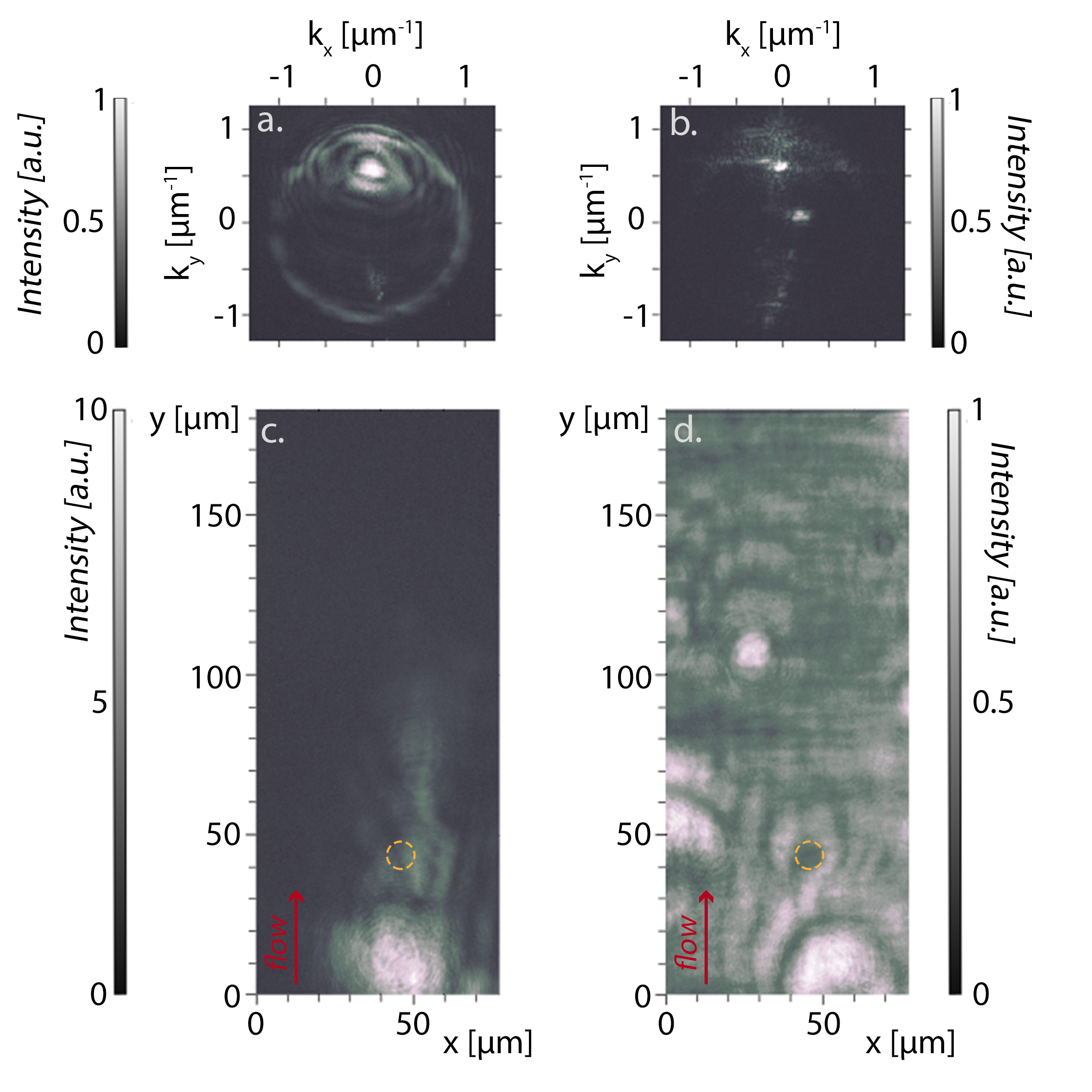}
    \caption{\textbf{Seed and support beams separately}. The left column presents the momentum and density maps of the seed alone, while the right one corresponds to the maps of the support beam. The momentum maps show the excitation spot of both beams at the same position. The support being collimated on the sample, its momentum space shows a very localized spot on panel b. The Rayleigh scattering ring appears due to sample inhomogeneities. The density maps display the range of the areas excited by both beams: on panel c., the seed is localized on the lower part of the image whereas the support on panel d. is everywhere. Note that the intensity scale of the picture c is ten times higher than the one of d. The defect used to generate vortices is circled in orange: in both cases, no topological excitations are observable.}
    \label{fig:VortSeedSuppkR}
\end{figure}

The first step of the experiment is to align the seed and the support beam independently. Figure \ref{fig:VortSeedSuppkR} shows the momentum and density maps of each field: the seed is on the left and the support on the right. 
On top are presented the \textbf{k}\textsubscript{x}-\textbf{k}\textsubscript{y} images, centered around $\mathbf{k} = 0$. Both excitation spots are at the same position, the beams enter with the same in-plane wavevector in the cavity ($|\mathbf{k}_{y}| = 0.6 $ \textmu m\textsuperscript{-1} and $|\mathbf{k}_{x}| = 0 $ \textmu m\textsuperscript{-1}). The support beam is however collimated on the sample while the seed is focused: in the momentum space, the support is therefore more localized than the seed.

In addition to the excitation spots, the images also show a circle, centered in $\mathbf{k} = 0$, and with a radius slightly larger than the excitation vector. 
It is the Rayleigh scattering ring due to the sample structural disorder and the elastic scattering on defects. The detuning between the excitation and the lower polariton branch, here $\delta E = 0.26$ meV is visible in the figure.

\paragraph{}
The density maps show us the size and location of each beam on the sample. The flow goes from bottom to top, at a velocity $v_{f} = 0.9$ \textmu m/ps. The seed in picture c. is very localized and placed at the bottom of the detection field of view. The support is much more extended and is present in all the detection picture.
An important point to notice is the intensity scales. The one of the seed is ten times higher than the one of the support: as predicted, the seed has to be much more intense than the support.
The support itself is inside the bistability cycle (see figure \ref{fig:SeedSuppBist}), alone, it generates a linear fluid on the lower branch of the bistability, with low density.

The structural defect used to generate topological excitation is circled in orange: in any of those cases, nothing appears in its wake, no vortices are generated. Indeed, the seed alone illuminates a small area of the fluid where it fixes its phase, preventing any turbulence generation; while the support alone creates a linear fluid with a too low density to observe such phenomenon.

\begin{figure}[h]
    \centering
    \includegraphics[width=\linewidth]{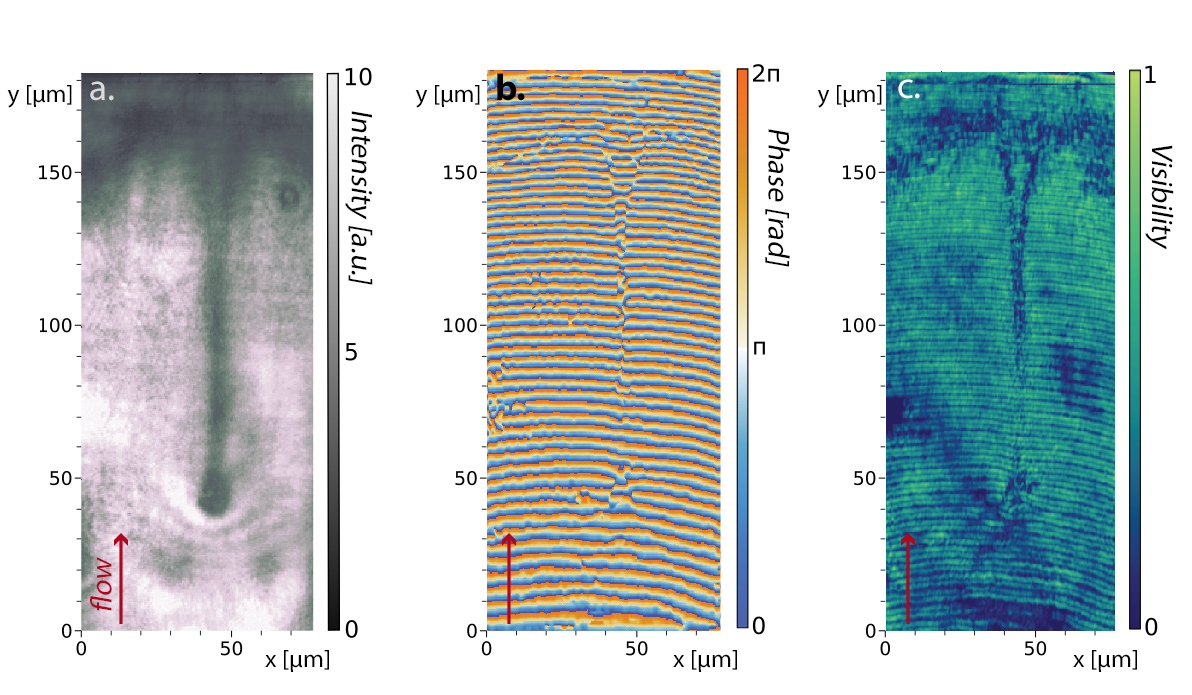}
    \caption{\textbf{Vortex stream generation}. a. Time integrated density map of a flow of vortex pairs generated in the wake of a defect. As the vortices move along the flow, the time integrated image results in a blurry density dip in the wake of the defect. b. Interferogram of the previous image, showing phase irregularities along the vortices propagation. c. Visibility map extracted from the interferogram, displaying a lower visibility along the vortex stream.}
    \label{fig:VortSeedSuppIphVis}
\end{figure}

\paragraph{}
The interesting situation happens when both seed and support are sent together. Figure \ref{fig:VortSeedSuppIphVis} presents an intensity map, an interferogram and a visibility map of this case.
As predicted, a shadow appears in the wake of the defect, which corresponds exactly to the time integrated image numerically calculated and presented in figure \ref{fig:VortSnapInt}. The pairs of vortex-antivortex are generated around the defect and follow the flow, leading to a decrease of density along their propagation path on the time integrated image.

\paragraph{}
To confirm that this shadow is indeed due to the presence of vortices, an interferogram has been realized as well as a visibility map, displayed on figures \ref{fig:VortSeedSuppIphVis}b. and c.
As the vortices are moving along the flow, one can not expect to spot forks, the typical signature of vortices, as they would move by the time the picture is taken (as a reminder, the order of magnitude of the integration time is the millisecond, while the one of the flow speed is 1 \textmu m/ps).
However, the presence of the vortices is visible on the phase map as a blur of the fringes along their path. It is in particular more visible after a few micrometers, while just after the defect the visibility dip is not so evident: this can be due to the fact that the flow is accelerated around and just after the defect \cite{Frisch1992}, which leads to a lower density of vortices.

In order to further insure that the vortex generation is responsible for the shadow, a visibility map as been extracted from the interferogram and shown in figure \ref{fig:VortSeedSuppIphVis}c. If the irregularities of the phase pattern are indeed a phase blur due to the the flow of vortices, the fringes visibility should decrease. This is exactly what is observed: the shadow and the phase variations coincide with a dip in visibility, confirming its attribution to the vortex stream.

\paragraph{}
Another interesting feature to observe in those pictures is the fact that at the top end of the fluid, the stream separates into two thinner and darker lines.
This is due to the fact that in this region, vortices merge together and become gray solitons. 
Indeed, the vortex core size is controlled by the hydrodynamic properties of the system, and in particular its healing length $\xi$, defined as \cite{Carusotto2013a}:

\begin{equation*}
\xi = \dfrac{\hbar}{\sqrt{2m^{*}\hbar gn}}
\end{equation*}

with $m^{*}$ the effective mass of the polariton and the product $gn$ the polariton density. 
On the pictures of figure \ref{fig:VortSeedSuppIphVis}, one can see that we are at the edge of the polariton fluid, and that the dark region on top of the density map corresponds to a linear fluid.
But before reaching it, the non linear fluid on the upper bistability branch sees its density slowly decreasing. This density decrease has a direct impact on the healing length which inversely increases, leading to vortex cores bigger and bigger. 
At some point, vortices on one hand and antivortices on the other hand reach their respective neighbor, and merge together into a gray soliton.
A precise phase jump is indeed visible at the corresponding position of the interferogram.
It can also be explained it terms of Mach number, as the decrease of density means also a decrease of the sound speed and therefore an increase of the Mach number: the flow becomes supersonic there.

\begin{figure}[h]
    \centering
    \includegraphics[width=\linewidth]{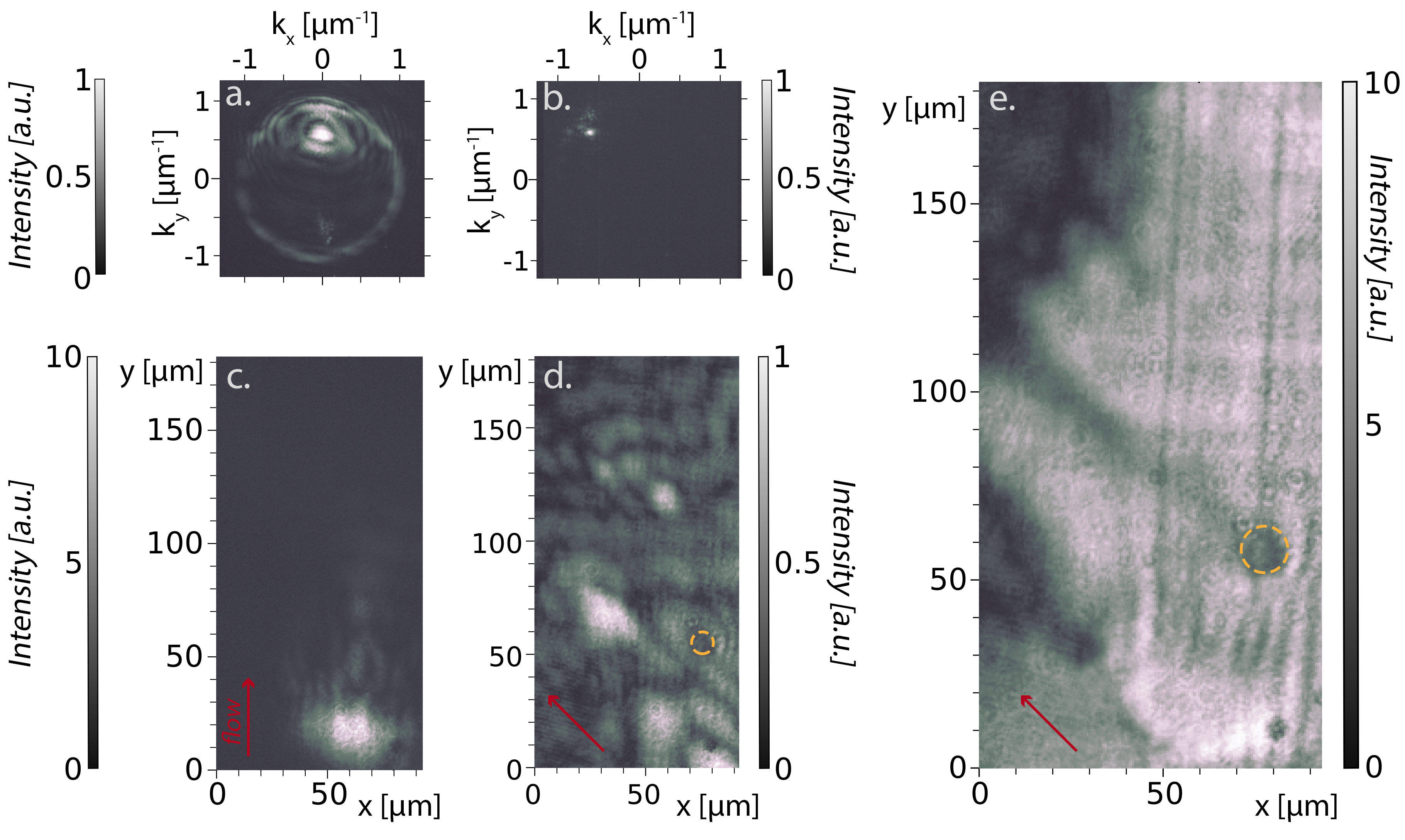}
    \caption{\textbf{Vortex stream propagation direction}. The first column presents the results of the seed alone, sent with a vertical wavevector. a. corresponds to the momentum space image and c. is the density map. The second column is the support, tilted of about 40\textdegree\ compared to the seed; again, b. is the \textbf{k}\textsubscript{x}-\textbf{k}\textsubscript{y} image and d. is the real space. Note that the scales of c. and d. have a factor 10 between them. The position of the defect is circled by the dashed orange line. e. shows the density map resulting in the superposition of seed and support. The vortex stream line is again visible, this time tilted in the direction of the flow.}
    \label{fig:VortExpTiltedFlow}
\end{figure}

\paragraph{}
We have tried another configuration reported in figure \ref{fig:VortExpTiltedFlow}. 
This time, the two beams where not sent with the same wavevector, as shown in pictures a. and b. 
While the seed in a. still has only one component in \textbf{k}\textsubscript{y}, the support (picture \ref{fig:VortExpTiltedFlow}b.) is displaced of roughly 40\textdegree\ compared to the seed, and has components in both \textbf{k}\textsubscript{x} and \textbf{k}\textsubscript{y} directions.

Density maps of each beams alone are presented in images c. (seed) and d. (support). As previously, the seed is localized, and induces a light vertical flow of polariton in its wake.
The support is much more extended, and due to its \textbf{k}\textsubscript{x} component, it creates a diagonal polariton flow. The considered defect is positioned through the orange dashed circle.
Again, the scales of the density maps are different by a factor ten.

\paragraph{}
The density map obtained when both beams are sent together is presented in picture \ref{fig:VortExpTiltedFlow}e. As before, the surface of the nonlinear fluid is greatly enhanced, and the vortex stream is generated after the defect. 
This time however, the stream direction has changed and follows the flow imposed by the support orientation. The vortex direction can thus be easily controlled by tuning the excitation wavevector of the support beam.

\begin{figure}[h]
    \centering
    \includegraphics[width=0.8\linewidth]{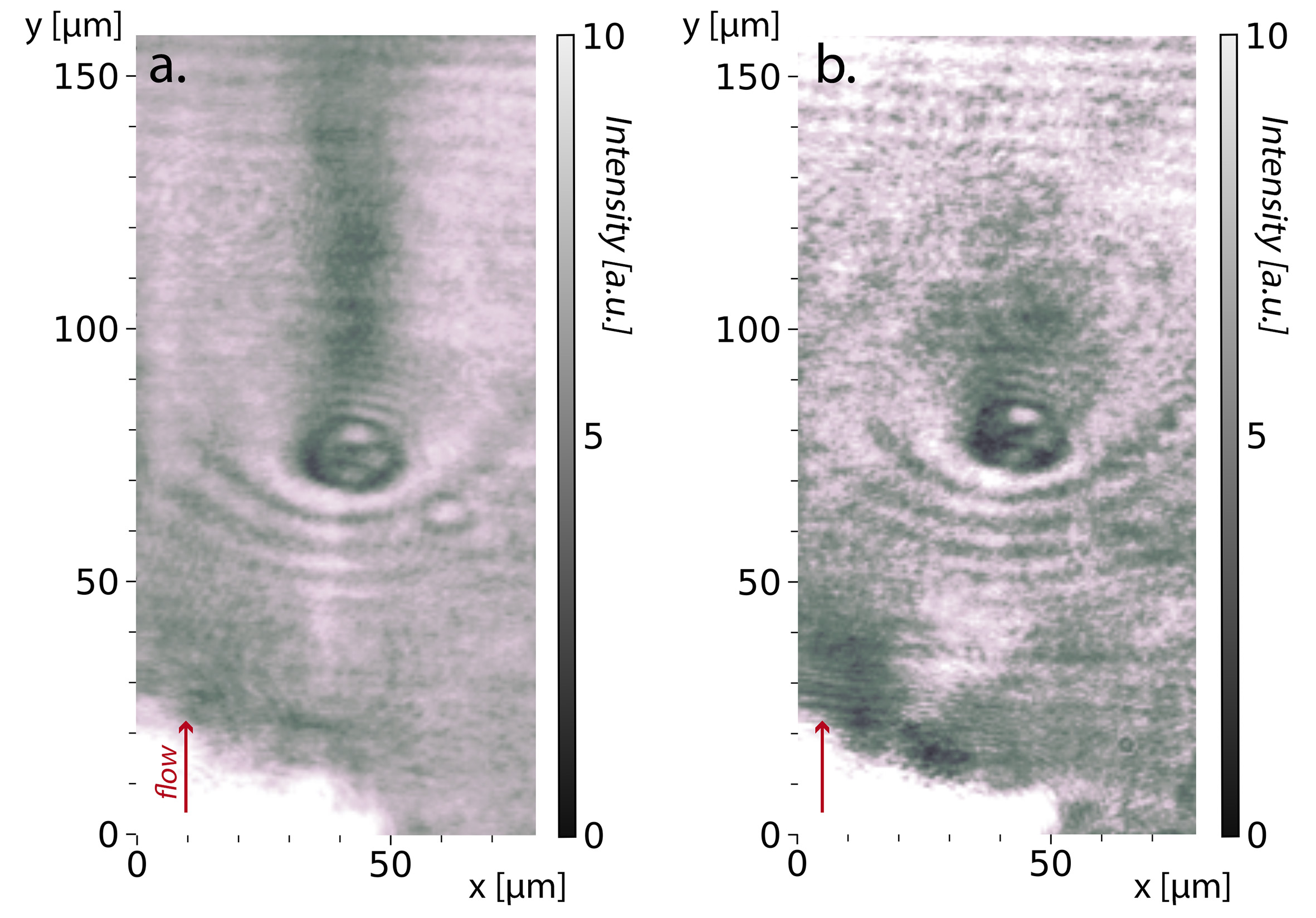}
    \caption{\textbf{Vortex stream suppression}. a. Vortex stream in the wake of a defect under 5.8 W/mm\textsuperscript{2} support power density. b. Suppression of the vortex stream for a support power density of 6 W/mm\textsuperscript{2}}
    \label{fig:VortExpSupprHD}
\end{figure}

\paragraph{}
A last theoretical prediction was experimentally realized: the confirmation that the vortices are generated only if the support intensity is on the left side of the bistability cycle.
Figure \ref{fig:VortExpSupprHD} displays two images between which only the support intensity varies: on a., the support power density is 5.8 W/mm\textsuperscript{2}, while it increased to 6 W/mm\textsuperscript{2} on picture b. 

The first picture exhibits the typical dark shadow associated with the vortex stream generation. But in the second one, the support intensity has become too high to allow the topological excitation to generate and propagate: the phase is fixed by the driving field and therefore flat. 
No shadow appears, the vortex stream is inhibited.

\section{Supersonic flow: parallel dark soliton pair generation}
\label{sec:SpontSolitons}

\paragraph{}
The subsonic case inducing the generation of vortices and their sustainability for macroscopic distances, the supersonic configuration is now investigated. According to previous studies in polaritons superfluid \cite{Amo2011}, higher Mach numbers, around or above 1, allow for the generation of dark solitons. 
However, the system configuration did not allow for long propagation: the quasi-resonant pump fixing the phase of the fluid, it was sent only upstream of the defect and the solitons were observed within the exponential decay of the solitons propagation. Therefore, they could only be seen for short distances (around 30 micrometers, limited by the polariton lifetime) and with changing conditions along their propagation as the density exponentially decreased.
The goal of this part is thus to use the previous configuration of seed-support excitation, and to generate dark solitons sustained over long distances to be able to study their hydrodynamic behaviour. 
After a brief summary of the physics of solitons, experimental results and theoretical description are presented.

\subsection{State of the art}

\paragraph{}
Solitons are nonlinear solitary waves which have the property of conservation of their shape and velocity during their propagation.
The first reported observation of such phenomenon dates from 1834 and was realized by the Scottish engineer John Russel who described it as such \cite{Russel1844}:

\paragraph{}
\textit{
I was observing the motion of a boat which was rapidly drawn along a narrow channel by a pair of horses, when the boat suddenly stopped—not so the mass of water in the channel which it had put in motion; it accumulated round the prow of the vessel in a state of violent agitation, then suddenly leaving it behind, rolled forward with great velocity, assuming the form of a large solitary elevation, a rounded, smooth and well-defined heap of water, which continued its course along the channel apparently without change of form or diminution of speed. I followed it on horseback, and overtook it still rolling on at a rate of some eight or nine miles an hour [14 km/h], preserving its original figure some thirty feet [9 m] long and a foot to a foot and a half [30-45 cm] in height. Its height gradually diminished, and after a chase of one or two miles [2–3 km] I lost it in the windings of the channel. Such, in the month of August 1834, was my first chance interview with that singular and beautiful phenomenon which I have called the Wave of Translation.
}

\paragraph{}
Even though Russel passed many years focusing on the study of such solitary waves, the scientific community did not realized at the time the significance of his work.
The first theoretical understanding of their dynamics was made by Korteweg and de Vries \cite{Korteweg1895} in 1895 who derived a simple equation of propagation of wave on water surface traveling only in one direction, the KdV equation. 
However, it is only in the 1960's that researchers understood that this equation's solutions had the form of solitary waves, which exhibit properties of particles, as their shape is conserved during propagation or even after collision with each other. From that came the name \textit{solitons} as "particles of a solitary wave".

\paragraph{}
For long, the term "solitons" was only used to describe such nonlinear stationary shape-maintaining waves, but as their study expanded, it now refers to a wider range of phenomena which can evolve during propagation, sometimes accelerate, decompose or form coupled states by interacting \cite{Maimistov2010}. 
Two families of solitons can be easily distinguished, depending on their impact on the system they evolve in.
If the system is not impacted by the crossing of the wave, and keeps the same energetic state before and after it, we talk about non topological solitons, described by the KdV equation.
If however the system has a different state on both side of the wave, it is a topological solitons which can be described by different equations as for instance the non-linear Schr\"{o}dinger equation or the Sine Gordon equation, as it is the case in our system with a phase jump across the soliton \cite{Maimistov2010}. 

\paragraph{}
In the last few decades, solitons have been studied in many different field of physics. Obviously, the study started on water surface and shallow water for fluid dynamics \cite{Benjamin1967, Yuen1975, Melville1982, Denardo1990}. Later on, solitons were created as temporal pulses in optical fibers \cite{Tai1986, Emplit1987} or as spatial structures in waveguides \cite{Swartzlander1991}.
They were also studied in thin magnetic films \cite{Chen1993, Shinjo2000} as well as in semiconductor microcavities \cite{Barland2002a}. Liquid Helium \cite{Williams1999} or complex plasma \cite{Heidemann2009} were also able to sustain those features. 
Theoretical works were realized starting from the 70s, in particular in the context of Bose-Einstein condensates \cite{Tsuzuki1971, Zakharov1973, Pitaevskii1961, Gross1961, Gredeskul1989}, which were followed a few decades later by experimental observations after the first realization of atomic condensates \cite{Anderson1995a, Bradley1995, Davis1995}.
Research focused much on matter-wave solitons \cite{Burger1999, Denschlag2000, Dutton2001, Anderson2001b, Bongs2001} as one of the first purely nonlinear states experimentally realized in BECs \cite{Frantzeskakis2010b}.

\paragraph{}
In any cases, all those phenomena result in the balance between the wave dispersion, which tends to spread it out, and the nonlinear properties of the medium, which on the contrary compresses the wave. Nonlinearity is therefore essential for the solitons to be sustained.
In optics in particular, two cases need to be separated: the self-focusing and the self-defocusing nonlinearity. Self-focusing results from a positive correction to the refractive index, proportional to the intensity: the beam's divergence is suppressed and the beam collapses. This phenomenon allows for the generation of bright solitons, characterized by an intensity peak above a continuous background \cite{Kivshar2000}. On the other hand, self-defocusing supports the propagation of dark solitons as an intensity dip on a continuous background.

\paragraph{}
On polariton fluids, the first observation of solitons was realized in our group ten years ago \cite{Amo2011}. Sending the polariton flow toward a structural defect with suitable size, solitons were generated in its wake. 
The system was resonantly pumped: as previously explained, a strong localized pump fixes the phase of the fluid and prevents the formation of topological excitations. Adopting a theoretical proposal \cite{Pigeon2011}, a half circle shaped pump was implemented just upstream of the defect and the solitons were observed along the free propagation of the polaritons.

In the case of a spontaneous formation in an unpumped region, the two generated solitons are oblique: they are formed just after the defect, then propagate straight with an aperture angle $\alpha$ between them.
This angle can be linked to the Mach number of the system  through the theory of dark solitons \cite{Pitaevskii2003a, Carusotto2013a}: it predicts that solitons are well sustained for supersonic speeds \cite{Kamchatnov2008a}. 
At subsonic speeds, they are subject to snake instabilities that can lead to their decay into vortices \cite{Anderson2001b}, even though their propagation is kinematically not prohibited.
Indeed, the experiment of Amo \textit{et al.} presents stable solitons at subsonic speeds: it can be explained as a result of the finite lifetime of the polaritons \cite{Kamchatnov2012}. 

\paragraph{}
The main restriction of this configuration is the fact that the solitons propagate in an unpumped region, and consequently in a region where the polariton density exponentially decays. 
Not only does it limits the propagation length of the solitons, it also limits their study as the hydrodynamic conditions are constantly changing along the propagation.
Even though, many researches have been pursued since then to understand better the kinematic of solitons in polaritons fluid, like the study of their stability \cite{Grosso2011, Yulin2015}, the generation of half solitons in spinor condensates \cite{Flayac2011, Hivet2012a}, the observation of bright solitons \cite{Sich2012} or the formation of solitons train \cite{Goblot2016b, Caputo2019}.
Different experimental configurations have been implemented to obtain all those results, however none of them was able to generate dark solitons and to sustain their propagation for macroscopic distances.

\paragraph{}
The idea that led to our work was to use the seed support configuration at higher Mach number than before. Combining the hydrodynamic conditions of dark solitons generation with a support field inside the bistability cycle should greatly enhance the solitons propagation distance and allow for a deeper study of their properties.

\subsection{Parallel dark solitons observation}

\paragraph{Experimental implementation}
The implementation of this experiment is similar to the one presented in section \ref{sec:ExpVortSpont}. The experimental setup is the same, with two colinear beams sent to the cavity: the seed, localized and intense, and the support, extended and with an intensity within the bistability cycle. 

\begin{figure}[h]
    \centering
    \includegraphics[width=0.6\linewidth]{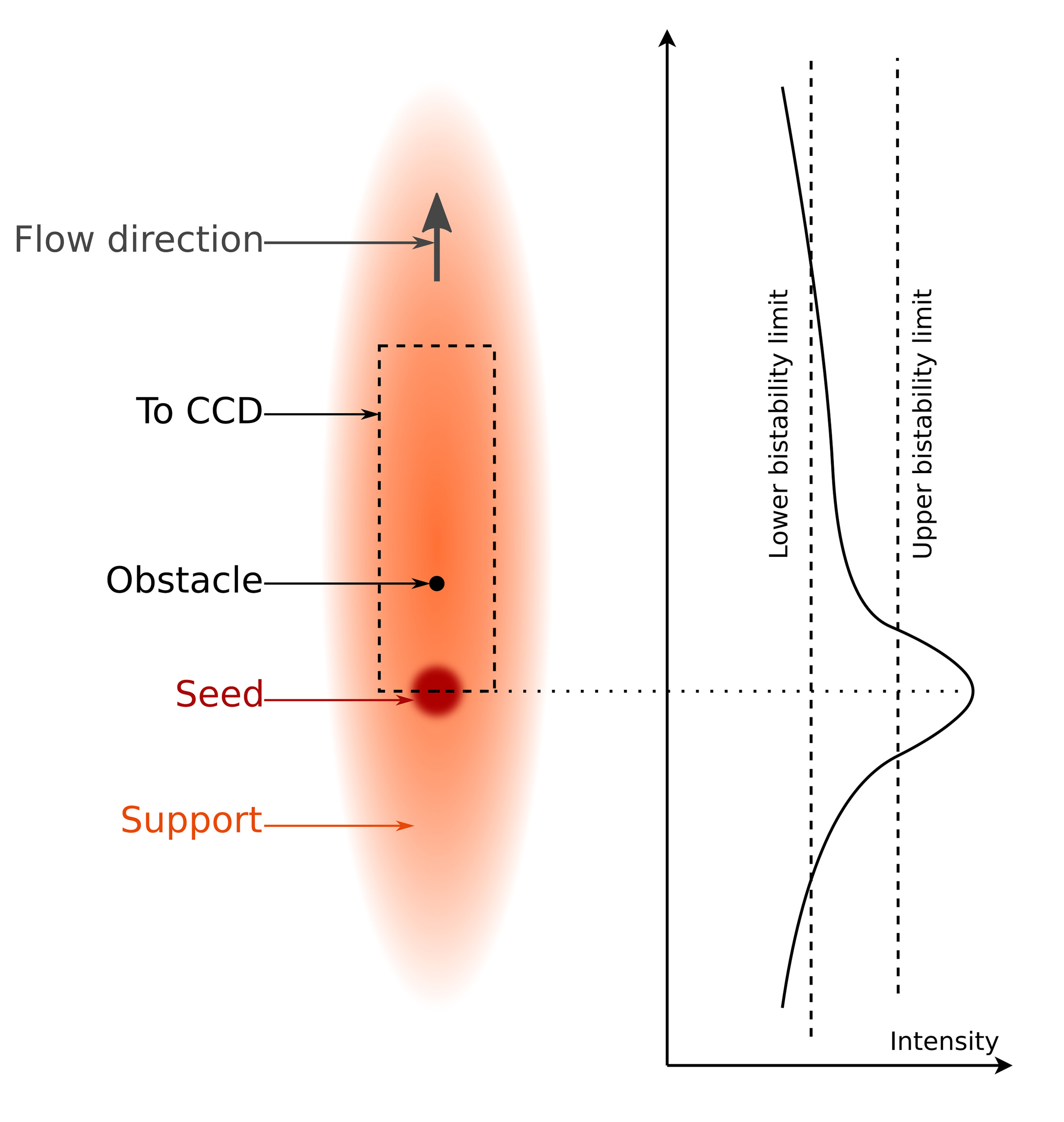}
    \caption{\textbf{Seed and support beams distribution}. The support beam is elongated along the vertical direction and its intensity is within the bistability cycle. The flow is from bottom to top, and a structural defect is placed in the center of the support. Upstream to it is sent the seed, localized and with an intensity above the bistability cycle. The distribution of the total intensity along the vertical axis is displayed on the right as well as the upper and lower bistability limits.}
    \label{fig:SolBeams}
\end{figure}

\paragraph{}
The position of the seed and the support beams on the sample are shown in figure \ref{fig:SolBeams}. The seed is shifted compared to the support center, to be placed just upstream of the considered structural defect - the flow is from bottom to top. 
The curve on the right shows the distribution of the intensity in the cavity along the vertical axis. The region close to the seed is above the upper bistability limit, while the main part of the support is inside the cycle. The combination of the two beams ensures the fluid to be bistable and on the upper branch of the cycle.

\paragraph{Dark soliton pairs generation}
To generate dark solitons in the wake of a defect, supersonic conditions need to be created. The in-plane wave vector is chosen to be high ($k = 1.2$ \textmu m\textsuperscript{-1}) in order to ensure a high velocity of the fluid: in this case, $v_{f} = 1.52$ \textmu m/ps. 
The sound of speed is extracted through the energy renormalization and is measured to be $c_{s} = 0.4$ \textmu m/ps: the supersonic conditions are reached.

\begin{figure}[h]
    \centering
    \includegraphics[width=0.6\linewidth]{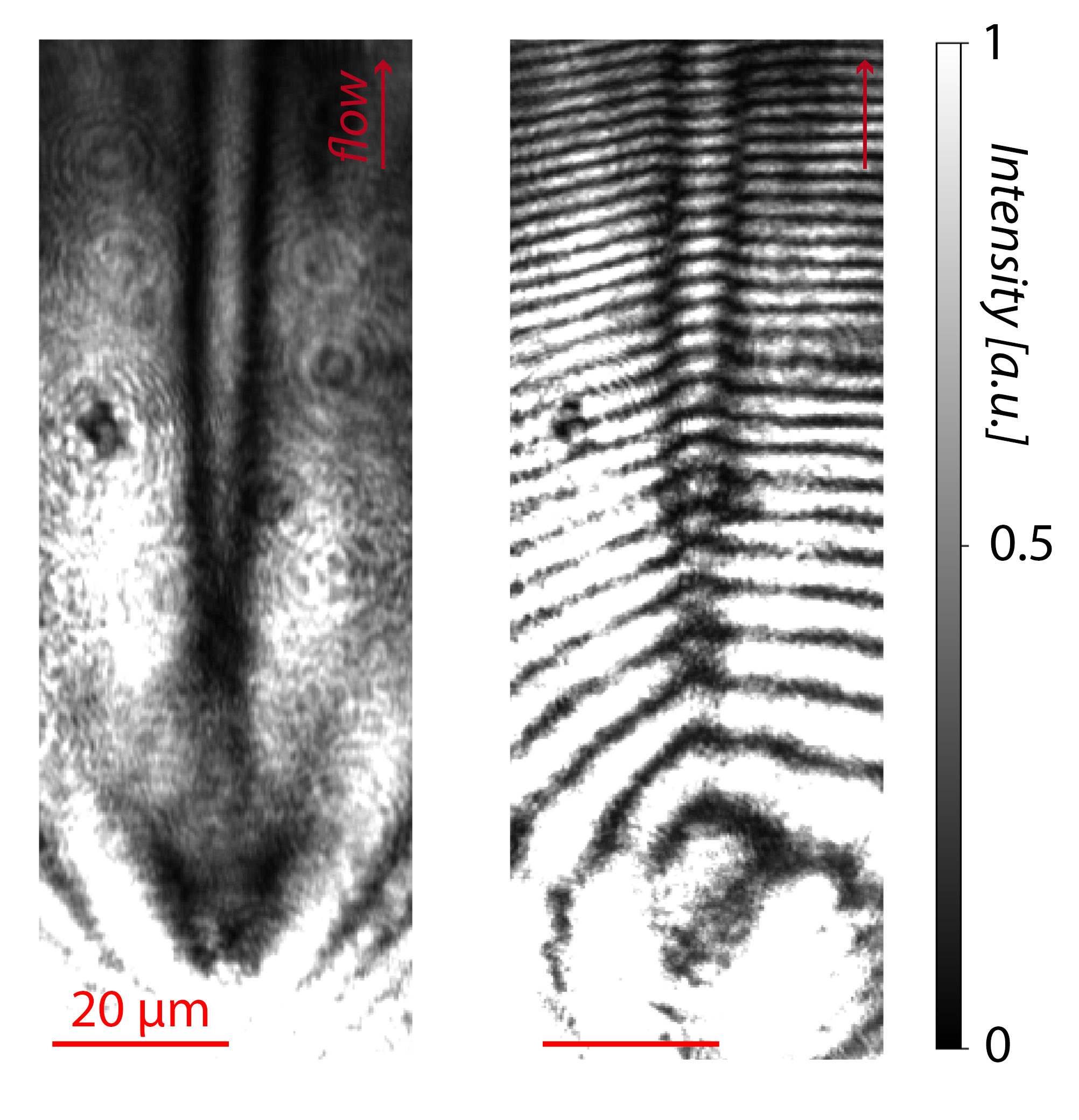}
    \caption{\textbf{Spontaneous generation of a pair of parallel dark solitons}. Intensity (left) and interferogram (right) of a dark soliton pair spontaneously generated in the wake of a structural defect. They propagate away from each other for a short distance, before aligning and staying parallel for over a hundred microns.}
    \label{fig:SolParrExp}
\end{figure}

\paragraph{}
The figure \ref{fig:SolParrExp} presents the observation of a spontaneous generation of dark solitons in the wake of a defect: intensity on the left and interferogram on the right.
The first conclusion that can be done from those results is that the propagation length is indeed greatly enhanced. The scale bar illustrates 20 microns: the solitons themselves are sustained for more than a hundred microns, one order of magnitude more than previously reported.

Moreover the solitons have a surprising behaviour. They are generated at the same position and propagate away from each other for a few microns; but eventually, they reach an equilibrium separation distance of about 8 microns, from which they align and stay parallel for as long as they can be sustained. This is quite unexpected to observe such a bound state of dark solitons as they have repulsive interactions and usually constantly repel each other \cite{Zakharov1973}.

The observed solitons are fully dark: not only does the intensity dip goes to zero once the background is removed, but the phase jump associated with the solitons is very close to \textpi\ all along the propagation, confirming their transverse velocity is close to zero.

\begin{figure}[h]
    \centering
    \includegraphics[width=0.9\linewidth]{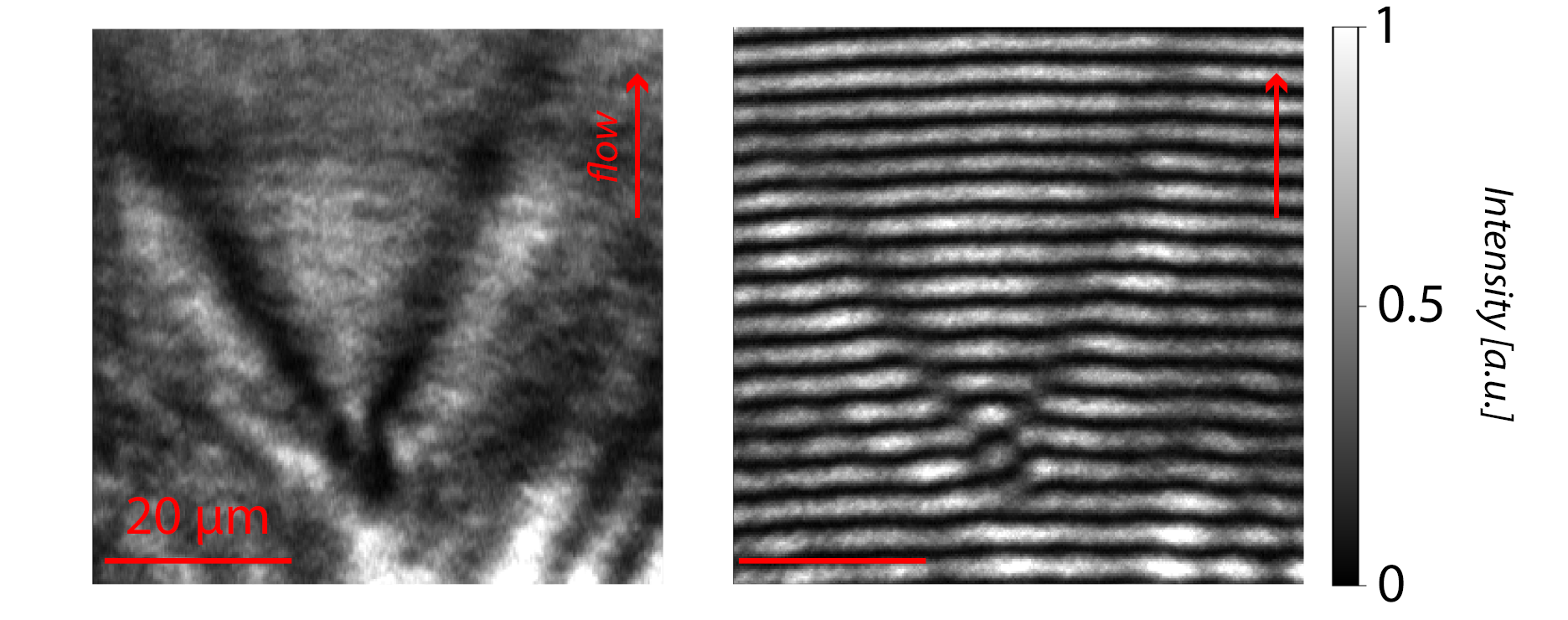}
    \caption{\textbf{Spontaneous generation of a pair of gray solitons}. Intensity (left) and interferogram (right) of the hydrodynamic generation of oblique gray solitons in the wake of a structural defect. The strong and localized pump is placed about 20 microns upstream of the defect, outside of the field of view, hence the good contrast of the picture.}
    \label{fig:SolOblExp}
\end{figure}

\paragraph{Oblique grey solitons generation}
In order to have a compared analysis of the different cases, we performed an experiment in a configuration close to the one of Amo \textit{et al.} \cite{Amo2011}, with a strong pumping upstream of the defect. The results are presented in figure \ref{fig:SolOblExp}: again, the left image corresponds to the intensity map while the right one is the phase interferogram.
In this case, only the seed is sent: it is localized about twenty microns before the defect, in order to surely avoid the phase fixing of a strong coherent pumping. It is therefore outside of the field of view of figure \ref{fig:SolOblExp} in order to enhance contrast in the picture.

\paragraph{}
The generated solitons have a different shape than in the previous configuration, as they stay oblique for their whole propagation. They are generated dark with a phase jump of \textpi\, but do not stay that way. They vanish along the propagation, and with it their phase jump gradually reduces.
Their propagation length is also much shorter than in the presence of the support field: they have completely vanished after 40 microns propagation. 
Finally, their width is large: about 10 \textmu m full width at half maximum (FWHM) for the oblique solitons while the parallel one was fitted at 2.8 \textmu m FWHM.

\begin{figure}[h]
    \centering
    \includegraphics[width=0.6\linewidth]{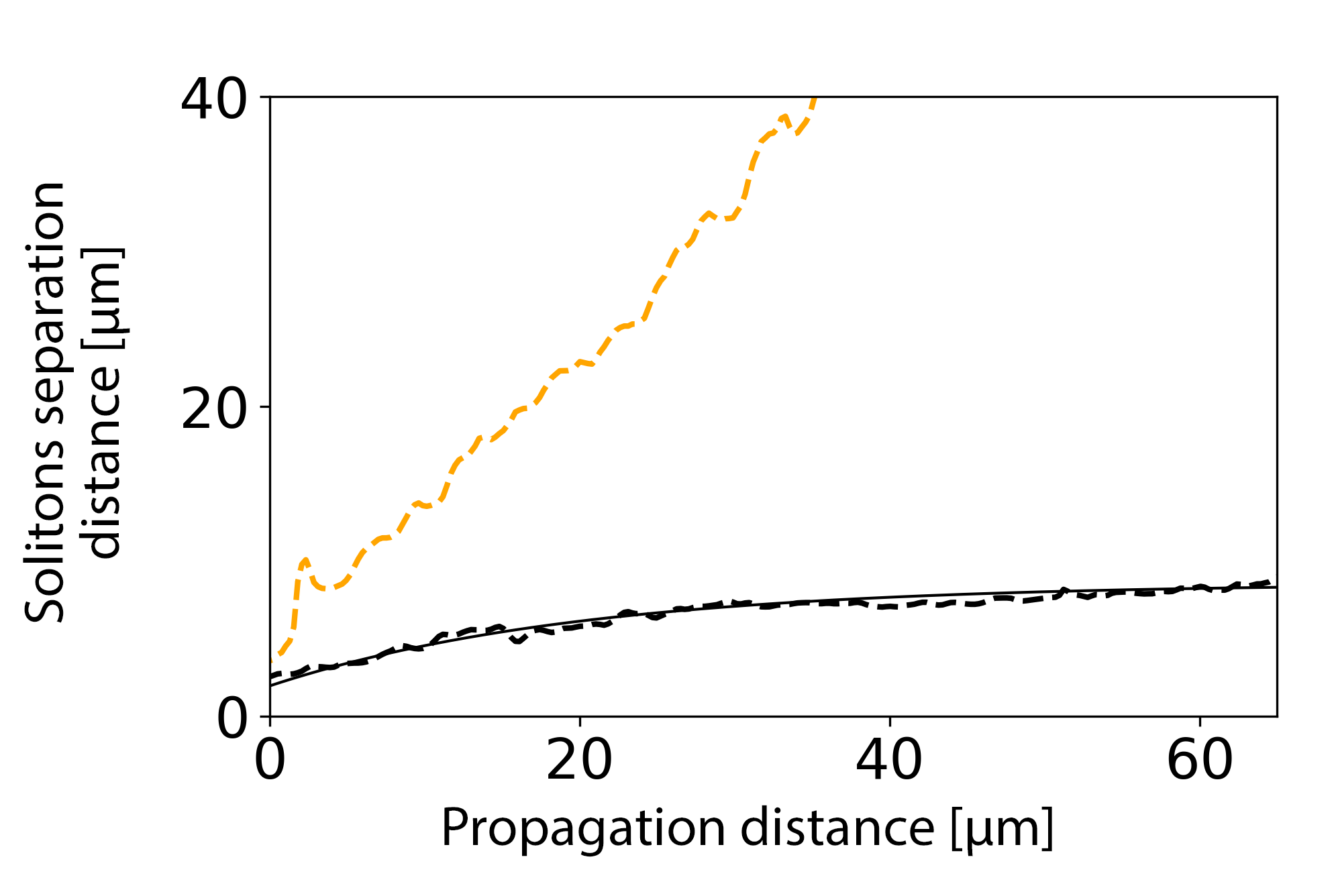}
    \caption{\textbf{Spontaneous solitons separation distance}. The dotted lines show the experimental values of the separation distance between the solitons along their propagation: the orange line corresponds to the oblique solitons presented in figure \ref{fig:SolOblExp}, while the black one is extracted from the parallel solitons in figure \ref{fig:SolParrExp}. The solid black line is a fit of the parallel solitons data.}
    \label{fig:SolPropDistComp}
\end{figure}

\paragraph{}
A direct comparison between the two cases previously presented has been done, in particular about the separation distance.
Figure \ref{fig:SolPropDistComp} presents the evolution of the separation distance between the solitons along their propagation, for both cases.

The dotted lines show the experimental results: in black the parallel solitons with both seed and supports fields, and in orange the oblique case with only the seed.
The black solid line is a fit of the experimental points, in order to get a value of the equilibrium separation distance. The fitting function is chosen to be $2 d_{0} (1 - exp(-x/d_{l}))$ in order to get a transient and a stationary region. The transient region corresponds to the first part of the propagation where the solitons are oblique, with a asymptotic length of $d_{l} = 14$ \textmu m. The solitons then reach their stationary region where they propagate parallel, at an equilibrium separation distance of $2 d_{0} = 8$ \textmu m. 

The oblique solitons always follow the same trend: they continuously go further to one another, before completely vanishing after 40 microns propagation.

\subsection{Numerical approach}

\paragraph{}
To fully understand the phenomenon taking place in the system, numerical simulations in collaboration with the group of Guillaume Malpuech in Clermont-Ferrand have been realized to reproduce the solitons behaviour.
They used the coupled equations of the excitons and cavity photons fields, $\psi_{X}$ and $\psi_{\gamma}$:

\begin{gather}
    i \hbar \dfrac{\partial \psi_{\gamma}(\mathbf{r}, t)}{\partial t} =
    \bigg[ - \dfrac{\hbar^{2} \nabla^{2} }{2m_{\gamma}^{*}} + V(\mathbf{r}) - i \Gamma_{cav} \bigg] \psi_{\gamma}(\mathbf{r}, t) + V \psi_{X}(\mathbf{r}, t) 
    + ( S(\mathbf{r}) + R(\mathbf{r})) e^{-i\omega_{0}t}
    \\
    i \hbar \dfrac{\partial \psi_{X}(\mathbf{r}, t) }{\partial t} = 
    [V(\mathbf{r}) + g_{X} |\psi_{X}(\mathbf{r}, t)|^{2} - i \Gamma_{X} - \Delta_{X} ]
    \psi_{X}(\mathbf{r}, t) + V \psi_{\gamma} (\mathbf{r}, t)
\end{gather}

with $m_{\gamma}^{*}$ the effective cavity photon mass, $\omega_{0}$ the laser energy, $\Gamma_{cav}$ and $\Gamma_{X}$ the photon and exciton lifetimes respectively,  $V$ the half Rabi splitting and $\Delta_{X}$ the cavity-exciton detuning.
All these parameters have been taken the same as the experimental ones.

$V(\mathbf{r})$ is a 10 meV potential barrier that reproduces the presence of the structural defect, using a Gaussian shape of 10 \textmu m width.
The spatial profiles of the seed and support are modeled through $R(\mathbf{r})$ and $S(\mathbf{r})$ respectively, with adjustable magnitudes.

\begin{figure}[h]
    \centering
    \includegraphics[width=0.6\linewidth]{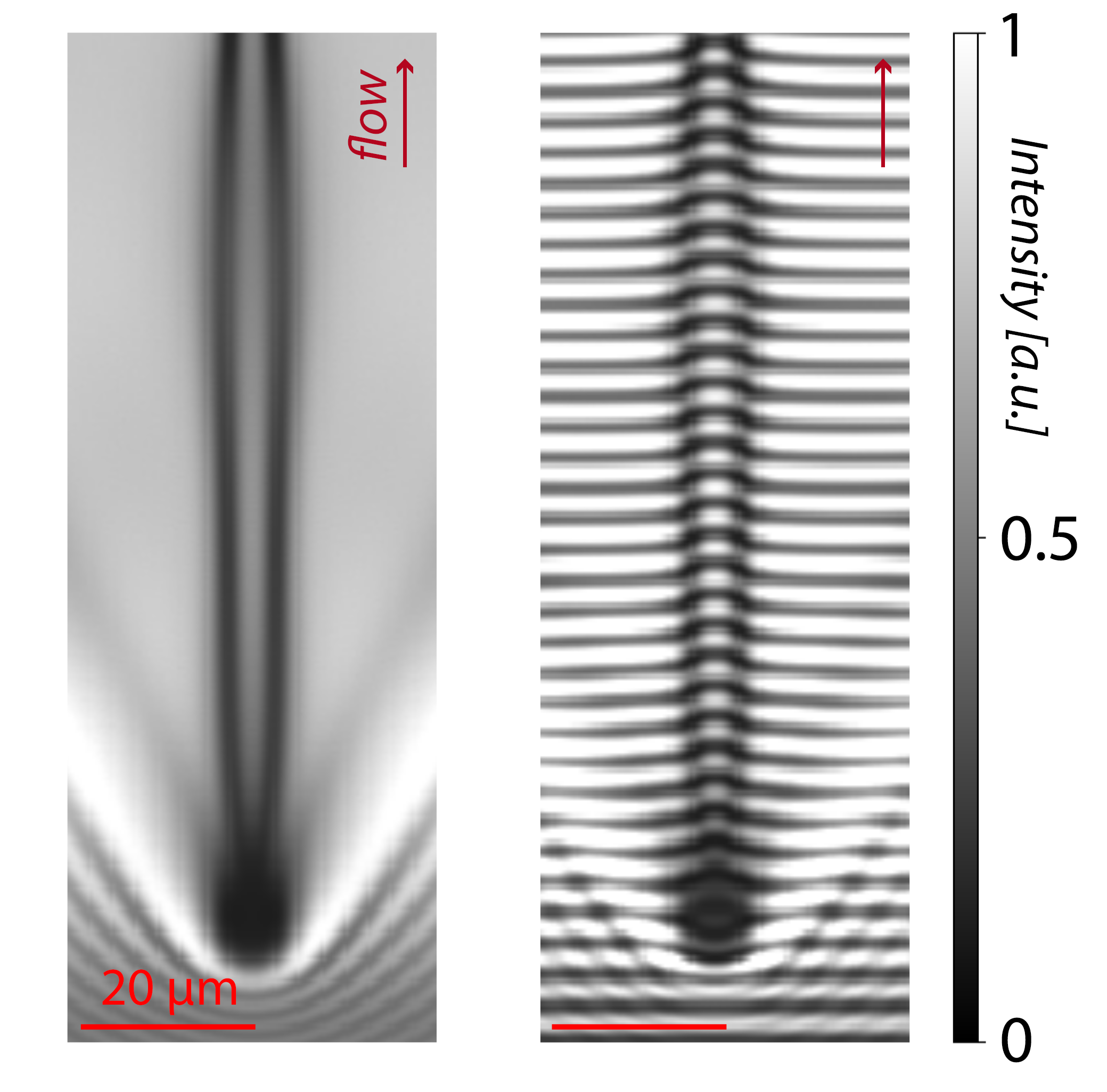}
    \caption{\textbf{Simulation of the spontaneous generation of dark parallel solitons}. Using the driven-dissipative Gross-Pitaevskii equation, intensity (left) and interferogram (right) simulation of a dark soliton pairs spontaneously generated in the wake of a defect. }
    \label{fig:SolParrSimu}
\end{figure}

\paragraph{}
The results of those simulations are presented in figure \ref{fig:SolParrSimu}. They offer a perfect agreement with the experimental pictures of figure \ref{fig:SolParrExp}. The left image is the intensity map, switched on to the upper branch of the bistability cycle by combination of both the localized intense seed and the bistable extended support. 
The interference pattern on the right exhibits a clear phase jump all along the propagation, confirming the solitonic nature of the intensity dips. 
As expected, the solitons are sustained for a macroscopic distance and stay parallel to one another during their propagation.

\paragraph{}
The use of the numerical simulations allows for a better resolution in time, and in particular to understand the establishment of the steady state.
The seed and the support need to be sent simultaneously, but it does not matter which one is turned on first. Once the two beams enter the system, the flow created by the seed is strong enough to propagate to all the support illuminated area and switch it to the upper branch of the bistability.
During this expansion phenomenon, the high density region is delimited by a domain wall, separating it from the low density region. 
When this wall hits the defect, it got split into a V shape around the defect, still pushed by the expansion of the high density area which tends to reduce the angle between its two branches. 
Different cases can occur depending on the velocity of the walls. The first possibility is that they spread away and vanish soon, leaving only a blurry gray region in the wake of the defect.
They can also move quickly toward one another, and switch everything to the upper branch of bistability, which means a flat high intensity everywhere.
Finally, if they go toward the center but with a lower velocity, this one can be compensated by the repulsive interaction between the nonlinear shock waves associated with the domain walls. The stabilization of the soliton pair can therefore be seen as a balance between the repulsion of the dark solitons and the pressure of the domain walls sending them toward one another.

\paragraph{}
Another important result to notice is the fact that the solitons come as pair.
Indeed, the phase jump across one soliton is \textpi, and the total phase jump across the fluid must remain null. The total number of solitons have consequently to be even.

\paragraph{}
Numerically, the propagation distance of the dark solitons is not limited, as the support beam is modeled to be everywhere. It does not correspond exactly to the experimental configuration, as the wedge of the cavity and the support finite extension should be taken into account.

%\bibliographystyle{unsrt}
%\bibliography{bibs/LKB-bibs-bib_thesis-SpontaneousChap}  

%\end{document}
%\documentclass[a4paper,11pt]{book}
%\usepackage[utf8]{inputenc}
%\usepackage[T1]{fontenc} 
%\usepackage{lmodern} 
%\usepackage[margin=28mm,includeheadfoot,bindingoffset=5mm]{geometry}[2010/03/13]

%\usepackage{graphicx}
%\usepackage{amsmath}
%\usepackage{bbold}
%\usepackage{amssymb} % pour le signe \lesssim
%\usepackage{textcomp} % \textdegree
%\usepackage[most]{tcolorbox} 
%\usepackage{enumitem} 
%\usepackage{xcolor}
%\usepackage{float}
%\usepackage{lscape}
%\usepackage{physics} 
%\usepackage{stmaryrd}
%\usepackage{wasysym} 
%\usepackage{tikz}
%\usepackage{cite}
%\usepackage{hyperref}
%\usepackage{textgreek}
%\newcommand*\circled[1]{\tikz[baseline=(char.base)]{
%          \node[shape=circle,draw,inner sep=2pt] (char) {#1};}}
%\renewcommand{\thesubsubsection}{\roman{subsubsection}}
%\tcbset{enhanced,colback=red!5!white, colframe=red!75!black,fonttitle=\bfseries}
%\graphicspath{{figures/}} %Setting the graphicspath
%\setcounter{tocdepth}{3}
%\setcounter{secnumdepth}{3}

%\begin{document}

%\tableofcontents

\setcounter{chapter}{3}

\chapter{Impression of bound soliton pairs in a polariton superfluid}
\chaptermark{Impression of bound soliton pairs}
\label{chap:Impression}

\paragraph{}
We described in the previous chapter how to greatly enhance the propagation length of a polariton superfluid, and simultaneously getting rid of the phase constraint of the pump. 
This allowed us to observe the spontaneous generation of topological excitations and their propagation for over a hundred microns. However, their generation was not controlled, and depended on parameters out of our reach; in particular, the necessary presence of a structural defect to induce the turbulence leading to the topological excitations \cite{Lerario2020, Lerario2020a}.

\paragraph{}
The starting point of this chapter is to get rid of this limitation and to be able to artificially create solitons in the fluid. 
It is realized by imprinting a phase modulation on the system, leading to the formation of dark solitons that can evolve freely on the nonlinear fluid. 
Such impression of solitons offers many tools for the detailed study of this phenomenon, as their shape and position can be tuned on demand. 
It results once again in the unexpected binding mechanism between the imprinted solitons, leading to the propagation of a dark soliton molecule \cite{Maitre2020}.

\paragraph{}
After an introduction of the experimental implementation and of different solitons configurations, the hydrodynamic behaviour of these structures is studied, as well as their connection with the driven-dissipative nature of the system.

\newpage

\section{Experimental implementation}
\label{sec:expimpl}

\paragraph{}
To be able to properly imprint solitons on the polariton fluid, some specific arrangements have been used in the excitation part. 
In particular, the pump beam has been designed to simultaneously allow a phase impression and a free evolution of the system to sustain the solitons propagation.

After a detailed description of the setup, the experimental results will be presented, as well as the different shapes and configurations tried on the fluid.

\subsection{Beam shaping}

\subsubsection{Intensity and bistability}

\paragraph{}
The main difficulty of this experiment is to combine an impression of a phase modulation, i.e. a region of the system whose phase is imposed by the pump; and a free propagation area, \textit{i.e.} a region where the phase of the pump beam is flat but where the system can still sustain topological excitations.

\begin{figure}[h]
    \centering
    \includegraphics[width=0.9\linewidth]{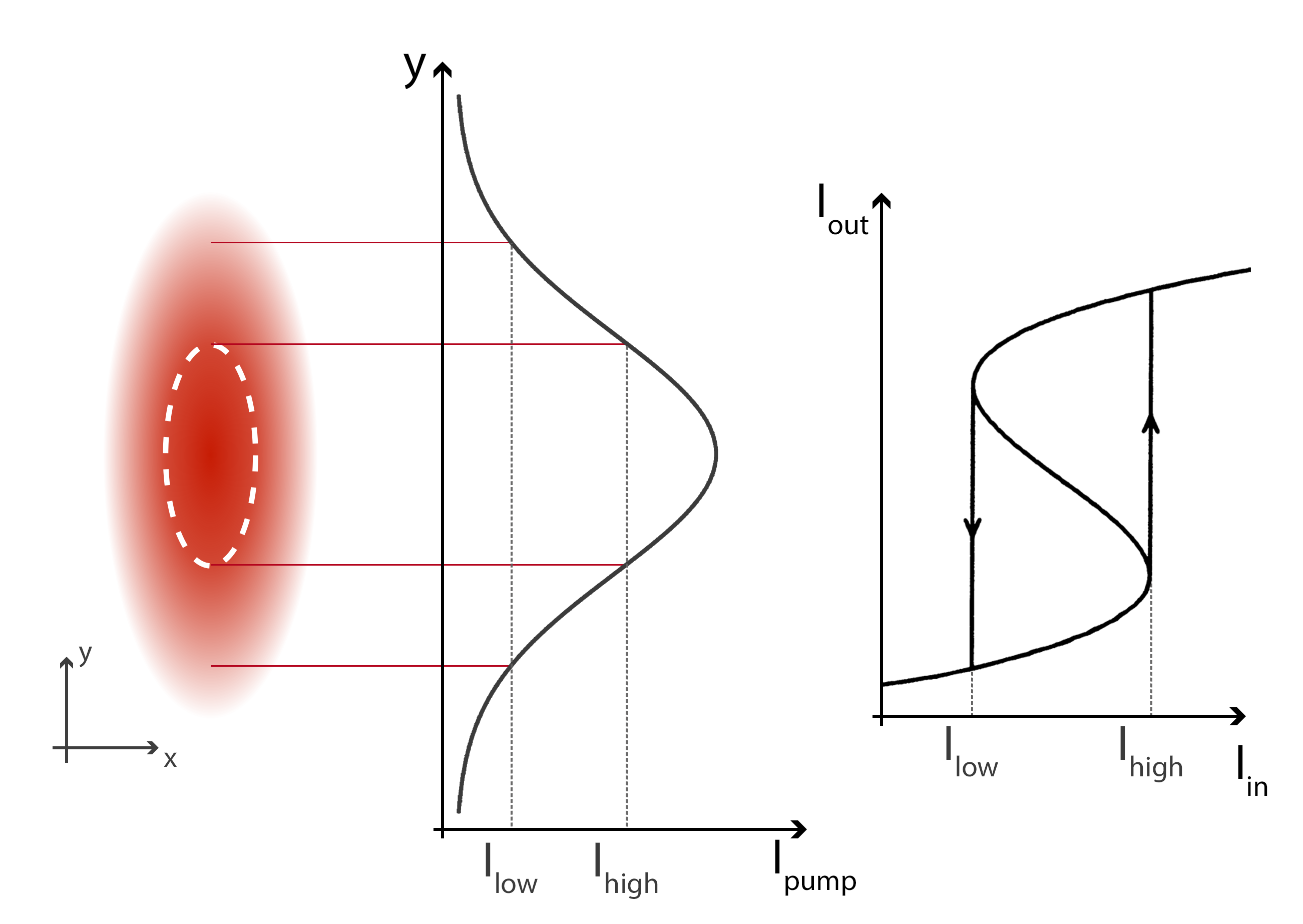}
    \caption{\textbf{Beam intensity and bistability}. Left, spatial distribution of the excitation beam. The white dashed line illustrates the position of I\textsubscript{high}, upper intensity threshold of the bistability cycle, as shown in the intensity profile on the center image. On the right, a reminder of the theoretical bistability cycle with the definition of I\textsubscript{high} and I\textsubscript{low}}
    \label{fig:BeamInt}
\end{figure}

It is achieved by using the properties of the system's optical bistability (see section \ref{sec:OpticalBist}). Indeed, the input intensities above the bistability cycle impose their phase on the fluid, while this constraint is released for the input intensities within the bistability cycle.
The two situations are obtained through the gaussian shape of the excitation beam, as illustrated in figure \ref{fig:BeamInt}. 
The spatial distribution of the beam is plotted on the left, where the white dashed line indicates the position of the high intensity threshold of the bistability $I_{high}$ (see right picture). 
All the area inside this circle is above the bistability cycle, as shown in the profile in the center. Therefore, it fixes the phase of the fluid and corresponds to the effective impression region.
Outside of this circle however, most of the beam intensities are within the bistability cycle: the phase is not imposed anymore and the system is able to readjust it and to sustain the free propagation of topological excitations.
The beam is elongated in the \textit{y} direction in order to flatten its profile and extend the bistable region where the solitons free propagation will be studied.

\subsubsection{Phase profile design}

\paragraph{}
As the solitons induce a phase jump on the system, their implementation can be done by modeling the phase of the excitation. To do so, we use a Spatial Light Modulator, a liquid-crystal based device that can shape the wavefront of an incident light (see section \ref{sec:SLM}). 

\paragraph{}
The phase modulation induced by dark solitons is a phase jump of \textpi: the phase profile corresponding to a pair of dark solitons is thus an elongated region \textpi-shifted compared to the background beam. 
The solitons must not be imprinted everywhere but only in the impression region: a rectangular shape in the center of the beam should be created.

\begin{figure}[h]
    \centering
    \includegraphics[width=0.6\linewidth]{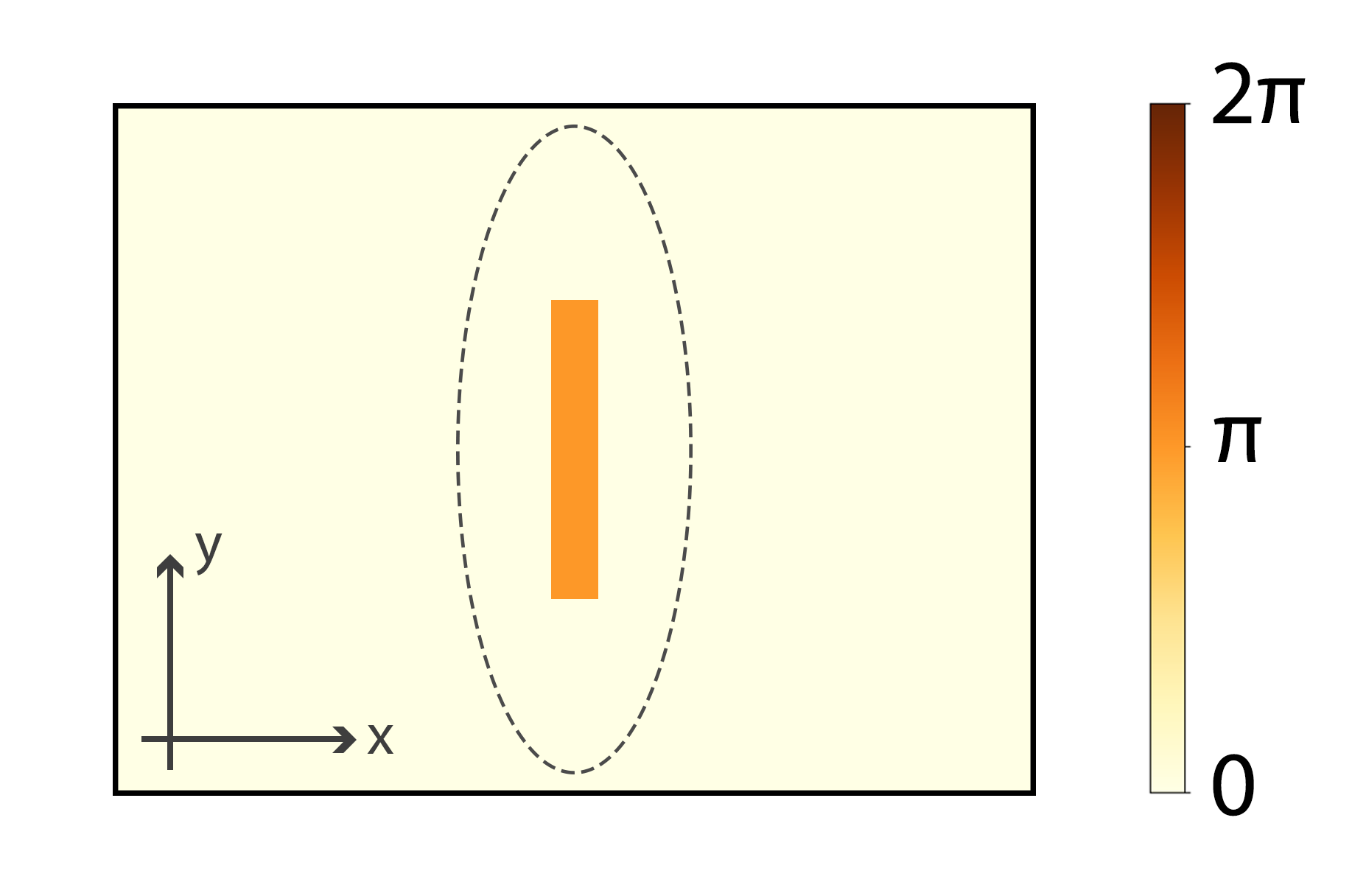}
    \caption{\textbf{SLM profile}. Typical image sent to the SLM to create a \textpi-shifted rectangular shape on the excitation beam. Its position and shape are easily tunable. The dashed line corresponds to the position of the incident beam on the SLM screen.}
    \label{fig:SLMpatt}
\end{figure}

The SLM is very convenient for the design of the soliton phase profile. Furthermore, as it is entirely controlled via a software, the position of the solitons can be finely tuned to match ideally with the excitation beam.
The figure \ref{fig:SLMpatt} shows a typical image sent to the SLM to create the solitons pattern. 
The SLM reproduces it on its screen which transfers it to the beam wavefront: it results, as desired, in a rectangular phase jump of \textpi\ which consequently induces an intensity dip. The dashed line illustrates the incident beam position on the screen of the SLM.

\paragraph{}
The efficiency of the SLM is not perfect (90\% ): in order to get rid of the residual light not shaped by the device, we implement a Fourier filtering. In addition to the desired phase pattern, we add to the image a grating, so that the light effectively shaped by the SLM is diffracted in the first order, while the rest stays in the zeroth one. 
In the Fourier plane of the SLM, all the orders of the grating focus in different spots: only the first order signal is kept, which cleans the beam of all impurities due to the reflection on the SLM.

\subsubsection{Spatial filtering}
\label{sec:SpatialFilt}

\paragraph{}
The filtering in the Fourier space is actually not only for cleaning the beam from SLM impurities: we also use it to finely tune the shape of the phase jumps. Indeed, we want to imprint the solitons vertically, along the \textit{y} axis: we do want the horizontal phase jumps to be as sharp as possible, in order not only to induce dark solitons well imprinted on the system, but also such that they can freely continue their propagation along \textit{y} through the bistable region of the fluid. 
However, the rectangular shape created by the SLM also induces short phase jumps along the \textit{y} direction, that we want to avoid in the fluid. 
Indeed, they join the two horizontal phase jumps, limiting the region in phase opposition: on the fluid, if they are too well defined, they prevent the \textit{y}-solitons to follow their path and connect them together.

\paragraph{}
To avoid this situation, the Fourier filtering is used to blur this vertical phase jump. Practically, the filter is a slit of tunable width along \textit{y}.
It thus cuts the frequency components in the \textit{y} direction, while leaving the \textit{x} ones unchanged.

\begin{figure}[h]
    \centering
    \includegraphics[width=0.95\linewidth]{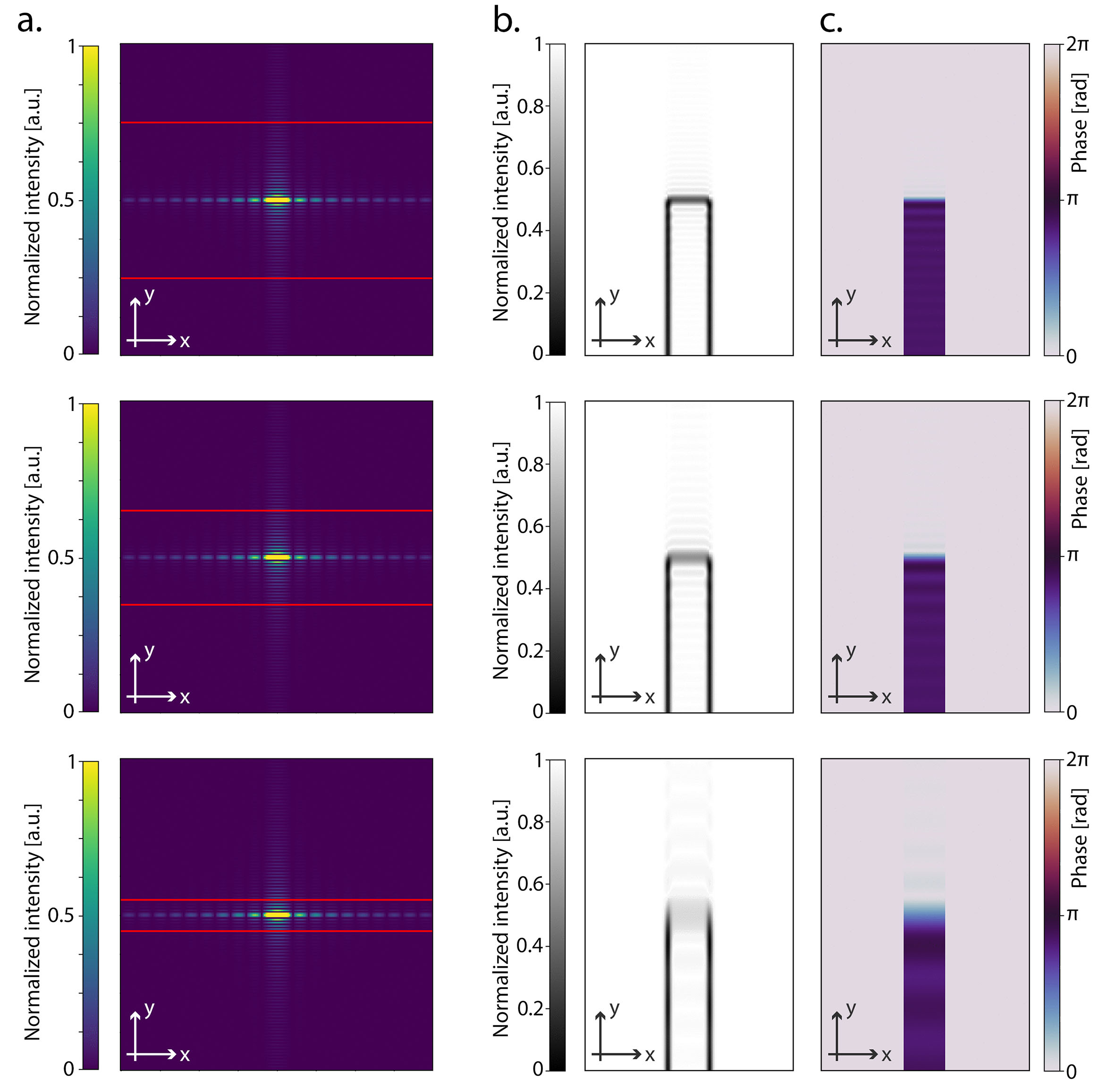}
    \caption{\textbf{Filtering in the SLM Fourier space}. Fourier space of the SLM (a.) and the different filtering indicated by the slit opening in red. The top part of the corresponding filtered signal is shown in intensity (b.) and phase (c.). The vertical phase jump definition is defined by the slit opening, while the horizontal ones are left unchanged.}
    \label{fig:Filtering}
\end{figure}

\paragraph{}
The figure \ref{fig:Filtering} presents the effect of the slit on the filtered signal. The left column (figure \ref{fig:Filtering}.a) shows the signal in the Fourier space of the SLM (zoomed on the first order of the grating, \textit{i.e.} the one we keep), and the opening of the slit is marked by the red horizontal lines. The second and third columns show the top part of the corresponding real space signal after the slit, in intensity (fig. \ref{fig:Filtering}.b) and phase (fig. \ref{fig:Filtering}.c).

\paragraph{}
The Fourier space of the SLM has some frequency components in the \textit{x} and in the \textit{y} directions, corresponding to the horizontal and vertical phase jumps.
On the upper line, the slit is quite open and only a few components of the vertical phase jump are removed: the filtered real space is therefore very similar to the signal designed by the SLM, and both the intensity dip and the phase jump along the \textit{y} direction are well defined.

\paragraph{}
By reducing the slit opening (see medium line), more components of the vertical phase jump are cut, which blurs its definition in intensity as well as in phase.

Finally, on the bottom line, the slit is almost closed: in that case, most of the vertical components are filtered out, which leads to an extended and shallow intensity dip and a smooth phase jump. Meanwhile, all of the horizontal components are kept: therefore the horizontal phase jump keep their good definition and are unchanged by the filtering.

\begin{figure}[h]
    \centering
    \includegraphics[width=0.6\linewidth]{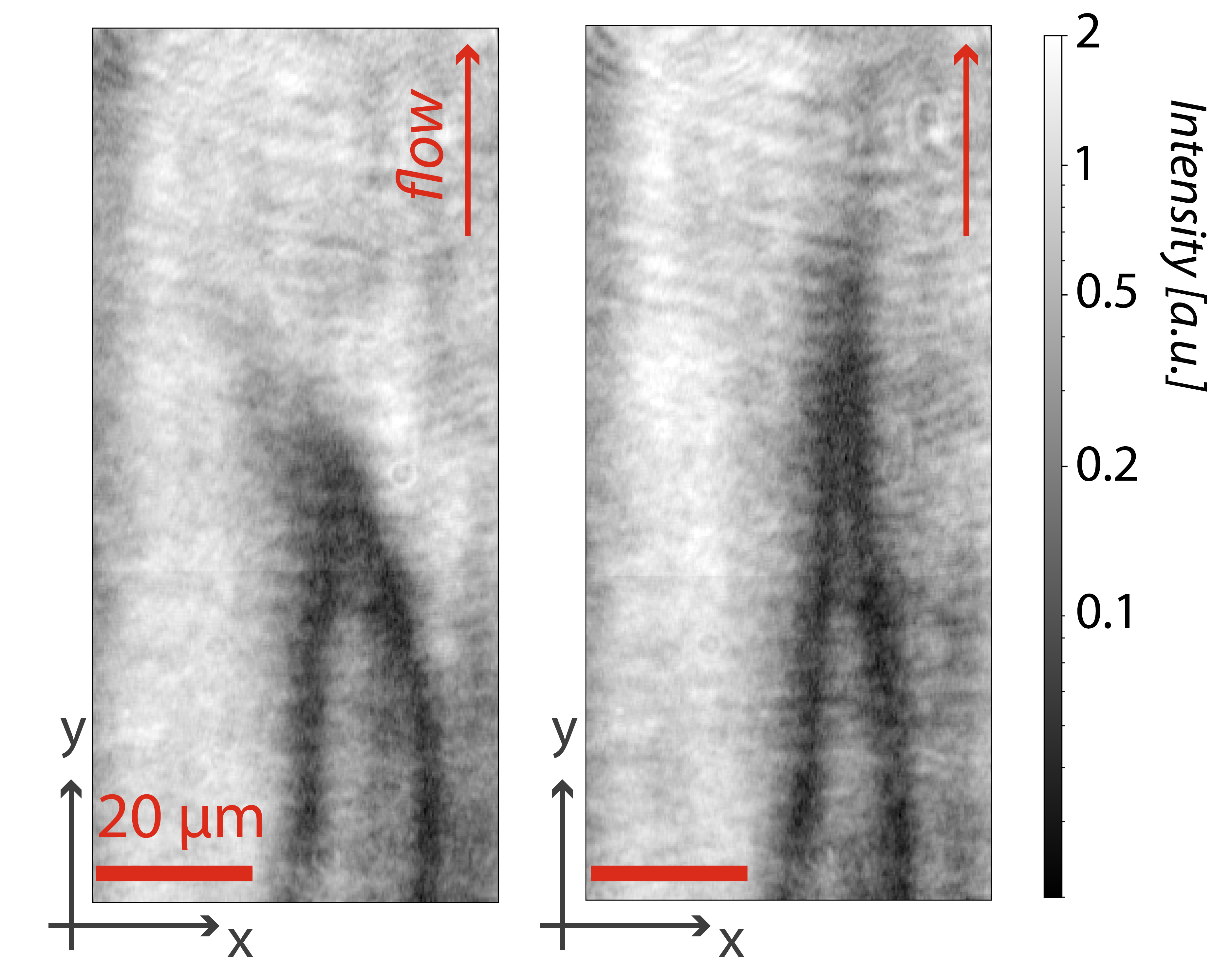}
    \caption{\textbf{Effect of the slit on the fluid}. Images if the top end of the imprinted solitons on the fluid, at high pump intensity preventing any free propagation. On the left, the solitons obtained with a wide slit. Because of the presence of the flow from bottom to top, the solitons are slightly stretched, but join together quickly. On the right, the slit is much more closed: the solitons are very elongated and their respective transverse velocity is low when the join together.}
    \label{fig:SlitImgSamp}
\end{figure}

\paragraph{}
The result of the filtering procedure on the fluid impression is illustrated in figure \ref{fig:SlitImgSamp}.
The images are a density map of the cavity plan in logarithmic scale, in a high power configuration: all the fluid is above bistability and gets its phase and properties from the pump, preventing any free propagation of the solitons. 

They are centered on the top part of the imprinted solitons.
The fluid is created with a flow, from bottom to top on the pictures.

On the left, the slit is open, filtering out the grating components but keeping most of the ones of the soliton pattern.
Due to the presence of the flow, the horizontal soliton is stretched, but join the vertical solitons together on a short distance along \textit{y}.

On the right, the slit is much more closed, filtering out most of the components of the vertical phase jump of the pump. As this one is therefore sent very smooth to the sample, it induces in the fluid an extension of the vertical solitons along \textit{y}: they use all the large \textit{y} range of the blurred phase jump to join, finally meeting each other with a low transverse velocity.
That configuration should benefit the further propagation of the solitons.

\subsubsection{Global experimental setup}

\begin{figure}[h]
    \centering
    \includegraphics[width=\linewidth]{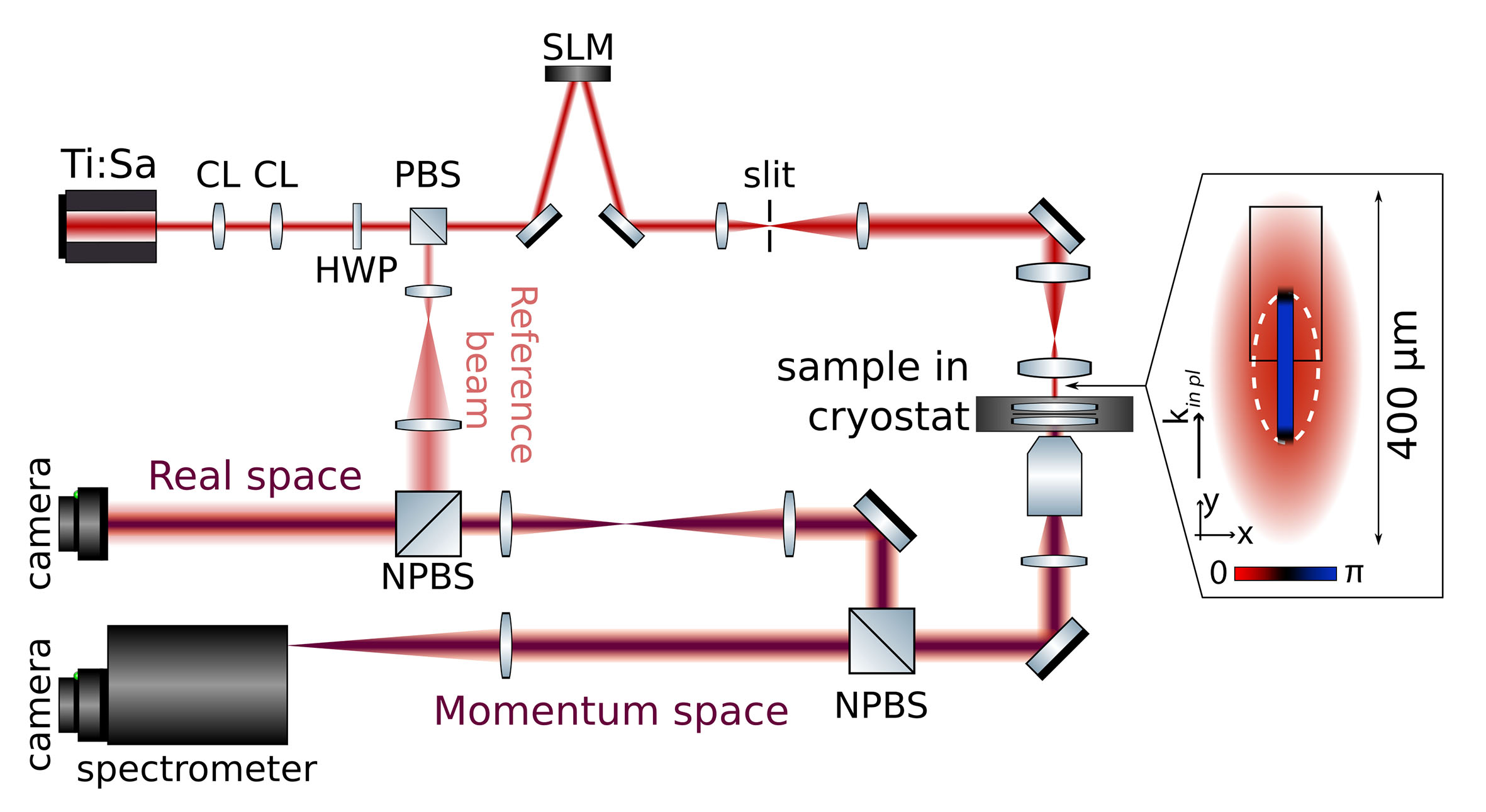}
    \caption{\textbf{Experimental setup}. The excitation beam is designed by the SLM and filtered through the slit. It is collimated on the sample, imaging the SLM plan. The inset illustrates the beam configuration: the solitonic pattern is in the center, where the intensity is above the bistability limit, located at the white dashed line. The detection field of view is delimited by the black rectangle, shifted from the center in order to focus on the bistable region and the solitons free propagation. The detection is done in real and momentum space, so that the experimental conditions can be associated with the corresponding intensity and phase maps.}
    \label{fig:SetupImpr}
\end{figure}

\paragraph{}
The experimental setup is sketched in figure \ref{fig:SetupImpr}. The laser source is a Titanium Sapphire Matisse, and its spot is elongated in the \textit{y} direction by two cylindrical lenses (CL). The beam is then split in two by a polarizing beam splitter (PBS) preceded by a half-wave plate (HWP). 
The main beam is shaped as desired: the SLM draws the phase front, later filtered by the slit. It is sent collimated to the cavity, imaging the SLM plan so that the phase jumps are well defined on the sample.
The inset illustrates the excitation beam configuration on the sample: the solitonic pattern is placed in the center, where the intensity is above the bistability limit, delimited by the white dashed line. 
The beam enters the cavity with an upward in-plane wave vector that gives a flow to the polaritons.
The black rectangle is the detection field of view: it is shifted on top of the illuminated region to observe the solitons free propagation through the bistable area. 

\paragraph{}
As usual, the detection is done in both real and momentum space. The real space gives the intensity map of the cavity plan, as well as information on the phase pattern through the interferences with the reference beam previously separated from the laser beam.
The experimental conditions of the system are extracted from the momentum space images (see section \ref{sec:DataAnalysis}).

\subsection{Solitons impression}

\subsubsection{Parallel propagation}

\begin{figure}[h]
    \centering
    \includegraphics[width=\linewidth]{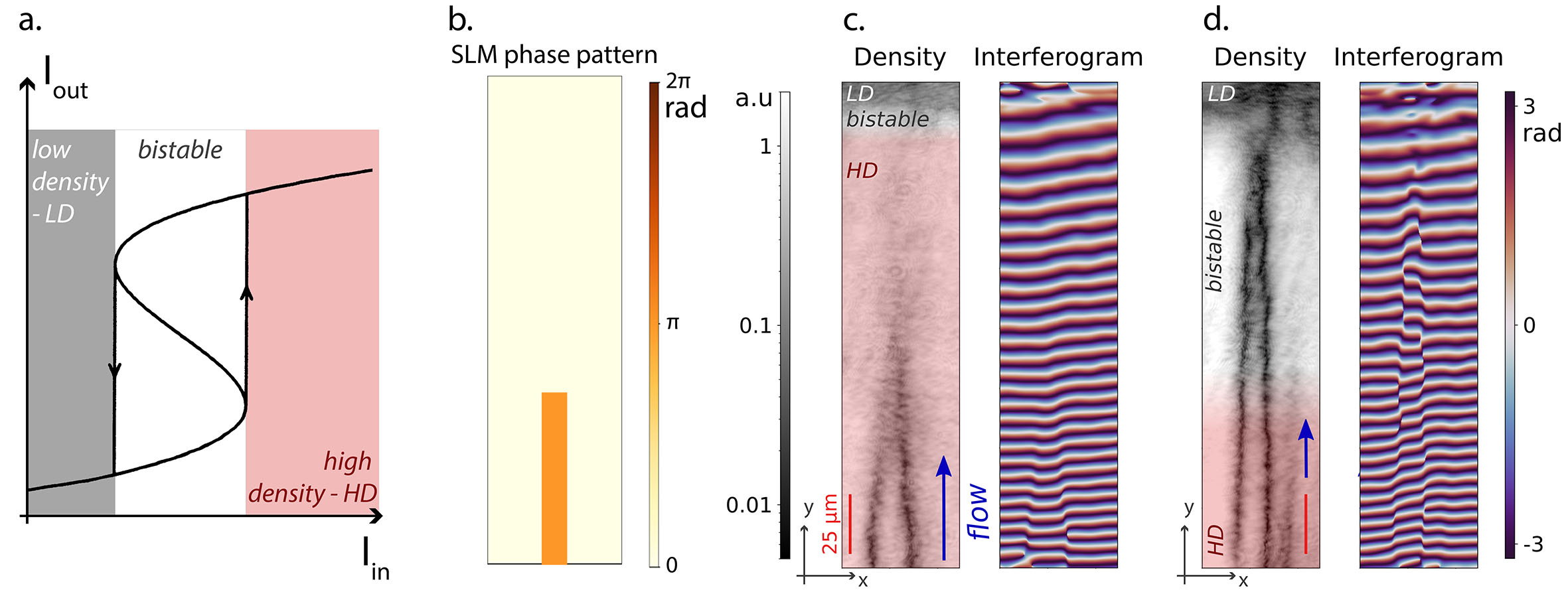}
    \caption{\textbf{Impression of dark parallel solitons}. a. Theoretical bistability profile and the three associated intensity ranges. The color code is used all along this chapter. b. SLM phase pattern with the same field of view as the detection: the phase modulation are only sent on the bottom part of the images.
    c. High power density and phase maps. Almost all of the illuminated area is in the HD regime: the phase is fixed by the driving field and replicates the pattern designed by the SLM. d. Same configuration as c. at lower power. The bistable region has extended toward the beam center, and reached the top part of the solitons. The system has readjusted its phase to let the solitons propagate through the bistable area, until the low density region where the nonlinear interactions are too low to sustain dark solitons.}
    \label{fig:ImprParallel}
\end{figure}

\paragraph{}
The essential role of the bistability in the propagation of the solitons is explained in figure \ref{fig:ImprParallel}.
The figure a. reminds the S shape of the bistability curve
and the three associated intensity regions: below the cycle, the low density region in grey, denoted as LD; above the cycle, the high density region highlighted in red and denoted as HD; and the bistable cycle left blank.

\paragraph{}
On b. is plotted the phase pattern designed by the SLM, with the same field of view as the detection: we can see that the phase modulation is only present on the bottom part of the images.
Figures c. and d. were realized in the exact same conditions except for the total intensity of the excitation. In c., the total laser power is high, which puts almost all the illuminated area above the bistability cycle: the red HD region expands above the major part of the picture. The high density region fixes the properties of the fluid: it is therefore a replica of the driving pump field. 
Indeed, the solitons are artificially created only in the bottom part of the picture, while on top, phase and intensity are flat.

Figures d. are obtained from the c. ones by gradually decreasing the input intensity. The bistable region expands toward the center of the beam, and eventually reaches the top part of the imprinted solitons.
The fluid then readjusts its phase and lets the solitons propagate through the bistable region, even though the region between the solitons is out of phase with the driving field.
Indeed, the dark solitons in d. are clearly visible within the bistable region, inducing a phase jump of \textpi\ all along their propagation. The propagation is sustained as long as the system is in the bistable regime. As the illuminated region is finite, the solitons will reach its border: in the low density region, the nonlinear interaction are too low to sustain dark solitons.

\subsubsection{Influence of the intensity}

\paragraph{}
To find the good configuration for the solitons to propagate through the bistable region, several parameters need to be finely tuned. In particular, the total intensity of the pump has an important impact on the soliton propagation, as they need a bistable fluid to readjust the phase of the fluid. 

\begin{figure}[h]
    \centering
    \includegraphics[width=\linewidth]{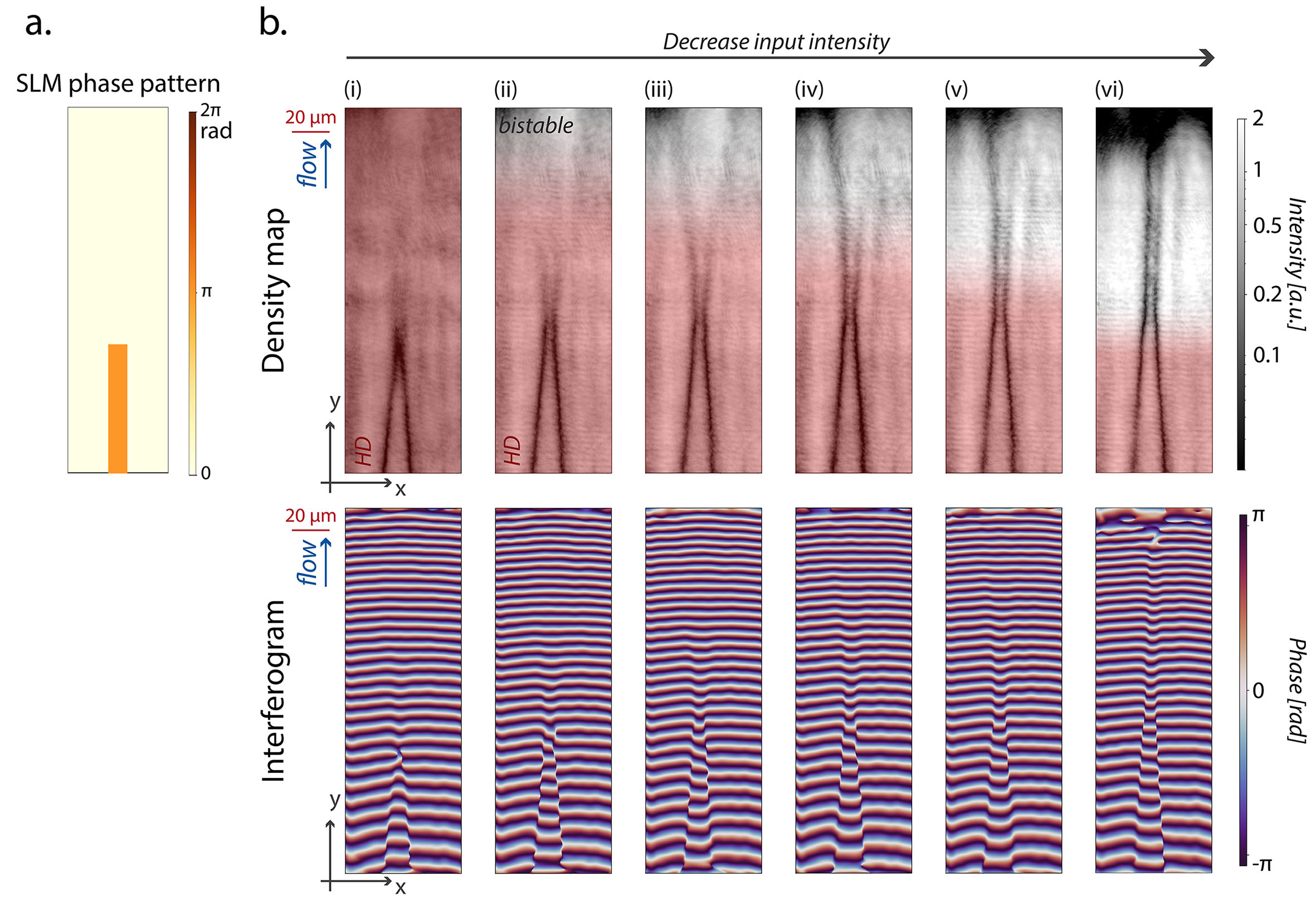}
    \caption{\textbf{Scan of the input power}. a. Phase pattern designed by the SLM with the same field of view as the detection.
    b. Density (top) and phase (bottom) maps of a scan of the total input intensity. On (i), the power is maximum and the phase is fixed everywhere. The intensity gradually decreases and with it the size of the HD region in red: the solitons propagate and open, until the bistable part reaches the imprinted solitons in (vi) where they align and propagate parallel.}
    \label{fig:ScanInt}
\end{figure}

\paragraph{}
To qualitatively study the influence of the pump intensity, a scan is realized and the results are presented in the figure \ref{fig:ScanInt}. The SLM phase pattern is presented on a., again with the same field of view as the detection images, and the experimental images are shown in b.
The top line shows the density maps and the bottom one the interferograms.
The input power is gradually decreased from picture (i) to (vi). The flow is from bottom to top, and the red colored regions indicates the area above the bistability cycle.

On picture (i), the total power is high: the fluid is above bistability on the whole picture. Its phase is therefore fixed, which shows that the solitons are sent only on the bottom part of the image.
From picture (ii) to (v), as the power decreases, the bistable region expands. The solitons propagate further and further but the system still can not perfectly sustain them: they are grey as their phase jump is lower than \textpi\, and they open and vanish along the flow. 
The phase maps confirm as well that the phase modulation induced by the solitons vanishes with them.

Finally, on picture (vi), the bistable area joins the top part of the imprinted solitons. They are then able to align to each other, and to remain dark and parallel all along their propagation. This time, their phase is \textpi\ and stays constant.
They are sustained through the whole bistable region, and vanish only at its edge, where the system jumps to the low density regime.

\paragraph{}
This set of measurement scans the bistability cycle, and confirms the necessity to be in this regime to achieve the free propagation of dark solitons in a resonantly pumped polariton fluid.

\subsubsection{Influence of the filtering}

\paragraph{}
A second crucial parameter for the good propagation of the solitons is the spatial filtering in the Fourier plan of the SLM.
As explained in section \ref{sec:expimpl}.\ref{sec:SpatialFilt}, a slit is placed along the \textit{x} axis in order to clean the beam after the SLM, but also to smooth the phase jump along \textit{y} to stimulate the solitons propagation.

\begin{figure}[h]
    \centering
    \includegraphics[width=0.95\linewidth]{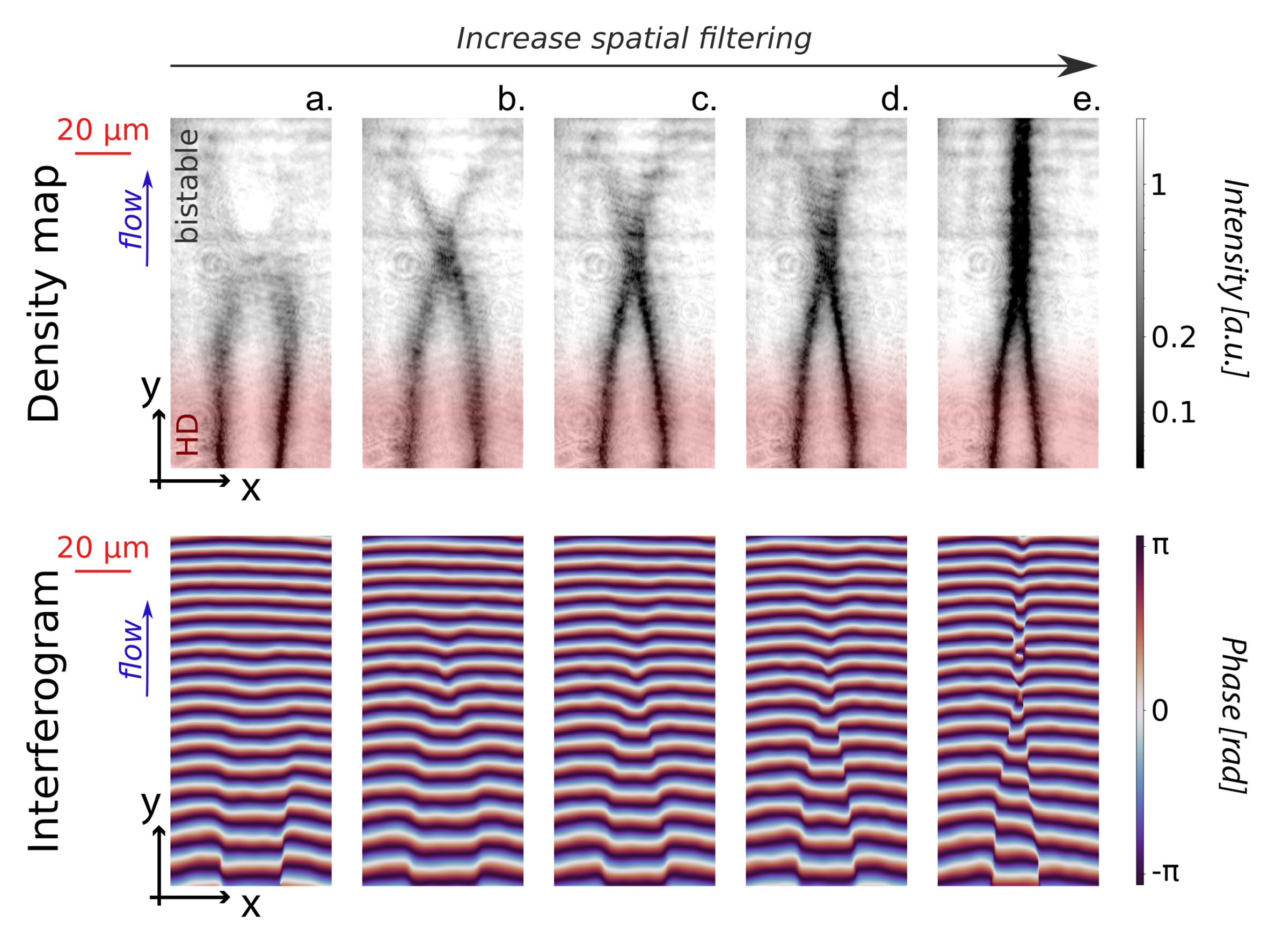}
    \caption{\textbf{Scan of the spatial filtering}. Density (top) and phase (bottom) maps of a scan of the spatial filtering: on a., the slit is widely open, while its width is gradually reduced from b. to e. It reduces the respective lateral velocity of the solitons, which align only on e. where they meet with a small enough angle.}
    \label{fig:ScanSlit}
\end{figure}

\paragraph{}
The qualitative study of the slit width influence is shown in figure \ref{fig:ScanSlit}. From a. to e., the slit is progressively closed, while all other parameters remain the same. In particular, the pump intensity is unchanged: the imprinting region highlighted in red is the same in all the pictures. 

\paragraph{}
The images a. present the density and phase maps corresponding to a fully opened slit. The phase pattern imprinted in the fluid has therefore a sharp profile. The soliton along the \textit{x} axis is however not visible: the polariton flow, from bottom to top, prevents a well defined impression. 
The solitons propagation is anyway inhibited: they vanish soon after the HD region and the phase jump is quickly erased.

Figures b., c. and d. show intermediate configurations for different slit openings. In those three cases, the solitons start to propagate by getting closer to one another. However, they bounce on each other and reopen, eventually vanishing: they have a too high relative lateral velocity to align. 
Remarkably, we observe that they conserve their relative lateral velocity through the collision: the angle is the same before and after the bouncing in all three figures (27\textdegree, 17\textdegree\ and 13\textdegree\ respectively).
This could also allow for a systematic study of collisions between solitons.

On figure e., the filtering is further increased, that elongates the imprinted solitons which meet with an even smaller angle of 11\textdegree.
This time, the angle is small enough for the solitons to align: their propagation is maintained for over a hundred microns, and they remain dark with a \textpi\ phase jump all along.

\subsection{Different shape configurations}

\paragraph{}
The setup previously implemented gives a large flexibility on the shape of the imprinted phase pattern.
Even though the first goal of the experiment was to reproduce the generation of a dark solitons pair, the SLM pattern is easily tuned and some other configurations have been tried, in order to study the solitons behaviour with different initial conditions.

\subsubsection{Four solitons}

\begin{figure}[h]
    \centering
    \includegraphics[width=0.9\linewidth]{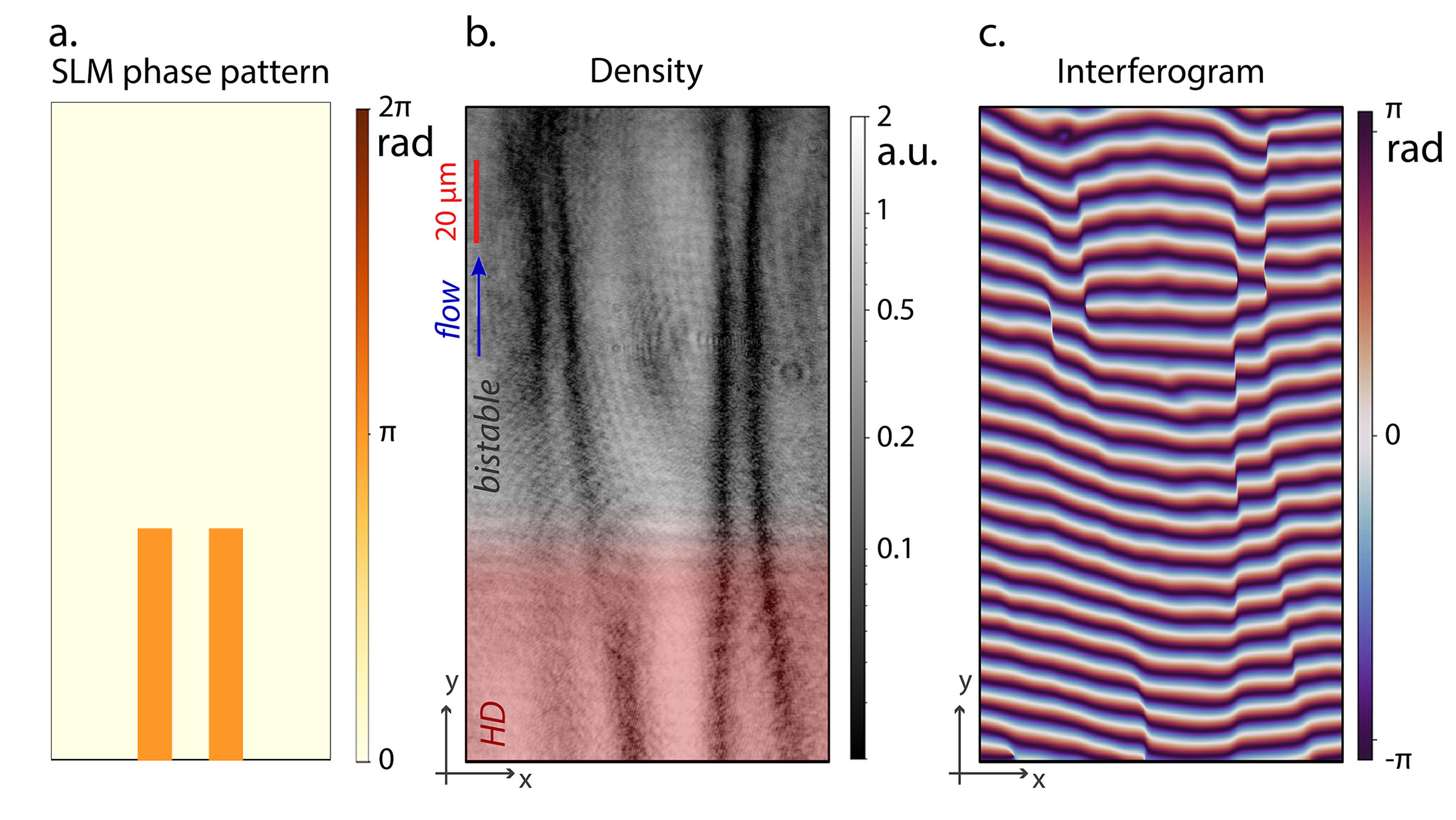}
    \caption{\textbf{Four solitons imprinting}. a. SLM phase pattern corresponding to the detection region. b. Density map of the fluid. The two solitons pairs are imprinted on the red region and propagate through the bistable white one. c. Interferogram of the fluid. The phase jump propagates with the solitons.}
    \label{fig:Impr4Sol}
\end{figure}

\paragraph{}
The first tested pattern, shown in figure \ref{fig:Impr4Sol}, is a double pair of solitons.
On the left, figure a. is a scheme of the SLM phase pattern. 
It does not represent the whole pattern, but only the one corresponding to the top part of the beam, in order to coincide with the corresponding detection pictures. 
Those ones are plotted on figure b. and c. and show the density map and the interferogram, respectively.

\paragraph{}
The double pair of solitons is realized by sending two rectangular shapes in phase opposition with the background thanks to the SLM. The imprinted phase pattern is designed so that each of the four solitons is equidistant from its neighbor. It stays like that in the high density region of the fluid, where the solitons are imprinted (red part of figure \ref{fig:Impr4Sol}.b.), but during their free evolution, they get closer to their respective pair, so that the area in phase with the driving expands while the one in phase opposition is reduced. However, the solitons do propagate and maintain a phase jump on the fluid.

This configuration gives an overview of the scalability of our method, and opens the way of the study of soliton lattices in polariton superfluids.

\subsubsection{Opening solitons}

\paragraph{}
As the tendency of the solitons seems to be to get closer to one another, we tried to compensate this effect by sending them with an opposite direction and a positive respective lateral velocity. 

\begin{figure}[h]
    \centering
    \includegraphics[width=0.9\linewidth]{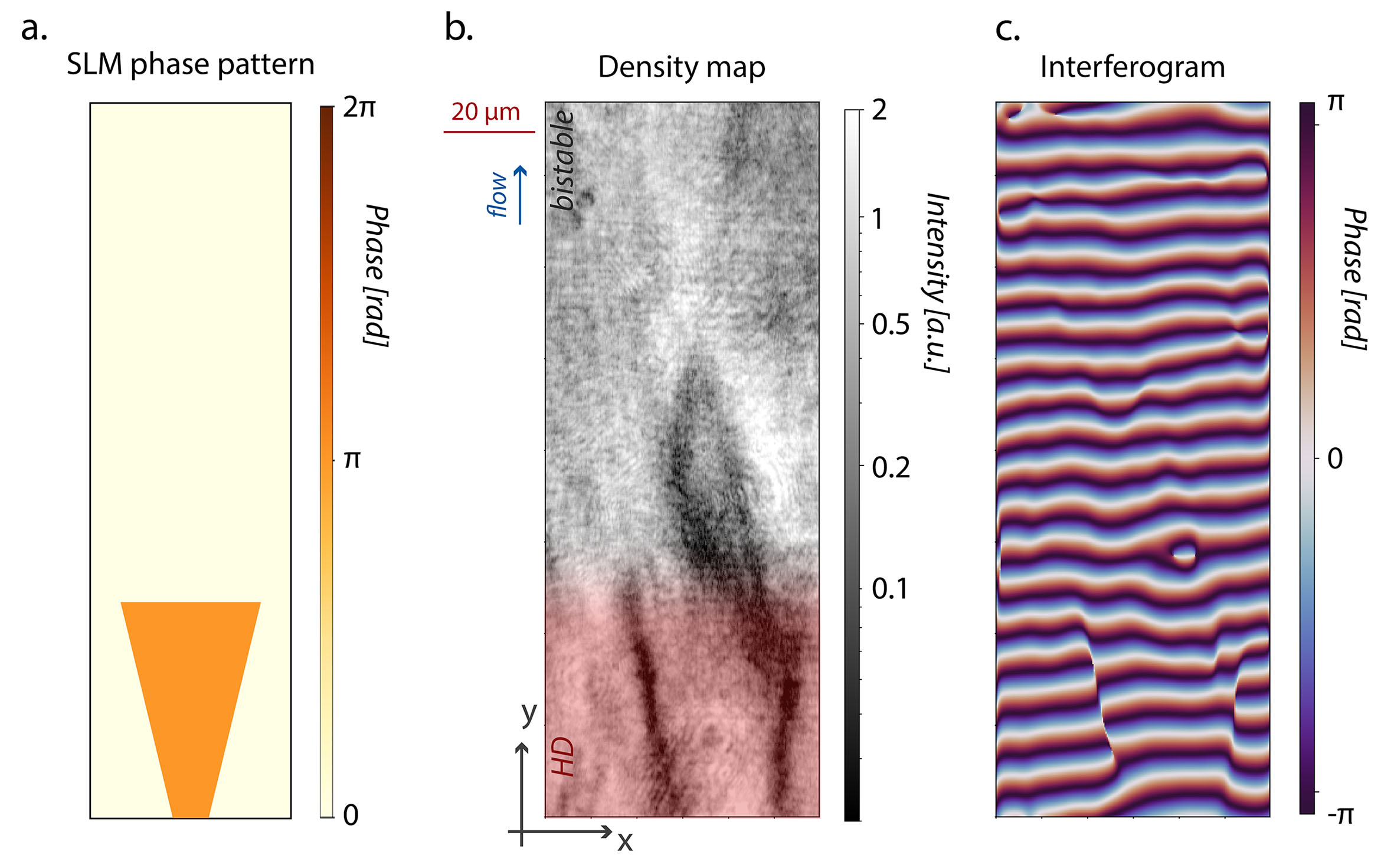}
    \caption{\textbf{Opening solitons imprinting}. a. Imprinted phase pattern inducing solitons moving away from each other. b. and c. Fluid density map and interferogram of the impression.}
    \label{fig:ImprOpenSol}
\end{figure}

\paragraph{}
The results are presented in figure \ref{fig:ImprOpenSol}: the left figure shows the SLM phase pattern corresponding to the fluid density and phase maps plotted in figures b. and c.
The solitons are well defined in the imprinting region, at the bottom of the figure. However, when they enter the bistable fluid, they are not properly sustained. Indeed, they are send with a velocity going away from each other, while the driving field pushes them toward one another, preventing their continuous propagation.
They become grey and still tend to come closer to one another, even though it induces an important change of direction compared to their imprinted part. They still produce a phase modulation but smaller than \textpi, in agreement with their color.
They vanish when they meet, as grey solitons are not stable through collision \cite{Zakharov1973, Maimistov2010}.

\subsubsection{Closing solitons}

\begin{figure}[h]
    \centering
    \includegraphics[width=0.9\linewidth]{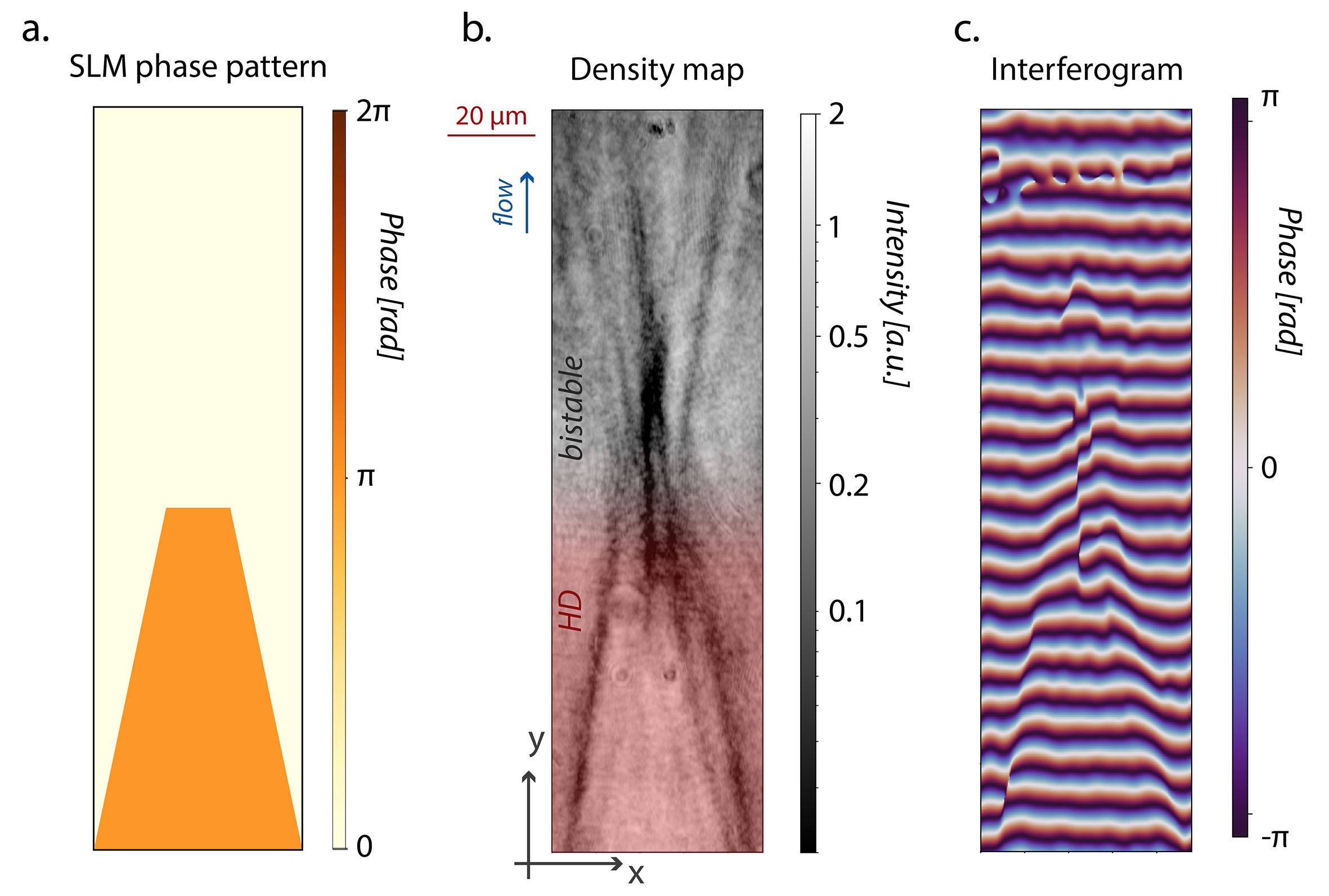}
    \caption{\textbf{Closing solitons imprinting}. Imprinted phase pattern (a.), density map (b.) and interferogram (c.) of solitons sent toward each other.}
    \label{fig:ImprClosSol}
\end{figure}

\paragraph{}
Figure \ref{fig:ImprClosSol} presents the opposite case where the solitons are sent toward each other. The solitons are sent oblique with a relatively high opposite velocity. As before, the three pictures illustrate the phase pattern designed by the SLM, the density map of the fluid and the corresponding interference pattern. The flow is from bottom to top and the impression is realized in the high density (HD) region highlighted in red.

\paragraph{}
The collision of the solitons induces a breaking of the pair into several solitons which then spread though the fluid. 
They indeed meet with a too high velocity to smoothly align or bounce, and split into different pieces, each of them inducing a phase modulation. However, they are not completely black, and not perfectly sustained by the system: they vanish along their propagation.

\subsubsection{Single soliton}

\paragraph{}

\begin{figure}[h]
    \centering
    \includegraphics[width=0.9\linewidth]{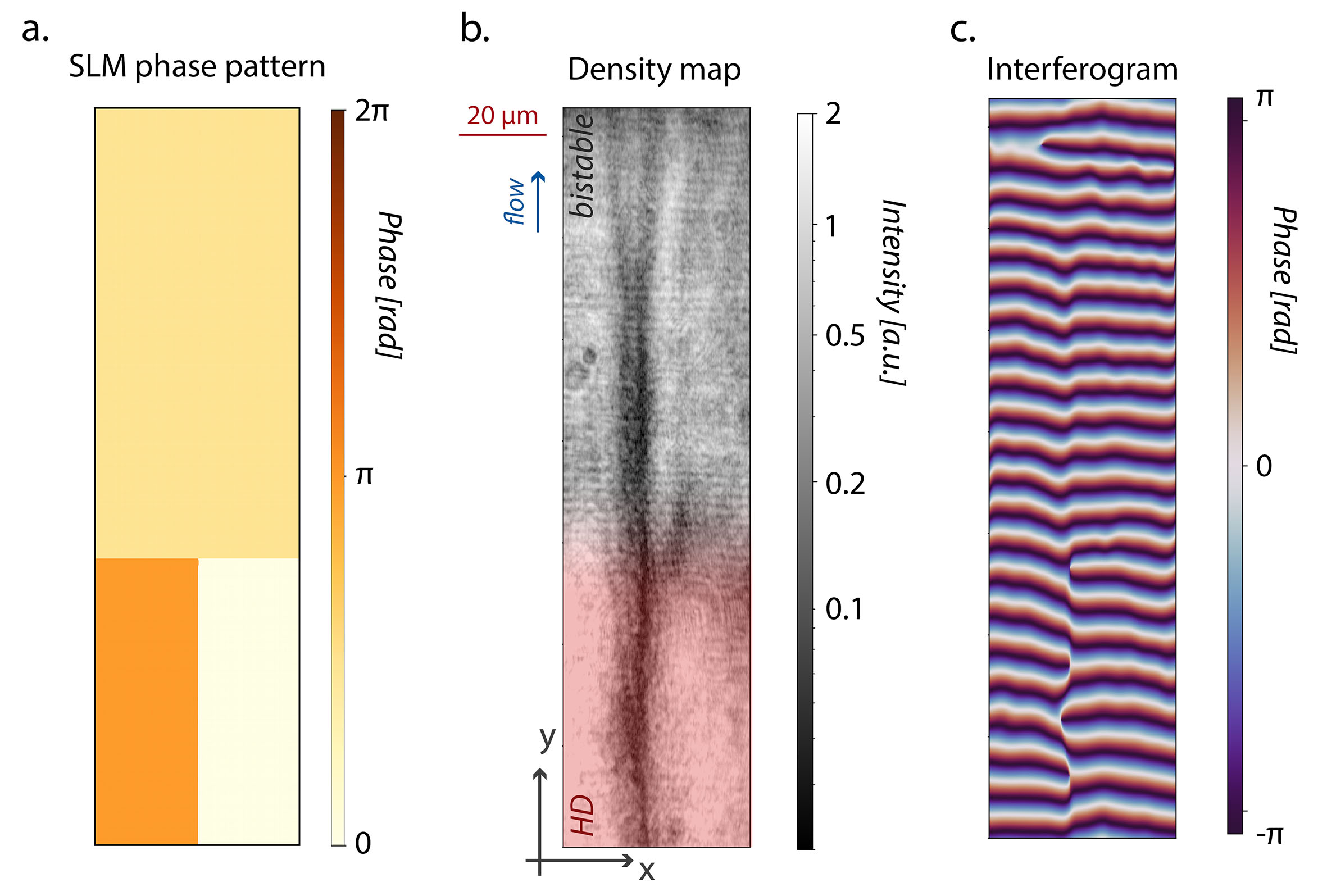}
    \caption{\textbf{One soliton imprinting} a. The soliton is imprinted on the bottom part of the images, inducing a single delimitation between two area in phase opposition. The bistable part of the beam is sent with a \textpi/2 phase to ensure the symmetry. b. On the fluid, the single soliton is not sustained in its initial form and splits into two (b.), as a phase switch of half the fluid is not supported by the system. c. The two resulting grey solitons have a phase jump between them and stay quite close to each other.}
    \label{fig:Impr1half}
\end{figure}

A different configuration has been tested in figure \ref{fig:Impr1half}, where a single soliton is imprinted on the fluid.
The phase pattern designed by the SLM is now different than the previous ones. The impression region is split into two parts in phase opposition, between which the soliton will appear. 
In order to have a completely symmetrical pattern, the bistable part  of the beam is chosen to be at a \textpi/2 phase difference from both domains of the high density region, so that none of them is favored by the system.

\paragraph{}
The fluid density is presented in figure b. The impression is done in the HD region: the single soliton is well imprinted, with a clear phase jump. But when it enters the bistable region, the soliton is not supported anymore and splits into two.
A single soliton indeed implies a phase difference between the two half-parts of the system, so compared to the driving field, either half of the fluid is in total phase opposition, or all the fluid has a \textpi/2 phase difference.
This seems to be energetically too expensive for the system, even in the bistable regime.

\paragraph{}
The splitting of the soliton is anyway interesting as it shows that its propagation is not totally inhibited, and even preferred to its halt. 
The propagation can be separated into two parts. First, the solitons are quite dark and stay close to each other, inducing an important phase dip.
After about 50 microns however, they are not perfectly supported anymore and start to open, getting grey and vanishing along the way.
This can be due to irregularities in the sample or to the intensity gradient which changes the hydrodynamic conditions.

\section{Parallel alignment: equilibrium separation distance}

\paragraph{}
The study of those different configurations indicates that the dark parallel solitons are the most stable in the polariton fluid in the bistable regime. It is thus the configuration that is going to be further studied in the next section. 
In order to better understand the characteristic behaviour of such phenomenon, several runs of measurements have been performed varying the initial inter-soliton distances as well as the hydrodynamic conditions. The results of this analysis will be reported in this section.

\subsection{Scan of the imprinting separation distance}

\paragraph{}
The most intriguing interrogation about the solitons alignment is to identify the different parameters that play a role in the establishment of such an equilibrium distance.
The first idea is to try to modify the initial conditions of the system to see if it has any effect on its behaviour. 
This has been realized through the use of the SLM, changing the shape of the phase pattern to modify the solitons imprinted on the fluid. 
The first parameter we decided to study is the separation distance between the solitons.

\begin{figure}[h]
    \centering
    \includegraphics[width=0.4\linewidth]{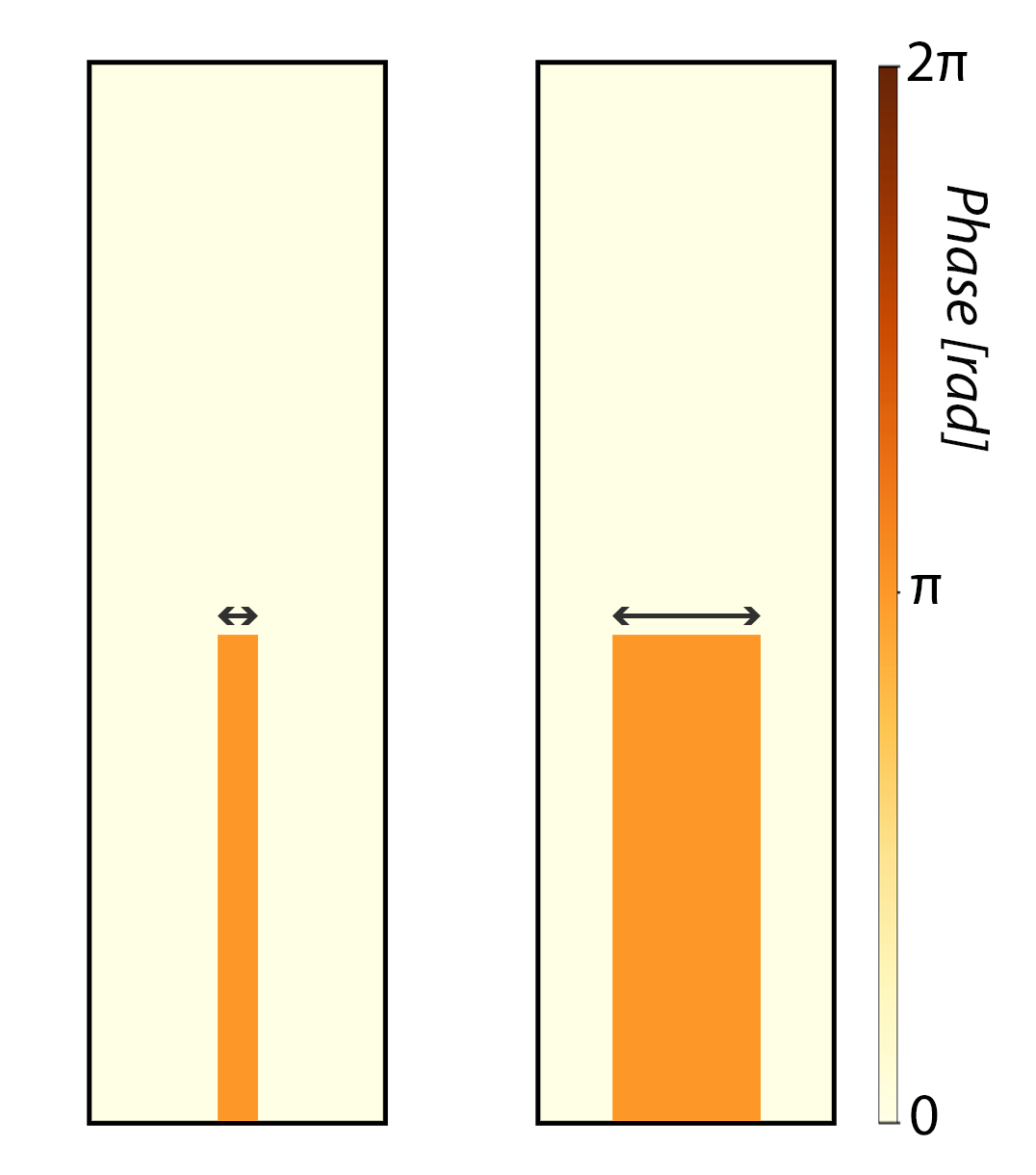}
    \caption{\textbf{SLM patterns corresponding to different separation distances} The distance between the imprinted solitons is tuned by changing the width of the rectangular \textpi-shifted rectangle.}
    \label{fig:ImprSLMpattdsep}
\end{figure}

\paragraph{}
It can be easily controlled by the shape designed on the driving beam by the SLM. By enlarging the \textpi-shifted rectangular shape of the pattern (see figure \ref{fig:ImprSLMpattdsep}), the solitons are imprinted further apart.

\begin{figure}[h]
    \centering
    \includegraphics[width=\linewidth]{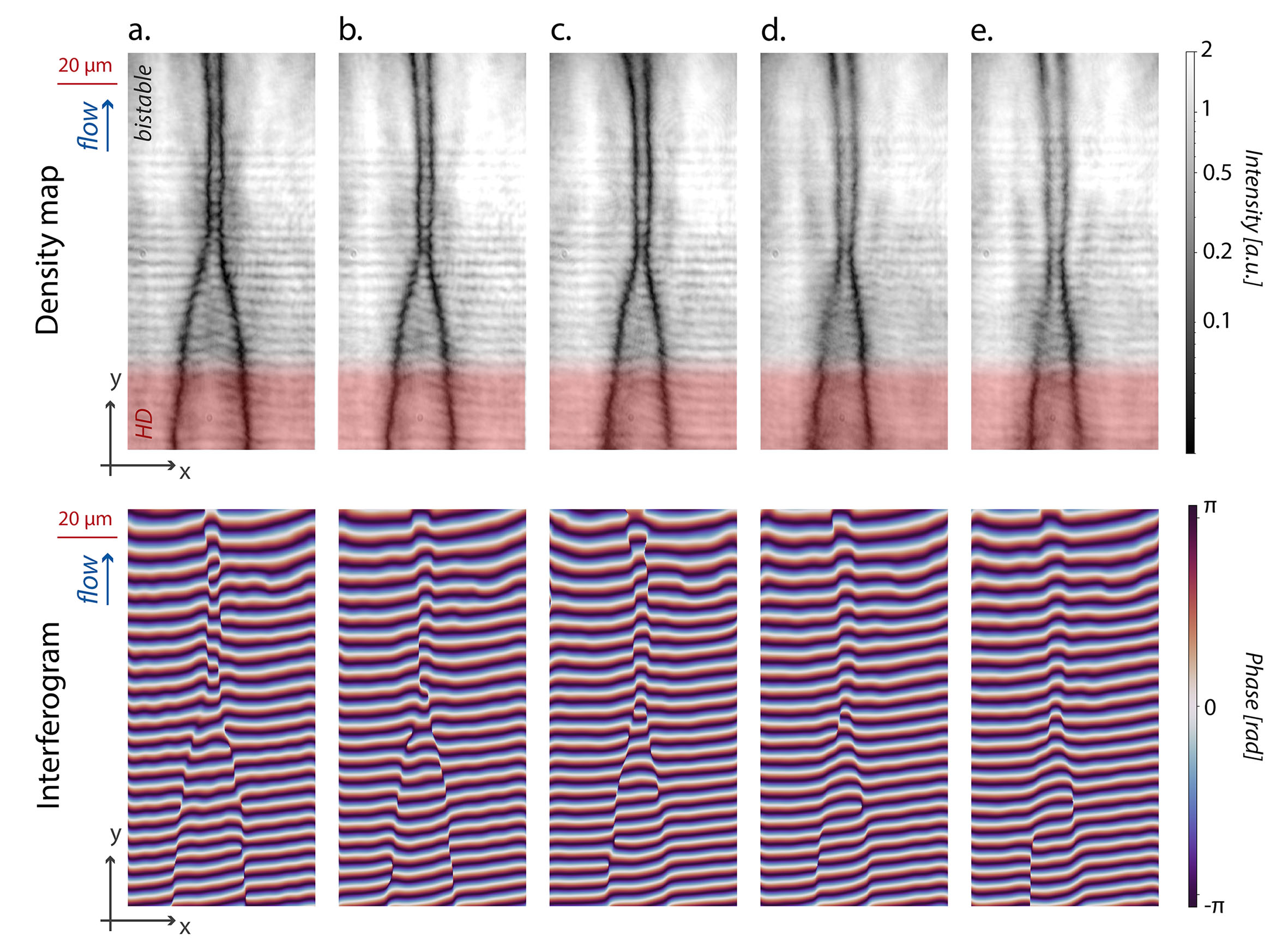}
    \caption{\textbf{Experimental scan of the initial separation distance}. The distance between the imprinted solitons is tuned from 24 \textmu m in a. to 15 \textmu m in e. by changing the width of the rectangular \textpi-shifted rectangle. When they enter the bistable fluid, the solitons propagation is maintained, but not their separation distance: they get closer to one another until reaching an equilibrium separation distance from which they align.}
    \label{fig:ImprScandsepExp}
\end{figure}

The results of a typical run of measurements are exposed in figure \ref{fig:ImprScandsepExp}. The top images present the intensity of the fluid while the bottom ones show the corresponding interferograms. For this set of experiments, the fluid velocity is 1.00 \textmu m/ps, the sound velocity $c_{s} = 0.67$ \textmu m.ps, the effective polariton mass $m^{*} = 1.24 \cdot 10^{-34}$ kg and the energy detuning $\Delta E_{lasLP} = 0.34$ meV.
Surprisingly, even though the initial separation distance is reduced from 24 \textmu m in figure a. to 15 \textmu m in figure e., we observe that the behaviour of the solitons is identical in all the cases. When they enter the bistable part of the fluid and are not artificially imprinted anymore, the separation distance between the solitons reduces until reaching the same equilibrium value, from which the solitons align and stay parallel.

\paragraph{}
In the case of figure \ref{fig:ImprScandsepExp}, the equilibrium separation distance is about 5 \textmu m for all the plotted figures. After reaching this configuration, they keep this separation distance as well as their associated phase jump and continue their propagation for more than 50 \textmu m, until they can not be sustained by the system anymore.

\paragraph{}
The experimental results have been compared with numerical simulations based on the generalized Gross Pitaevskii equation which describes the polariton system, to check if the observed behaviour was compatible with the model. The simulations were based on a split-step method in the exciton-cavity photon basis, as presented in \cite{Pigeon2011}.
The model uses two coupled equations for the excitons ($\psi_{X}$) and the cavity photon ($\psi_{\gamma}$) fields respectively:

\begin{gather}
    i \hbar \dfrac{\partial \psi_{\gamma} }{\partial t} = 
    \Bigg[ - \dfrac{\hbar^{2} \nabla^{2} }{2m_{\gamma}^{*}} - i \Gamma_{cav}\Bigg] \psi_{\gamma}
    + V \psi_{X} + F(\mathbf{r})e^{-i\omega_{0}t} \\
    i \hbar \dfrac{\partial \psi_{X}}{\partial t} = 
    \big[ g_{X} |\psi_{X}|^{2} - i \Gamma_{X} - \Delta E_{Xcav} \big] \psi_{X}
    + V \psi_{\gamma}
\end{gather}

where $m_{\gamma}^{*}$ is the effective cavity photon mass, $\omega_{0}$ the laser energy, $\Gamma_{cav}$ the photon lifetime,$g_{X}$ the exciton interaction constant, $V$ the half Rabi splitting and $\Delta E_{Xcav}$ the cavity-exciton energy detuning.
In order to best match the experimental configuration, all those parameters have been taken the same as in the experiment (the exciton lifetime $\Gamma_{X}$ was set equal to 150 ps).
$F(\mathbf{r})$ represents the pump and contains both its amplitude and phase.
\begin{figure}[h]
    \centering
    \includegraphics[width=\linewidth]{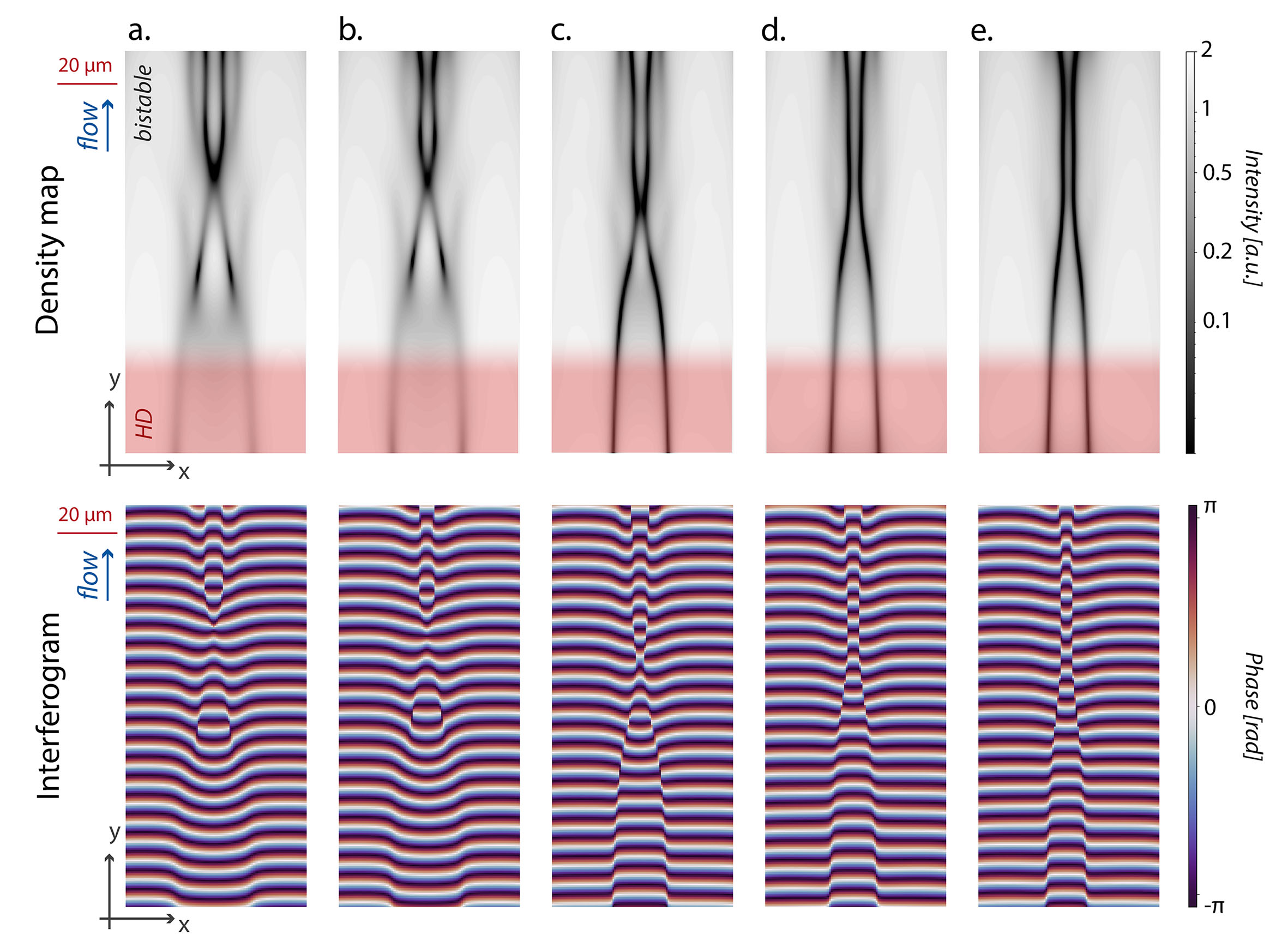}
    \caption{\textbf{Numerical simulations of the scan of the initial separation distance}. As for the experimental set (figure \ref{fig:ImprScandsepExp}), the imprinted separation distance is numerically scanned from 24 \textmu m (a.) to 15 \textmu m (e.). The same behaviour is observed: when the solitons enter the bistable fluid, their separation distance reduces until the solitons align at a specific equilibrium distance from each other. }
    \label{fig:ImprScandsepSimu}
\end{figure}
The steady-state solutions are presented in figure \ref{fig:ImprScandsepSimu}. As usual, the flow is from bottom to top, the top images show the density map of the fluid while the bottom images are the associated interferograms, giving access to the fluid phase.

\paragraph{}
The driving field used in the simulations has been chosen to reproduce the experimental one: it has been taken as a Gaussian beam whose center is shifted towards the bottom of the images plotted on figure \ref{fig:ImprScandsepSimu}, so that the region where the intensity is above the bistability cycle is highlighted in red.
In this high density (HD) region, the solitons are imprinted: a \textpi\ phase jump is imposed; the experimental filtering operated by the slit is modeled here through a smoothing of the phase jump along the \textit{y} direction.

\paragraph{}
The numerical simulations are in good agreement with the experiment: the solitons follow exactly the same behaviour. No matter how far away from each other they are imprinted, as soon as they are released within the bistable fluid, they get closer to one another before aligning at a specific equilibrium separation distance.
This alignment takes place for solitons at around 5 \textmu m from each other, which reproduce accurately the experimental configuration.

\begin{figure}[h]
    \centering
    \includegraphics[width=0.7\linewidth]{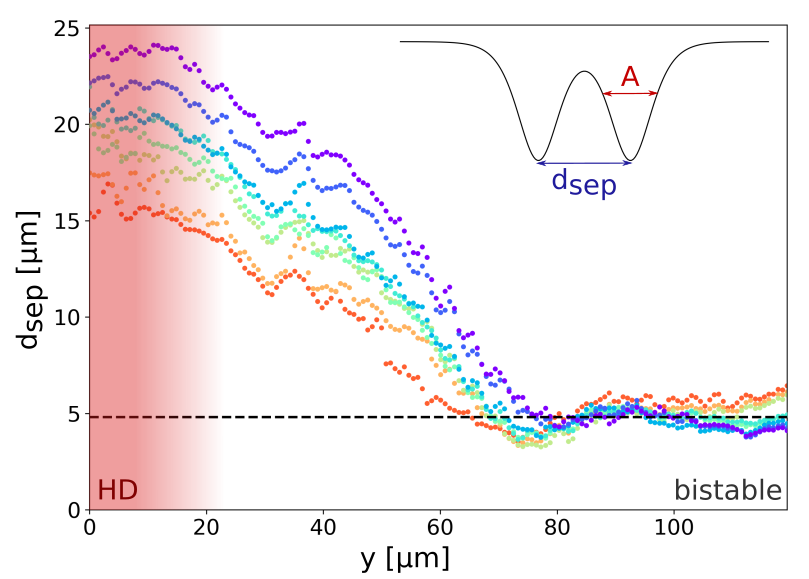}
    \caption{\textbf{Experimental scan of the initial separation distance}. Each color line corresponds to one image of the experimental set presented in figure \ref{fig:ImprScandsepExp}. Each dot illustrates the fitted separation distance at the corresponded position along the propagation axis \textit{y}. No matter the initial separation distance, they all converge to the same value of 4.8 \textmu m in this case, illustrated by the black dashed line.}
    \label{fig:ImprdsepProp}
\end{figure}

\paragraph{}
To get a quantitative idea of the separation distance evolution, it has been studied all along the solitons propagation. To do so, the transverse solitons profile has been fitted by its expression \cite{Pitaevskii2003a}:

\begin{equation}
\label{eq:SolitonsProfile}
    |\psi(x)|^{2} = \tanh^{2} \Bigg( \dfrac{x-d_{sep}/2}{A} \Bigg)
    \tanh^{2} \Bigg( \dfrac{x+d_{sep}/2}{A} \Bigg)
\end{equation}

where $A$ corresponds to the full width at half-maximum (FWHM) of one soliton and $d_{sep}$ to their separation distance, as illustrated in the inset of figure \ref{fig:ImprdsepProp}.

\paragraph{}
The fit results of the experimental set are reported in figure \ref{fig:ImprdsepProp}. The hydrodynamic conditions are kept constant during the whole set, the only parameter which is varied between each picture is the initial separation distance. 
The color lines correspond to the different images, and show the evolution of the fitted separation distance $d_{sep}$ along the propagation axis \textit{y}.
As usual, the red HD region illustrates the impression area where the input intensity is above the bistability, while in the white bistable region, the solitons propagates freely through a flat driving field. 

\paragraph{}
The graph confirms the tendency: despite the change of the initial separation distance, the freely propagating solitons get closer to another, and align at a specific separation distance. This distance is the same for all the cases of the set, at 4.8 \textmu m illustrated by the black dashed line.
It confirms the existence of a equilibrium between two opposite forces, which push the solitons in opposite directions and only equilibrate when they are at a specific distance from each other. 

On hand hand, dark solitons have repulsive interactions \cite{Frantzeskakis2010b, Walker2017a}, which pushed them away from each other. 
However, they are also propagating within a pumped fluid, even if it is a bistable one. So even if phase readjustments are possible, all the area of the fluid with a different phase than the pump one remain effectively unpumped. Sustaining a large out-of-phase region has therefore a high energetic cost for the system, which tends to reduce this area and pushes the solitons toward one another.

\paragraph{}
By imprinting the solitons far away from each other, their propagation is first dominated by the force induced by the driving, and they propagate towards each another. However by getting closer to one another, their repulsion increases, until reaching an equivalent value to the driving-induced force. Prevented to go closer together or further apart, the solitons propagate parallel.

\paragraph{}
This parallel propagation of a soliton pair corresponds to the one observed in the spontaneous case, reported in section \ref{sec:SpontSolitons}, but remains in full opposition with the previously expected \cite{Frantzeskakis2010b} and reported \cite{Amo2011} soliton behaviour.
However, this artificial implementation has high control and scalability: we will use those advantages to lead for a deeper study of the phenomenon.

\subsection{Role of the hydrodynamic conditions}

\paragraph{}
The first interrogation from those results is the origin of this equilibrium separation distance, and on which parameters it depends.
The size of a vortex core is known to be of the order of the healing length $\xi$ of the system \cite{Pitaevskii2016}.
By analogy, the FWHM of the solitons ($A$ in equation \ref{eq:SolitonsProfile}) should also be connected to $\xi$. The equilibrium separation distance seems also to be the minimal one between the solitons, and therefore should be connected to their width so that both solitons can be effectively distinguished.
The healing length is thus the parameter that will be considered in the following.

\paragraph{}
The healing length of a polariton fluid is defined by the hydrodynamic conditions of the system, and in particular by the effective polariton mass $m^{*}$, the polariton interaction constant $g$ and by the polariton density $n_{0}$ \cite{Pitaevskii2016}. 
An approximation can however be made, as we are in the bistable regime, and in particular in the left part of the bistability cycle. Indeed, at the turning point of the cycle, on the upper bistability branch at the low intensity threshold, the laser-polariton detuning verifies the relation $\Delta E_{lasLP} = g n_{0}$.
We can thus write the healing length as:

\begin{equation}
    \xi = \sqrt{\dfrac{\hbar^{2}}{2 m^{*} g n_{0}}} = \dfrac{\hbar}{\sqrt{2 m^{*} \Delta E_{lasLP} }}
\end{equation}

\paragraph{}
The hydrodynamic parameters can be tuned experimentally: 

\begin{itemize}
    \item the \textbf{effective mass of the polaritons $m^{*}$} is defined through the curvature of the dispersion, and can be tuned by changing the detuning between the exciton and the cavity photon $\Delta E_{Xcav}$. Experimentally, this is done by changing the working point on the sample: the wedge between the two DBR modifies the length of the cavity and thus the cavity photon resonance. The induced modification of the dispersion consequently changes the polariton mass and influences the hydrodynamic behaviour of the system.
    \item the \textbf{detuning between the laser and the lower polariton branch $\Delta E_{lasLP}$} is directly tunable by shifting the laser frequency. It also affects the bistability curve of the system: a different detuning leads to a different intensity profile on the fluid and has a direct influence on the impression and propagation of the solitons. The hydrodynamic conditions must be adjusted in order to find a favourable configuration again, by tuning the fluid and sound speeds.
    \item the \textbf{fluid velocity} $v_{fluid}$ is the derivative of the dispersion: 
    \begin{equation}
        v_{fl} = \dfrac{1}{\hbar} \dfrac{\partial E}{\partial k}
    \end{equation}
    which changes with the in-plane wavevector of the driving field. 
    Therefore, the resonance of the lower polariton branch is also affected, as well as the detuning laser-lower polariton $\Delta E_{lasLP}$ and with it the bistability cycle.
    \item finally, the \textbf{speed of sound} $c_{sound}$ depends on the square root of both the detuning laser - lower polariton $\Delta E_{lasLP}$ and on the effective mass $m^{*}$: 
    \begin{equation*}
        c_{sound} = \sqrt{\dfrac{\Delta E_{lasLP}}{m^{*}}}
    \end{equation*}
\end{itemize}

\paragraph{}
All those parameters are difficult to tune separately in the experiment, as they are all connected and a modulation of one of them would lead to a modification of the system bistability, key element of the soliton propagation.
An interesting way to combine them all is to consider the \textbf{Mach number}:
\begin{equation}
    M = \frac{v_{fluid}}{c_{sound}}
\end{equation}
which contains not only $m^{*}$ and $\Delta E_{lasLP}$ through the speed of sound, present in the healing length definition too, but also the fluid velocity $v_{fluid}$.
It is thus the parameter chosen to compare the different results.

\paragraph{}
The influence of the hydrodynamic parameters on the equilibrium separation distance between the solitons is investigated by analyzing the propagation of imprinted solitons on polariton fluids with different Mach numbers.
In the experiment, the cavity is spatially explored, and on several positions which presents a low enough disorder, solitons are imprinted. Their free propagation is then achieved by alternatively playing with the wavevector, the laser energy and the pump power to reach an intensity pattern where the bistable region coincide with the top part of the imprinted solitons.

\begin{figure}[h]
    \centering
    \includegraphics[width=0.8\linewidth]{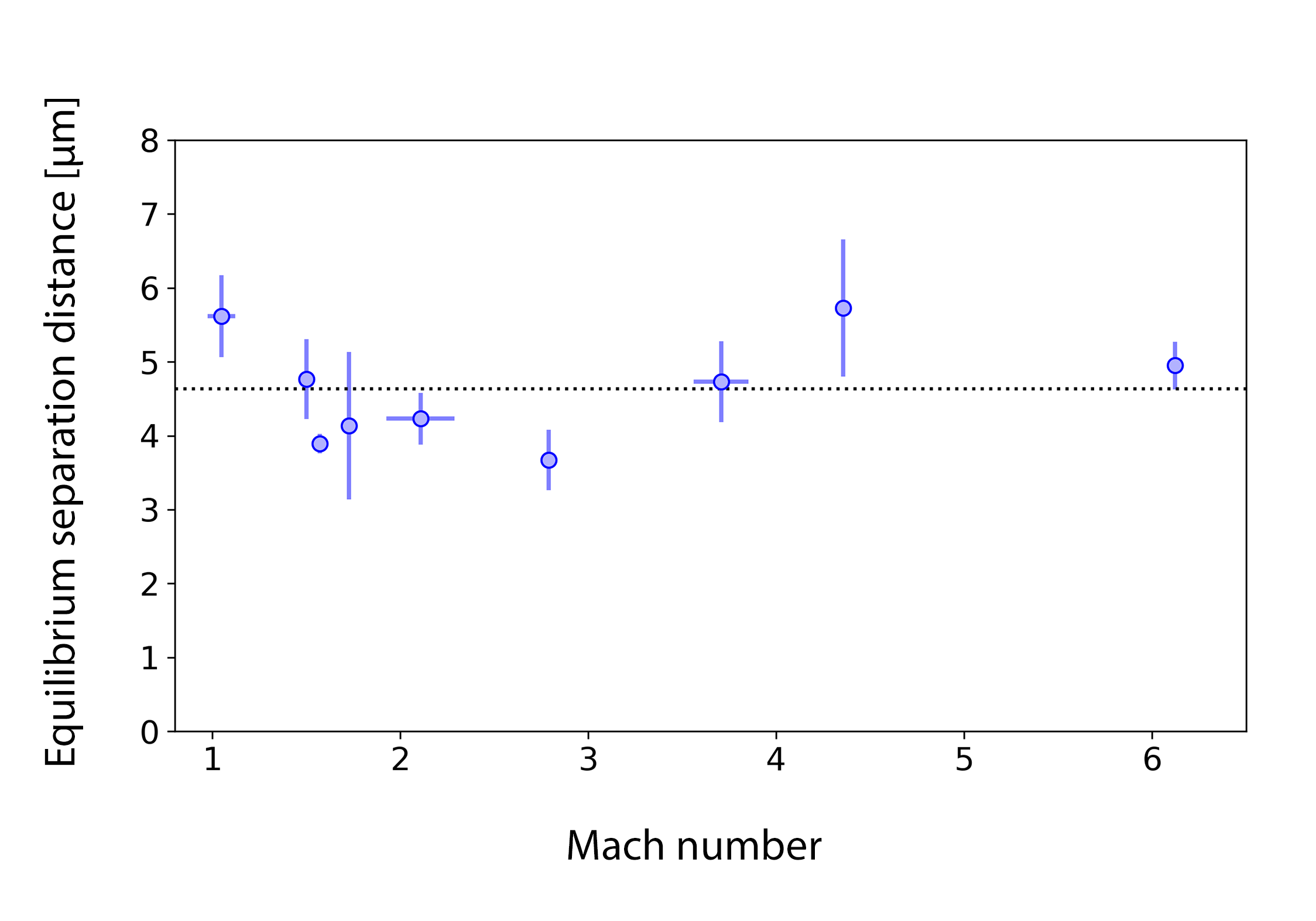}
    \caption{\textbf{Experimental equilibrium separation distances as a function of the Mach number}. Each spot corresponds to the mean equilibrium distance for one set of measurements with the same hydrodynamic conditions. Even though the Mach number range is quite large, the equilibrium distance is almost constant around a mean value of 4.75 \textmu m, illustrated by the black dashed line.}
    \label{fig:ImprMdsepExp}
\end{figure}

\paragraph{}
The experimental results have all been gathered in figure \ref{fig:ImprMdsepExp}, where each spot corresponds to one set of measurements: it shows the mean equilibrium separation distance of all the solitons pictures taken in the same hydrodynamic conditions, but changing only the initial separation distance. 
They have been plotted as a function of the Mach number: the conditions have been widely scanned as it goes from $M=1$ to more than 6.
All the parameters have also been reported in table \ref{tab:ImprSetParam}, where are listed the detuning laser - lower polariton branch $\Delta E_{lasLP}$, the speed of sound $c_{sound}$, the effective polariton mass $m^{*}$ and the Mach number $M$.

\begin{table}[h]
\centering
\begin{tabular}{||cccccc||}
\hline
\hline
$\Delta E_{lasLP}$ [meV]&
$c_{sound}$ [\textmu m/ps]&
$m^{*}$ [kg]&
$v_{fluid}$ [\textmu m/ps]&
M&
$d_{sep}$ [\textmu m]\\

\hline

0.24 & 0.60 & 1.05$\cdot 10^{-34}$ & 0.62 & 1.05 & 5.62 \\
0.34 & 0.67 & 1.24$\cdot 10^{-34}$ & 1.00 & 1.50 & 4.77 \\
0.11 & 0.42 & 1.04$\cdot 10^{-34}$ & 0.66 & 1.57 & 3.89 \\
0.33 & 0.60 & 1.47$\cdot 10^{-34}$ & 1.03 & 1.73 & 4.14 \\
0.08 & 0.33 & 1.11$\cdot 10^{-34}$ & 0.70 & 2.11 & 4.23 \\
0.22 & 0.42 & 1.95$\cdot 10^{-34}$ & 1.17 & 2.79 & 3.67 \\
0.16 & 0.39 & 1.75$\cdot 10^{-34}$ & 1.43 & 3.71 & 4.73 \\
0.07 & 0.31 & 1.25$\cdot 10^{-34}$ & 1.33 & 4.36 & 5.73 \\
0.06 & 0.23 & 1.84$\cdot 10^{-34}$ & 1.42 & 6.12 & 4.95 \\

\hline
\hline
\end{tabular}
\caption{\label{tab:ImprSetParam} \textbf{Parameters of the experimental sets plotted in figure \ref{fig:ImprMdsepExp}}.
}
\end{table}

\paragraph{}
The results shown in figure \ref{fig:ImprMdsepExp} are very surprising.
Indeed, the expected dependence of the separation distance on the hydrodynamic of the system is not observed at all: on the whole range of conditions, the equilibrium separation of the solitons is almost constant, slightly fluctuating around the mean value of 4.75 \textmu m shown by the black dashed line.
This result is quite unexpected, as the behaviour of dark solitons is usually governed by the hydrodynamic conditions of the system where they are generated.
On the other hand, as explained earlier, in this experimental study all the parameters are connected. By trying to modify the global conditions, all of the parameters are changed between each points in order to find each time a favorable situation for the solitons propagation. They could eventually compensate each other, leading to a similar configuration.

\paragraph{}
The scan of one single parameter is experimentally very difficult to realize, as they are all connected. In order to independently study the influence of each parameter, accurate numerical simulations have been done, presented in the next section.

\subsection{Numerical investigations}

\paragraph{}
An important advantage of the numerical simulations is the possibility to tune the system parameters independently.
It has been very useful for the verification of our assumption that the equilibrium separation distance is set by the hydrodynamic parameters through the healing length.
To do so, several sets of simulations have been realized, where the system conditions are kept constant, and only one parameter is tuned. This way, the healing length is also gradually tuned and its influence on the separation distance should be clearly visible.

\paragraph{}
The chosen tunable parameter is the detuning between the laser and the lower polariton branch $\Delta E_{lasLP}$, while the other ones are left unchanged, similar to the typical experimental ones.
In the definition of the Mach number, $\Delta E_{lasLP}$ is present only in the sound velocity; all others parameters being kept constant, the Mach number is therefore tuned as the inverse of $\sqrt{ \Delta E_{lasLP}}$. 
As the healing length $\xi$ also follows a relation on $1/\sqrt{\Delta E_{lasLP}}$, $M$ and $\xi$ are linearly connected.

\begin{figure}[h]
    \centering
    \includegraphics[width=0.8\linewidth]{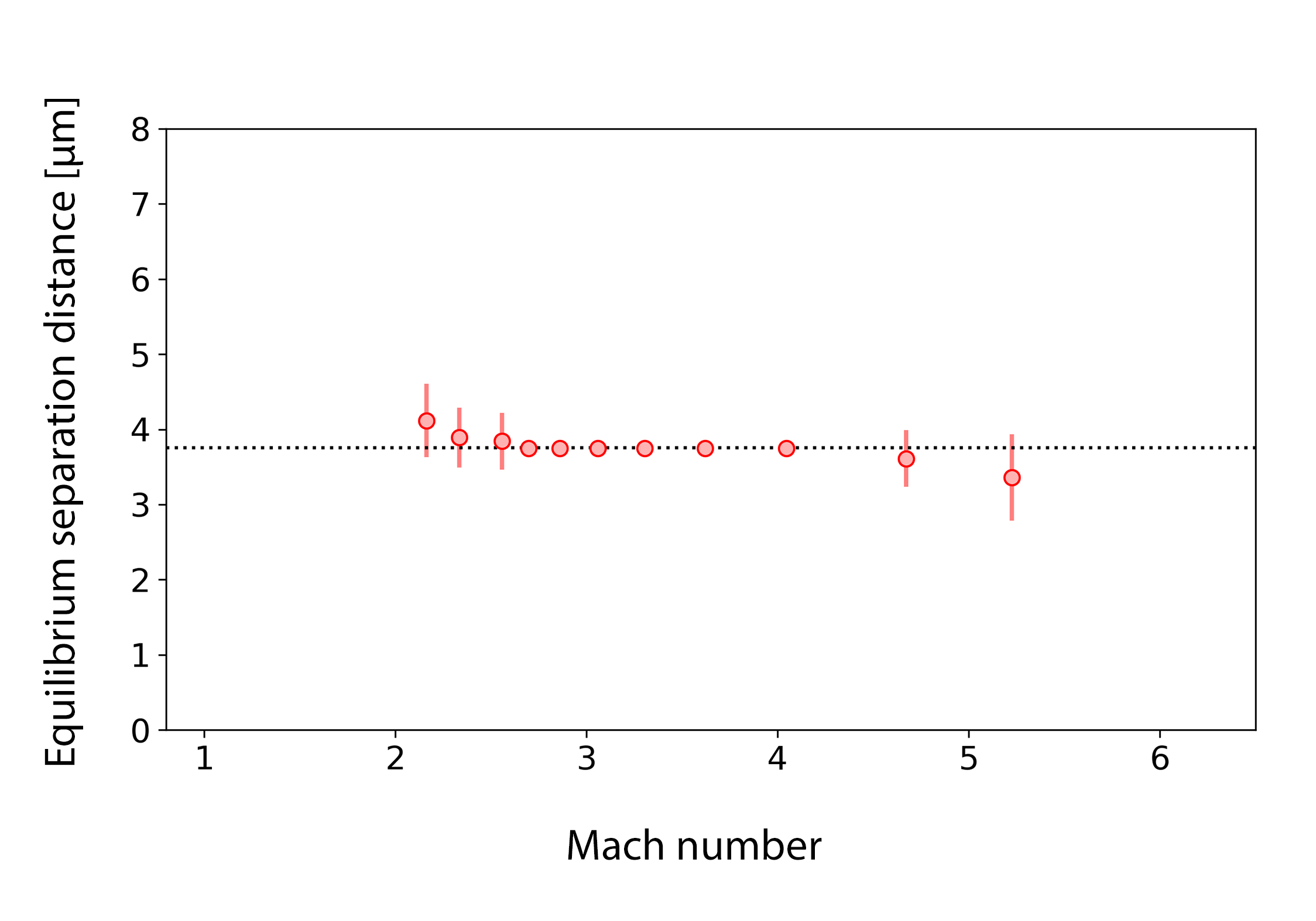}
    \caption{\textbf{Numerical equilibrium separation distances as a function of the Mach number}. Each spot corresponds to the equilibrium distance between two parallel solitons for a different detuning $\Delta E_{lasLP}$; all the other parameters are kept constant.}
    \label{fig:ImprMdsepSimu}
\end{figure}

\paragraph{}
The results of such simulations are presented in figure \ref{fig:ImprMdsepSimu}: as for the experimental ones, no trend appears in the equilibrium separation distance, it stays the same independently of the hydrodynamic conditions.
The black dashed line shows the mean value of the different points, of 3.76 \textmu m in this case.

\paragraph{}
Some remarks need to be done concerning the figure \ref{fig:ImprMdsepSimu}. First of all, the Mach number range is shorter than the experimental one. 
It is due to the fact that only the detuning is changed on this scan: also numerically, a large enough detuning is necessary to implement a bistability ($\Delta E_{lasLP} > \sqrt{3} \hbar \gamma$, with $\gamma$ the lower polariton decay rate), itself required for the propagation of the solitons through the fluid. On the other hand, a too high detuning leads to unstable solitons \cite{Kong2010b, Koniakhin2019a}: instead of staying parallel and dark, they oscillate and blur, making impossible the definition of a separation distance.

\begin{figure}[h]
    \centering
    \includegraphics[width=0.6\linewidth]{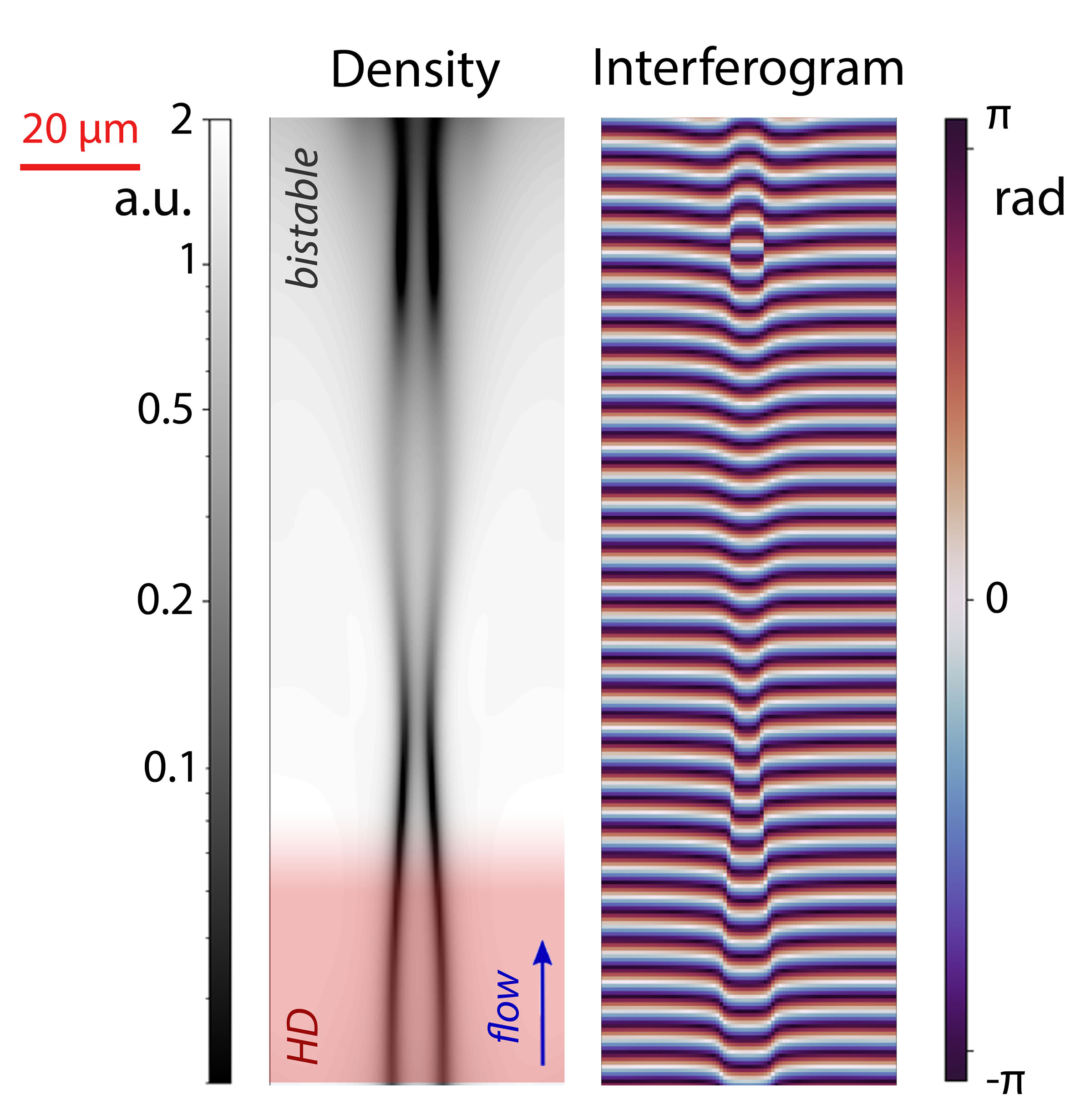}
    \caption{\textbf{Numerical simulation of the impression of unstable solitons.}. The conditions do not allow a parallel propagation of the soliton pair, which induces a breathing behaviour and an uncertainty on the separation distance definition.}
    \label{fig:SimuUnst}
\end{figure}

The second remark, connected with this behaviour, is the presence of error bars in numerical results. They come from the previously mentioned oscillations: the dots of figure \ref{fig:ImprMdsepSimu} are the mean value of the separation distance between the solitons over 50 \textmu m propagation. If the conditions are not exactly favourable to the solitons propagation, they slightly oscillate, which is represented through the error bars.
Such behaviour is presented on figure \ref{fig:SimuUnst}, where we can see a breathing along the propagation of the soliton pair.

\paragraph{}
Even though unexpected at first, those numerical results confirm the experimental ones: the equilibrium distance between the parallel dark solitons is independent of the hydrodynamic conditions. But the question remains: what does it depend on ?

\paragraph{}
Another specificity of our system that could explain this unusual behaviour is its driven-dissipative nature. In our configuration, the pump is sent everywhere; that is exactly what led the solitons to align in the spontaneous configuration (see section \ref{sec:SpontSolitons}). 
This driven-dissipative character is imposed by the decay rate of the polariton $\gamma_{polaritons}$, itself defined by the cavity mirrors reflectivity.
Therefore, the equilibrium distance could be governed by the dissipation of the system. From the experimental point of view, a scan of the polariton decay rate would mean a collection of different samples, with exactly the same properties except for their DBR thickness. As we do not have them in the lab, the verification of the effect of the decay rate has been studied numerically.

\paragraph{}
To do so, specific simulations have been made, where this time all the hydrodynamic conditions were kept constant. Only the decay rate of the polaritons has been tuned, so that any observed trend could be associated with the dissipation.
The parameters have been chosen close to the experimental ones: $\Delta E_{Xcav} = E_{X} - E^{0}_{cav} = 1$ meV, $2 \hbar\Omega_{R} = 2.7$ meV, $\Delta E _{lasLP} = 0.3$ meV, $v_{fluid} = 1$ \textmu m/ps.

\begin{figure}[h]
    \centering
    \includegraphics[width=0.85\linewidth]{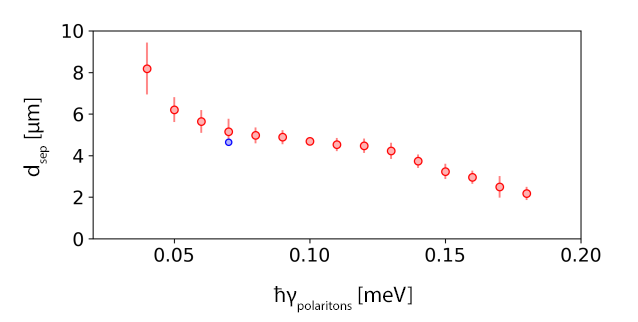}
    \caption{\textbf{Numerical simulations of the equilibrium separation distance as a function of the decay rate of the polaritons}. The red dots correspond to the equilibrium separation distance of the parallel solitons for different polariton decay rate, while fluid and sound speeds are kept constant at typical experimental values. The blue dot illustrates our experimental results: the mean value of all the extracted separation distances for the decay rate of our cavity.}
    \label{fig:ImprgmadsepSimu}
\end{figure}

\paragraph{}
The results are presented in figure \ref{fig:ImprgmadsepSimu}. The red dots correspond to the mean separation distance of each simulation realized for a specific polaritons decay rate. 
The error bars come again from the slight breathing of the solitons in the region where the separation distance is averaged.
This time, a clear trend is observable: the separation distance strongly depends on the polaritons decay. A higher decay rate, \textit{i.e.} a shorter polariton lifetime, induces a smaller separation distance.

\paragraph{}
This is indeed concordant with the fact that the inner region of the solitons is effectively unpumped by the driving field because it is out of phase: the refilling is only done through tunneling across the solitons.
Therefore, for a certain distance between the solitons, a shorter polariton lifetime leads to a lower density in between the solitons: the equilibrium takes place for solitons closer together.

\paragraph{}
The linewidth of our cavity is $\hbar \gamma_{polaritons} = 0.07$ meV, which corresponds to a lifetime of 10 ps. 
We illustrated our experimental results through the blue dot, which corresponds to the mean separation distance of all our results, the 4.75 \textmu m extracted from figure \ref{fig:ImprMdsepExp}.
It stands within the error bar of the numerical results and allows to conclude for a good agreement between the theory and the experiment.

\paragraph{}
The work of this chapter has resulted in the implementation of a new technique of phase pattern impression. 
The phase constraint imposed by the quasi-resonant pumping has been released using the bistability properties, and the design of the driving field through the SLM allows to artificially generate dark solitons on demand. 
The presence of the driving field along the propagation of the solitons enhances its distance, but is also responsible for an interesting binding mechanism between the solitons. 
They propagate parallel, as a form of dark-soliton molecule in a local nonlinear medium.
The deeper study of this behaviour lead us to notice that the characteristic separation distance between the soliton is governed by the driven-dissipative nature of our system.
This technique is very promising for the control and manipulation of such collective excitations, and could for instance be used in the generation of multiple solitonic pattern, or the study of the solitons mutual interaction.

%\bibliographystyle{unsrt}
%\bibliography{bibs/LKB-bibs-bib_thesis-ImprintingChap}  

%\end{document}
%\documentclass[a4paper,11pt]{book}
%\usepackage[utf8]{inputenc}
%\usepackage[T1]{fontenc} 
%\usepackage{lmodern} 
%\usepackage[margin=28mm,includeheadfoot,bindingoffset=5mm]{geometry}[2010/03/13]

%\usepackage{graphicx}
%\usepackage{amsmath}
%\usepackage{bbold}
%\usepackage{amssymb} % pour le signe \lesssim
%\usepackage{textcomp} % \textdegree
%\usepackage[most]{tcolorbox} 
%\usepackage{enumitem} 
%\usepackage{xcolor}
%\usepackage{float} 
%\usepackage{physics}
%\usepackage{stmaryrd}
%\usepackage{wasysym} 
%\usepackage{tikz}
%\usepackage{hyperref}
%\usepackage{cite}
%\usepackage{textgreek}
%\newcommand*\circled[1]{\tikz[baseline=(char.base)]{
%          \node[shape=circle,draw,inner sep=2pt] (char) {#1};}}
%\renewcommand{\thesubsubsection}{\roman{subsubsection}}
%\tcbset{enhanced,colback=red!5!white, colframe=red!75!black,fonttitle=\bfseries}
%\graphicspath{{figures/}} %Setting the graphicspath
%\setcounter{tocdepth}{3}
%\setcounter{secnumdepth}{3}

%\begin{document}

% \tableofcontents
%\setcounter{chapter}{4}

\chapter{Generation of solitonic patterns in static polariton fluid}
\chaptermark{Solitons in static fluid}

\paragraph{} 
In the two previous chapters, we studied the generation and propagation of parallel dark solitons in a supersonic polariton fluid. However, we did not focus on the stability of those solitons, which was assured by the high speed of the polariton flow. 
Indeed, solitons are unidimensional objects, which, when placed in bi- or tridimensional environments, develop transverse modulations known as "snake instabilities" \cite{Kivshar2000}. 
Those ones have been studied theoretically and experimentally in different systems, such as self-defocusing nonlinear media \cite{Kuznetsov1995, Tikhonenko1996, Mamaev1996} or ultracold bosonic and fermionic gases \cite{Anderson2001b, Dutton2001, Cetoli2013}.

\paragraph{}
Snake instabilities lead solitons to decay into quantized vortex-antivortex pairs, resulting into quantum vortex streets which can be seen as quantum equivalent of the von Karman vortex streets.
They were not observed in our driven dissipative quantum fluids, as under resonant excitation, solitons were always generated within a flow as we saw, and under non-resonant excitation, the very fast relaxation of the solitonic structures should prevent the observation of those instabilities.

\paragraph{}
This chapter is based on a theoretical proposal made as part of a collaboration with the group of Guillaume Malpuech in Clermont-Ferrand \cite{Koniakhin2019a}.
It follows the previous results on bistability and suggests a new configuration to generate solitonic pattern, but this time within a static polariton fluid. 
It uses a transverse confinement within an intensity channel to create a pair of dark solitons, which decays into vortex streets due to the disorder of the system.

\newpage

\section{Theoretical proposal}
\label{sec:TheoryChannel}

\paragraph{}
Before presenting our experimental results in the next section, we give here an overview of the theory \cite{Koniakhin2019a} and of the first numerical realizations.

\subsection{Bistability and domain walls}

\paragraph{}
We use again in this section a polariton fluid pumped quasi-resonantly, but the pump is not spatially uniform and can be decomposed into two parts.
A low intensity pump is sent everywhere, and is called the support, labeled as $S$. 
On top of it, some regions are also pumped by a strong power, named the pump and labeled $P(\mathbf{r})$.

This configuration is very similar to the one we presented in chapter \ref{chap:SpontChap}, except that the support intensity is this time scanned also below the bistability cycle. 
The position of the support intensity plays an important role on the behaviour of the illuminated fluid, as we will discover.

The dynamics of the fluid is still described by the driven-dissipative Gross-Pitaevskii equation (equation \ref{eq:ddGPE}), that we can rewrite in our context, considering only the lower polariton branch:

\begin{equation}
    i \dfrac{\partial \Psi}{\partial t} = \Big(
    - \dfrac{\hbar^{2}}{2 m^{*}} \nabla^{2} - i \gamma
    + g |\Psi|^{2} \Big) \Psi
    + (S + P) e^{-i \omega_{0}t}
\end{equation}

where $m^{*}$ is the effective mass of the polaritons, $\gamma$ their decay rate, $g$ their interaction constant and $\Delta E_{lasLP} = \hbar \omega_{0}$ the detuning between the pump energy and the lower polariton resonance.

\paragraph{}
It is important to notice at this point that both S and P are sent at normal incidence: there is no in-plane wavevector, and consequently no velocity of the fluid which is at rest.

\begin{figure}[h]
    \centering
    \includegraphics[width=0.65\linewidth]{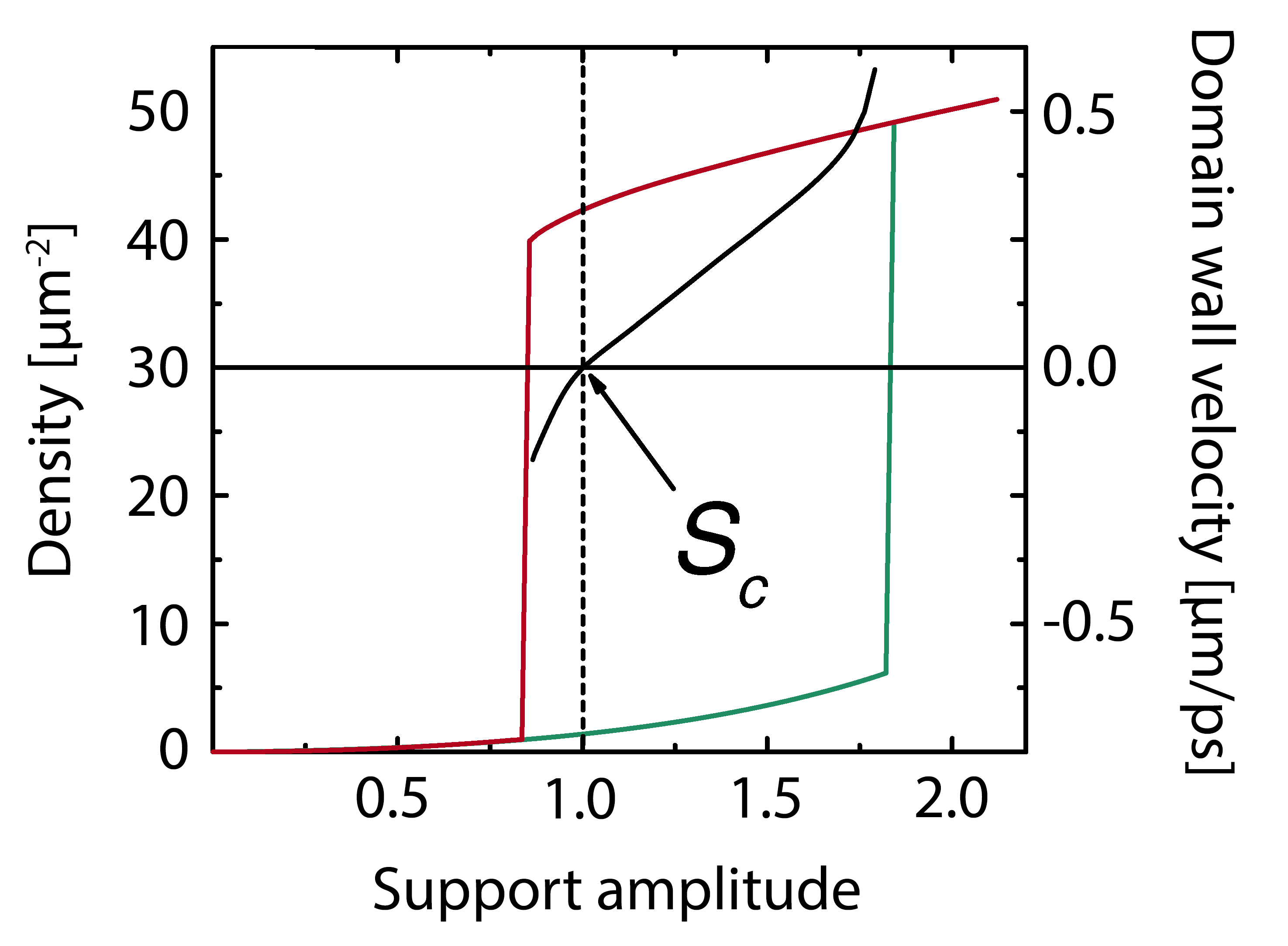}
    \caption{\textbf{Polariton bistability and domain wall velocity}. Polariton density as a function of the support amplitude, in green for increasing support and red for the decreasing one. The black line illustrates the domain wall velocity (\textit{y} axis on the right). The typical support amplitude $S_{c}$ corresponds to zero velocity of the domain wall.
    From \cite{Koniakhin2019a}}
    \label{fig:TheorBist}
\end{figure}

\paragraph{}
If we first consider only the support, the polariton fluid is homogeneously pumped, and a bistability takes place.
This one is illustrated on figure \ref{fig:TheorBist} in green and red: the green line shows the evolution of the polariton density for an increasing support field, and the red one for a decrease of the support, which shows different behaviours.

\begin{figure}[h]
    \centering
    \includegraphics[width=0.7\linewidth]{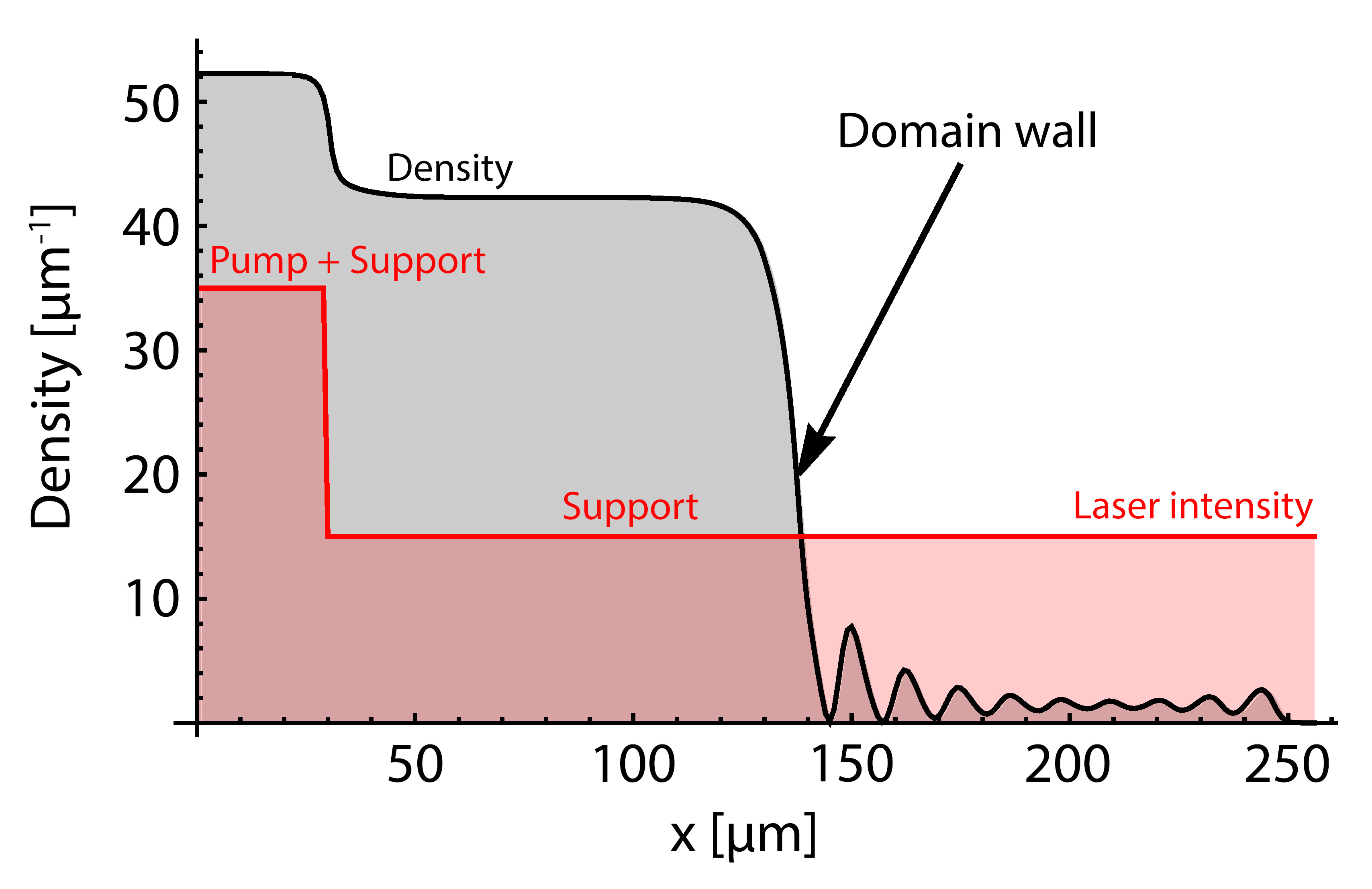}
    \caption{\textbf{Pump-support configuration and domain wall}. Pumping profile in red: the weak support is everywhere while the strong pump is sent only for $x < 30$ \textmu m. The black line shows the polariton density: it remains high even in region where only the support is sent, before abruptly dropping around $x = 135$ \textmu m, creating a domain wall.
    From \cite{Koniakhin2019a}}
    \label{fig:DomainWall}
\end{figure}

\paragraph{}
Now we can add the pump on a specific region of the fluid: figure \ref{fig:DomainWall} illustrates in red the laser profile. 
The support is sent everywhere, while the pump $P(\mathbf{r})$ is only present for $x<30$ \textmu m.
The polariton density is plotted in black. As in chapter \ref{chap:SpontChap}, we see that it does not follows the same profile as the pump.
Indeed, the pump sets the polariton density in the upper branch of the bistability, which stays there even in region where only the support is sent ($30<x<140$ \textmu m in figure \ref{fig:DomainWall}).
At some point however, the high density is not sustained anymore, and the polariton density abruptly drops.
As a discrete symmetry has been spontaneously broken, this sudden drops can be defined as a domain wall \cite{Weinberg1996}, and is followed by small density oscillations (from $x \approx 140$ \textmu m).
The decay of the density takes place within one healing length $\xi = \hbar / \sqrt{2 m^{*} g n}$ ($m^{*}$ the polariton effective mass, $g$ their interaction constant and $n = |\Psi|^{2}$ their density).

\paragraph{}
The domain wall can propagate along the \textit{x} direction, and is stable against instabilities along the \textit{y} one.
The propagation of such domain walls have been previously studied in polaritons \cite{Liew2008, Amo2010} and, more largely, in optics \cite{Rosanov2002, Ganne2001, Odent2016}.

\paragraph{}
The velocity of this domain wall is of particular interest as it gives the direction of the moving wall and the condition of stability. It has been computed numerically as a function of the support intensity, and plotted in black line in figure \ref{fig:TheorBist}. 
We can see the particular support intensity $S_{c}$ for which the domain wall velocity is zero. Its value can be evaluated by the Maxwell construction \cite{Clerk-Maxwell1875, Koniakhin2019a} for $\gamma \rightarrow 0$.
It is done by considering the stationary driven-dissipative Gross-Pitaevskii equation in which we neglect the loss term, resulting in:

\begin{equation}
    \bigg( - \hbar \omega_{0} - \dfrac{\hbar^{2}}{2m^{*}} \nabla^{2} 
    + g |\Psi|^{2} \bigg) \Psi + S = 0
\end{equation}

Since all coefficients are real, the wave function $\Psi$ and the pumping $S$ can only take real values as well. The previous equation can thus be rewritten as a Newton's equation of motion for a material point ($x = \Psi$) with a mass $m_{0} = \hbar^{2}/2m^{*}$:

\begin{equation}
    m_{0} \dfrac{d^{2}x}{dt^{2}} = F(x)
\end{equation}

This material point is thus governed by a position-dependent force $F(x) = gx^{3} - \hbar \omega_{0} x +S$, to which can be defined a potential $U$, which the two maxima correspond to the two stable domain of the system, \textit{i.e.} the high and low density branch of the bistability. 
The system remains stationary only if these two maxima are equal, otherwise the domain wall propagates. 
The analytical solution of this equation leads finally to the critical value of the support intensity $S_{c}$:

\begin{equation}
    S_{c} \approx \dfrac{2(\hbar \omega_{0})^{3/2}}{3 \sqrt{3g}}
\end{equation}

\paragraph{}
The velocity of the domain wall evolves as a function of $S - S_{c}$. Indeed, for a support intensity higher than $S_{c}$, the wall propagates to the right with a speed proportional to $S-S_{c}$:

\begin{equation}
    v \approx 2 \dfrac{S - S_{c}}{S_{c}} \dfrac{\xi}{\tau}
\end{equation}

\paragraph{}
Then, the high density region expands and eventually fills up the whole space.
The healing length $\xi$ has been chosen to be 1.8 \textmu m for these calculations.
On the contrary, for support intensity lower than $S_{c}$, the domain wall move backward and reach the limit of the pump region.
Finally, if the support intensity is close to $S_{c}$, the speed of the domain wall is close to zero and the domain wall is localized in space. In these conditions, the Gross-Pitaevskii solutions bifurcate to dark soliton multiplets, as we will see in the next section.

\paragraph{}
The moving property of this domain walls explains the results we presented in chapter \ref{chap:SpontChap}. At that time, we wanted to extend the region of the fluid with high polariton density, which is why we placed the system above the critical support intensity $S_{c}$, placing the whole illuminated area on the upper branch of bistability. In this chapter, on the contrary, we want to keep a low intensity on the support illuminated area and use the pressure of the neighbour high density regions to create solitonic pattern. Therefore, the chosen support intensity will mostly been chosen below the critical intensity $S_{c}$.

\subsection{Guided solitons instabilities}

\paragraph{}
The next step is now to add another high density region in front of the other. In the bidimensional fluid, it results in a central channel where only the support is present, while the lateral regions are illuminated by both the support and the pump. The width of the channel is fixed constant at $L = 23$ \textmu m, which corresponds to $13 \xi$ of the high density region.

\paragraph{}
The opposition of the two domain walls can lead to the development of solitons. Indeed, with an adequate ratio of the pump/support intensities, their interaction can create a phase jump and thus two regions in phase opposition, resulting in two parallel solitons guided by the low density channel.
The channel is indeed in the lower branch of the bistability cycle, where stationary phase defects exists: the phase is not fixed as the particles diffuse from the high density regions \cite{Koniakhin2019a}.

\begin{figure}[h]
    \centering
    \includegraphics[width=0.99\linewidth]{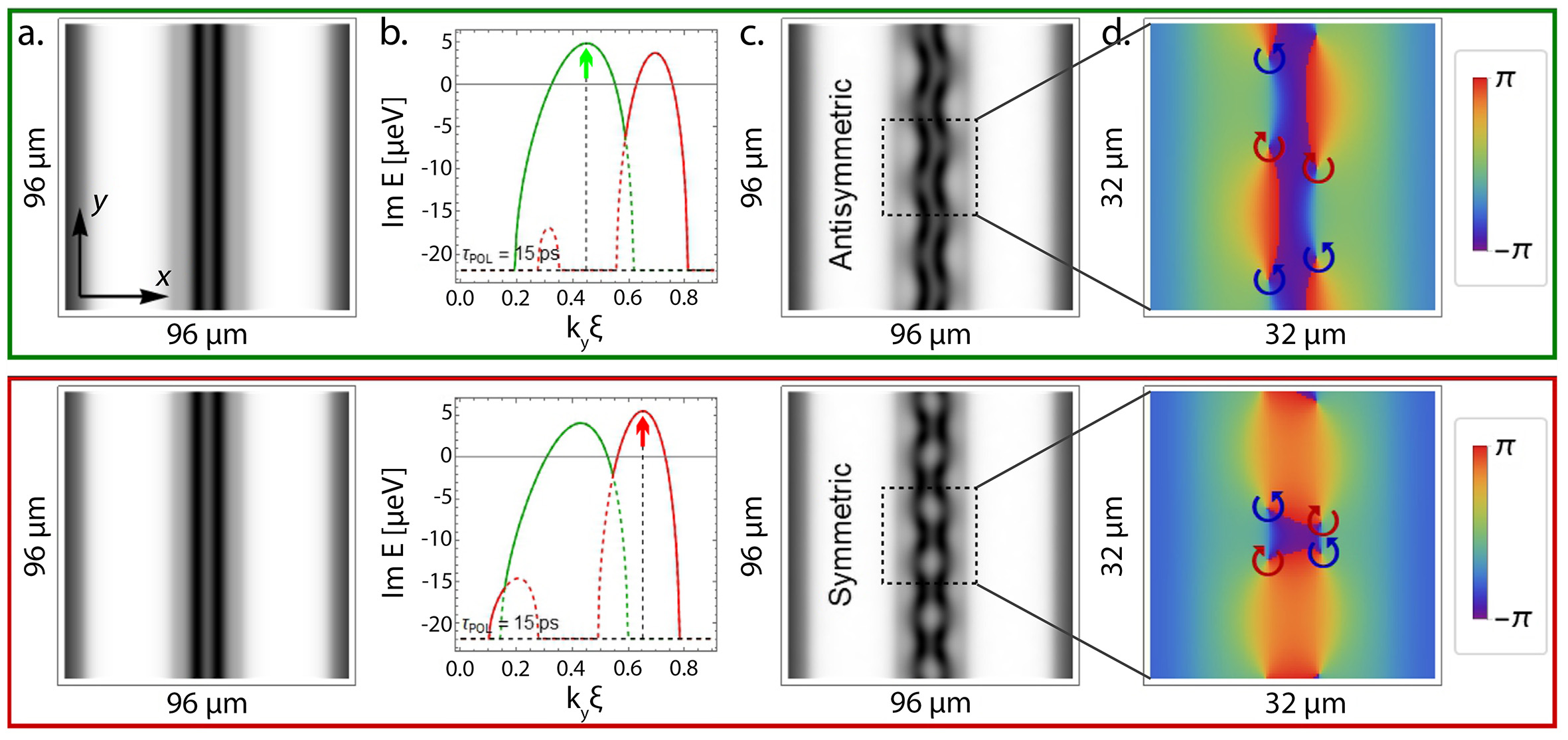}
    \caption{\textbf{Instabilities of guided solitons}. Solitons generation inside a channel of low intensity. The channel has a width $L = 25$ \textmu m and the support intensity within the channel is $s = 0.25 S_{c}$. The pump on the lateral regions is $P = 1.25 S_{c}$ for the upper line and $P = 2 S_{c}$ for the lower one.
    a. Intensity map of the ideal case of stable parallel dark solitons generation. b. Imaginary part of the energy of weak excitations of the stationary solutions of a. The green and red colors represent the symetric and antisymetric modes. 
    c. Intensity map of the stationary solutions with weak disorder, leading to the generation of modulational instabilities. d. Phase map corresponding to the bordered region of picture c.
    From \cite{Koniakhin2019a}}
    \label{fig:GuidedInst}
\end{figure}

\paragraph{}
The numerical results of this configuration are presented in figure \ref{fig:GuidedInst}, for a support intensity $S = 0.25 S_{c}$. 
Images a. present the ideal case density map: the presence of two domain walls in front of each other gives rise to two dark solitons in the channel. The system is invariant along \textit{y} so is effectively 1D.
One can notice that, as the high density regions impose their identical phase to the fluid, only a even number of solitons can be sustained inside the channel.

\paragraph{}
These solitons are however created in a static fluid: they are subject to instabilities along the \textit{y} direction.
They can be studied through the imaginary part of the energy of weak excitations, obtained through the Bogoliubov-de Gennes equations \cite{Morgan2013, Carusotto2004a, Solnyshkov2008}.
This one is plotted in figure \ref{fig:GuidedInst}.b, for two different pump intensities: the upper line corresponds to the case of $P = 1.25 S_{c}$, while for the lower one $P = 2 S_{c}$. The green and red colors show the symmetric and antisymmetric modes.

Modulational instabilities, known as \textit{snake instabilities} and well known in conservative condensates \cite{Kuznetsov1988}, are linked to the presence of a positive imaginary part of the energy.
Those ones are plotted in figure \ref{fig:GuidedInst}.b: we can see that they can be decomposed into two components (in red and green) which correspond to modes with different symmetries. 
The highest one defines the symmetry of the instabilities: on the upper line of figure \ref{fig:GuidedInst}, the final instabilities have an antisymmetric pattern, while on the lower line, the highest mode is the red one, which leads to symmetric instabilities.
More details about the modes competition are given in \cite{Koniakhin2019a}.

\paragraph{}
The instabilities in the simulations are generated by adding small noise or fluctuations as this breaks the translational symmetry along
the \textit{y} axis.
The results of figure \ref{fig:GuidedInst} have been obtained with a weak Gaussian disorder with a correlation length of 2 \textmu m and amplitude $\gamma = 0.01$ meV.

\paragraph{}
Figure \ref{fig:GuidedInst}.c. and d. show the density and phase map of the instable solitonic pattern. Figure d. corresponds to a zoom on the central region of figure c. The position of the instabilities along \textit{y} are determined by the disorder, however it does not affect the pattern shape if the disorder is low enough ($\gamma \ll \hbar \omega_{0}$).
In both cases, the dark parallel solitons break into two stationary vortex streets.
Further analysis \cite{Koniakhin2019a} show the stability of these final patterns: the snake instability has been frozen by the confining potential.

\begin{figure}[h]
    \centering
    \includegraphics[width=0.85\linewidth]{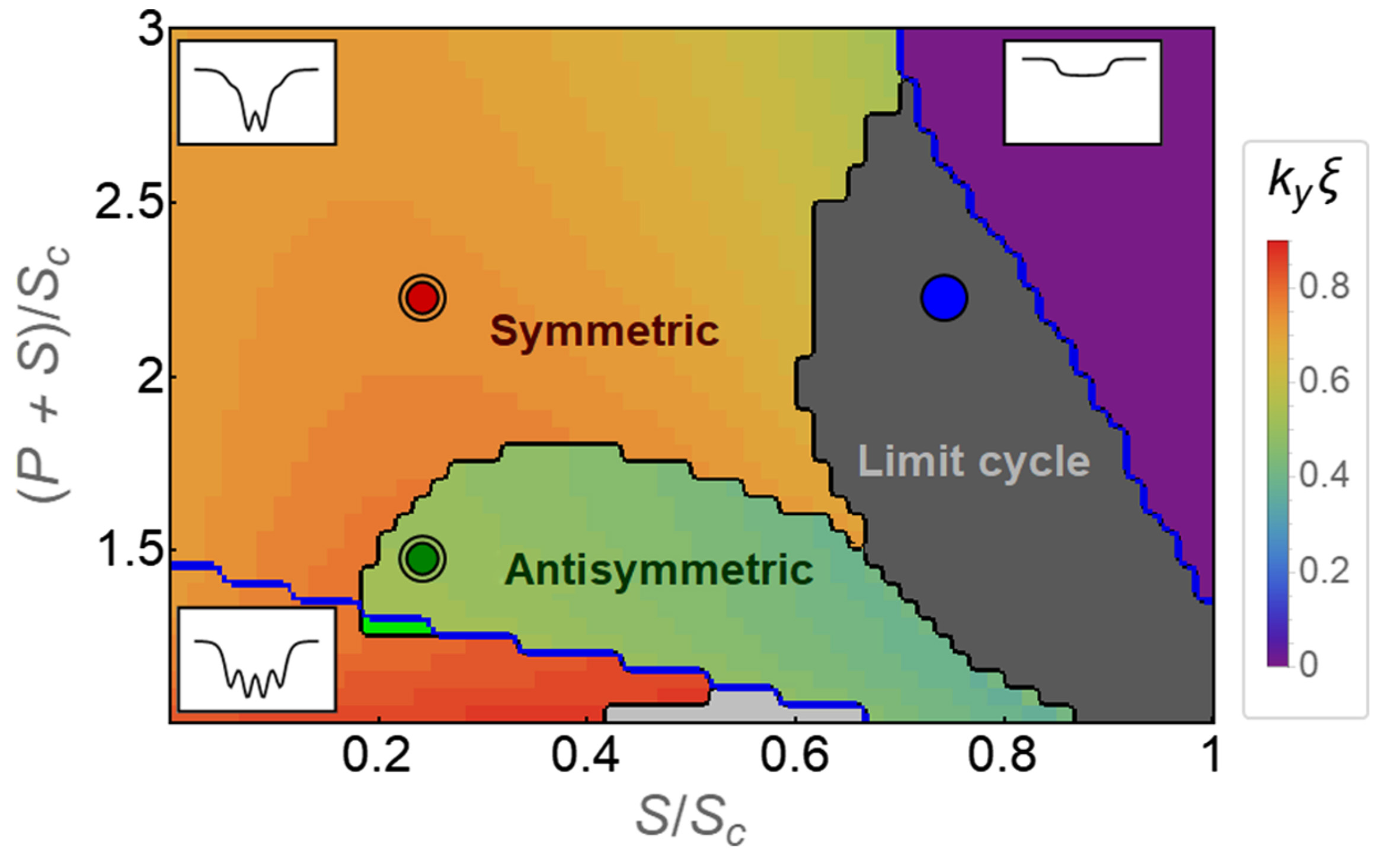}
    \caption{\textbf{Phase diagramm as a function of $S$ and $P$ intensities}. The color scale indicates the maximal instability wavevector, and therefore the configuration of the instabilities (orange-red is symmetric and green is antisymmetric). The blue lines separates the regions with different numbers of solitons.
    On the right, the purple region possess a high intensity inside the channel and therefore no solitons, while the dark grey region delimits a non-stationary steady state.
    The red and green dots correspond to the conditions shown in figure \ref{fig:GuidedInst}, while the blue one indicates the conditions of the maze solving of figure \ref{fig:Maze}.
    From \cite{Koniakhin2019a}}
    \label{fig:PhaseDiag}
\end{figure}

\paragraph{}
We have seen that depending on the pump intensity, the instability symmetry can change. In order to summarize the different configuration conditions, a phase diagram have been realized and is presented in figure \ref{fig:PhaseDiag}.
The colorscale corresponds to the value of the maximal instability wavevector, while the  x and y axis represent respectively the support intensity and the total intensity, normalized by $S_{c}$.
The length of the channel $L$ and the detuning laser-LP $\hbar \omega_{0}$ have been kept constant for the whole diagram.

\paragraph{}
Several regions can be delimited, with different color shades: each of them corresponds to a different configuration within the channel, whose profile is plotted in the insets. 
The whole diagram is divided in three regions delimited by two blue lines. They highlight the number of solitons: on the bottom left angle, four solitons are generated; two solitons can be observed in the center of the diagram, in between the two lines, while no solitons are sustained on the top right angle, where the support intensity is high.

\paragraph{}
The four solitons are sustained for lower pump intensity. The large majority of it corresponds to the symmetric configuration in orange, but a tiny light green region also shows the possibility of four solitons in the antisymmetric configuration.
The small elongated light gray region shows the position where a pattern of four solitons is generated.
It happens for low pump and high support, which therefore induces only a weak transverse polariton flow towards the channel which favors the stability of solitons pattern.

\paragraph{}
The two solitons region is subdivided in three parts. The orange part corresponds to the symmetric instabilities defined in figure \ref{fig:GuidedInst}, while the green one indicates the antisymmetric case.
The dark grey region highlights a non-stationary steady state (limit cycle). This happens in the particular conditions of the simulations, \textit{i.e.} without energy relaxation and with low disorder.
In that case, a pair of breathing solitons is generated which oscillates in time, as shown in the Supplemental Material Video of \cite{Koniakhin2019a}.

\paragraph{}
On the right, the purple region illustrates a high density in the channel, without any solitons nor vortices.
This is comparable to the polariton neuron picture depicted in \cite{Liew2008}, where logic gates are implemented in polariton fluids using similar intensity channels.

\paragraph{}
Another configuration can also be observed by working with no support. In that case, it is possible to tune the phase between the two high intensity regions. In particular, by putting them in opposition of phase, the number of solitons created in the channel is odd, in order to conserve the phase \cite{Koniakhin2019a}.
This can be seen as a 2D generalization of the phase conservation described in \cite{Goblot2016b}.

\subsection{Maze solving}

\paragraph{}
This channel configuration can now be extended in more complex geometries, and in particular to a complicated maze. 
Maze solving studies have started with the work of Leonhard Euler \cite{Euler1736} which is now considered as the starting point of topology. 
More recently, the problem is commonly solved using the potential method \cite{Adamatzky2017}, where a potential is assigned to the destination, and the choices are determined by its gradient. This method is used in many different fields, as in robotics \cite{Shannon1951, Adamatzky2020}, in biology through the motion of biological organisms \cite{Nakagaki2000, Tweedy2020}, in microfluidics \cite{Fuerstman2003} or in plasma physics \cite{Reyes2002}.
Some chemical methods based on the velocity map of the reaction have also been developed \cite{Steinbock1995}, as well as optical ones using the diffusion of a wavepacket \cite{Caruso2016}.
In our case, we suggest an algorithm based on the dead-end filling of instabilities, using an all optical control and a picosecond time resolution.

\begin{figure}[h]
    \centering
    \includegraphics[width=0.85\linewidth]{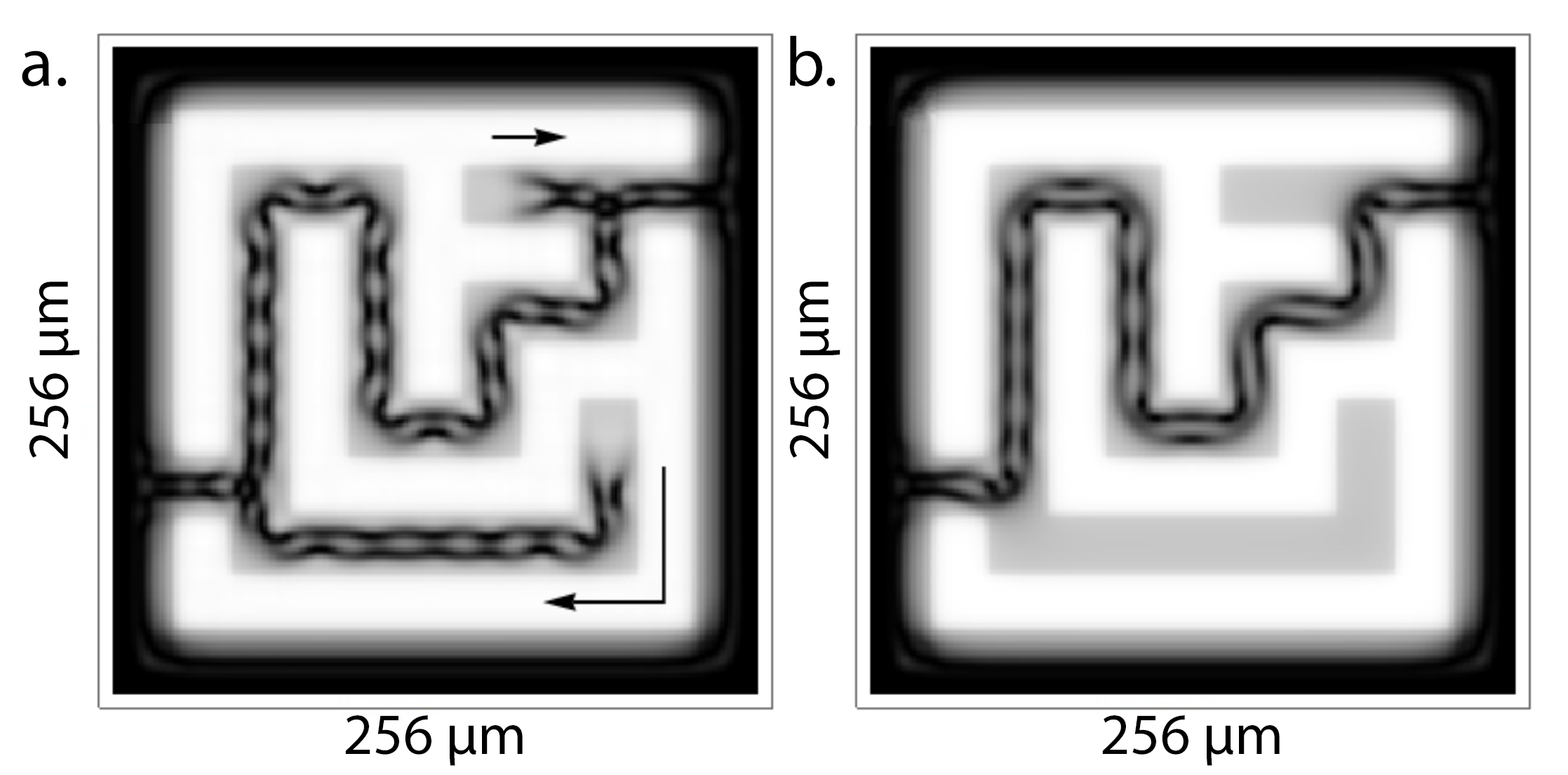}
    \caption{\textbf{Maze resolution}. Intensity maze for $S$ and $P$ conditions indicated by the blue dot in figure \ref{fig:PhaseDiag}. a. Snapshot of the system 20 ps after the pump and support have been switched on: the heads of the solitons start to be repelled by the dead-ends. b. Stationary final distribution after 1 ns: the solitons fill only the path providing the maze solution.
    From \cite{Koniakhin2019a}}
    \label{fig:Maze}
\end{figure}

\paragraph{}
The previous configuration is now developed into an arrangement of low intensity channels forming a maze, as pictured in figure \ref{fig:Maze}. 
The $S$ and $P$ intensities correspond to the blue dot of figure \ref{fig:PhaseDiag}.
All the channels do not have the same configuration: some of them are open-ends, connected to the outside of the maze, while the others are dead-ends, surrounded by high density walls.

\paragraph{}
If an appropriate ratio of $S$ and $P$ is set, just after the switching of the high density regions on the upper bistability branch, the corridors of the maze get filled by solitons.
However, the solitons present in the dead-end channels quickly start to withdraw: figure \ref{fig:Maze}.a shows the system distribution 20 ps after the support and the pump have been switched on.
The heads of the solitons move on the opposite direction of the channel end, as indicated by the black arrows.

\paragraph{}
Eventually, they reach the open-end channels: instead of following their movement until all channels are empty, those ones stay fixed. Figure \ref{fig:Maze}.b shows the configuration of the system 1 ns after being switched on: the only filled corridors are the one solving the maze, \textit{i.e.} the open end ones.

\begin{figure}[h]
    \centering
    \includegraphics[width=0.6\linewidth]{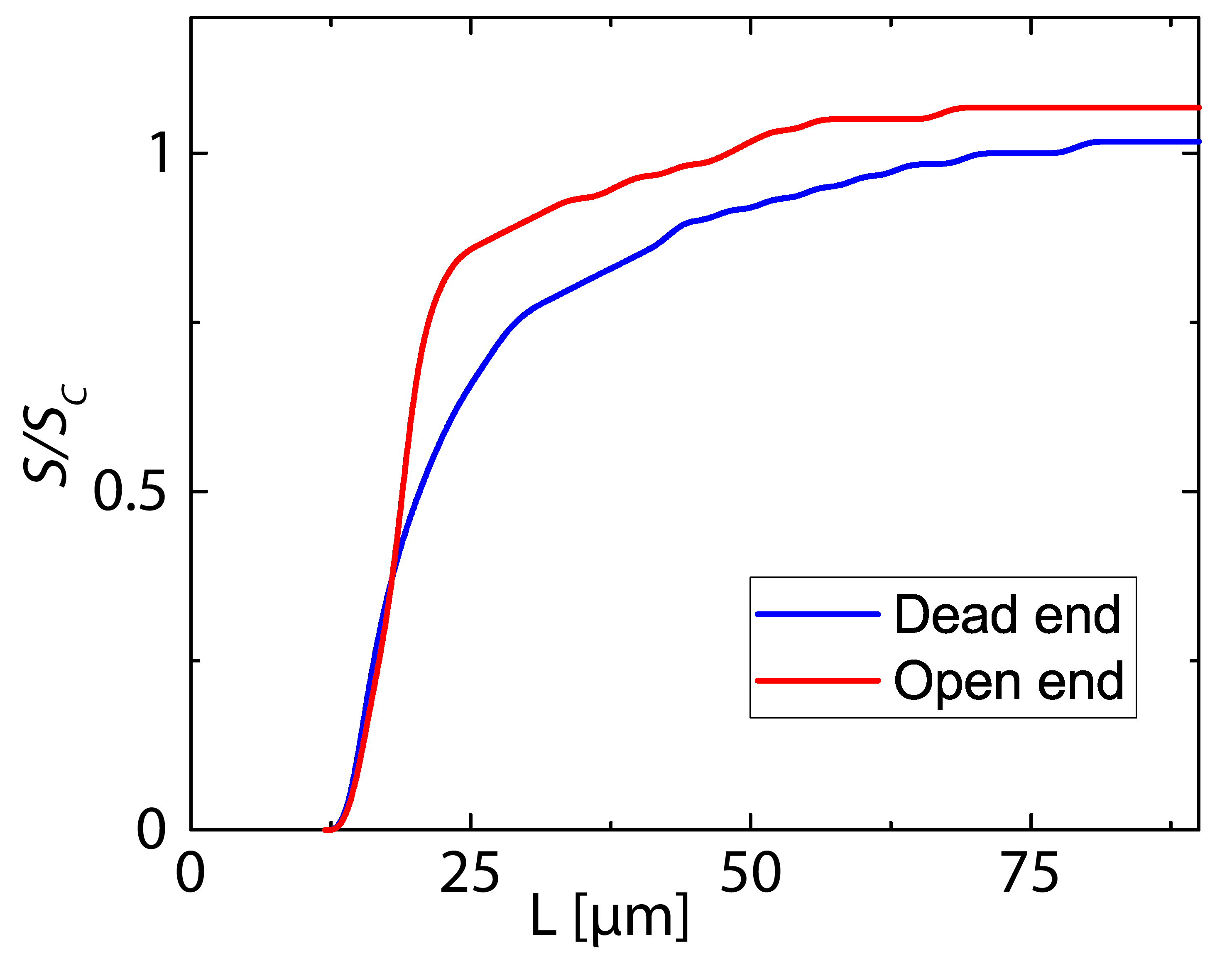}
    \caption{\textbf{Threshold support intensities}. Critical support intensities for the motion of the two types of solitons, the dead ends in blue and the open ones in red, as a function of the width of the channels. The dead ends start to move for a lower intensity than the open ones.
    From \cite{Koniakhin2019a}}
    \label{fig:DeadOpenS}
\end{figure}

\paragraph{}
The dead end channels have indeed a different behaviour than the open channels as their head can be seen as a domain wall by itself \cite{Koniakhin2019a}. However, the motion conditions are not the same as for the lateral domain wall that we studied previously. 
Figure \ref{fig:DeadOpenS} shows the thresholds of the support intensities that start the motion of the domain walls in both cases, dead ends in blue and open ends in red, as a function of the channel width $L$.

\paragraph{}
Below $L = 14$ \textmu m, the channels are to thin to be filled, hence the sharp decrease of the critical support intensities.
Above 14 \textmu m however, the open ends always need a higher support intensity to be emptied of their solitons than the dead ends. Therefore, as long as we choose the support intensity to be in the range between those critical intensities, the dead end solitons will move and thus be removed of the channels, while the open end ones are stable: the maze is solved.

\section{Experimental realization}

\paragraph{}
This theoretical proposal has been followed by its experimental realization. 
After a brief description of the setup, we report the results we obtained for the different configurations we tried.

\subsection{Implementation}

\begin{figure}[h]
    \centering
    \includegraphics[width=\linewidth]{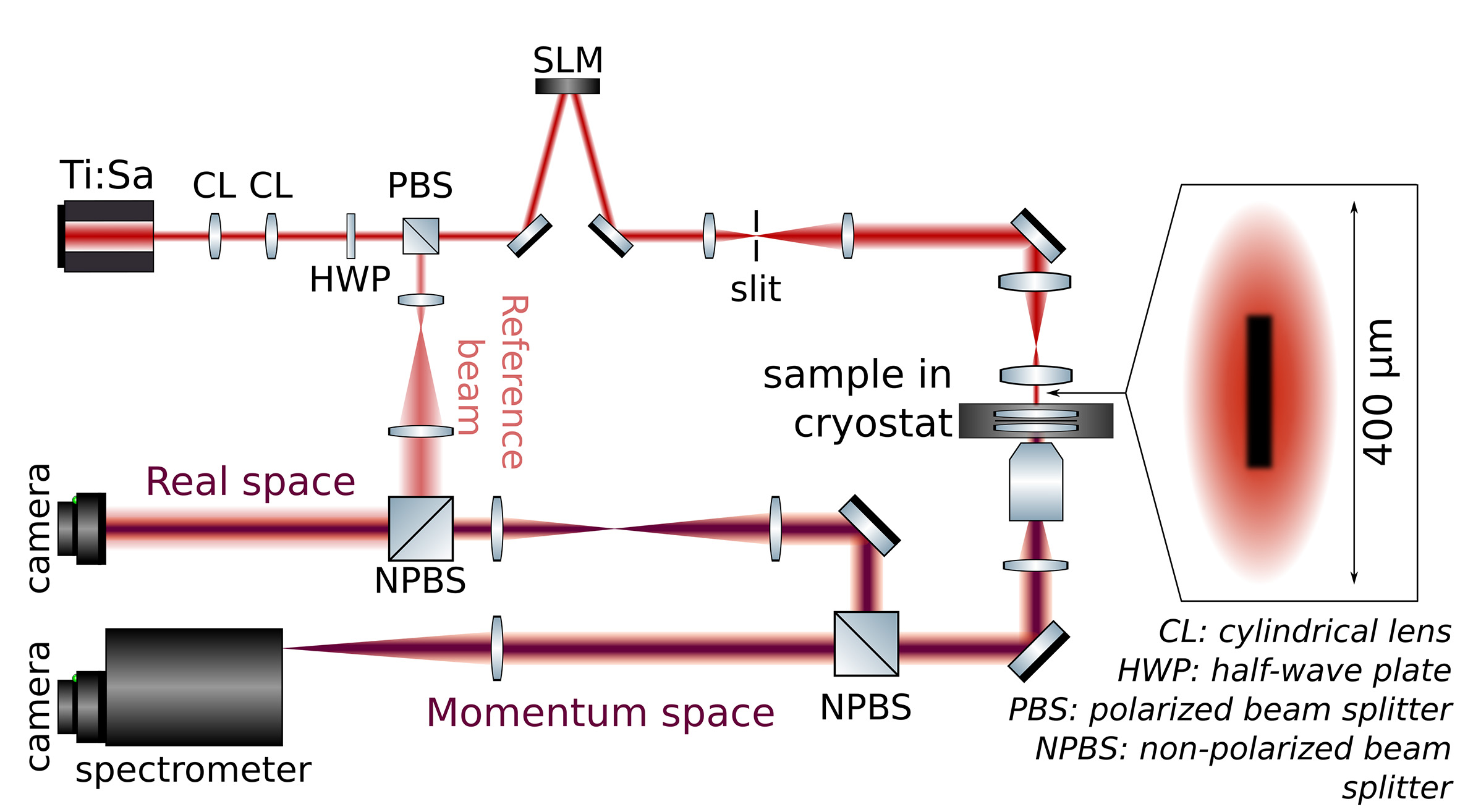}
    \caption{\textbf{Experimental setup}. As for the soliton impression, the beam is shaped through the SLM, this time in intensity: a darker channel is designed in its center. The beam is sent at normal incidence to the sample, and the signal is detected in real and momentum space.}
    \label{fig:SetupChannel}
\end{figure}

\paragraph{}
The setup we used for this experiments is shown in figure \ref{fig:SetupChannel} and is quite similar to the one of the soliton impression, presented in chapter \ref{chap:Impression}.
Two main differences have yet to be noticed.
First of all, to facilitate the development of the instabilities, the experiment needs a static fluid: the excitation is therefore sent at normal incidence, and the in-plane wavevector is zero.
Then, the solitons are generated through intensity channels: the excitation beam is shaped in intensity and not in phase anymore.

\paragraph{}
As we saw in chapter \ref{chap:Setup} and used in chapter \ref{chap:Impression}, the Spatial Light Modulator (SLM) can shape the wavefront of the incoming light beam. 
However, a trick can be employed to use it as a intensity modulator, and thus design an intensity pattern on the beam.
To do so, we work again with a grating from which we keep only the first order, but with a specific, controlled contrast.
Indeed, if the full range of the gray scale is used in the SLM pattern, all the light is diffracted to the first order.
However, by using only part of the scale, only a portion of light is diffracted, resulting in a darker region.
This way, by tuning the grating contrast of the SLM screen and filtering out the first order in the Fourier plane, we can shape a beam with a designed intensity pattern.
Again, as it is managed through a computer, the shape is easily tunable.

\paragraph{}
The detection is the same as before, realized in both real and momentum space. The excitation conditions are extracted from the momentum images, while the real space gives us the density map as well as the phase from interferences with a reference beam.

\subsection{Intensity analysis}

\paragraph{}
The starting point of the experiment is to explore the numerical phase diagram presented in figure \ref{fig:PhaseDiag}.
To do so, we should scan the intensity along both axis, \textit{i.e.} vertically by tuning the total intensity, and horizontally by tuning only the support one.

\begin{figure}[h]
    \centering
    \includegraphics[width=\linewidth]{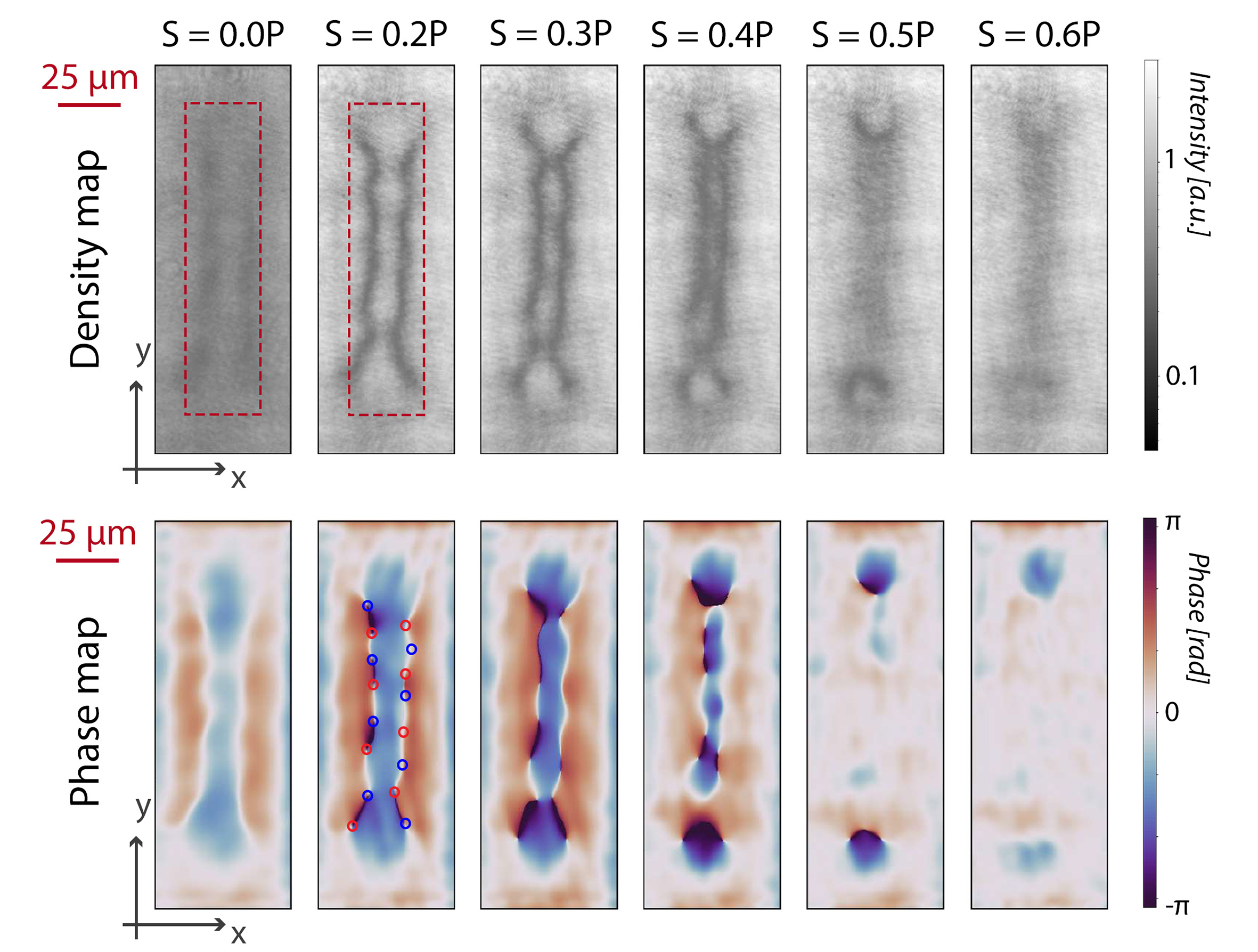}
    \caption{\textbf{Scan of the support intensity $\mathbf{S}$}. The density within the channel increases from left to right: it starts completely dark ($S = 0.0P$) and raises up to $S=0.6P$. The red dashed rectangle indicates the channel position. The symmetric solitonic pattern appears at $S=0.2P$, elongates a bit and vanishes with the intensity increase. The phase maps show the appearance of vortex streets, where vortex and antivortex are marked by the blue and red circles for $S = 0.2P$.}
    \label{fig:ScanS}
\end{figure}

\paragraph{}
Figure \ref{fig:ScanS} illustrates the evolution of the system with a tuning of the support intensity $S$ only. The top line shows the density maps while the corresponding phase maps are plotted on the bottom line.
The support density increases from left to right: the first picture corresponds to a completely dark channel. Nothing happens then as the vortex streets need a support to develop.

On the second picture however, for $S = 0.2P$, the expected vortex streets are clearly visible: the symmetric pattern is clear, and the pinned vortex-antivortex are spotted on the phase map by the red and blue circles, respectively.

\paragraph{}
By increasing gradually the intensity within the channel, the pattern elongates ($S = 0.3P$ then $S = 0.4P$) while looses in definition.
At $S=0.5P$, it has already vanished in the center, only remains the pinned ends of the vortex street.

This behaviour is quite surprising as it does not correspond to the numerical simulations, where the solitons vanished from the end of the channels towards the center. However, it can be explained by the gaussian shape of our excitation beam, while the previous simulations were realized for a flat intensity. 
The conditions are therefore not the same along the channel, the density in particular is lower at the ends, which explains why the instabilities remain there longer.

\paragraph{}
On the last picture, where $S = 0.6P$, the instabilities have completely vanished. The difference between the support and the pump intensities are not high enough for the domain walls to be generated, which therefore prevents the generation of the vortex streets.

\paragraph{}
Those first results coincide well with the numerical simulations, but only the symmetric pattern has been observed. In the phase diagram however, the antisymmetric region has an elongated shape toward the horizontal axis: in order to see them, we then tried to tune the total intensity $S+P$ to scan the diagram vertically.

\begin{figure}[H]
    \centering
    \includegraphics[width=\linewidth]{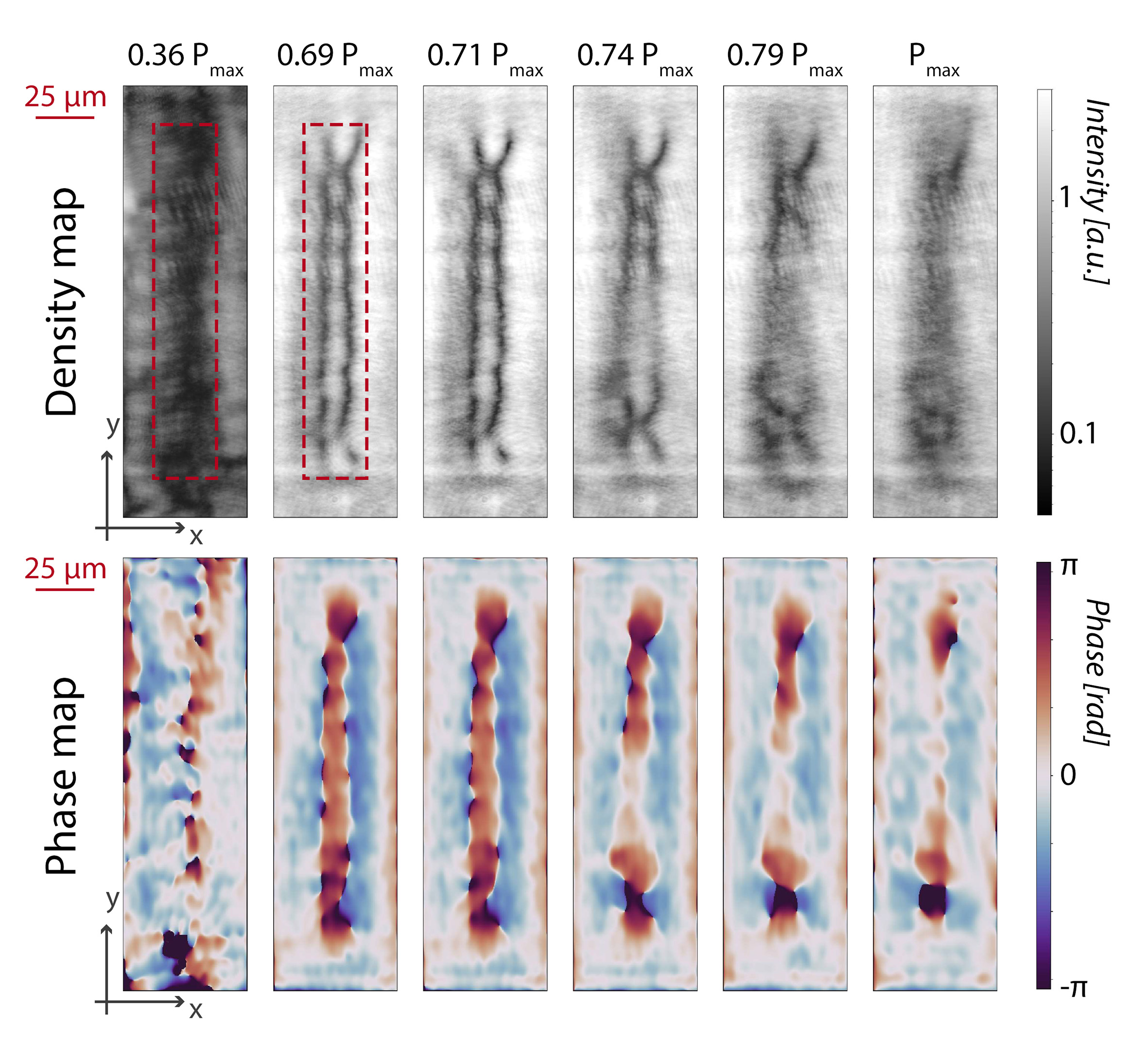}
    \caption{\textbf{Scan of the total intensity $S+P$}. Density (upper line, logarithmic scale) and phase (lower line) maps for different values of the total intensity $S+P$. The laser power increases from left to right until the maximum value $P_{max}$ (0.36$P_{max}$, 0.69$P_{max}$, 0.71$P_{max}$, 0.79$P_{max}$ and $P_{max}$, respectively). At low intensity, the whole fluid is linear. When the pump regions jump into the nonlinear regime, the solitonic pattern appears, best defined for $P = 0.71P_{max}$, then vanishes for higher intensities.}
    \label{fig:ScanSP}
\end{figure}

\paragraph{}
Figure \ref{fig:ScanSP} presents the images of the system for several values of the total intensity $S+P$. 
Again, the top line shows the density maps, and the phase maps are presented in the bottom line.
The intensity increases from left to right until a maximal total power $P_{max}$. 
The red dashed rectangle indicates the position of the low intensity channel: it is centered on the figure, with a horizontal width of 15 \textmu m and a vertical length of 150 \textmu m.

\paragraph{}
The first picture, where $P = 0.36 P_{max}$, shows a linear fluid. Indeed even the regions of the pump do not reach the threshold intensity value for the fluid to jump into the upper nonlinear branch of the bistability.

\paragraph{}
When the intensity becomes high enough so that the high density regions jump into the nonlinear regime, the vortex streets appear ($P = 0.69 P_{max}$). The pattern is again clearly symmetric. 
The instabilities are maintained over a small range of intensities, then gradually vanish from the center, starting from $P = 0.74P_{max}$. 
At the maximum laser power $P_{max}$, the instabilities are gone and only remains the grey shadow of the channel.

\paragraph{}
This time again, only the symmetric configuration has been observed: no clear transition from one symmetry to another is visible. The experiment was repeated several times with different conditions ($S/P$ ratio, laser-LP detuning, channel length...) but the antisymmetric instabilities were never found.

In order to understand our experimental difficulties, new simulations have been made, to plot a new phase diagram corresponding more accurately to our sample properties.

\paragraph{}
The main difference between the previous theory and the experiment comes from the disorder of the sample.
The numerical results presented in section \ref{sec:TheoryChannel} were done in the quasi-ideal case: a slight disorder is necessary to generate the instabilities, but it was chosen to be small, of 0.01 meV amplitude. This way, the solitonic pattern is not affected by the disorder position and thus get a regular shape.

\begin{figure}[h]
    \centering
    \includegraphics[width=0.8\linewidth]{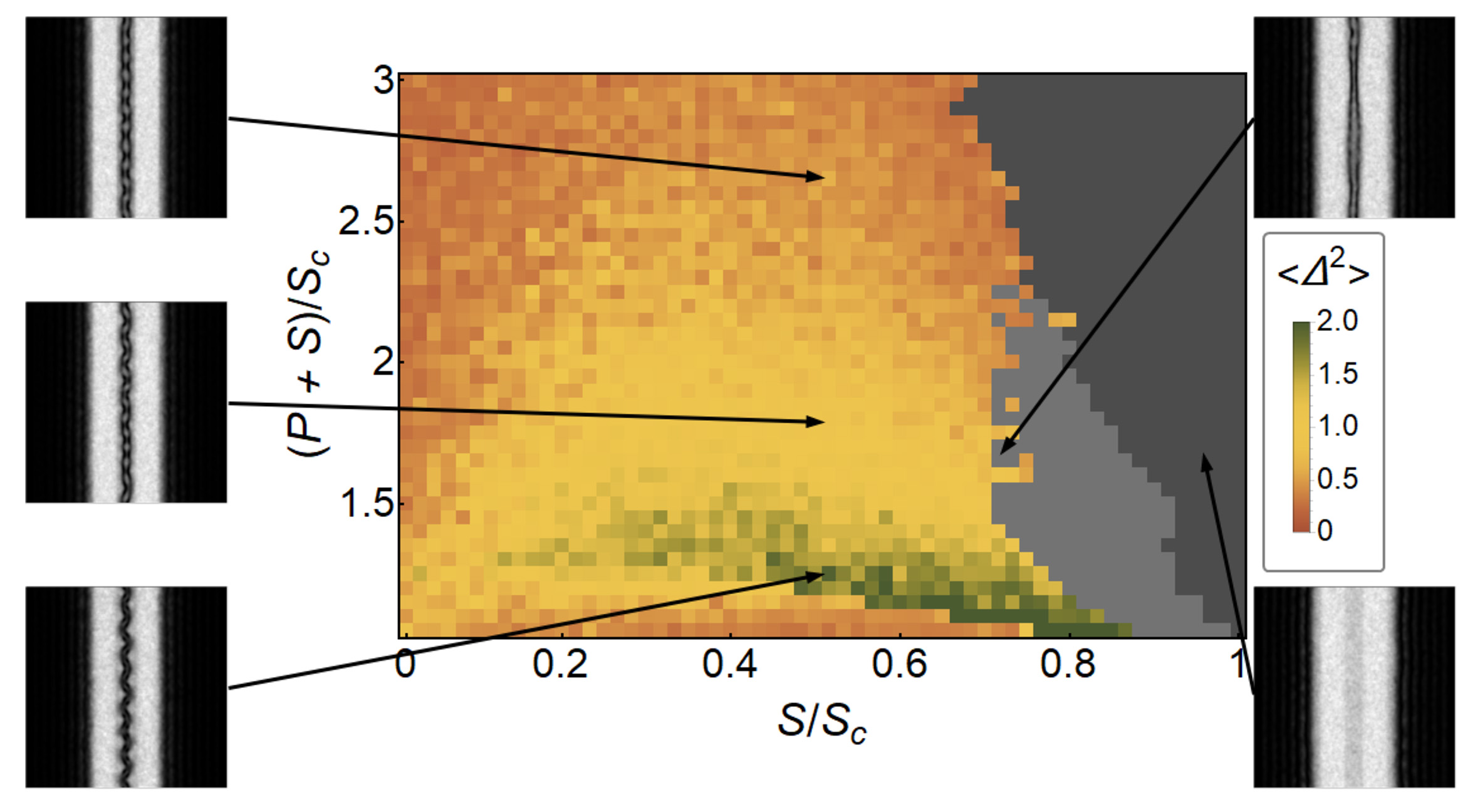}
    \caption{\textbf{Phase diagram in the presence of disorder}. The color scale illustrates the standard deviation of the center between the solitons $\langle \Delta^{2} \rangle$. The different instability configurations are plotted in the insets. The antisymmetric pattern corresponds to the green area, much smaller than previously. The light grey region indicates where the solutions are not stationary in time, while the dark grey area shows where no solitons are visible and the channel is entirely filled with polaritons.
    From \cite{Koniakhin2019a}}
    \label{fig:PhaseDiagSample}
\end{figure}

\paragraph{}
Experimentally, our sample has a more important disorder amplitude: it was evaluated around 0.1 meV. In this case, it plays a role in the shape of the instabilities as some vortex/antivortex can be pinned by the disorder speckle, leading to irregularities in the vortex street pattern.
Therefore, a new phase diagram has been realized, plotted in figure \ref{fig:PhaseDiagSample}

\paragraph{}
The impact of a stronger disorder to the instability shape also makes it more difficult to numerically determine the dominant mode, and therefore the symmetry of the instabilities. 
The new color scale of figure \ref{fig:PhaseDiagSample} phase diagram is thus based on the soliton mass center $X_{c}(y)$, numerically easy to define. 

\paragraph{}
Indeed, in the symmetric configuration, the soliton mass center stays quite aligned along the \textit{y} axis, while in the antisymmetric case, it follows a sinusoidal trajectory. 
Its standard deviation $\big \langle \Delta^{2} \big \rangle = \Big \langle \big( X_{c}(y) - \langle X_{c}(y) \rangle_{y} \big)^{2} \Big \rangle_{y}$ can therefore give us useful information: the higher it is, the more antisymmetric the pattern is.

\paragraph{}
Even though the color scale parameter has been modified compared to figure \ref{fig:PhaseDiag}, it still corresponds to the same symmetry pattern, as illustrated by the insets: orange areas shows symmetrical instabilities while the antisymmetrical ones are located on the green regions.

\paragraph{}
The dark grey region shows where no solitons are generated in the channel, which is entirely filled by the polariton fluid. The lighter grey corresponds to solutions that are not stationary in time, and thus not clearly observable experimentally.

The clear antisymmetric pattern corresponds to the greener pixels: the area is actually much smaller than previously, hence the difficulty to experimentally observe it.

\paragraph{}
Now that we have tuned the intensities of the system, we can also use the properties of the SLM to study the influence of the shape of the channel on the instability behaviour.

\subsection{Shape tuning}

\begin{figure}[H]
    \centering
    \includegraphics[width=\linewidth]{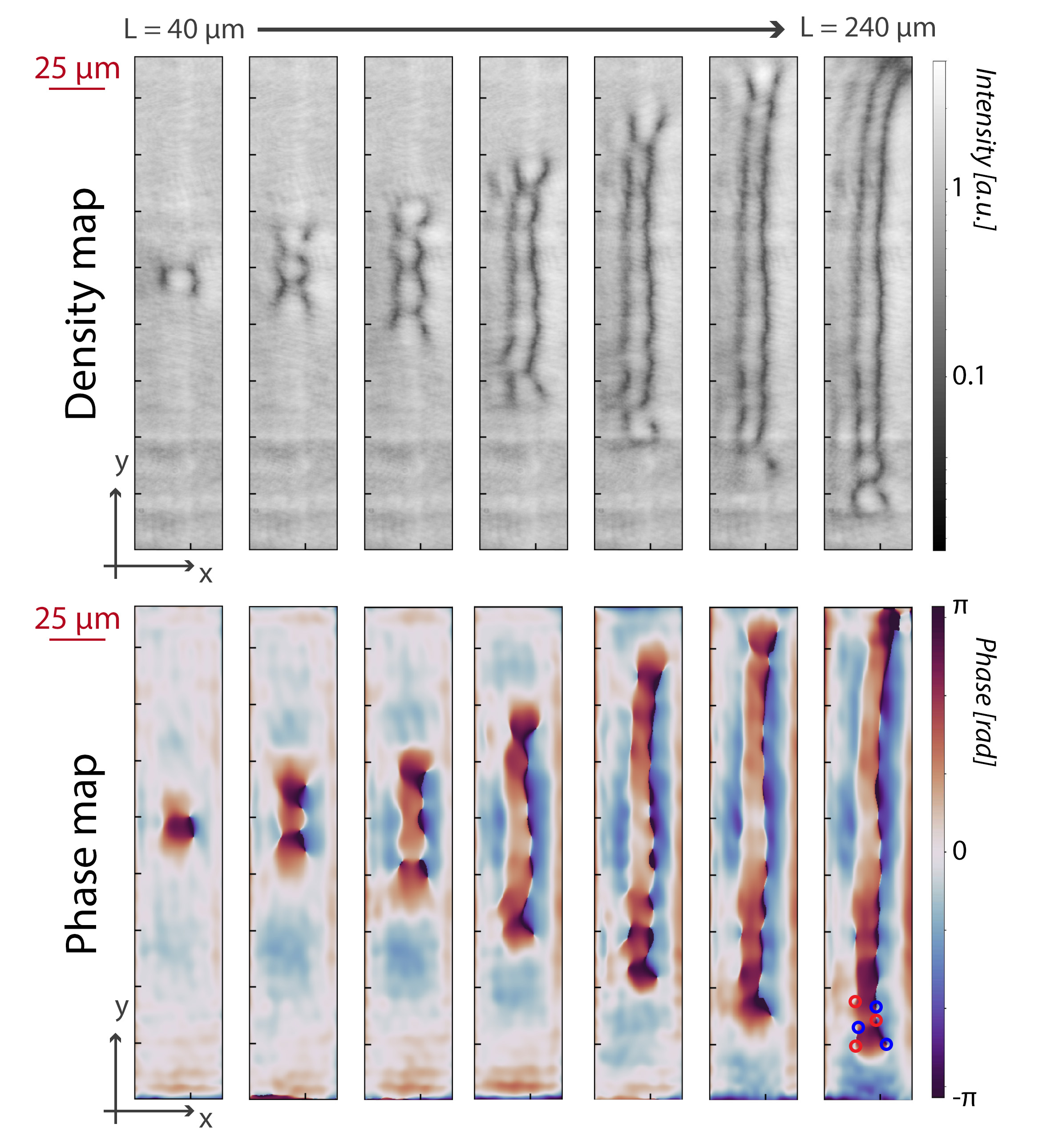}
    \caption{\textbf{Scan of the length $L$ of the channel}. Density (upper line) and phase (bottom line) maps of the system for different lengths $L$ of the channel. The length increases from 40 \textmu m (left) to 240 \textmu m (right). The symmetric instabilities are well defined for short channels, while they blurry in the center of the long ones. In the last picture on the right, the top end of the channel has reached the low density regions.}
    \label{fig:ScanL}
\end{figure}

\paragraph{}
The first dimension of the channel that we have scanned is its length $L$. Again, the operation is very easy as it consists in changing the image send on the SLM screen. 
The results are presented in figure \ref{fig:ScanL}.

\paragraph{}
This set of experiment has been realized for a channel with a horizontal width of 15 \textmu m.
It expands vertically along the images, from 40 \textmu m length up to 240 \textmu m in the last image on the right.

\paragraph{}
When the channel is short, the instabilities are very clear, as well as their symmetric pattern. Indeed, the dimension of the channel imposes boundary conditions on their spatial period and position \cite{Koniakhin2019a}, and therefore on the number of vortex-antivortex pairs that can appear in a channel of finite length.
Besides, one can see that in all the images, the instabilities are better defined at the ends of the channel, where the boundary conditions are strong.
In the center, however, they have a relative translational symmetry along the channel axis and therefore get a larger flexibility.

\paragraph{}
A particularly interesting image in this set of measurements is the last one on the right. Here, the top end of the channel is connected to the low density region, while the lower one is still surrounded by high density fluid. 
The instability pattern is therefore different. On the bottom, as before, the presence of the horizontal wall breaks the symmetry along \textit{y} and pins the instabilities: the vortex streets are very clear in the phase map, indicated there by the red and blue circles.
On the top end however, the connection to the low density region maintains the translational symmetry, allowing the instabilities to slightly move along the \textit{y} axis. In this time integrated image, it results in blurred parallel solitons.

\begin{figure}[H]
    \centering
    \includegraphics[width=\linewidth]{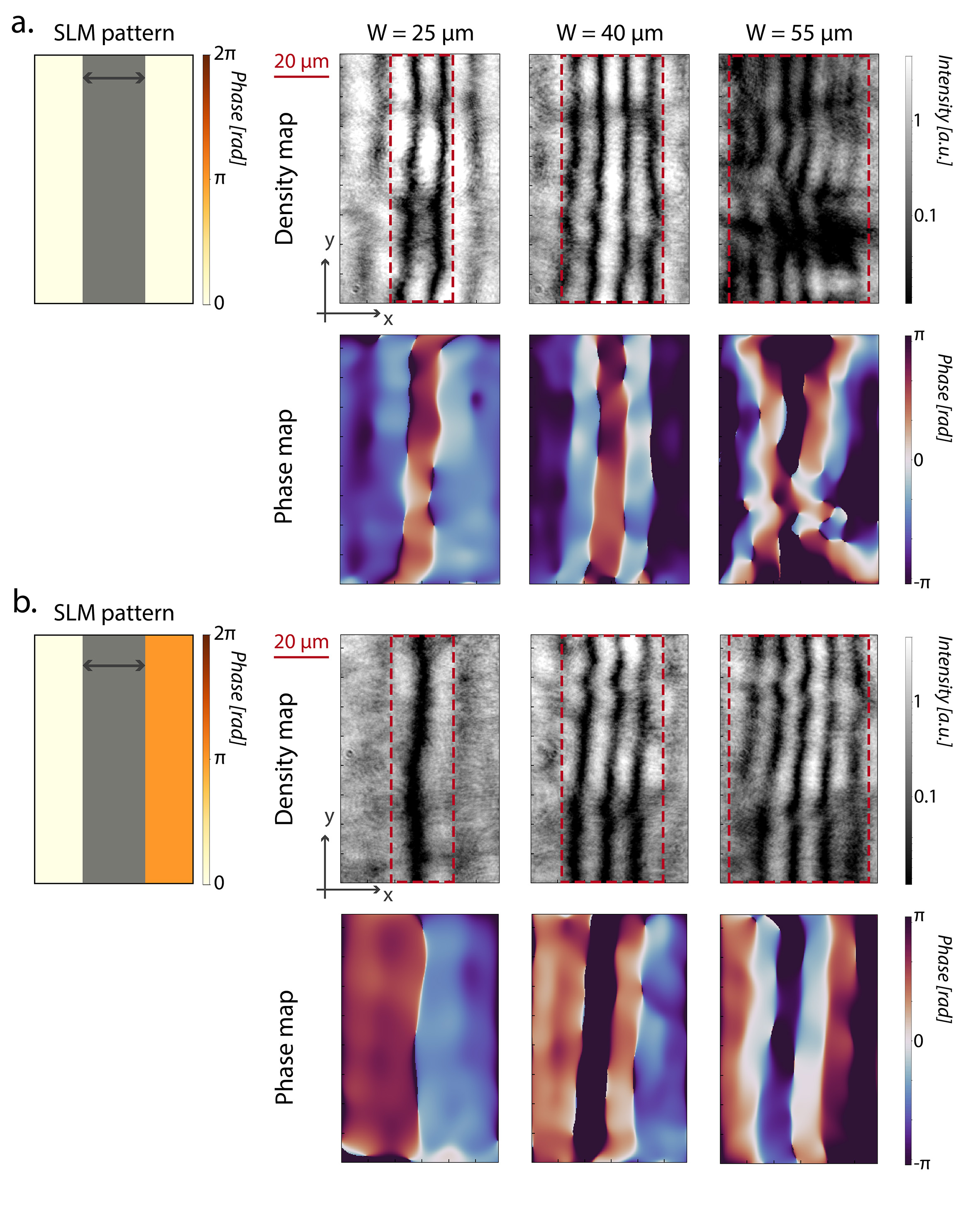}
    \caption{\textbf{Scan of the width $W$ of the channel}. Density and phase maps of the system for different width of the channel (25 \textmu m, 40 \textmu m and 55 \textmu m, respectively). The solitons fill the entire channel and therefore increase their number. On a., the two high density regions have the same phase: the soliton number is even; while on b, the high density regions are in opposition of phase, which leads to an odd number of solitons.}
    \label{fig:ScanW}
\end{figure}

\paragraph{}
The horizontal dimension of the channel has also been tuned, as presented in figure \ref{fig:ScanW}.
This time, the channel is sent over the whole beam, so is connected to the low density region on both ends.
Three different widths have been studied: 25 \textmu m, 40 \textmu m and 55 \textmu m. The dimensions of the channel are represented by the red dashed rectangles.

\paragraph{}
For each channel width, we see that the instability patterns fill all the surface of the channel: the soliton number increases with the width.
In a. for instance, we observe 2, then 4 and eventually up to an array of 6 solitons.
We only have an even number of solitons, as the wavefront of the pump has a flat phase and the total phase jump must be conserved.

\paragraph{}
Now we can also combine the wavefront shaping property of the SLM with the intensity modulation: on panel \ref{fig:ScanW}.b, the two high density regions have a phase shift of \textpi\ between them. Therefore, the number of solitons inside the channel has to be odd, so that the total phase jump in conserved: we can see one, three and five solitons depending on the total width.
Those results are analog to the one presented in \cite{Goblot2016b} in the unidimensional case, with in our case a translational symmetry along \textit{y}.

This one is responsible for the inhibition of the modulational instability. However, for wide channels, we can see the breaking of the solitons: the confinement induced by the high density region is not strong enough to sustain stable solitons, hence the development of instabilities.

\paragraph{}
Now that we have studied the single channel configurations, we can study the effect of a modification of the shape of the corridor, as a starting point of a maze realization. 
Experimentally however, the optical impression is limited by the cavity bandwith: in the momentum space, the bandwidth of the cavity filters out the signal with too high components in k-space. 
Therefore, if the shape of the incoming beam is too large in momentum space, the high spatial frequency components will not enter the cavity and the signal will be blurred out on the fluid, hence the difficulty to imprint complicated patterns.

\begin{figure}[h]
    \centering
    \includegraphics[width=\linewidth]{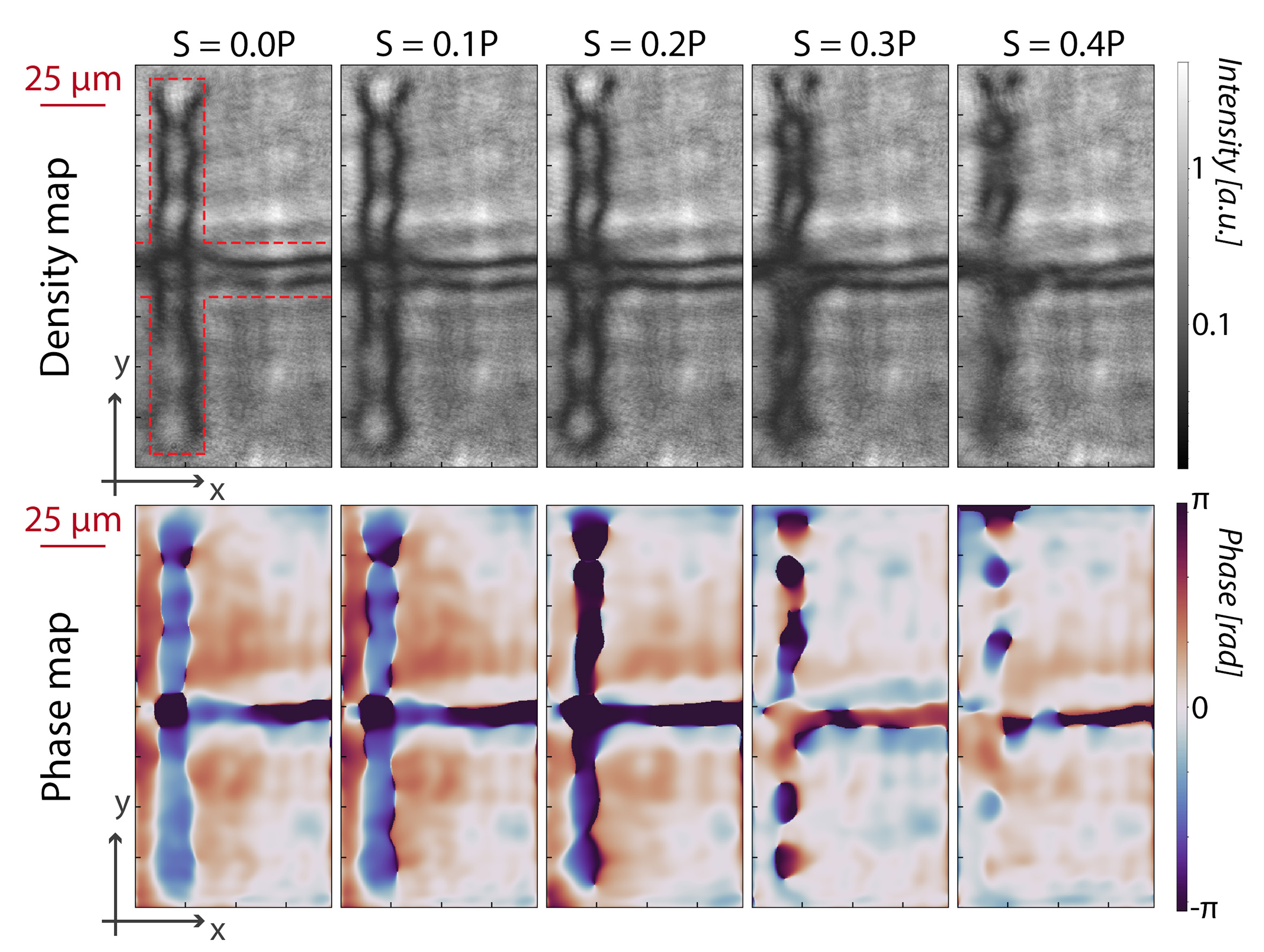}
    \caption{\textbf{Cross shaped channel}. Scan of the support intensity of a cross shaped channel, from $S = 0.0P$ to $S=0.4P$. The vertical corridor have dead-ends, surrounded by high density walls, while the horizontal corridor is connected to the low density region. For $S = 0.4P$, the solitons still fill the horizontal channel while they have vanished in the vertical one: the maze is solved.}
    \label{fig:Cross}
\end{figure}

\paragraph{}
This is why, as a first observation, we considered a cross, where the vertical corridor possess dead-ends surrounded by high density fluid, but where the horizontal channel is open and connected to the low density region.
The results are presented in figure \ref{fig:Cross}, for a channel width of 25 \textmu m and for different intensity ratios, from $S = 0.0P$ to $S = 0.4P$.
The red dashed line indicates the shape and position of the corridors.

\paragraph{}
The instabilities fill the corridors for low support intensities. However, looking at the phase maps, we can see that their best definition is around $S = 0.2P$, where the phase jumps are really clear. 
The shape of the instabilities also confirms what was already observed: the dead end channels pin the instabilities and the symmetric pattern is visible, while they appear parallel on the open horizontal channel.

Increasing again the support intensity, we reach the value $S = 0.4P$, where the dead-end channel has exceeded the critical support intensity $S_{C}$, contrary to the open-end one. 
Therefore, the solitons have been pushed away from the vertical corridor, but remains in the horizontal one, as we can clearly see in the phase map.
A few instabilities can however still be guessed in the upper part of the vertical channel due to disorder of the sample which locally pin them.
These results demonstrate that the proof of the resolution of this simplified maze is experimentally achieved.

\paragraph{}
Those results are very promising on the possibility of an actual maze solving with polaritonic devices. For now, the experimental limitation stays the implementation of the maze shape, which in our all optical configuration and with our kind of sample can not be develop to complicated shapes. It would indeed require better quality sample and higher laser power to implement a larger scale maze.
Other configurations could also be considered, as the etching of the maze directly on the sample, or the use of shaped mask.

%\bibliographystyle{unsrt}
%\bibliography{bibs/LKB-bibs-bib_thesis-OutlookChap}   

%\end{document}
%\documentclass[a4paper,11pt]{book}
%\usepackage[utf8]{inputenc}
%\usepackage[T1]{fontenc} 
%\usepackage{lmodern} 
%\usepackage[margin=28mm,includeheadfoot,bindingoffset=5mm]{geometry}[2010/03/13]

%\usepackage{graphicx}
%\usepackage{amsmath}
%\usepackage{bbold}
%\usepackage{amssymb} % pour le signe \lesssim
%\usepackage{textcomp} % \textdegree
%\usepackage[most]{tcolorbox} 
%\usepackage{enumitem} 
%\usepackage{xcolor}
%\usepackage{float} 
%\usepackage{physics}
%\usepackage{stmaryrd}
%\usepackage{wasysym} 
%\usepackage{tikz}
%\usepackage{hyperref}
%\usepackage{cite}
%\newcommand*\circled[1]{\tikz[baseline=(char.base)]{
%          \node[shape=circle,draw,inner sep=2pt] (char) {#1};}}
%\renewcommand{\thesubsubsection}{\roman{subsubsection}}
%\tcbset{enhanced,colback=red!5!white, colframe=red!75!black,fonttitle=\bfseries}
%\graphicspath{{figures/}} %Setting the graphicspath
%\setcounter{tocdepth}{3}
%\setcounter{secnumdepth}{3}

%\begin{document}

%\tableofcontents

%\setcounter{chapter}{5}

\chapter*{Conclusion}

\paragraph{}
The starting point of this work came from the study of the polariton optical bistability and its related properties \cite{Pigeon2017}. 
The ability to release the phase constraint imposed by the quasi-resonant pump particularly caught our attention, as well as the proposal of the seed-support configuration, leading to an extended high-density bistable fluid.

The experimental implementation showed to be in excellent agreement with the announced theory \cite{Lerario2020, Lerario2020a}.
Indeed, topological excitations, such as vortex-antivortex pairs and dark solitons, were spontaneously generated in the wake of a structural defect, within the high density quasi-resonantly pumped fluid. 
Therefore, the system did not only get rid of the pump phase constraint, but were also able to sustain those topological excitations over hundreds of microns, greatly enhancing their propagation length compared to the previous observations \cite{Amo2011}.

This experiment also revealed a very unexpected behaviour of the solitons. Indeed, the presence of the driving field imposed by the bistable pump prevents the dark solitons to propagate away from each other, as they usually do in an undriven system \cite{Amo2011, Hivet2012a}. 
In our configuration, they tend to align to each other and propagate parallel as a binding mechanism.

\paragraph{}
Following those results, we tried to get rid of the last constraint we had on the soliton generation, namely their spontaneous formation in the wake of structural defect.
To do so, we decided to artificially imprint them on the fluid and observe their further free propagation. This was implemented by wisely designing the excitation beam using a Spatial Light Modulator, and we managed to generate on demand dark solitons on a polariton fluid \cite{Maitre2020}.
Once again, due to the presence of the driving field, we observed this binding between the solitons, propagating parallel as a dark soliton molecule.

The flexibility of our method allowed us to explore the system parameters in order to better understand this surprising behaviour. After a careful study, we realized it is directly connected to the driven-dissipative nature of our system, and that the separation distance between the solitons is governed by the decay rate of the microcavity.
The easy design and implementation of this technique opens the way to deeper studies of quantum turbulence phenomena.

\paragraph{}
It was done in the last part of this thesis work, where the previous technique was developed to implement intensity modulations on the polariton fluid. 
This time, the idea was to generate solitonic structures in guided low-density channel in a static polariton fluid \cite{Koniakhin2019a}.
The absence of flow destabilizes the solitons, which break into vortex streets due to transverse snake instabilities. Those instabilities present different symmetries that we were able to study, as well as their different behaviour depending on the dead- or open-end nature of the channel.
This property was then applied to offer an all-optical fast analog maze solving algorithm, even though some technical limitations still prevents us to implement too complicated shapes.

\paragraph{}
This thesis work has been focused on the study of topological excitations, namely vortices and dark solitons.
This topic is an important subject of research in the field of exciton-polaritons and more generally quantum fluids, and we have concentrated our investigations on their control. By implementing novel techniques to generate and sustain them at will, we were able to greatly facilitate their control and the one of their environment parameters. 
It offers many possibilities of developments, going from the study of interaction in multiple solitons pattern or their robustness against modulational instabilities, the improvement of ultrafast all-optical maze solving algorithms and more generally the controlled and quantitative study of quantum turbulence in driven-dissipative fluids of light.

%\bibliographystyle{unsrt}
%\bibliography{bibs/LKB-bibs-bib_thesis-Conclu}  

%\end{document}
%\documentclass[a4paper,11pt]{book}
%\usepackage[utf8]{inputenc}
%\usepackage[T1]{fontenc} 
%\usepackage{lmodern} 
%\usepackage[margin=28mm,includeheadfoot,bindingoffset=5mm]{geometry}[2010/03/13]

%\usepackage{graphicx}
%\usepackage{amsmath}
%\usepackage{bbold}
%\usepackage{amssymb} % pour le signe \lesssim
%\usepackage{textcomp} % \textdegree
%\usepackage[most]{tcolorbox} 
%\usepackage{enumitem} 
%\usepackage{xcolor}
%\usepackage{float} 
%\usepackage{physics}
%\usepackage{stmaryrd}
%\usepackage{wasysym} 
%\usepackage{tikz}
%\usepackage{hyperref}
%\usepackage{cite}
%\newcommand*\circled[1]{\tikz[baseline=(char.base)]{
%          \node[shape=circle,draw,inner sep=2pt] (char) {#1};}}
%\renewcommand{\thesubsubsection}{\roman{subsubsection}}
%\tcbset{enhanced,colback=red!5!white, colframe=red!75!black,fonttitle=\bfseries}
%\graphicspath{{figures/}} %Setting the graphicspath
%\setcounter{tocdepth}{3}
%\setcounter{secnumdepth}{3}

%\begin{document}

%\tableofcontents

%\setcounter{chapter}{5}

\chapter*{Communications}

\section*{Conferences}

\begin{itemize}
    \item \textbf{ICSCE10}, Melbourne, Australia, January 2020, oral presentation
    \item \textbf{TeraMetaNano-4}, Lecce, Italy, May 2019, oral presentation
    \item \textbf{Optique Toulouse 2018}, Toulouse, France, July 2018, poster presentation
    \item \textbf{QFLM}, Les Houches, France, June 2018, poster presentation
\end{itemize}

\section*{Publications}

\begin{itemize}
    \item Giovanni Lerario, Sergei V. Koniakhin, \textbf{Anne Maître}, Dmitry Solnyshkov, Alessandro Zilio, Quentin Glorieux, Guillaume Malpuech, Elisabeth Giacobino, Simon Pigeon, and Alberto Bramati. Parallel dark soliton pair in a bistable two-dimensional exciton-polariton superfluid. \textit{Physical Review Research}, \textbf{2}, 042041, 2020.
    
    \item Ferdinand Claude, Sergei V Koniakhin, \textbf{Anne Maître}, Simon Pigeon, Giovanni Lerario, Daniil D Stupin, Quentin Glorieux, Elisabeth Giacobino, Dmitry Solnyshkov, Guillaume Malpuech, and Alberto Bramati. Taming the snake instabilities in a polariton superfluid. \textit{Optica}, 7:1660-1665, 2020.
    
    \item \textbf{Anne Maître}, Giovanni Lerario, Adrià Medeiros, Ferdinand Claude, Quentin Glorieux, Elisabeth Giacobino, Simon Pigeon, and Alberto Bramati. Dark-soliton molecules in an exciton-polariton superfluid. \textit{Physical Review X}, \textbf{10}, 041028, 2020.
    
    \item Thomas Boulier, Maxime J. Jacquet, \textbf{Anne Maître}, Giovanni Lerario, Ferdinand Claude, Simon Pigeon, Quentin Glorieux, Alberto Amo, Jacqueline Bloch, Alberto Bramati, and Elisabeth Giacobino. Microcavity Polaritons for Quantum Simulation. \textit{Advanced Quantum Technologies}, page 2000052, 2020.
    
    \item M. J. Jacquet, T. Boulier, F. Claude, \textbf{A. Maître}, E. Cancellieri, C. Adrados, A. Amo, S. Pigeon, Q. Glorieux, A. Bramati, and E. Giacobino. Polariton fluids for analogue gravity physics. \textit{Philosophical Transactions of the Royal Society A: Mathematical, Physical and Engineering Sciences}, \textbf{378}(2177), 2020.
    
    \item Giovanni Lerario, \textbf{Anne Maître}, Rajiv Boddeda, Quentin Glorieux, Elisabeth Giacobino, Simon Pigeon, and Alberto Bramati. Vortex-stream generation and enhanced propagation in a polariton superfluid. \textit{Physical Review Research}, \textbf{2}, 023049, 2020.
    
    \item S. V. Koniakhin, O. Bleu, D. D. Stupin, S. Pigeon, \textbf{A. Maitre}, F. Claude, G. Lerario, Q. Glorieux, A. Bramati, D. Solnyshkov, and G. Malpuech. Stationary Quantum Vortex Street in a Driven-Dissipative Quantum Fluid of Light. \textit{Physical Review Letters}, \textbf{123}, 215301, 2019.
\end{itemize}

%\bibliographystyle{unsrt}
%\bibliography{bibs/LKB-bibs-bib_thesis-Conclu}  

%\end{document}
\backmatter

\bibliographystyle{unsrt}
\bibliography{bibs/LKB-bibs-bib_thesis}

%\backcover
\end{document}